\documentclass[prb,aps,twocolumn,superscriptaddress]{revtex4-2}
\usepackage{graphicx}
\usepackage{amsmath}
\usepackage{amssymb}
\usepackage{bm}
\usepackage{dsfont}
\usepackage[colorlinks]{hyperref}
\usepackage{multirow}
\usepackage{appendix}
\usepackage{color}
\usepackage{placeins}
\usepackage{color}

\usepackage{float}

\newcommand{\ket}[1]{\left| #1 \right\rangle}

\DeclareMathOperator{\sgn}{sgn}

\begin{document}

\title{Strong and Fragile Topological Dirac Semimetals with Higher-Order Fermi Arcs}

\author{Benjamin J. Wieder$^\star$}
\affiliation{Department of Physics,
Princeton University,
Princeton, NJ 08544, USA
	}
	
\author{Zhijun Wang}
\affiliation{Beijing National Laboratory for Condensed Matter Physics, 
and Institute of Physics, Chinese Academy of Sciences, Beijing 100190, China}
\affiliation{University of Chinese Academy of Sciences, Beijing 100049, China}
	
\author{Jennifer Cano}
\affiliation{Department of Physics and Astronomy, Stony Brook University, Stony Brook, New York 11974, USA}
\affiliation{Center for Computational Quantum Physics, The Flatiron Institute, New York, New York 10010, USA}

\author{Xi Dai}
\affiliation{Physics Department, Hong Kong University of Science and Technology, Clear Water Bay, Hong Kong}

\author{Leslie M. Schoop}
\affiliation{Department of Chemistry, Princeton University, Princeton, New Jersey 08544, USA}

\author{Barry Bradlyn$^\star$}
\affiliation{Department of Physics and Institute for Condensed Matter Theory, University of Illinois at Urbana-Champaign, Urbana, IL, 61801-3080, USA}
\thanks{Permanent Address}
\affiliation{Donostia International Physics Center, P. Manuel de Lardizabal 4, 20018 Donostia-San Sebasti\'{a}n, Spain}

\author{B. Andrei Bernevig$^\star$}
\affiliation{Department of Physics,
Princeton University,
Princeton, NJ 08544, USA
	}
\affiliation{Dahlem Center for Complex Quantum Systems and Fachbereich Physik,
Freie Universit{\"a}t Berlin, Arnimallee 14, 14195 Berlin, Germany
	}
\affiliation{Max Planck Institute of Microstructure Physics, 
06120 Halle, Germany
}

\begin{abstract}
Dirac and Weyl semimetals both exhibit arc-like surface states.  However, whereas the surface Fermi arcs in Weyl semimetals are topological consequences of the Weyl points themselves, the surface Fermi arcs in Dirac semimetals are not directly related to the bulk Dirac points, raising the question of whether there exists a topological bulk-boundary correspondence for Dirac semimetals.  In this work, we discover that strong and fragile topological Dirac semimetals exhibit 1D higher-order hinge Fermi arcs (HOFAs) as universal, direct consequences of their bulk 3D Dirac points.  To predict HOFAs coexisting with topological surface states in solid-state Dirac semimetals, we introduce and layer a spinful model of an $s-d$-hybridized quadrupole insulator (QI).  We develop a rigorous nested Jackiw-Rebbi formulation of QIs and HOFA states.  Employing \emph{ab initio} calculations, we demonstrate HOFAs in both the room- ($\alpha$) and intermediate-temperature ($\alpha''$) phases of Cd$_{3}$As$_2$, KMgBi, and rutile-structure ($\beta'$-) PtO$_2$.
\end{abstract}

\maketitle

 {
 $\\$
\centerline{\bf Introduction}
\vspace{0.05in}
}

Since the realization that the Fermi surface of graphene is characterized not only by its bulk 2D Dirac cones, but also by 1D arc-like states along zigzag edges~\cite{GrapheneReview}, there has been an ongoing effort to identify bulk-gapless systems with topological boundary modes. This effort has yielded a wide variety of 3D nodal semimetals with topological states on their 2D faces, including systems with bulk Weyl~\cite{AshvinWeyl1,AndreiWeyl,SYWeyl} and unconventional~\cite{DDP,NewFermions,RhSiArc,CoSiArc,KramersWeyl} fermions.  Despite the presence of bulk gapless points in these semimetals, bands are still generically gapped in momentum space away from the nodal points, allowing for topological invariants to be defined along closed surfaces in the Brillouin zone (BZ)~\cite{AshvinWeyl1,NagaosaDirac,YoungkukLineNode}.  Nontrivial values of these invariants necessitate the presence of topological surface bands. Examples include the surface Fermi arcs in Weyl~\cite{AshvinWeyl1,AndreiWeyl,SYWeyl} and unconventional chiral semimetals~\cite{NewFermions,RhSiArc,CoSiArc,KramersWeyl,AlPtObserve,CoSiObserveJapan,CoSiObserveHasan,CoSiObserveChina}, and topological boundary polarization modes, such as the solitons in the Su-Schrieffer-Heeger (SSH) and Rice-Mele chains~\cite{SSH,RiceMele}, the aforementioned Fermi arcs in graphene~\cite{GrapheneReview}, and the drumhead surface states in centrosymmetric nodal-line semimetals~\cite{YoungkukLineNode}.  Researchers have also identified 3D Dirac semimetals with arc-like surface states that resemble the Fermi arcs of Weyl semimetals~\cite{ZJDirac,NagaosaDirac,SYDiracSurface,ZJSurface}.  However, unlike the surface states of Weyl, nodal-line, and unconventional chiral semimetals, the surface Fermi arcs in Dirac semimetals can be disconnected and removed without breaking a symmetry or closing a gap~\cite{KargarianDiracArc1}, and therefore are not topological consequences of the bulk Dirac points themselves.  It has thus remained an open question as to whether 3D Dirac points can actually exhibit robust, nontrivial topology with spectroscopic consequences.

In this work, we exploit the theory of Topological Quantum Chemistry (TQC)~\cite{QuantumChemistry} and recent advances in higher-order~\cite{multipole,WladTheory,HigherOrderTIBernevig,HigherOrderTIChen,HigherOrderTIPiet,DiracInsulator,ChenRotation,AshvinTCI,HOTIBismuth,BernevigMoTe2} and fragile~\cite{AshvinFragile,ArisInversion,JenFragile1,BarryFragile,WiederAxion} topology to discover a large family of 3D Dirac semimetals that exhibit intrinsic, polarization- (quadrupole-) nontrivial higher-order Fermi-arc (HOFA) states on their 1D hinges as direct, topological consequences of their bulk Dirac points, definitively diagnosing condensed matter Dirac fermions as higher-order topological.  The HOFA states introduced in this work therefore represent a robust manifestation of a topological bulk-hinge correspondence in experimentally established 3D solid-state semimetals, and may be observable through experimental probes such as scanning tunneling microscopy (STM) and nonlocal quantum oscillation measurements.  We support our findings with extensive analytic, tight-binding, and first-principles calculations.

{
$\\$
\vspace{0.05in}
\centerline{\bf Results}
}

\begin{figure}[t]
\centering
\includegraphics[width=0.5\textwidth]{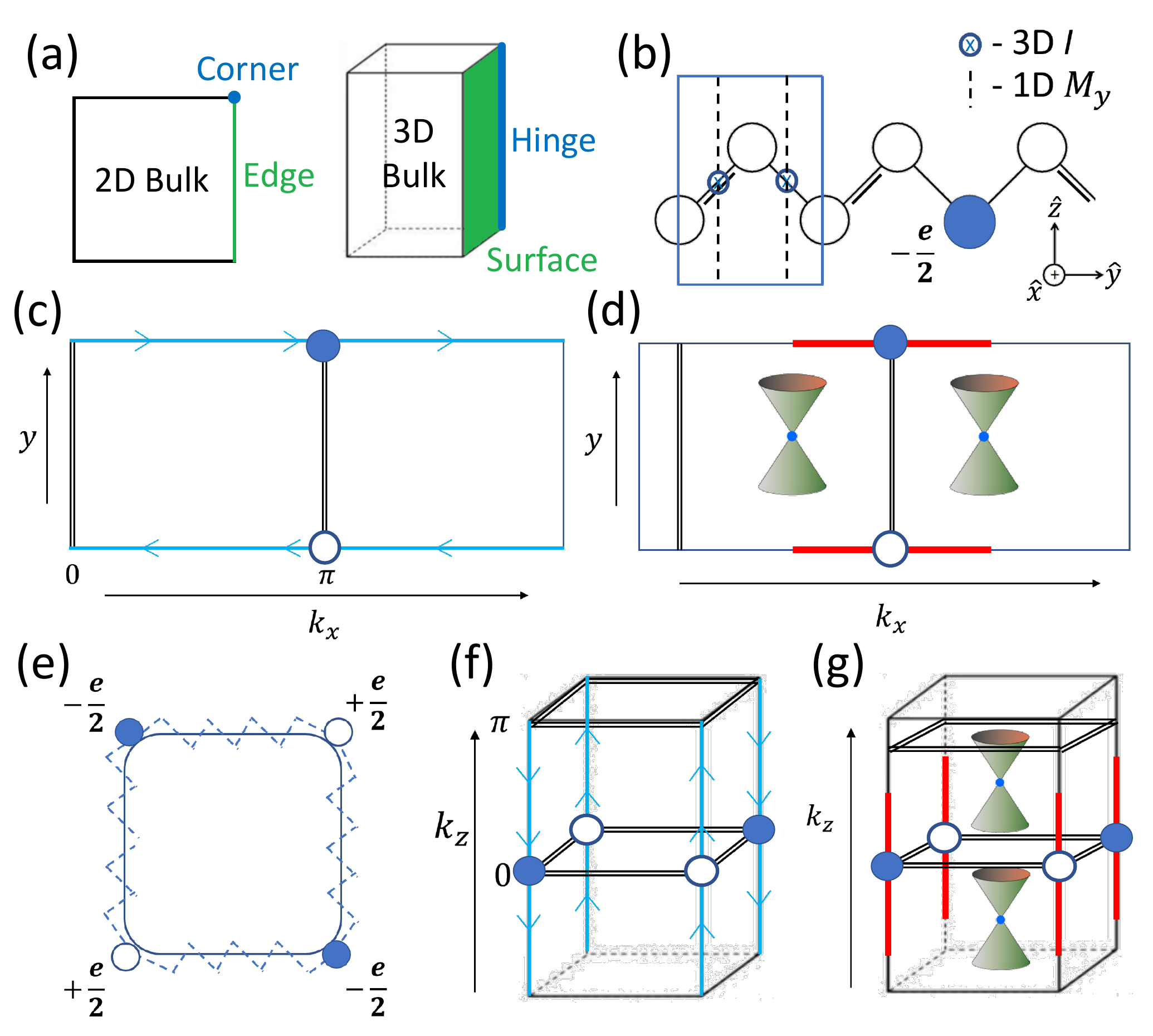}
\caption{Tuning cycles of 1D and 2D insulators with 0D boundary states.  (a) Terminology for the bulk and boundary of 2D and 3D systems. (b) A $y$-directed 1D SSH chain with quantized polarization, enforced by either 3D inversion $\mathcal{I}$ or 1D mirror symmetry along the chain (\emph{e.g.}, the operation $M_{y}$, which takes $y\rightarrow -y$)~\cite{SSH,RiceMele}.  (c) In a 2D crystal with $\mathcal{I}$ symmetry, $k_{x}$ can be treated as a parameter that periodically tunes between a $y$-directed SSH chain (double black lines in (c) and (d)) with zero polarization ($k_x=0$) and another with $e/2$ polarization ($k_x=\pi$), yielding a Chern insulator~\cite{ThoulessWannier,RiceMele} with chiral edge modes (blue lines). (d) In a 2D crystal with $M_{y}$ symmetry instead of $\mathcal{I}$, the Hamiltonian at each value of $k_{x}$ is equivalent to that of a $y$-directed SSH chain with a quantized polarization of $0$ or $e/2$; because the polarization cannot change continuously, a periodic tuning cycle indexed by $k_{x}$ between SSH polarizations $0$ and $e/2$ must pass through a pair of gapless points.  This yields a 2D band-inverted semimetal with topological polarization modes (red lines) analogous to those in zigzag-terminated graphene~\cite{GrapheneReview,GrapheneEdgeMullen}.  (e) A $\mathbb{Z}_{2}$ quantized quadrupole insulator (QI)~\cite{multipole} invariant under wallpaper group $p4m$. (f) A $C_{4z}$-broken, mirror-preserving pumping cycle of a QI (double black lines in (f) and (g)) is equivalent to a 3D 2nd-order Chern insulator~\cite{WladTheory,HigherOrderTIBernevig,HigherOrderTIChen,HigherOrderTIPiet} with chiral hinge modes (blue lines), whereas (g) a $p4m$-preserving cycle is equivalent to a 3D Dirac semimetal with higher-order Fermi arcs (HOFAs) on its 1D hinges (red lines).}
\label{fig:pumpMain}
\end{figure}

\textit{Boundary modes in topological tuning cycles} --  To provide context for the analysis performed in this work, we first review the crucial distinctions between topological polarization boundary modes and the surface states of topological insulators (TIs).  Whereas in topological (crystalline) insulators~\cite{AndreiTI,CharlieTI} the Bloch wavefunctions do not admit a description in terms of symmetric, exponentially localized Wannier functions~\cite{ThoulessWannier,AlexeyVDBTI,QuantumChemistry}, insulating phases with only quantized electric polarization conversely do admit a Wannier description; the quantized polarization leads to a nontrivial Berry phase indicating the positions of the electronic Wannier centers relative to the ionic positions~\cite{VDBpolarization,QuantumChemistry}.  In these insulators, such as the SSH chain~\cite{SSH,RiceMele}, the Berry phase is quantized by the presence of a crystal symmetry, typically mirror reflection $M$ or spatial inversion $\mathcal{I}$ (Fig.~\ref{fig:pumpMain}(b)). Correspondingly, the boundary between insulators with differing polarizations forms a domain wall that binds a topological soliton of fractional charge~\cite{SSH,RiceMele}, though the energy of this mode may float away from zero if particle-hole symmetry is broken. Nevertheless, as observed in polyacetylene~\cite{SSH,SSHExp}, zigzag-terminated graphene~\cite{GrapheneEdgeMullen}, and nodal-line semimetals~\cite{YoungkukLineNode,LeslieLine2,LineNodeExp}, topological polarization boundary modes can still frequently lie near the Fermi energy in real materials.

By reinterpreting one of the momenta as an external parameter, a subset of topological semimetals and (crystalline) insulators, can be reexpressed as the adiabatic, cyclic tuning of an insulator with quantized electric polarization in one fewer dimension~\cite{ThoulessWannier}.  For example, the $y$-directed (hybrid) Wannier centers of a Chern insulator exhibit spectral flow as a function of the momentum $k_x$, which can be indicated by the product of parity eigenvalues if $\mathcal{I}$ symmetry is present~\cite{ThoulessWannier,FuKaneInversion}.  Reinterpreting $k_x$ as an external tuning parameter, we can recast the Chern number, $C$, as a nontrivial tuning cycle (Thouless pump) of a 1D SSH chain; as $k_x$ is tuned from $0$ to $2\pi$, charge $eC$ is pumped across the unit cell of the crystal.  If the cycle is $\mathcal{I}$-symmetric, then $C\text{ mod }2$ can be detected by (twice) the change in quantized polarization between effective $y$-directed SSH chains at $k_x=0$ and~\cite{RiceMele,FuKaneInversion} $k_x=\pi$.  We show this schematically in Fig.~\ref{fig:pumpMain}(c).  In a crystalline semimetal, the presence of additional symmetries in the tuning cycle can force the gap to close at certain values of $k_x$. For instance, adding (spinless) time-reversal ($\mathcal{T}$) symmetry to the $\mathcal{I}$-symmetric Thouless pump obstructs the presence of a nonzero Chern number; in order for the polarization to change by $e/2$ from $k_x=0$ to $k_x=\pi$, there must be a gapless point~\cite{YoungkukLineNode} at some $k_x^*\in(0,\pi)$, with a time-reversed partner at $-k_x^*$. A similar gapless point occurs when the polarization of an SSH chain in line group $pm$ is periodically tuned (Fig.~\ref{fig:pumpMain}(d)).  There, taking the mirror to be $M_{y}$, each value in parameter space indexed by the periodic tuning parameter $k_{x}$ corresponds to a $y$-directed SSH chain with a quantized polarization indicated by the mirror eigenvalues of the occupied bands~\cite{ArisInversion}; each time the polarization jumps between $0$ and $\pi$, a robust gapless point forms because the crossing bands carry different mirror eigenvalues.

Recently these arguments were generalized to higher electric multipole moments.  In Ref.~\onlinecite{multipole}, the authors demonstrated the theoretical existence of spinless insulators with threaded flux that exhibit trivial dipole moments, but topologically quantized electric quadrupole and octupole moments, and which host boundary (corner) modes in two and three dimensions fewer than the bulk, respectively (Fig.~\ref{fig:pumpMain}(a,e)). Many of these corner-mode phases~\cite{multipole,WladTheory,HigherOrderTIChen} can be identified by their bulk symmetry eigenvalues, exploiting the theory of band representations~\cite{QuantumChemistry,JenFragile1,BarryFragile,AshvinIndicators,ChenTCI,SlagerKanePrx}.  As shown in recent independent proposals, imposing combinations of rotational, rotoinversion, and $\mathcal{T}$ symmetries allows for 3D topological insulating crystals that are equivalent to nontrivial pumping cycles of quantized quadrupole insulators (QIs)~\cite{WladTheory,HigherOrderTIBernevig,HigherOrderTIChen,HigherOrderTIPiet,PhononQuadrupole}, or other 2D phases with corner modes~\cite{ChenTCI,AshvinTCI,ChenRotation,BernevigMoTe2,WiederAxion}.  These 3D higher-order TIs~\cite{WladTheory,HigherOrderTIBernevig,HigherOrderTIChen,HigherOrderTIPiet,DiracInsulator,ChenRotation,AshvinTCI,HOTIBismuth,BernevigMoTe2} host chiral or helical modes not on their 2D faces, but instead on their 1D hinges (Fig.~\ref{fig:pumpMain}(a,f)).

\textit{Summary of results} --  In this work, we present the discovery of higher-order (polarization) topology and HOFA states in a large family of previously identified Dirac semimetals, completing the set of interrelated (higher-order) topological insulators and semimetals shown in Fig.~\ref{fig:pumpMain}.  We demonstrate the intrinsic, topological nature of the HOFA states by performing several extensive calculations that bridge the significant gap between previously established theoretical concepts and the candidate real-material HOFA Dirac semimetals identified in this work.  First, we use TQC~\cite{QuantumChemistry} to formulate a new, spinful model of a QI derived from $s-d$-orbital hybridization in a magnetic layer group, and show that it is topologically equivalent to the spinless model with staggered magnetic flux proposed in Ref.~\onlinecite{multipole} (see Supplementary Appendices, Section~\ref{sec:equivalence} (Appendix~\ref{sec:equivalence})). This puts the QI ($s-d$ hybridization) on the same physical foundation as previous dipole insulators, such as the SSH chain~\cite{SSH,RiceMele} ($s-p$ hybridization). We then prove using band representations that the QI is an obstructed atomic limit with localizable Wannier functions~\cite{QuantumChemistry} (Appendix~\ref{sec:bandrep}).  Next, we use crystal symmetry to develop an extensive, angular-momentum-based, nested Jackiw-Rebbi~\cite{JackiwRebbi} formulation of intrinsic corner modes in order to analytically obtain the bound states of the $s-d$-hybridized QI (Appendix~\ref{sec:boundary}) and to relate them to SSH (anti)solitons (Figs.~\ref{fig:pumpMain}(e) and Appendix~\ref{sec:boundary}).  Because our construction employs an isotropic (\emph{i.e.} cylindrical) boundary, it uniquely represents an analytic formulation of the QI in which the presence of intrinsic 0D boundary modes can be separated from the extrinsic effects of the singular curvature of sharp corners.  Furthermore, because our construction is explicit, general, and rigorous, it can also be employed to predict and analyze other corner-mode phases~\cite{BernevigMoTe2,WiederAxion}.  Through our TQC-based model of a QI and our Jackiw-Rebbi analysis, we discover a fragile topological phase~\cite{AshvinFragile,ArisInversion,JenFragile1,BarryFragile,WiederAxion} that exhibits the same corner charges as a QI modulo $e$; because these charges are a property of the fragile bands closest to the Fermi energy, they persist even when the valence manifold of the fragile phase is trivialized by additional (trivial) bands (Appendices~\ref{sec:fragile} and~\ref{sec:TCIBoundary}), as is expected to occur in real materials.  Stacking our spinful, TQC-based model of a QI in 3D, we construct both $\mathcal{T}$-symmetric and $\mathcal{T}$-breaking realizations of Dirac semimetals with higher-order Fermi arcs (HOFAs) on their 1D hinges (Fig.~\ref{fig:pumpMain}(a,g)), \emph{i.e.} in two fewer dimensions than their bulk.  Furthermore, unlike the surface Fermi arcs in Dirac semimetals, which can be removed by symmetry- and bulk-band-order-preserving potentials~\cite{KargarianDiracArc1}, HOFA states represent a direct, topological boundary consequence of the bulk Dirac points.

Crucially, because our analysis is derived from TQC, atomic orbitals, and symmetry-based (nested) Jackiw-Rebbi domain walls, it allows the immediate connection to real materials, unlike recent toy models with HOFA states (\emph{i.e.}, the flux-lattice and particle-hole symmetric semimetallic models in Refs.~\onlinecite{TaylorToy,VladHOFA}, respectively) that appeared while we were expanding our material search to fragile and experimentally favorable structural phases of established topological Dirac semimetals.  Specifically, while one can naively stack the spinless QI and obtain a toy-model HOFA state, without the careful symmetry- and orbital-based analysis developed in this work, the resulting HOFA states bear no clear connection to 2D TIs (Appendices~\ref{sec:TIboundary} and~\ref{sec:TItoTrivial}), TCIs (Appendix~\ref{sec:TCIBoundary}), fragile phases (Appendices~\ref{sec:fragile} and~\ref{sec:TCIBoundary}), or to 3D Dirac points in real materials (Appendices~\ref{sec:symmetry}  and~\ref{sec:DFT}).  Furthermore, in this work, we explicitly relax particle-hole symmetry, which numerous other works, such as Ref.~\onlinecite{VladHOFA}, centrally exploit.  Because particle-hole is not generically a symmetry of real materials, it can protect corner (and thus HOFA) states that appear in toy models, but which are not observable in real materials.  This can be understood by making an analogy to the SSH model of polyacetylene~\cite{SSH,RiceMele}.  Specifically, while real polyacetylene exhibits only $\mathbb{Z}_{2}$ polarization topology~\cite{SSH,RiceMele}, the particle-hole symmetric toy-model SSH chain exhibits strong, $\mathbb{Z}$-valued topology (Class AIII in the nomenclature of Ref.~\onlinecite{KitaevClass}).

We predict previously unidentified HOFAs and related fragile-phase corner charges (Appendix~\ref{sec:TCIBoundary}) in established candidate Dirac semimetals.  We present \emph{ab initio} and tight-binding calculations demonstrating the presence of HOFAs in the intermediate-temperature ($\alpha''$) phase of the well-studied Dirac semimetal Cd$_{3}$As$_2$ in space group (SG) 137 ($P4_{2}/nmc1'$)~\cite{ZJDirac,NagaosaDirac,CavaDirac2,StableCadmium} and in the candidate Dirac semimetals KMgBi in SG 129 ($P4/nmm1'$)~\cite{KMgBi2,KMgBi3} and rutile-structure ($\beta'$-) PtO$_2$ in SG 136 ($P4_{2}/mnm1'$)~\cite{PtO21,PtO22} (here and throughout this work, we follow Ref.~\onlinecite{MagneticBook} in using primes to denote antiunitary group elements).  We also use symmetry arguments to predict that the archetypal room-temperature ($\alpha$) phase of Cd$_3$As$_2$ in SG 142 ($I4_{1}/acd1'$) exhibits a related variant of HOFA states that derive from relaxing the reflection symmetries of the QI phase (Appendix~\ref{sec:unpinnedHOFAs}).  Finally, we also demonstrate that, in the presence of an external electric field, the topological Dirac semimetal phase of $\beta'$-PtO$_2$ can be converted into a previously uncharacterized variant of fragile topological Dirac semimetal that displays HOFA states coexisting with fractionally charged corner (hinge) states.

\begin{figure}[t]
\centering
\includegraphics[width=0.44\textwidth]{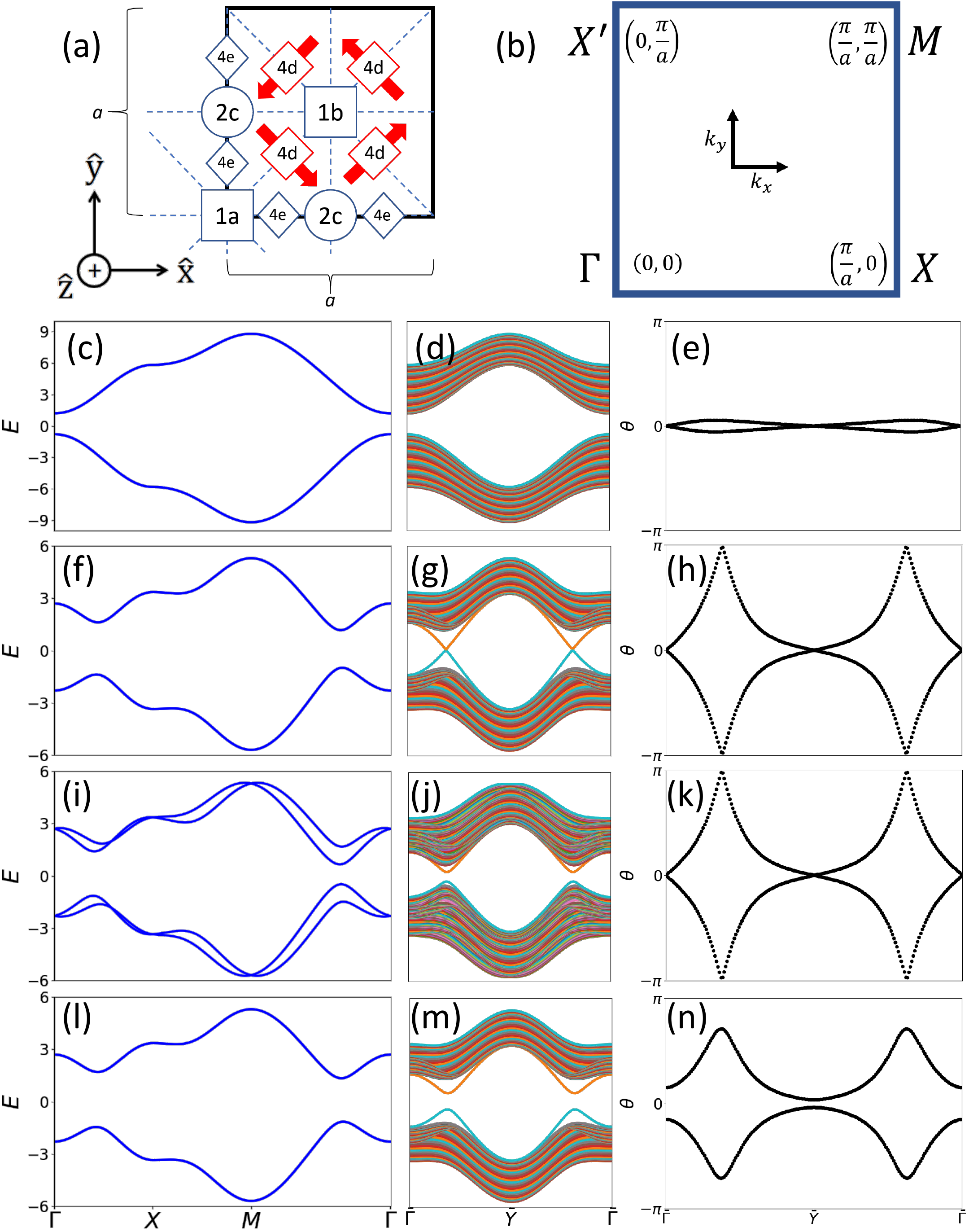}
\caption{TCI, fragile, and QI phases in 2D insulators with $p4m$ symmetry.  (a) $\mathcal{T}$-symmetric layer group $p4/mmm1'$ reduces to type-III magnetic layer group~\cite{MagneticBook} $p4/m'mm$ under the application of a magnetic potential with no net magnetic moment in each unit cell (Appendix~\ref{sec:bandrep}); this can be achieved by placing spin-$1/2$ magnetic moments (red arrows in (a)) at the $4d$ Wyckoff position with orientations related by $C_{4z}$ and $M_{x,y}$.  (b) BZ and (c) bulk bands for a model (Eq.~(\ref{eq:my2DquadMain})) that respects $p4/mmm1'$ (Table~\ref{tb:symsMain}), which has $M_{z}$ and $\mathcal{T}$ symmetries, as well as the symmetries of $p4/m'mm$ in (a); this model is constructed from $s$ and $d_{x^{2}-y^{2}}$ orbitals at the $1a$ Wyckoff position.  Eq.~(\ref{eq:my2DquadMain}) can be tuned between a trivial and a mirror TCI phase (f), distinguished by their (d,g) ribbon edge spectra and (e,h) $x$-directed Wilson loops plotted as functions of $k_{y}$ (Eq.~(\ref{eq:wilsoncontMain})). (i) Relaxing $M_{z}$ while preserving $C_{2z}$ and $\mathcal{T}$ by introducing Eq.~(\ref{eq:breakIMain}) to Eq.~(\ref{eq:my2DquadMain}), (k) we realize a four-band model with the same Wilson loop winding as the 2D TCI phase in (f-h), but (j) without topological edge states.  The Wilson loop in (k) can either be trivialized by the addition of more orbitals to the model (Appendix~\ref{sec:fragile}), or gapped by the magnetic potential depicted in (a). (l) Upon gapping (m) the surface and (n) Wilson bands with magnetism that breaks $M_{z}$ while preserving magnetic wallpaper group~\cite{MagneticBook,DiracInsulator} $p4m$, which we accomplish by adding the potential in  Eq.~(\ref{eq:quadMain}), the Wannier centers of the topological phases in (f-k) localize at the $1b$ position (Appendix~\ref{sec:bandrep}), realizing an insulator topologically equivalent to the QI introduced in Ref.~\onlinecite{multipole} (Appendix~\ref{sec:equivalence}).}
\label{fig:2Dmain}
\end{figure}

\textit{2D TCIs, fragile TIs, and QIs in $p4m$} -- We begin by providing a more physical formulation of the 2D QI introduced in Ref.~\onlinecite{multipole} using atomic orbitals, which clarifies the connection with the SSH chain.  We place spin-$1/2$ $s$ and $d_{x^{2}-y^{2}}$ orbitals at the center ($1a$ Wyckoff position) of a square unit cell in 2D (Fig.~\ref{fig:2Dmain}(a)) and then, following Ref.~\onlinecite{multipole}, impose the symmetries of wallpaper group $p4m$, which is generated by $M_{x}$ and $C_{4z}$ about the $1a$ position in Fig.~\ref{fig:2Dmain}, as well as 2D square lattice translations (For the distinctions between wallpaper and layer groups and their relationship to topological semimetals and insulators, see Ref.~\onlinecite{DiracInsulator}).  In addition to the symmetries of $p4m$, we will first additionally impose $M_{z}$ and $\mathcal{T}$ symmetries to explore 2D phases with spin-orbit coupling (SOC), and then subsequently relax $M_{z}$ and $\mathcal{T}$ with magnetism to induce the QI.  Eliminating all non-essential symmetries and degeneracies, we form the Hamiltonian:
\begin{eqnarray}
\mathcal{H}({\bf k})&=&t_{1}\tau^{z}[\cos(k_{x}) + \cos(k_{y})] + t_{2}\tau^{x}[\cos(k_{x}) - \cos(k_{y})] \nonumber \\
&+& v_{m}\tau^{z} + t_{PH}\mathds{1}_{\tau\sigma}[\cos(k_{x}) + \cos(k_{y})]  \nonumber \\
&+& v_{s}\tau^{y}\sigma^{z}\sin(k_{x})\sin(k_{y}),
\label{eq:my2DquadMain}
\end{eqnarray}
where $\tau$ ($\sigma$) indexes the $s,d$-orbital (spin) degree of freedom and $\mathds{1}_{\tau\sigma}$ is the $4\times 4$ identity.  Here, $v_{m}$ produces on-site orbital splitting, $t_{1}$ ($t_{2}$) is first-neighbor hopping between the same (opposite) orbital, $t_{PH}$ is spin- and orbital-independent first-neighbor hopping that explicitly breaks particle-hole symmetry, and $v_{s}$ represents second-neighbor SOC (Appendix~\ref{sec:parameters}).  Eq.~(\ref{eq:my2DquadMain}) is invariant under the symmetries of layer group $p4/mmm1'$ (Table~\ref{tb:symsMain}).  Since $\mathcal{I}=M_{x}M_{y}M_{z}$ is given by the identity matrix and $\{M_{x},M_{y}\}=0$ in the representation in Table~\ref{tb:symsMain}, our model with four spinful orbitals (Eq.~(\ref{eq:my2DquadMain})) exhibits the same bulk symmetry eigenvalues and symmetry algebra as the original, spinless QI model in Ref.~\onlinecite{multipole}.  The bulk bands of Eq.~(\ref{eq:my2DquadMain}), due to the presence of $\mathcal{I}\times\mathcal{T}$ symmetry, are twofold degenerate (Fig.~\ref{fig:2Dmain}(b,c)).  In Appendices~\ref{sec:altHOFAsem} and~\ref{sec:boundary}, we additionally introduce and analyze models of QIs and HOFA semimetals with $p-d$ hybridization.

\begin{table}[t]
\begin{tabular}{|c|c|}
\hline
\multicolumn{2}{|c|}{Symmetries of 2D Hamiltonians $\mathcal{H}(k_{x},k_{y})$}  \\
\hline
$g$ & $g\mathcal{H}(gk_{x},gk_{y})g^{-1}$ \\
\hline
\hline
$M_{x}$ & $\sigma^{x}\mathcal{H}(-k_{x},k_{y})\sigma^{x}$ \\
\hline
$M_{y}$ & $\sigma^{y}\mathcal{H}(k_{x},-k_{y})\sigma^{y}$ \\
\hline 
$C_{4z}$ & $\tau^{z}\left(\frac{\mathds{1}_{\sigma} - i\sigma^{z}}{\sqrt{2}}\right)\mathcal{H}(k_{y},-k_{x})\tau^{z}\left(\frac{\mathds{1}_{\sigma} + i\sigma^{z}}{\sqrt{2}}\right)$ \\
\hline
$M_{z}$ & $\sigma^{z}\mathcal{H}(k_{x},k_{y})\sigma^{z}$ \\
\hline
$\mathcal{I}$ & $\mathcal{H}(-k_{x},-k_{y})$  \\
\hline
$\mathcal{T}$ & $\sigma^{y}\mathcal{H}^{*}(-k_{x},-k_{y})\sigma^{y}$ \\
\hline
\multicolumn{2}{c}{ }  \\
\hline
\multicolumn{2}{|c|}{Symmetries of 3D Hamiltonians $\mathcal{H}(k_{x},k_{y},k_{z})$}  \\
\hline
$g$ & $g\mathcal{H}(gk_{x},gk_{y},gk_{z})g^{-1}$ \\
\hline
\hline
$M_{x}$ & $\sigma^{x}\mathcal{H}(-k_{x},k_{y},k_{z})\sigma^{x}$ \\
\hline
$M_{y}$ & $\sigma^{y}\mathcal{H}(k_{x},-k_{y},k_{z})\sigma^{y}$ \\
\hline 
$C_{4z}$ &  $\tau^{z}\left(\frac{\mathds{1}_{\sigma} - i\sigma^{z}}{\sqrt{2}}\right)\mathcal{H}(k_{y},-k_{x},k_{z})\tau^{z}\left(\frac{\mathds{1}_{\sigma} + i\sigma^{z}}{\sqrt{2}}\right)$ \\
\hline
$M_{z}$ & $\sigma^{z}\mathcal{H}(k_{x},k_{y},-k_{z})\sigma^{z}$ \\
\hline
$\mathcal{I}$ & $\mathcal{H}(-k_{x},-k_{y},-k_{z})$   \\
\hline
$\mathcal{T}$ & $\sigma^{y}\mathcal{H}^{*}(-k_{x},-k_{y},-k_{z})\sigma^{y}$\\
\hline
\end{tabular}
\caption{The symmetry representation of the 2D and 3D $s-d$-hybridized models in the main text (Eqs.~(\ref{eq:my2DquadMain}),~(\ref{eq:breakIMain}),~(\ref{eq:quadMain}),~(\ref{eq:magHingeMain}), and~(\ref{eq:hingeMain})).  These models derive from Eq.~(\ref{eq:my2DquadMain}), which contains the symmetries (wallpaper group~\cite{DiracInsulator} $p4m$) of a QI~\cite{multipole,WladTheory} (Appendix~\ref{sec:bandrep}), as well as $M_{z}$, $\mathcal{I}=M_{x}M_{y}M_{z}$, and $\mathcal{T}$.}
\label{tb:symsMain}
\end{table}

To diagnose the topology of Eq.~(\ref{eq:my2DquadMain}), we examine the $x$-directed Wilson loop (holonomy) matrix~\cite{Fidkowski2011,ArisInversion,BarryFragile}, a bulk quantity defined by:
\begin{equation}
\mathcal{W}_{(k_{x0},k_y)} \equiv P e^{i\int_{k_{x0}}^{k_{x0}+2\pi} dk_x A_x(k_x,k_y)},
\label{eq:wilsoncontMain}
\end{equation}
where $P$ indicates that the integral is path-ordered and $A_x(k)_{ij} \equiv i\langle u^i(k)|\partial_{k_x} u^j(k)\rangle$ is the matrix-valued Berry connection. The eigenvalues $\theta(k_{y})$ of $\mathcal{W}$ are gauge invariant and form bands in one fewer dimension than that of the bulk, with connectivity and degeneracy constrained by the symmetries of the $x$-projected edge symmetry group~\cite{DiracInsulator}, as well as by the representations of bulk symmetries~\cite{ArisInversion}.  At half filling, Eq.~(\ref{eq:my2DquadMain}) exhibits two topologically distinct insulating phases (Fig.~\ref{fig:2Dmain}(c,f)), indicated by the relative ordering of the Kramers pairs of $C_{4z}$ eigenvalues of the occupied bands at $\Gamma$ and $M$ (Fig.~\ref{fig:2Dmain}(b))~\cite{multipole}.  In Fig.~\ref{fig:2Dmain}, we show the Wilson spectra computed from the lower two bands of Eq.~(\ref{eq:my2DquadMain}) in the uninverted (trivial) phase (e) and in the inverted (nontrivial) phase (h).  As we will detail below, we then also calculate the Wilson spectrum of Eq.~(\ref{eq:my2DquadMain}) in the presence of potentials that break $M_{z}$ symmetry while either preserving (k) or breaking (n) $\mathcal{T}$ symmetry; we also compare the Wilson loop spectra to the surface states of tight-binding Hamiltonians calculated in a ribbon geometry (d,g,j,m).

Using Eq.~(\ref{eq:wilsoncontMain}), we identify the nontrivial phase of Eq.~(\ref{eq:my2DquadMain}) as a topological crystalline insulator (TCI)~\cite{LiangOriginalTCI,TeoFuKaneTCI,DiracInsulator} with mirror Chern number $C_{M_{z}} = 2$ (Fig.~\ref{fig:2Dmain}(g,h)).  By introducing a term that breaks $M_{z}$ and $\mathcal{I}$ while preserving the symmetries of wallpaper group $p4m1'$ (generated by $M_{x,y},\ C_{4z},$ and $\mathcal{T}$)~\cite{DiracInsulator,MagneticBook}:
\begin{equation}
V_{M_{z}}({\bf k}) = v_{M_{z}}\left[\tau^{z}\sigma^{y}\sin(k_{x}) - \tau^{z}\sigma^{x}\sin(k_{y})\right],
\label{eq:breakIMain}
\end{equation}
we can gap the edge states of this TCI (Fig.~\ref{fig:2Dmain}(j)). However, its two-band $x$-directed Wilson loop still winds (Fig.~\ref{fig:2Dmain}(k)). This phenomenon is related to recently identified fragile topological phases~\cite{AshvinFragile,ArisInversion,JenFragile1,BarryFragile,WiederAxion}, whose Wilson loops can be rendered topologically trivial by the introduction of trivial bands.  In Appendix~\ref{sec:fragile}, we show how the topological Wilson connectivity of this four-band model is unstable to the addition of spinful $s$ orbitals at the $2c$ position of $p4m1'$.  In both the TCI (Eq.~(\ref{eq:my2DquadMain})) and fragile (Eqs.~(\ref{eq:my2DquadMain}) and~(\ref{eq:breakIMain})) phases, $\mathcal{T}$ symmetry obstructs the presence of singly degenerate corner modes; however we find that the $M_{z}$-broken fragile phase, when the overall system is kept at a constant half filling, still exhibits three-quarters-filled Kramers pairs of corner modes that can float into the bulk gap (Appendices~\ref{sec:fragile} and~\ref{sec:TCIBoundary}).  We show in Appendix~\ref{sec:TCIBoundary} that, as this fragile phase can be connected to a QI by restoring $\mathcal{T}$ symmetry without closing a bulk or edge gap, its corner modes still exhibit the same charges as the QI modulo $e$.  $C_{M_{z}}=2$ TCI phases in layer group $p4/mmm1'$ have been proposed in XY (X=Sn, Te; Y=S, Se, Te) monolayers~\cite{LiangTCIMonolayer}.  However because band inversion in these XY monolayers occurs at the $X$ and $X'$ points (Fig.~\ref{fig:2Dmain}(b)) between bands with different $\mathcal{I}$ eigenvalues, rather than at the $\Gamma$ or $M$ points between bands with different pairs of $C_{4z}$ eigenvalues, XY monolayers will realize a different insulating phase than the fragile phase of Eqs.~(\ref{eq:my2DquadMain}) and~(\ref{eq:breakIMain}) when $M_{z}$ is broken with a substrate or an external field (Appendix~\ref{sec:TCIBoundary}).

To induce the QI phase (Fig.~\ref{fig:2Dmain}(l)), we first set $v_{M_{z}}=0$ in Eq.~(\ref{eq:breakIMain}); this restores $M_{z}$ and $\mathcal{I}$ symmetries.  We then instead add to Eq.~(\ref{eq:my2DquadMain}) a term that anticommutes with $\mathcal{H}({\bf k})$ in its particle-hole symmetric limit ($t_{PH}\rightarrow 0$):
\begin{equation}
U({\bf k})=u[\tau^{y}\sigma^{y}\sin(k_{x}) + \tau^{y}\sigma^{x}\sin(k_{y})].
\label{eq:quadMain}
\end{equation}
Eq.~(\ref{eq:quadMain}) breaks $M_{z}$, $\mathcal{I}$, and $\mathcal{T}$ while preserving the unitary symmetries of $p4m$ and the magnetic antiunitary symmetries $M_{z}\times\mathcal{T}$ and $\mathcal{I}\times\mathcal{T}$, the latter of which continues to enforce a twofold band degeneracy (Fig.~\ref{fig:2Dmain}(l)).  The new Hamiltonian $\mathcal{H}({\bf k}) + U({\bf k})$ (Eqs.~(\ref{eq:my2DquadMain}) and~(\ref{eq:quadMain})) therefore has the symmetry of magnetic layer group $p4/m'mm$, a supergroup of $p4m$.  We note that because $U({\bf k})$ preserves two orthogonal mirrors, $M_{x,y}$, it cannot be induced by a constant Zeeman field alone, and must instead come from several internal magnetic moments or applied quadrupolar magnetism.  An example of a configuration of spin-$1/2$ magnetic moments is shown in Fig.~\ref{fig:2Dmain}(a) that, like $U({\bf k})$, lowers the symmetry of $p4/mmm1'$ to $p4/m'mm$.  When Eq.~(\ref{eq:quadMain}) is added to Eq.~(\ref{eq:my2DquadMain}), the surface states and Wilson spectrum gap (Fig.~\ref{fig:2Dmain}(m,n)), but gapped, SSH-like states remain bound to the 1D edges~\cite{multipole,WladTheory} (Appendix~\ref{sec:boundaryNumbers}). By projecting onto one of the eigenstates of $\mathcal{W}$ (for example the lower Wilson band in Fig.~\ref{fig:2Dmain}(n)), a second, nested Wilson loop can be computed in the $y$ direction, and displays a nested Berry phase $\theta_{2}$ of $0$ ($\pi$) if this magnetic insulator is in a trivial (quadrupole) phase~\cite{multipole}.  For all nonzero values of $u$ in Eq.~(\ref{eq:quadMain}), transitions between QI and trivial phases occur when the bulk gap closes at $\Gamma$($M$) for $2t_{1} = - (+)v_{m}$, with $|\frac{v_{m}}{t_{1}}| < (>) 2$ characterizing the QI (trivial) phase. As we show in Appendix~\ref{sec:equivalence}, the Hamiltonian $\mathcal{H}({\bf k}) + U({\bf k})$ (Eqs.~(\ref{eq:my2DquadMain}) and~(\ref{eq:quadMain})) is topologically equivalent to the quadrupole model introduced in Ref.~\onlinecite{multipole}.  We can also choose to reintroduce $V_{M_{z}}({\bf k})$ (Eq.~(\ref{eq:breakIMain})) to Eqs.~(\ref{eq:my2DquadMain}) and~(\ref{eq:quadMain}), which, as it is invariant under $p4m$, will preserve the QI phase if it does not close a bulk or edge gap, even though it breaks the combined magnetic symmetries $M_{z}\times\mathcal{T}$ and $\mathcal{I}\times\mathcal{T}$ in $p4/m'mm$, the (magnetic) layer group of Eqs.~(\ref{eq:my2DquadMain}) and~(\ref{eq:quadMain}).  For weak $v_{M_{z}}$ this therefore results in a QI phase in $p4m$ with singly degenerate bands, and for stronger values, it can induce a crystalline semimetallic phase (Appendix~\ref{sec:parameters}).  It also follows from the theory of band representations~\cite{QuantumChemistry,JenFragile1,BarryFragile} that the QI phase of Eqs.~(\ref{eq:my2DquadMain}) and~(\ref{eq:quadMain}) with (without) Eq.~(\ref{eq:breakIMain}) is an obstructed atomic limit~\cite{QuantumChemistry} with the two occupied Wannier orbitals shifted to the $1b$ Wyckoff position of $p4m$ ($p4/m'mm$) (Appendix~\ref{sec:bandrep}).

In Appendices~\ref{sec:TIboundary} through~\ref{sec:noMirror}, we construct a microscopic picture of the phase transitions between the TCI, fragile, and QI phases of the tight-binding Hamiltonians given by Eq.~(\ref{eq:my2DquadMain}) with the potentials in Eqs.~(\ref{eq:breakIMain}) and~(\ref{eq:quadMain}).  We also analytically examine the phase transition between a tight-binding model of a $p_{z}-d_{x^{2}-y^{2}}$-hybridized 2D TI (Appendix~\ref{sec:pd}) and an additional model of QI in $p4/m'mm$ that is distinct from (but topologically equivalent to) Eqs.~(\ref{eq:my2DquadMain}) and~(\ref{eq:quadMain}).  Specifically in Appendices~\ref{sec:TIboundary} and~\ref{sec:TItoTrivial}, we derive the low-energy $k\cdot p$ theories of $p_{z}-d_{x^{2}-y^{2}}$- and $s-p_{z}$-hybridized 2D TIs whose atoms lie at the $1a$ position of $p4/mmm1'$, and analytically solve for the bound states on their corners when their edge states are gapped with $p4m$-symmetric magnetism.  We find that the $p-d$- ($s-p$-) hybridized TI evolves into a QI (trivial insulator) when $p4m$-symmetric magnetism is introduced, precisely because the inverted bands exhibit different (the same) Kramers pairs of $C_{4z}$ eigenvalues, such that the symmetry eigenvalues of the occupied bands (do not) match those of a QI in $p4m$ (Appendix~\ref{sec:bandrep}).  We then show in Appendix~\ref{sec:TCIBoundary} that the edge states of a $C_{M_{z}}=2$ TCI (such as the $s-d_{x^{2}-y^{2}}$-hybridized TCI phase of Eq.~(\ref{eq:my2DquadMain})) can gap under an $M_{z}$-breaking, $\mathcal{T}$-symmetric potential (such as Eq.~(\ref{eq:breakIMain})) into four Kramers pairs of corner modes that, if the total system filling is fixed at $1/2$, are quarter- (three-quarters-), half-, or fully filled depending on the $C_{4z}$ eigenvalues of the inverted bulk bands.  We then demonstrate that the quarter-filled and three-quarters-filled cases evolve into QIs under $p4m$-preserving magnetism, also indicating that the $s-d$-hybridized TCI phase of Eq.~(\ref{eq:my2DquadMain}), like the $p-d$-hybridized 2D TI in Appendices~\ref{sec:pd} and~\ref{sec:TIboundary}, can transition into a QI when its edge states are gapped with $p4m$-symmetric, $M_{z}$-breaking magnetism.  In Appendix~\ref{sec:EBRsforTCI}, we explain this by using TQC~\cite{QuantumChemistry,JenFragile1,BarryFragile} to show that the $s-d$-hybridized TCI phase of Eq.~(\ref{eq:my2DquadMain}) exhibits the same quadrupole moment (modulo $e$) as a $p_{z}-d_{x^{2}-y^{2}}$-hybridized 2D TI (when their edge states are gapped by breaking $M_{z}$ and $\mathcal{T}$).

\begin{figure}[t]
\centering
\includegraphics[width=0.43\textwidth]{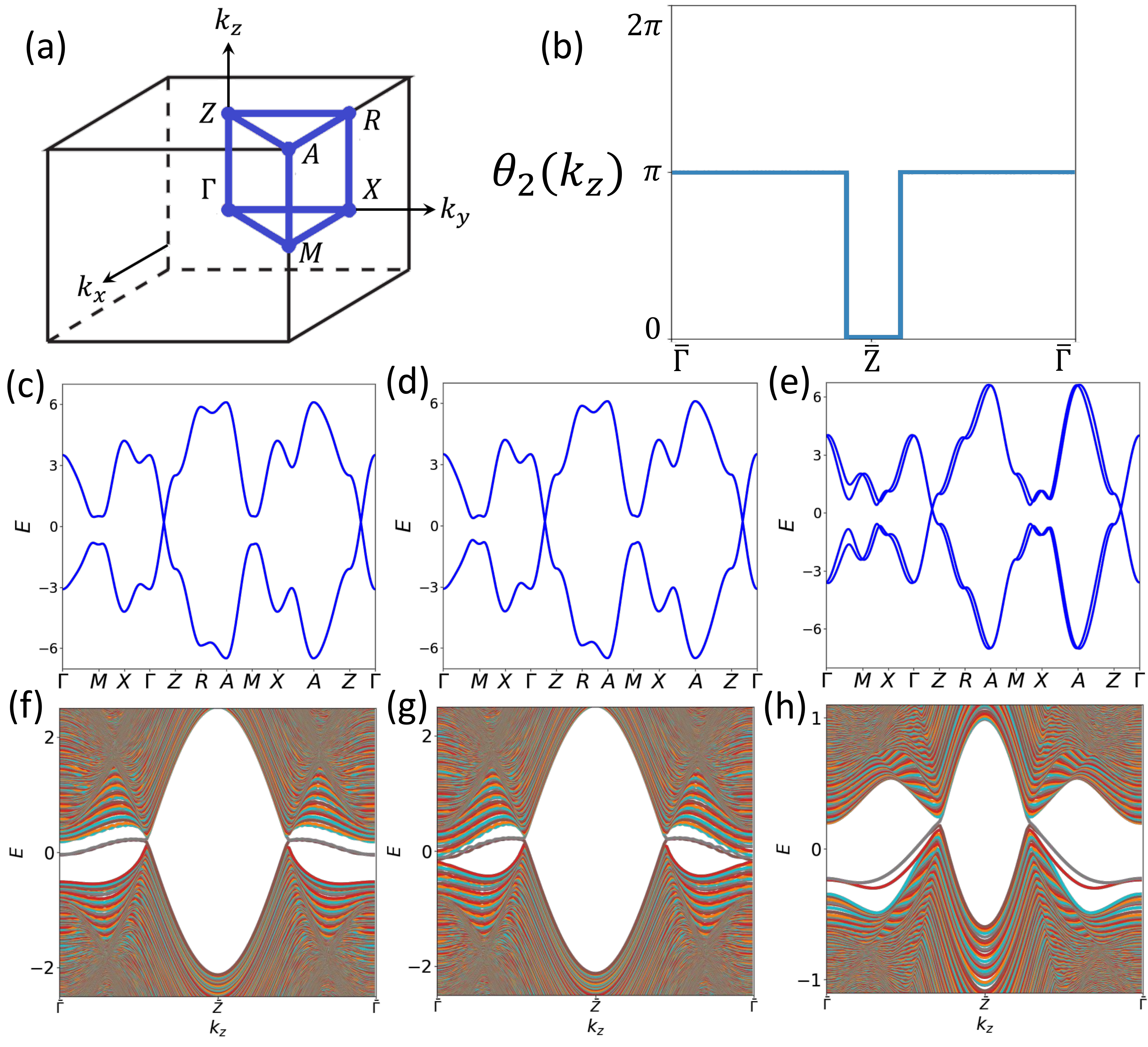}
\caption{HOFA states in magnetic, nonmagnetic, and fragile topological Dirac semimetals.  (a) The BZ, (c) bulk bands, and (f) hinge bands of a $z$-directed rod of a $\mathcal{T}$-broken 3D Dirac semimetal with HOFAs in magnetic SG $P4/m'mm$ (Eq.~(\ref{eq:magHingeMain})).  The Hamiltonian of each $k_{z}$ slice describes either a trivial insulator or a QI, and the bulk Dirac points occur at the quantum critical points between the two phases. (d)  HOFAs can also be realized in a closely related $\mathcal{T}$-symmetric Dirac semimetal in SG 123 $P4/mmm1'$ (Eq.~(\ref{eq:hingeMain})).  (g) Here, the Hamiltonian of the $k_{z}=0$ plane is in the 2D TCI phase~\cite{NagaosaDirac} shown in Fig.~\ref{fig:2Dmain}(f,g), and therefore there is no gap in the hinge-projected surface states at $\bar{\Gamma}$.  (e) Upon breaking $M_{z}$ and $\mathcal{I}$ symmetries by adding Eq.~(\ref{eq:breakIMain}) to Eq.~(\ref{eq:hingeMain}), which reduces the symmetry to SG 99 ($P4mm1'$), (h) the face TCI cones gap as they did previously in Fig.~\ref{fig:2Dmain}(i-k), resulting in a noncentrosymmetric Dirac semimetal without surface states~\cite{KargarianDiracArc1} whose only topological boundary modes are HOFA states.  In (h), two sets of weakly split HOFAs meet in Kramers pairs at $\bar{\Gamma}$; if the system filling is fixed to $1/2$ (\emph{i.e.}, to the filling of the Dirac points), then one set of HOFA states in (h) is half-filled and carries a topological quadrupole moment, and the other set is fully filled, and is topologically trivial.  At $\bar{\Gamma}$, this implies that the Kramers pairs of hinge states are three-quarters filled, and that they exhibit the same quadrupole moment (modulo $e$) as a QI (Appendix~\ref{sec:TCIBoundary}).  The Hamiltonian of the $k_{z}=0$ plane in (e,h) exhibits the same fragile topology as the 2D model in Fig.~\ref{fig:2Dmain}(i-k) (Appendix~\ref{sec:fragile}), implying that the hinge states at $\bar{\Gamma}$ are equivalent to the fractionally charged corner modes of a 2D fragile TI.  The model (Eqs.~(\ref{eq:hingeMain}) and~(\ref{eq:breakIMain})) shown in (e,h) therefore represents a previously uncharacterized variant of topological semimetal that carries observable signatures of fragile topology.  (b) The 3D semimetallic phases of Eqs.~(\ref{eq:magHingeMain}) and~(\ref{eq:hingeMain}), whether or not Eq.~(\ref{eq:breakIMain}) is additionally present, exhibit the same quadrupole-quantized nested Berry phase~\cite{multipole} $\theta_{2}(k_{z})$.}
\label{fig:HingeSMmain}
\end{figure}

\textit{3D Dirac semimetals with HOFA states} -- We now stack the previous 2D models into 3D to create physically motivated Hamiltonians modeling solid state materials that are equivalent to tuning cycles of Eqs.~(\ref{eq:my2DquadMain}),~(\ref{eq:breakIMain}), and~(\ref{eq:quadMain}) (Fig.~\ref{fig:pumpMain}(f,g)).  In this work, we restrict focus to gapless tuning cycles, which are equivalent 3D topological semimetals.  We begin constructing 3D models by stacking Eq.~(\ref{eq:my2DquadMain}) in the $z$ direction, adding a term ($t_{H}\tau^{z}\cos(k_{z})$) that varies the gaps at $k_{x}=k_{y}=0,\pi$ as functions of $k_{z}$, and adding Eq.~(\ref{eq:quadMain}) with a modulation governed by one of two distinct interlayer coupling terms:
\begin{eqnarray}
\mathcal{H}_{H1}({\bf k}) &=& \mathcal{H}({\bf k}) + U({\bf k}) + t_{H}\tau^{z}\cos(k_{z}), \label{eq:magHingeMain} \\
\mathcal{H}_{H2}({\bf k}) &=& \mathcal{H}({\bf k}) + U({\bf k})\sin(k_{z}) + t_{H}\tau^{z}\cos(k_{z}).
\label{eq:hingeMain}
\end{eqnarray}
In addition to respecting the symmetries of magnetic SG $P4mm$ (number 99.163 in the BNS notation~\cite{MagneticBook}), the space group generated by adding translations in the $z$ direction to~\cite{MagneticBook} $p4m$, $H_{H1}({\bf k})$ respects the antiunitary symmetries $M_{z}\times\mathcal{T}$ and $\mathcal{I}\times\mathcal{T}$, whereas $H_{H2}({\bf k})$ individually respects $M_{z}$, $\mathcal{I}$, and $\mathcal{T}$ (Table~\ref{tb:symsMain}).  To tune $\mathcal{H}_{H1,2}({\bf k})$ into 3D Dirac semimetal phases, we invert bands by setting $v_{m}<0$, $t_{1}>0$, and tuning $t_{H}$.  When $|t_{H}|>2t_{1} + v_{m}$, a pair of Dirac points forms along the $\Gamma Z$ line (Fig.~\ref{fig:HingeSMmain}(c)). Viewing $\mathcal{H}(k_{x},k_{y})$ on each constant-$k_{z}$ slice of the 3D BZ as a 2D system, these Dirac points are equivalent to the critical point between trivial and QI phases (Appendix~\ref{sec:double}).  To see this, note that the Dirac points are formed by inverting bands with different pairs of $C_{4z}$ eigenvalues in a 3D BZ for which slices indexed by $k_{z}$ are invariant under magnetic supergroups of $p4m$ (Appendix~\ref{sec:altHOFAsem}).  For both Eqs.~(\ref{eq:magHingeMain}) and~(\ref{eq:hingeMain}), the QI-nontrivial BZ slices are identified by the bulk nested Wilson loop~\cite{multipole} parameterized as a function of $k_{z}$ (Fig.~\ref{fig:HingeSMmain}(b)).  When $|t_{H}|$ is further increased beyond $2t_{1} - v_{m}$, an additional pair of Dirac points forms along $MA$; we analyze the HOFA-state structure of this semimetal in Appendix~\ref{sec:double}.  We note that similar results were obtained in Ref.~\onlinecite{TaylorToy} in toy models of magnetic Dirac semimetals.  However, in this work, we also uniquely discover HOFA states in $\mathcal{T}$-symmetric Dirac and Weyl semimetals, allowing their prediction in real materials, which we will address shortly.

We first search for HOFA states in the 3D Dirac semimetal phase of $\mathcal{H}_{H1}({\bf k})$ (Eq.~(\ref{eq:magHingeMain})) that only exhibits a pair of Dirac points along $\Gamma Z$ (specific parameters are listed in Appendix~\ref{sec:parameters}).  In Eq.~(\ref{eq:magHingeMain}), $\mathcal{T}$ symmetry is broken, and therefore the 2D Hamiltonians of all $k_{z}$-indexed BZ planes (including $k_{z}=0,\pi$) describe either trivial insulators or QIs.  Calculating the hinge and surface states of the Dirac semimetal phase of Eq.~(\ref{eq:magHingeMain}) in a rod geometry that is finite in the $x$ and $y$ directions (Fig.~\ref{fig:HingeSMmain}(f)), we observe the absence of 2D surface states and the presence of HOFAs spanning the projections of the bulk 3D Dirac points along the 1D hinges.  If the bulk Dirac points are gapped by breaking $C_{4z}$ while preserving mirror symmetry, the HOFA states can evolve into the chiral hinge modes of a 3D (magnetic) higher-order TI (axion insulator)~\cite{WladTheory,BernevigMoTe2,WiederAxion}.  Though $\mathcal{H}_{H1}({\bf k})$ provides the simplest realization of a HOFA Dirac semimetal without surface states, it also requires the complicated mirror-preserving magnetism of magnetic SG $P4/m'mm$ (123.341 in the BNS notation~\cite{MagneticBook}), which cannot be realized in a constant external field or with ferromagnetism.  As the number of known magnetic structures is small compared to the number of known materials~\cite{BilbaoMagStructures}, it is difficult to identify material candidates for the magnetic HOFA Dirac semimetal phase of $\mathcal{H}_{H1}({\bf k})$.  However, we do find that the antiferromagnetic phase of the Dirac semimetal CeSbTe in magnetic space group $P_{c}4/ncc$ (130.432 in the BNS notation~\cite{MagneticBook}) is closely related~\cite{SchoopAFM}, and may exhibit HOFA states (Appendix~\ref{sec:pinnedHOFAs}).

Fortunately, we discover that topological HOFA states are also present in $\mathcal{T}$-symmetric Dirac semimetals.  To see this, we tune $\mathcal{H}_{H2}({\bf k})$ (Eq.~(\ref{eq:hingeMain})) into the parameter regime $2t_{1} + v_{m}<|t_{H}|<2t_{1} - v_{m}$ (specific parameters for Fig.~\ref{fig:HingeSMmain}(d,g) are detailed in Appendix~\ref{sec:parameters}) to realize a $\mathcal{T}$-symmetric Dirac semimetal in SG 123 ($P4/mmm1'$) with a time-reversed pair of Dirac points along $\Gamma Z$ and with mirror Chern number $C_{M_{z}}=2$ ($0$) at $k_{z}=0$ ($\pi$).  As with the magnetic Dirac semimetal phase of $\mathcal{H}_{H1}({\bf k})$ (Eq.~(\ref{eq:magHingeMain})), the bulk bands of $\mathcal{H}_{H2}({\bf k})$ (Fig.~\ref{fig:HingeSMmain}(d)) are twofold degenerate throughout the BZ as a consequence of $\mathcal{I}\times\mathcal{T}$ symmetry (Table~\ref{tb:symsMain}).  Crucially, while this 3D model ($\mathcal{H}_{H2}({\bf k})$ in Eq.~(\ref{eq:hingeMain})) is $\mathcal{T}$-symmetric, 2D planes of the BZ indexed by $k_{z}\neq 0,\pi$ are still invariant under magnetic layer group $p4/m'mm$, and thus can still be topologically equivalent to QIs.  In the Dirac semimetal phase of $\mathcal{H}_{H2}({\bf k})$, there are two kinds of topological boundary modes: mirror TCI cones on $M_{z}$-preserving 2D faces at $k_{z}=0$, and singly-degenerate HOFAs on each of the four 1D hinges connecting the projections of the TCI cones to those of the bulk Dirac points (Fig.~\ref{fig:HingeSMmain}(g)).  Furthermore, we recognize that $U({\bf k})\sin(k_{z})$ in Eq.~(\ref{eq:hingeMain}), which acts in each $k_{z}\neq 0,\pi$ BZ slice like the $p4m$-symmetric magnetism (Eq.~(\ref{eq:quadMain})) depicted in Fig.~\ref{fig:2Dmain}(a), is also equivalent to the bulk spin-orbit term previously introduced in Ref.~\onlinecite{KargarianDiracArc1} to destabilize the surface Fermi arcs of a Dirac semimetal.

As with the 2D TCI in Fig.~\ref{fig:2Dmain}(j), the $(100)$-surface states of the TCI-nontrivial plane at $k_{z}=0$ of the 3D Dirac semimetal phase of $\mathcal{H}_{H2}({\bf k})$ can be gapped by breaking $M_{z}$ while preserving $\mathcal{T}$.  We accomplish this by adding $V_{M_{z}}({\bf k})$ in Eq.~(\ref{eq:breakIMain}) to Eq.~(\ref{eq:hingeMain}); this breaks $M_{z}$ and $\mathcal{I}$ while preserving $p4m$, $\mathcal{T}$, and $z$-direction lattice translations, lowering the overall symmetry to SG 99 $P4mm1'$ (Table~\ref{tb:symsMain}).  In Fig.~\ref{fig:HingeSMmain}(e,h), we respectively plot the bulk and hinge bands of the noncentrosymmetric Dirac semimetal phase resulting from adding Eq.~(\ref{eq:breakIMain}) to Eq.~(\ref{eq:hingeMain}).  We observe that the previous mirror TCI surface states of Eq.~(\ref{eq:hingeMain}) have become split and, instead, there are four hinge-localized Kramers pairs of states at $\bar{\Gamma}$ in Fig.~\ref{fig:HingeSMmain}(h).  These eight states become weakly split into two sets of HOFA states at $k_{z}\neq 0$; as described in Appendix~\ref{sec:TCIBoundary}, if we fix the overall system filling to $1/2$ (\emph{i.e.}, to the filling of the Dirac points), then one of the sets of four HOFA states in Fig.~\ref{fig:HingeSMmain}(h) is half-filled and carries a topological quadrupole moment, and the other set is fully filled, and is topologically trivial.  This implies that the Kramers pairs of hinge states at $\bar{\Gamma}$ in Fig.~\ref{fig:HingeSMmain}(h) are three-quarters-filled and exhibit the same quadrupole moment (modulo $e$) as a QI (Appendix~\ref{sec:TCIBoundary}).  In this noncentrosymmetric Dirac semimetal phase (Eqs.~(\ref{eq:breakIMain}) and~(\ref{eq:hingeMain})), the Hamiltonian of the $k_{z}=0$ plane exhibits the same fragile topology as the 2D insulator in Fig.~\ref{fig:2Dmain}(l-n), and the anomalous, fractionally charged Kramers pairs of states on each hinge at $k_{z}=0$ represent an observable signature of the fragile bands  (or of an obstructed atomic limit that can be decomposed into the sum of fragile bands and unobstructed atomic limits~\cite{WiederAxion}) (Appendix~\ref{sec:TCIBoundary}).  Therefore, the noncentrosymmetric Dirac semimetal phase of Eqs.~(\ref{eq:breakIMain}) and~(\ref{eq:hingeMain}) represents a previously uncharacterized fragile topological variant of Dirac semimetal.  Like the 3D HOTIs (axion insulators) analyzed in Ref.~\onlinecite{WiederAxion}, we refer to this variant of Dirac semimetal (Eqs.~(\ref{eq:breakIMain}) and~(\ref{eq:hingeMain})) as fragile because its minimal realization is equivalent to a tuning cycle between a 2D fragile TI with anomalous corner modes and a 2D insulator with a trivial Wilson and corner spectrum.  Specifically, because the 3D Dirac semimetal phase of Eqs.~(\ref{eq:breakIMain}) and~(\ref{eq:hingeMain}) respects fourfold rotation and $\mathcal{T}$ symmetries, then the appearance of quarter-empty (or -filled) Kramers pairs of hinge states at $\bar{\Gamma}$ (where the overall system filling is fixed to the filling of the Dirac points) indicates that the occupied bands at $k_{z}=0$ contain the fragile valence bands of the 2D model described by Eqs.~(\ref{eq:my2DquadMain}) and~(\ref{eq:breakIMain}).  This occurs because the band inversion that creates the Dirac points along $\Gamma Z$ in Eqs.~(\ref{eq:breakIMain}) and~(\ref{eq:hingeMain}) also drives the Hamiltonian of the $k_{z}=0$ plane to exhibit the same $C_{4z}$ eigenvalues as a QI in $p4m$ (Appendix~\ref{sec:TCIBoundary}).  While not every fragile phase exhibits intrinsic (anomalous) corner modes (for example, two superposed copies of the $\mathcal{I}$-symmetric fragile TIs examined in Refs.~\onlinecite{BernevigMoTe2,WiederAxion} combine to form an insulator that is also fragile, but one without anomalous corner states), our results further imply that specific corner states (or state counting imbalances, as discussed in Appendix~\ref{sec:TCIBoundary}) can still represent a robust signature of a valence manifold that can be decomposed into the sum of unobstructed (trivial) atomic limits and fragile bands, when crystal symmetries and band connectivity are taken into account.

Alternatively, we can formulate a model of a 3D Dirac semimetal from hybridized layers of $p_{z}$ and $d_{x^{2}-y^{2}}$ orbitals in which the Hamiltonian of the $k_{z}=0$ plane instead characterizes a 2D TI~\cite{AndreiTI}, as occurs in the experimentally confirmed Dirac semimetals~\cite{ZJSurface,ZJDirac,NagaosaDirac,CavaDirac2,StableCadmium} Cd$_3$As$_2$ and~\cite{SYDiracSurface} Na$_{3}$Bi.  To realize HOFAs as the only boundary (surface and hinge) modes in a semimetal with TI surface cones, unlike with the Dirac semimetal phase of $\mathcal{H}_{H2}({\bf k})$ (Eq.~(\ref{eq:hingeMain})), one must break $\mathcal{T}$ symmetry (Appendix~\ref{sec:TIboundary}), or apply strain to drive additional band inversions (Appendix~\ref{sec:double}).  In Appendix~\ref{sec:pd}, we present a model of a $p-d$-hybridized Dirac semimetal with coexisting TI surface states and HOFA hinge states.  We also note that the three $\mathcal{T}$-symmetric semimetal models presented in this work -- the TCI-nontrivial~\cite{NagaosaDirac} Dirac semimetal phase of Eq.~(\ref{eq:hingeMain}), the fragile topological Dirac semimetal phase of Eqs.~(\ref{eq:breakIMain}) and~(\ref{eq:hingeMain}), and the $p-d$-hybridized Dirac semimetal in Appendix~\ref{sec:pd} -- all exhibit the same number of half-filled HOFA states at $k_{z}\neq 0$ (where the system filling is fixed to the filling of the bulk Dirac points), despite displaying differing numbers of gapped surface states at $k_{z}\neq 0,\pi$.  This reinforces the notion that the surface states of Dirac semimetals are not themselves a topological consequence of the bulk Dirac points, but rather only appear due to the topology of high-symmetry planes, and are not required to connect to the surface projections of the bulk Dirac points~\cite{KargarianDiracArc1}.

\begin{figure}[t]
\centering
\includegraphics[width=0.5\textwidth]{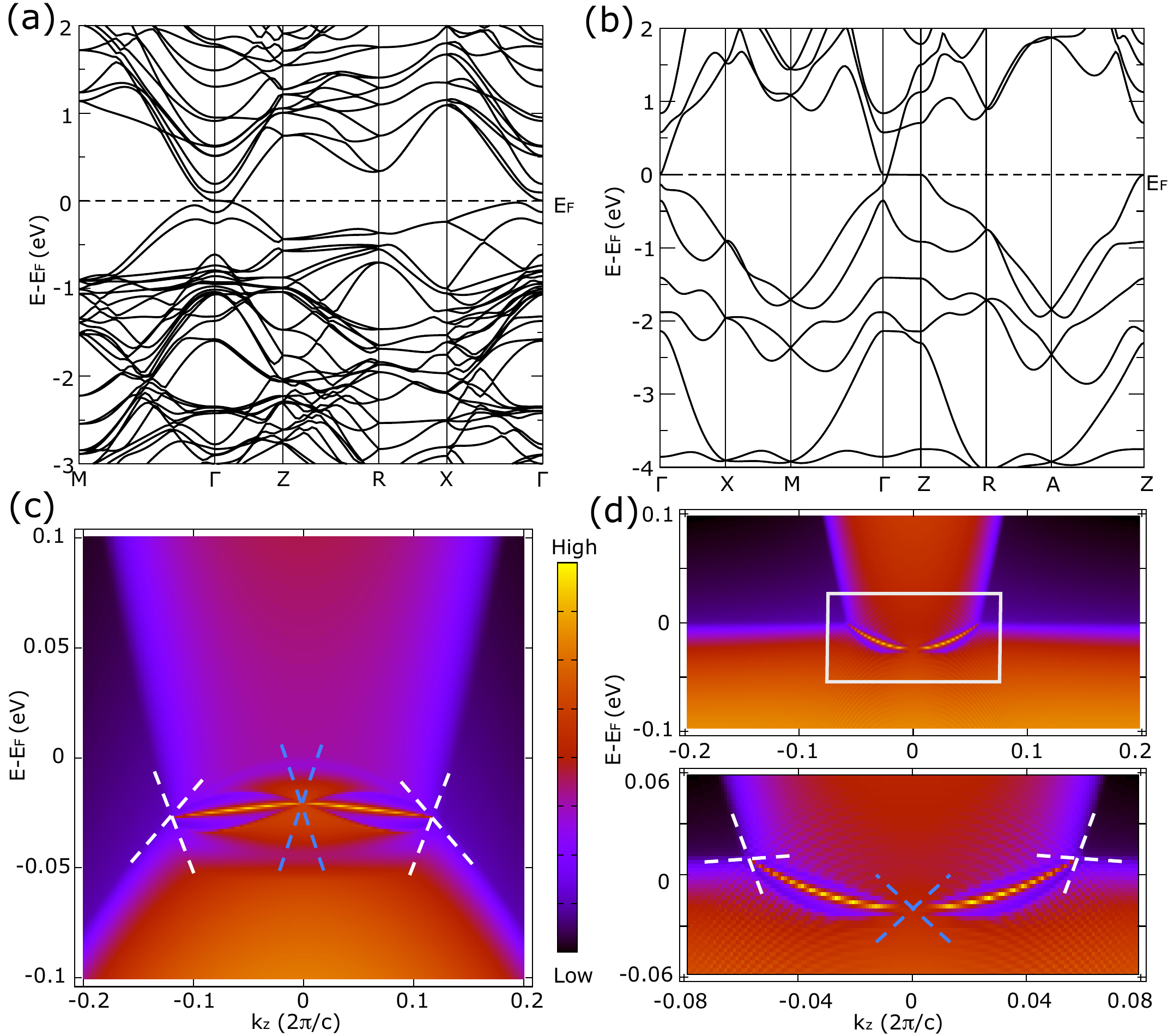}
\caption{HOFA states in $\alpha''$-Cd$_3$As$_2$ and KMgBi.  (a) Bulk bands incorporating the effects of SOC of $\alpha''$-Cd$_3$As$_2$ in SG 137 ($P4_{2}/nmc1'$)~\cite{ZJSurface,ZJDirac,NagaosaDirac,CavaDirac2,StableCadmium}.  This semimetal exhibits $\mathcal{T}$-reversed pairs of bulk 3D Dirac cones as well as 2D TI cones on its faces as a consequence of the nontrivial $\mathbb{Z}_{2}$ topology of the $k_{z}=0$ plane (Appendix~\ref{sec:cadmium})~\cite{ZJDirac,ZJSurface,SYDiracSurface,NagaosaDirac}.  (c) The hinge spectrum of the $k\cdot p$ model of $\alpha''$-Cd$_3$As$_2$ introduced in Ref.~\onlinecite{ZJDirac} exhibits previously undetected HOFA states connecting the projections of the bulk Dirac cones (white) to the hinge projection of the topological face cones (blue).  (b) Bulk bands incorporating the effects of SOC and (d) hinge states of the candidate tilted Dirac semimetal KMgBi in SG 129 ($P4/nmm1'$)~\cite{KMgBi2,KMgBi3}.  (d) Zooming into the green boxed region, HOFAs are clearly visible connecting the projections of the bulk, tilted 3D Dirac points (white) to the projections of surface 2D TI cones (blue) (Appendix~\ref{sec:kmgbi}).  The bulk band structures in (a) and (b) were obtained from first-principles, and then used to fit tight-binding models whose hinge Green's functions are shown in (c) and (d), respectively (Appendices~\ref{sec:cadmium} and~\ref{sec:kmgbi}, respectively).}
\label{fig:DFTmain}
\end{figure}

\vspace{0.15in}
\textit{Material realizations} -- Most surprisingly, the Dirac points in Fig.~\ref{fig:HingeSMmain} display the same $k\cdot p$ Hamiltonian as the bulk nodes~\cite{ZJDirac,NagaosaDirac} in the centrosymmetric structural ($\alpha$ and $\alpha''$) phases of the experimentally established Dirac semimetal~\cite{CavaDirac2,ZJDirac,ZJSurface} Cd$_3$As$_2$ (Fig.~\ref{fig:DFTmain}(a)).  This is because $\mathcal{H}_{H2}({\bf k})$ in Eq.~(\ref{eq:hingeMain}), which respects symmorphic SG 123 $P4/mmm1'$, and Cd$_{3}$As$_{2}$ in its room- (high-) temperature $\alpha$ ($\alpha''$) phase, which respects nonsymmorphic SG 137 $P4_{2}/nmc1'$ (SG 142 $I4_{1}/acd1'$), have little groups along their respective $k_{x}=k_{y}=0$ lines with isomorphic unitary subgroups (Fig.~\ref{fig:HingeSMmain}(a) and Appendices~\ref{sec:double} and~\ref{sec:pinnedHOFAs}).  Both the $\alpha$ and $\alpha''$ structural phases of Cd$_{3}$As$_{2}$ exhibit the same bulk topology -- they both host a time-reversed pair of Dirac points along $k_{x}=k_{y}=0$, and are equivalent at $k_{z}=0$ to 2D TIs due to a band inversion between the $5s$ orbitals of Cd and the $m_{j}=\pm 3/2$ subset of the $4p_{x,y}$ orbitals of As~\cite{ZJDirac}.  In terms of the $s-p$- and $s-d$-hybridized semimetals and TIs analyzed in Appendices~\ref{sec:pd},~\ref{sec:TIboundary},~\ref{sec:TItoTrivial}, and~\ref{sec:TCIBoundary}, the topology of the $\alpha$ and $\alpha''$ structural phases of Cd$_3$As$_2$ can be understood by noting that the $m_{j}=\pm 3/2$ subset of spinful $p_{x,y}$ orbitals exhibits the same parity eigenvalues as spinful $p_{z}$ orbitals and the same fourfold rotation eigenvalues as spinful $d_{x^{2}-y^{2}}$ orbitals~\cite{QuantumChemistry}.  This implies that the bulk topology of Cd$_3$As$_2$ (Appendix~\ref{sec:cadmium}) is equivalent to the superposition of an $s-p_{z}$-hybridized 3D TI and an $s-d_{x^{2}-y^{2}}$-hybridized topological Dirac semimetal with HOFA states (or equivalently, to the $p_{z}-d_{x^{2}-y^{2}}$ HOFA Dirac semimetal in Appendix~\ref{sec:pd}).  Using an analytic formulation of topological (intrinsic) HOFA states derived in Appendices~\ref{sec:TIboundary} through~\ref{sec:noMirror}, we find that the $k\cdot p$ theory and symmetries of $\alpha$-Cd$_{3}$As$_{2}$ imply the presence of HOFA states on the hinges of $(001)$- ($z$-) axis-directed samples, which have recently been synthesized in experiment~\cite{StemmerCadmiumGrowth}.  Though the $\alpha$ phase is body-centered and respects $x$- and $y$-normal glide reflections, instead of $M_{x,y}$ like Eq.~(\ref{eq:hingeMain}), we respectively provide proofs in Appendices~\ref{sec:bodycenter} and~\ref{sec:unpinnedHOFAs} demonstrating that body-centered and glide-symmetric Dirac semimetals also exhibit topological HOFA states like those in Fig.~\ref{fig:HingeSMmain}(f-h).

The symmetries and $k\cdot p$ theory of the Dirac points of Eq.~(\ref{eq:hingeMain}) additionally imply that the primitive tetragonal ($\alpha''$) phase of Cd$_3$As$_2$ should also exhibit topological HOFA states.  Although the $\alpha''$ phase naturally occurs at high temperatures (475 --600 $^\circ$C)~\cite{CavaDirac2}, it can be stabilized in single crystalline form at room temperature and below by 2\% zinc doping~\cite{StableCadmium}; as Zn is isoelectronic to Cd, this doping will not affect the Fermi level.  Calculating the hinge spectrum of the original $k\cdot p$ model introduced in Ref.~\onlinecite{ZJDirac} for $\alpha''$-Cd$_3$As$_2$, we confirm our prediction of previously overlooked HOFAs (Fig.~\ref{fig:DFTmain}(c) and Appendix~\ref{sec:cadmium}).  This suggests a clear route towards predicting additional candidate Dirac semimetals with HOFA states: using the low-energy theory of the QI (Appendix~\ref{sec:boundary}), we determine that strong-SOC Dirac semimetals with SGs that contain point group $4mm$ ($C_{4v}$) will exhibit HOFA states when they are cut into nanorods or exhibit step edge configurations that preserve fourfold axes (Table~\ref{tb:SGsMain}).  This is analogous to the helical hinge modes in the HOTI bismuth, which are only observable in samples that are cut into nanowires (or terminated with step edge configurations) that preserve bulk rotation and $\mathcal{I}$ symmetries~\cite{HOTIBismuth}.  A number of candidate Dirac semimetals have already been identified in the SGs in Table~\ref{tb:SGsMain}, including the aforementioned $\alpha$ and $\alpha''$ phases of Cd$_3$As$_2$, the rutile-structure ($\beta'$-) phase of PtO$_2$ in SG 136 ($P4_{2}/mnm1'$)~\cite{PtO21,PtO22}, and families of tilted Dirac semimetals related to VAl$_3$ in SG 139 ($I4/mmm1'$)~\cite{Type2Dirac1}, YPd$_2$Sn in SG 225 ($Fm\bar{3}m1'$)~\cite{Type2Dirac2}, and KMgBi in SG 129 ($P4/nmm1'$)~\cite{KMgBi2,KMgBi3}.

Of the candidate HOFA semimetals that we identified, we highlight KMgBi and $\beta'$-PtO$_{2}$ because of their simple geometries.  KMgBi has recently been identified as a topological semimetal with critically tilted bulk Dirac cones~\cite{KMgBi2}, and its electronic properties have been examined in experiment~\cite{KMgBi3}.  In Fig.~\ref{fig:DFTmain}, we plot the bulk bands (b) calculated from first principles, and the hinge spectrum (d) of a lattice tight-binding model of KMgBi fit to the bands in (b) (Appendix~\ref{sec:kmgbi}). We find that the $k_{z}=0$ plane of KMgBi exhibits the topology of a 2D TI (Appendix~\ref{sec:kmgbi}), in agreement with the surface-state calculation in Ref.~\onlinecite{KMgBi2}.  In the vicinity of $k_{z}=0$ (Fig.~\ref{fig:DFTmain}(c)), HOFAs are clearly visible connecting the hinge projections of the bulk 3D Dirac points (white) to the projections of 2D surface TI cones at $k_{z}=0$ (blue).  This boundary mode structure is captured by the model of a $p-d$-hybridized Dirac semimetal in Appendix~\ref{sec:pd}.

\begin{table}[t]
\begin{tabular}{|c|c|c|}
\hline
\multicolumn{3}{|c|}{Space Groups Admitting Dirac Points with HOFA States}  \\
\hline
Point Group Name &  Point Group Symbol & SG Numbers  \\
\hline
\hline
$C_{4v}$ & $4mm1'$ & 99 -- 110 \\
\hline
$D_{4h}$ & $4/mmm1'$ & 123 -- 142 \\
\hline
$O_{h}$ & $m\bar{3}m1'$ & 221 -- 230 \\
\hline
\end{tabular}
\caption{Space groups (SGs) that admit Dirac points with HOFA states derived from the QI introduced in Ref.~\onlinecite{multipole}.  All of these SGs have point groups that contain $C_{4v}$.  We obtain this list of SGs by combining the nested Jackiw-Rebbi formulation of the QI in Appendix~\ref{sec:boundary} with an analysis of the crystallographic rod groups in Appendix~\ref{sec:SGs}.  In all of these SGs, semimetals with Dirac points along lines with $4mm$ or $4/m'mm$ symmetry will exhibit intrinsic HOFA states when cut into nanorods that preserve fourfold axes and are thick compared to the in-plane lattice spacing.  This list is a complete enumeration of the SGs that permit Dirac semimetals with HOFA states directly related to the QI introduced in Ref.~\onlinecite{multipole}; alternative realizations of Dirac and Weyl semimetals with HOFA states derived from other 2D magnetic insulators with corner states are also possible (Refs.~\onlinecite{BernevigMoTe2,WiederAxion} and Appendix~\ref{sec:unpinnedHOFAs}).}
\label{tb:SGsMain}
\end{table}

\begin{figure*}[t]
\centering
\includegraphics[width=0.85\textwidth]{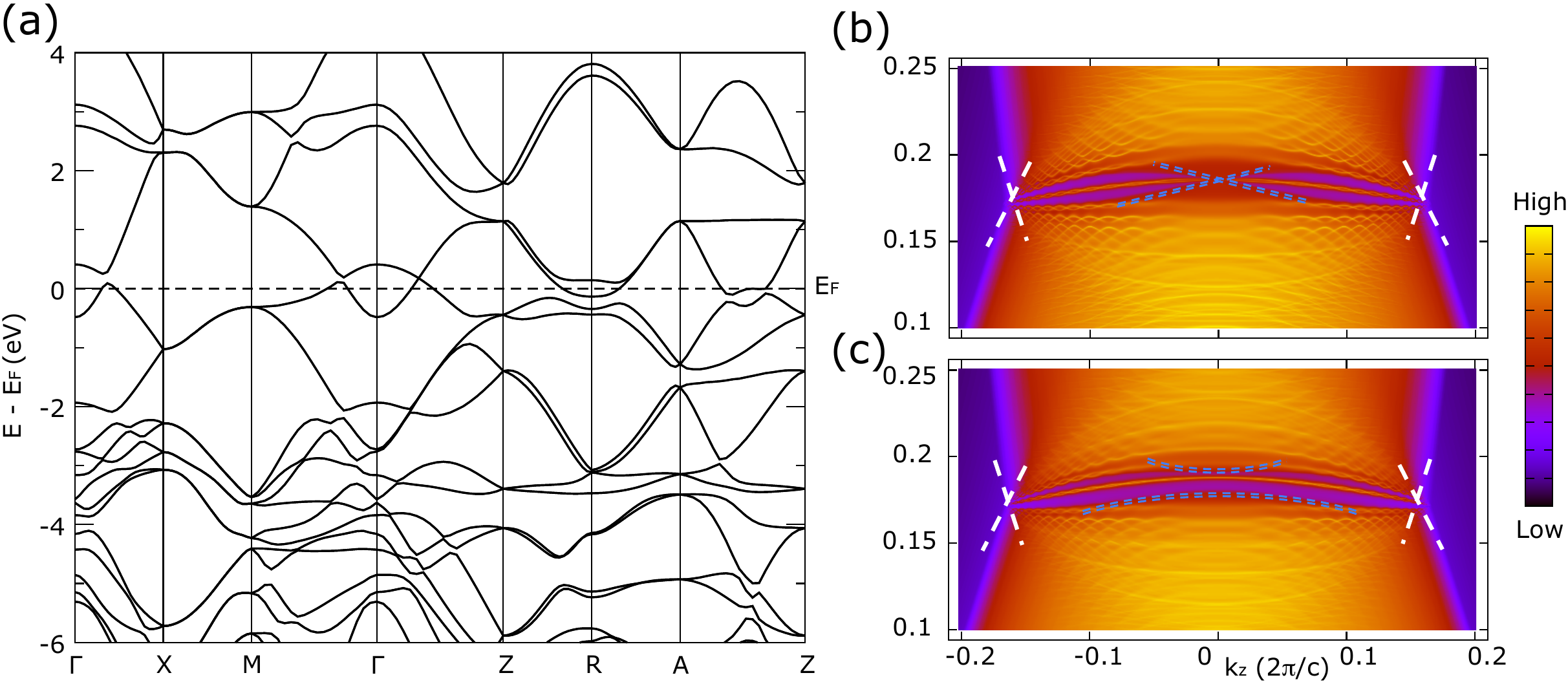}
\caption{HOFA states and fragile Dirac semimetal phase in $\beta'$-PtO$_2$.  (a) Bulk bands incorporating the effects of SOC of the candidate Dirac semimetal $\beta'$-PtO$_2$ in SG 136 ($P4_{2}/mnm1'$)~\cite{PtO21,PtO22}.  Unlike in the Dirac semimetals $\alpha''$-Cd$_3$As$_2$ and KMgBi examined in Fig.~\ref{fig:DFTmain}, the $k_{z}=0$ plane of $\beta'$-PtO$_2$ is equivalent to a 2D TCI with mirror Chern number $C_{M_{z}}=2$ (Ref.~\onlinecite{PtO22} and Appendix~\ref{sec:pto2}).  (b) The hinge spectrum of $\beta'$-PtO$_2$ exhibits two narrowly split HOFA states connecting the hinge projections of the bulk 3D Dirac points (white) to the projections of two surface TCI cones at $k_{z}=0$ (blue).  Fixing the system filling to that of the bulk Dirac points, we find that the lower HOFA state in energy in (b) is half-filled, and therefore exhibits a topological quadrupole moment, and the higher state is unoccupied, and is thus topologically trivial, as discussed in Appendices~\ref{sec:TCIBoundary} and~\ref{sec:pto2}.  (c)  In the presence of a $z$-directed external electric field, the surface TCI cones in $\beta'$-PtO$_2$ become gapped (blue), allowing for the two HOFA states to meet at $k_{z}=0$ in a quarter-filled Kramers pair of corner modes that is characteristic of the fragile phase introduced in this work (Fig.~\ref{fig:2Dmain}(i-k), Fig.~\ref{fig:HingeSMmain}(e,h), and Appendices~\ref{sec:fragile},~\ref{sec:TCIBoundary}, and~\ref{sec:pto2}).  The bulk band structure in (a) was obtained from first-principles, and then used to fit a tight-binding model whose hinge Green's functions in the absence and presence of an external electric field are shown in (b) and (c), respectively (Appendix~\ref{sec:pto2}).}
\label{fig:DFTfragileMain}
\end{figure*}

In Fig.~\ref{fig:DFTfragileMain}, we also examine the bulk and hinge spectra of the candidate Dirac semimetal $\beta'$-PtO$_2$.  Single crystals of PtO$_2$ in its rutile-structure ($\beta'$) phase have previously been prepared in experiment~\cite{PtO21}, and its bulk and surface electronic structure were examined in a previous theoretical work~\cite{PtO22}.  In $\beta'$-PtO$_2$, the Hamiltonian of the $k_{z}=0$ plane is equivalent to a 2D TCI with mirror Chern number $C_{M_{z}}=2$; therefore $\beta'$-PtO$_2$ is more closely related to the $s-d$-hybridized HOFA semimetal model introduced in this work (Eq.~(\ref{eq:hingeMain})) than it is to $\alpha''$-Cd$_3$As$_2$ and KMgBi, which at $k_{z}=0$ are instead equivalent to 2D TIs (Fig.~\ref{fig:DFTmain} and Appendix~\ref{sec:DFT}).  In Fig.~\ref{fig:DFTfragileMain}, we plot the bulk bands (a) and hinge spectrum (b) of $\beta'$-PtO$_2$ calculated from first principles, as detailed in Appendix~\ref{sec:pto2}.  In the spectrum of a single hinge (Fig.~\ref{fig:DFTfragileMain}(b)), we observe two narrowly split HOFA states connecting the hinge projections of the bulk 3D Dirac points to the projections of the surface 2D TCI cones.  Fixing the system filling to that of the bulk Dirac points, we observe that, similar to the hinge spectrum of the fragile topological Dirac semimetal in Fig.~\ref{fig:HingeSMmain}(h), only one of the HOFA states on each hinge of $\beta'$-PtO$_2$ is half-filled.  Specifically, we find that the lower HOFA state in energy in Fig.~\ref{fig:DFTfragileMain}(b) is half filled, and therefore carries a topological quadrupole moment (Appendix~\ref{sec:TCIBoundary}), and that the other HOFA state is unoccupied, and is therefore topologically trivial.

Because the TCI surface cones in $\beta'$-PtO$_2$ are only protected by $M_{z}$ symmetry, they can be gapped without breaking $\mathcal{T}$ symmetry, unlike the 2D TI surface cones in $\alpha''$-Cd$_3$As$_2$ and KMgBi (Appendix~\ref{sec:TIboundary}).  To preserve the bulk Dirac points and intrinsic HOFA states in $\beta'$-PtO$_2$ while gapping the surface TCI cones, we must break $M_{z}$ (and hence $\mathcal{I}$) symmetry while preserving the $4_{2}$ screw, $x$-directed $n$-glide reflection, $\mathcal{T}$, and lattice translation symmetries of SG 136 ($P4_{2}/mnm1'$).  Though this cannot be accomplished by uniaxial strain, which either manifests as symmetry-preserving stretching in the $z$ direction or as a translation-breaking strain gradient, these symmetry requirements can be satisfied in experiment by applying an external electric field that is spatially constant (or slowly varying on the scale of the lattice spacing) along the $z$- ($c$-) axis of a fourfold-symmetric $\beta'$-PtO$_2$ sample.  Implementing the effects of an external electric field into our Green's function calculation of the hinge states in $\beta'$-PtO$_2$ (Fig.~\ref{fig:DFTfragileMain}(c)), we observe that the TCI cones have become gapped, and that the HOFA states instead meet in a Kramers pair of quarter-filled corner modes at $k_{z}=0$.

Furthermore, because the $k_{z}=0$ plane of $\beta'$-PtO$_2$ both exhibits the topology of a $C_{M_{z}}=2$ TCI and carries the same bulk fourfold rotation eigenvalues as a QI in $p4m$ (Appendix~\ref{sec:bandrep}), then the quarter-filled corner modes that appear at $k_{z}=0$ in the hinge spectrum of $\beta'$-PtO$_2$ when its TCI surface states are gapped with an external electric field (Fig.~\ref{fig:DFTfragileMain}(c)) indicate that the valence manifold at $k_{z}=0$ can be separated into trivial bands and fragile bands with the same topology as the 2D fragile phase introduced in this work (Eqs.~(\ref{eq:my2DquadMain}) and~(\ref{eq:breakIMain})).  Specifically, $\beta'$-PtO$_{2}$ only differs from a trivial (unobstructed) atomic limit without Dirac points or hinge states by a single inversion at the $\Gamma$ point between bands with the same parity eigenvalues and different fourfold rotation eigenvalues (Fig.~\ref{fig:DFTfragileMain}(a) and Appendix~\ref{sec:pto2}).  Therefore, the band inversion in $\beta'$-PtO$_2$ drives the $k_{z}=0$ plane into the same $C_{M_{z}}=2$ TCI phase as that of Eq.~(\ref{eq:my2DquadMain}), which necessarily gaps into an insulator with fragile bands and fractionally charged Kramers pairs of corner modes when $M_{z}$ is relaxed while preserving fourfold rotation and $\mathcal{T}$ (Appendix~\ref{sec:TCIBoundary}).  Whether the entire valence manifold at $k_{z}=0$ is fragile or an obstructed atomic limit depends on the precise details of the bands below the Fermi energy, and for the case of the fragile phase introduced in this work (Eqs.~(\ref{eq:my2DquadMain}) and~(\ref{eq:breakIMain})), uniquely cannot be inferred from the symmetry eigenvalues of the occupied bands (Appendix~\ref{sec:fragile}), unlike the fragile phases examined in previous works~\cite{AshvinFragile,ArisInversion,JenFragile1,BarryFragile,WiederAxion}.  Nevertheless, like the $\mathcal{I}$-symmetric fragile phases with corner modes introduced in Refs.~\onlinecite{WiederAxion,BernevigMoTe2}, the fragile phase of Eqs.~(\ref{eq:my2DquadMain}) and~(\ref{eq:breakIMain}) still exhibits anomalous (intrinsic) corner modes when trivial bands (\emph{i.e.} unobstructed atomic limits without corner states) are introduced below the Fermi energy.  Therefore, because the $k_{z}=0$ plane of $\beta'$-PtO$_2$ can be decomposed into a set of trivial bands without corner states and the inverted bands at the Fermi energy, it still exhibits the fractionally charged corner states shown in Fig.~\ref{fig:DFTfragileMain}(c) when $M_{z}$ is relaxed while preserving $\mathcal{T}$ and fourfold rotation, whether or not the entire valence manifold at $k_{z}=0$ is fragile or an obstructed atomic limit.  We draw further connection between $\beta'$-PtO$_2$ and the model of an $s-d$-hybridized, noncentrosymmetric, fragile topological Dirac semimetal introduced in this work (Eqs.~(\ref{eq:breakIMain}) and~(\ref{eq:hingeMain}) and Fig.~\ref{fig:HingeSMmain}(e,h)) by noting that the quarter-filled corner modes at $k_{z}=0$ in Fig.~\ref{fig:DFTfragileMain}(c) represent the particle-hole conjugates of the three-quarters-filled fragile-phase corner modes observable at $k_{z}=0$ in Fig.~\ref{fig:HingeSMmain}(h) (Appendices~\ref{sec:TCIBoundary} and~\ref{sec:pto2}).

{ 
\vspace{0.1in}
\centerline{\bf Discussion}
}

The HOFA states introduced in this work may be detectable through transport and STM experiments~\cite{HOTIBismuth}.  Though our analysis has focused on nanowire geometries, HOFA states may also be observable through momentum-resolved probes of fourfold-symmetric arrangements of step edges on the surfaces of Dirac semimetals with the SGs in Table~\ref{tb:SGsMain}.  Nonlocal quantum oscillation experiments~\cite{AnalytisOscillation} and SQUID measurements~\cite{HOTIBismuth} performed on materials with HOFA states are likely to show interesting signatures reflecting the reduced dimensionality of the hinge modes.  By generalizing the analysis performed in this work, further examples of topological semimetals with HOFA states should be readily discoverable, including HOFA Dirac semimetals with sixfold symmetries and, as discussed in Appendices~\ref{sec:noMirror} and~\ref{sec:unpinnedHOFAs}, high-fold-rotation Weyl semimetals with coexisting surface Fermi arcs and HOFA states.  Additionally, our atomic-orbital description of QIs with $s-d$ hybridization suggests the possibility of quadrupolar generalizations of polyacetylene~\cite{SSH,SSHExp}.  Finally, because the analytic expression that we obtain for the bound (corner) states of the QI in Appendix~\ref{sec:TIboundary}, when the reflection symmetries of $p4m$ are relaxed, can be expressed as the superposition of $1+2n$ (\emph{i.e.} an odd number) of quadrupole moments whose direction is a free parameter but whose magnitude is fixed to $e/2$ (Appendix~\ref{sec:noMirror}), then it bears similarities with recent gauge-theory descriptions of fractons with anomalous tensor charges~\cite{Fracton}.

\vspace{0.05in}
\centerline{\bf Methods}

All tight-binding, surface state, hinge state, and Wilson loop calculations were performed using the standard implementation of the open-source~\textsc{PythTB} Python package~\cite{PythTB}.  Nested Wilson loop calculations were performed using an extension of~\textsc{PythTB} that is documented in Ref.~\onlinecite{WiederAxion}.

First-principles calculations were performed using the projector augmented wave (PAW)~\cite{paw1} method as implemented in the Vienna Ab initio Simulation Package (VASP)~\cite{vasp1,vasp2}.  The hinge states of $\alpha''$-Cd$_3$As$_2$, KMgBi, and $\beta'$-PtO$_2$ were obtained by mapping the bands closest to the Fermi energy to tight-binding models and then calculating the Green's function along a single 1D hinge of a slab that was infinite along the crystal axis parallel to the hinge and respectively finite and semi-infinite along the two perpendicular axes.  Further details of our first-principles and hinge Green's function calculations are provided in Appendix~\ref{sec:DFT}.

\vspace{0.05in}
\centerline{\bf Data Availability}
\vspace{0.03in}

The data supporting the findings of this study are available within the paper and other findings of this study are available from the corresponding authors upon reasonable request.  All first-principles calculations were performed using~\textsc{CIF} structure files with the experimental lattice parameters, which can be obtained from the Inorganic Crystal Structure Database (ICSD)~\cite{ICSD} using the accession numbers provided in Appendix~\ref{sec:DFT}.

\vspace{0.05in}
\centerline{\bf Acknowledgments}
\vspace{0.05in}

$^\star$Corresponding author: \url{bwieder@princeton.edu} (BJW), \url{bbradlyn@illinois.edu} (BB), \url{bernevig@princeton.edu} (BAB).  We thank Ady Stern, Ivo Souza, Maia G. Vergniory, Fan Zhang, Chen Fang, and Michael P. Zaletel for helpful discussions.  We further thank Kuan-Sen Lin for a close reading of the revised version of this work.  B. J. W., J. C., and B. A. B. acknowledge the hospitality of the Donostia International Physics Center, where parts of this work were carried out.  B. J. W. and B. A. B. were supported by the Department of Energy Grant No. DE-SC0016239, the National Science Foundation EAGER Grant No. DMR 1643312, Simons Investigator Grant No. 404513, ONR Grant No. N00014-14-1-0330, the Packard Foundation, the Schmidt Fund for Innovative Research, and a Guggenheim Fellowship from the John Simon Guggenheim Memorial Foundation.  Z. W. acknowledges support from the CAS Pioneer Hundred Talents Program.  J. C. acknowledges support from the Flatiron Institute, a division of the Simons Foundation.  L. M. S. was supported by a MURI grant on topological insulators from the Army Research Office, Grant No. ARO W911NF-12-1-0461.  B. J. W., L. M. S., and B. A. B. were additionally supported by the NSF through the Princeton Center for Complex Materials, a Materials Research Science and Engineering Center DMR-1420541.  As discussed in the main text, during the long preparation of this extensive work, simplified toy-models featuring variants of HOFA states were introduced in Refs.~\onlinecite{TaylorToy,VladHOFA}; further comparisons to the results of Ref.~\onlinecite{TaylorToy} are provided in Appendix~\ref{sec:boundaryNumbers}.  Corner modes in fragile phases were also recently recognized in Refs.~\onlinecite{BernevigMoTe2,WiederAxion}, and were connected in Ref.~\onlinecite{BernevigMoTe2} to a robust variant of spinless HOFA states distinct from those introduced in this work.  Finally, during the preparation of this work, Majorana HOFA states in nodal superconductors were analyzed in Ref.~\onlinecite{GhorashiTaylorHOFASC}.

\vspace{0.05in} 
\centerline{\bf Author Contributions}
\vspace{0.025in}

All authors contributed equally to the intellectual content of this work.  The existence of topological hinge states in layered quadrupole insulating (QI) Dirac semimetals was first recognized by B. A. B. and X. D. in consultation with B. J. W. and J. C.  The $s-d$-hybridized spinful QI, related 2D TCI and fragile phases, and layered 3D $\mathcal{T}$-symmetric HOFA Dirac semimetals introduced in this work were proposed by B. J. W.  HOFA states in doubly inverted $\mathcal{T}$-symmetric Dirac semimetals were proposed by J. C.  Equivalence between the spinless flux-threaded and spinful $s-d$-hybridized formulations of QIs was demonstrated by B. J. W. in consultation with J. C. and B. B.  Band representation analyses of the QI obstructed atomic limit and related topological phases were performed by B. J. W.,  J. C., and B. B.  The nested Jackiw-Rebbi formulation of QIs and HOFA states was introduced by B. J. W. under the supervision of B. B. and B. A. B.  The list of space groups supporting HOFA Dirac semimetals was obtained by B. J. W., Z. W., J. C., L. M. S., and B. B.  Analysis of HOFA states in body-centered semimetals was performed by B. J. W., J. C., and Z. W.  Tight-binding, surface, and hinge state calculations were performed by B. J. W. with assistance from J. C. and B. B.  Nested Wilson loop calculations were implemented by B. B., J. C., B. J. W., and Z. W.  The materials search was performed by B. J. W., Z. W., L. M. S., and X. D. with help from all authors.  First-principles and hinge Green's function calculations were performed by Z. W.  The manuscript was written by B. J. W. with help from all of the authors.  B. A. B. was responsible for the overall research direction.

\vspace{0.05in}
\centerline{\bf Competing Interests}
\vspace{0.05in}

The authors declare no competing interests.

\clearpage
\onecolumngrid
\begin{appendix}

\begin{center}
{\bf Appendices for ``Strong and Fragile Topological Dirac Semimetals with Higher-Order Fermi Arcs''}
\end{center}

\tableofcontents

\section{Tight-Binding Parameters for Figures \ref{fig:2Dmain} and \ref{fig:HingeSMmain} of the Main Text}
\label{sec:parameters}

Here, we list the specific model parameters used to generate the figures in the main text.  All plots shown in this paper were generated using the tight-binding, slab (ribbon), and Wilson loop functionality of the~\textsc{PythTB} package~\cite{PythTB}.  The nested Wilson loop shown in Fig.~\ref{fig:HingeSMmain}(b) of the main text was calculated by modifying the~\textsc{PythTB} functions relating to Wannier centers and Berry phase.  For convenience, we reproduce in this appendix the Hamiltonians and potentials introduced in the main text of this work.  First, in the main text, we formulated a 2D time-reversal- ($\mathcal{T}$-) symmetric Hamiltonian (Eq.~(\ref{eq:my2DquadMain}) of the main text):
\begin{eqnarray}
\mathcal{H}({\bf k})&=&t_{1}\tau^{z}[\cos(k_{x}) + \cos(k_{y})] + t_{2}\tau^{x}[\cos(k_{x}) - \cos(k_{y})] + v_{m}\tau^{z} \nonumber \\
&+& t_{PH}\mathds{1}_{\tau\sigma}[\cos(k_{x}) + \cos(k_{y})]  + v_{s}\tau^{y}\sigma^{z}\sin(k_{x})\sin(k_{y}),
\label{eq:my2Dquad}
\end{eqnarray}
that was invariant under layer group~\cite{WiederLayers,DiracInsulator,SteveMagnet,MagneticBook,subperiodicTables} $p4/mmm1'$, whose generating symmetries are represented by:
\begin{eqnarray}
M_{x}\mathcal{H}(k_{x},k_{y})M_{x}^{-1} &=& \sigma^{x}\mathcal{H}(-k_{x},k_{y})\sigma^{x},\ M_{z}\mathcal{H}(k_{x},k_{y})M_{z}^{-1} = \sigma^{z}\mathcal{H}(k_{x},k_{y})\sigma^{z}, \nonumber \\
C_{4z}\mathcal{H}(k_{x},k_{y})C_{4z}^{-1} &=& \tau^{z}\left(\frac{\mathds{1}_{\sigma} - i\sigma^{z}}{\sqrt{2}}\right)\mathcal{H}(k_{y},-k_{x})\tau^{z}\left(\frac{\mathds{1}_{\sigma} + i\sigma^{z}}{\sqrt{2}}\right),\ \mathcal{T}\mathcal{H}(k_{x},k_{y})\mathcal{T}^{-1} = \sigma^{y}\mathcal{H}^{*}(-k_{x},-k_{y})\sigma^{y}.
\label{eq:symsParams}
\end{eqnarray}
We also introduced the $M_{z}$- (and $\mathcal{I}$-) breaking, $\mathcal{T}$-symmetric potential $V_{M_{z}}({\bf k})$ (Eq.~(\ref{eq:breakIMain}) of the main text),
\begin{equation}
V_{M_{z}}({\bf k}) = v_{M_{z}}\left[\tau^{z}\sigma^{y}\sin(k_{x}) - \tau^{z}\sigma^{x}\sin(k_{y})\right],
\label{eq:breakI}
\end{equation}
and the $M_{z}$-, $\mathcal{I}$-, and $\mathcal{T}$-breaking, $M_{z}\times\mathcal{T}$- (and $\mathcal{I}\times\mathcal{T}$-) symmetric quadrupolar magnetic potential $U({\bf k})$  (Eq.~(\ref{eq:quadMain}) of the main text):
\begin{equation}
U({\bf k})=u[\tau^{y}\sigma^{y}\sin(k_{x}) + \tau^{y}\sigma^{x}\sin(k_{y})].
\label{eq:quad}
\end{equation}

\begin{figure}[h]
\centering
\includegraphics[width=1.0\textwidth]{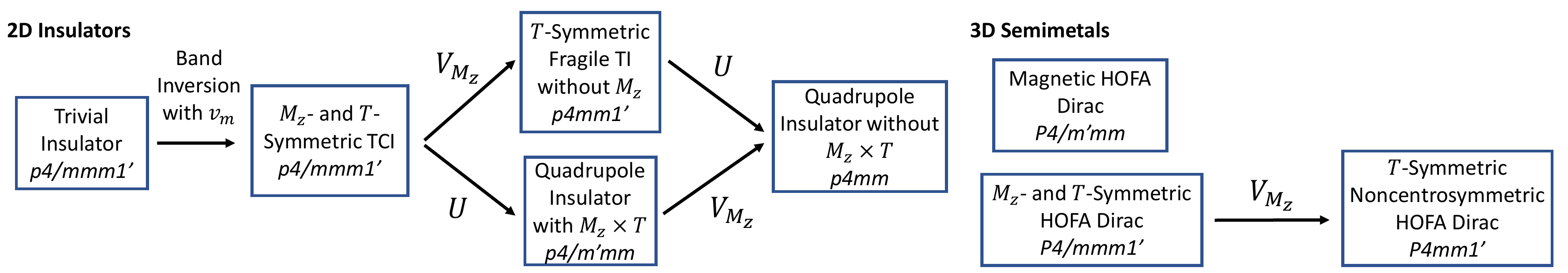}
\caption{Relationship between the phases of the 2D and 3D models introduced in this work.  The 2D $\mathcal{T}$-symmetric Hamiltonian in Eq.~(\ref{eq:my2Dquad}) can be tuned between trivial and and topological crystalline insulating (TCI) phases.  From the TCI phase, breaking $M_{z}$ and keeping time-reversal ($\mathcal{T}$), accomplished by the introduction of $V_{M_{z}}({\bf k})$ in Eq.~(\ref{eq:breakI}), results in a phase with ``fragile'' topology~\cite{AshvinFragile,JenFragile1,JenFragile2,AdrianFragile,BarryFragile,ZhidaBLG,AshvinBLG1,AshvinBLG2,AshvinFragile2,YoungkukMonopole,BernevigMoTe2,HarukiFragile,KoreanFragile,FragileKoreanInversion,ZhidaFragileAffine,KooiPartialNestedBerry,WiederAxion,WiederDefect}.  Conversely, starting in the TCI phase and breaking $\mathcal{T}$ while keeping the product $M_{z}\times\mathcal{T}$, accomplished through the introduction of $U({\bf k})$ in Eq.~(\ref{eq:quad}), results in a quadrupole insulating (QI) phase.  By introducing the magnetic potential $U({\bf k})$ into the fragile topological phase, or by introducing the $M_{z}\times\mathcal{T}$-breaking potential $V_{M_{z}}({\bf k})$ into the QI in $p4/m'mm$, a QI can be realized in the type-I magnetic group~\cite{BigBook,MagneticBook} $p4mm$.  In this work, this $M_{z}\times\mathcal{T}$-broken QI is realized in 2D Brillouin zone (BZ) planes indexed by $k_{z}\neq 0,\pi$ in Fig.~\ref{fig:HingeSMmain}(e,h) of the main text.  In 3D, an $M_{z}\times\mathcal{T}$-symmetric magnetic Dirac semimetal with higher-order Fermi arcs (HOFAs) is realized by $\mathcal{H}_{H1}({\bf k})$ (Eq.~(\ref{eq:magHinge})), and a $\mathcal{T}$- and $M_{z}$-symmetric centro- ($\mathcal{I}$-) symmetric Dirac semimetal with HOFAs is realized by $\mathcal{H}_{H2}({\bf k})$ (Eq.~(\ref{eq:hinge})).  Finally, the $\mathcal{T}$- and $M_{z}$-symmetric HOFA semimetal phase of $\mathcal{H}_{H2}({\bf k})$ (Eq.~(\ref{eq:hinge})) can be reduced to a noncentrosymmetric HOFA Dirac semimetal without surface states through the introduction of $V_{M_{z}}({\bf k})$ in Eq.~(\ref{eq:breakI}). For both the 2D and 3D phases, the layer and space groups are listed, respectively, using the expanded Shubnikov magnetic group notation~\cite{BigBook,MagneticBook}.  The specific parameters used to realize the 2D and 3D phases highlighted in this work are listed in Tables~\ref{tb:2D} and~\ref{tb:3D}, respectively.}
\label{fig:Phases}
\end{figure}

In Fig.~\ref{fig:Phases}, we show the relationships between the 2D insulating phases highlighted in Fig.~\ref{fig:2Dmain} of the main text, which are realized using the parameters listed in Table~\ref{tb:2D}.  Beginning with the trivial (uninverted) phase of the $M_{z}$- and $\mathcal{T}$-symmetric 2D Hamiltonian in Eq.~(\ref{eq:my2Dquad}), we tune $v_{m}$ to invert bands with different $C_{4z}$ eigenvalues and the same parity eigenvalues at $\Gamma$, while keeping the bulk bands uninverted at the other TRIM points.  This induces a 2D spinful topological crystalline insulator (TCI)~\cite{TeoFuKaneTCI,NagaosaDirac} with mirror Chern number~$C_{M_{z}}=2$.  This TCI phase can be reduced to an insulating phase with ``fragile'' topology~\cite{AshvinFragile,JenFragile1,JenFragile2,AdrianFragile,BarryFragile,ZhidaBLG,AshvinBLG1,AshvinBLG2,AshvinFragile2,YoungkukMonopole,BernevigMoTe2,HarukiFragile,KoreanFragile,FragileKoreanInversion,ZhidaFragileAffine,KooiPartialNestedBerry,WiederAxion,WiederDefect} through the introduction of nonzero $v_{M_{z}}$ in the $M_{z}$- and $\mathcal{I}$-breaking, $\mathcal{T}$-symmetric potential $V_{M_{z}}({\bf k})$ in Eq.~(\ref{eq:breakI}).  The $C_{M_{z}}=2$ TCI phase can also be reduced to a $\mathcal{T}$-broken quadrupole insulating (QI) phase~\cite{multipole,WladTheory,ZeroBerry,FulgaAnon,HigherOrderTIChen,EmilCorner,WladPhotonCorner,WladCorners,FrankCorners,EzawaCorner,TeoWladTCI,YoungkukBLG} through the introduction of nonzero $u$ in the $\mathcal{T}$-, $M_{z}$-, and $\mathcal{I}$-breaking potential $U({\bf k})$ in Eq.~(\ref{eq:quad}).  Crucially, although these fragile and QI phases are topologically distinct, we show in Appendix~\ref{sec:TCIBoundary} that they both exhibit quadrupolar 0D corner modes with charges $\pm e/2\mod e$ when half filled, \emph{even though the occupied bands of the fragile phase are not Wannierizable, \emph{i.e.}, they do not admit a description in terms of symmetric, exponentially localized Wannier functions}~\cite{ThoulessWannier,AlexeyVDBTI,QuantumChemistry,BarryFragile}.  By introducing the magnetic potential $U({\bf k})$ into the fragile topological phase, or by introducing the $M_{z}\times\mathcal{T}$-breaking potential $V_{M_{z}}({\bf k})$ into the QI in $p4/m'mm$, a QI can be realized in the type-I magnetic group~\cite{BigBook,MagneticBook} $p4mm$.  We do not explicitly show an $\mathcal{I}\times\mathcal{T}$-broken QI phase in 2D, but we do present in Fig.~\ref{fig:HingeSMmain}(e,h) of the main text a 3D model of a Dirac semimetal in which an $\mathcal{I}\times\mathcal{T}$-broken QI phase occurs in 2D Brillouin zone (BZ) planes indexed by $k_{z}\neq0,\pi$.

\begin{table}[h]
\begin{tabular}{|c|c|c|c|c|c|c|c|c|c|}
\hline
\multicolumn{10}{|c|}{Parameters for the 2D Models in Fig.~\ref{fig:2Dmain} of the Main Text} \\
\hline
Phase & Panels & Equations & $t_{1}$ & $t_{2}$ & $v_{m}$ & $t_{PH}$ & $v_{s}$ & $v_{M_{z}}$ & $u$ \\
\hline
\hline
Trivial Insulator & (c-e) & (\ref{eq:my2Dquad}) & $2$ & $1.5$ & $-5$ & $0.1$ & $1.3$ & $0$ & $0$ \\
\hline
$M_{z}$- and $\mathcal{T}$-Symmetric $C_{M_{z}}=2$ TCI & (f-h) & (\ref{eq:my2Dquad}) & $2$ & $1.5$ & $-1.5$ & $0.1$ & $1.3$ & $0$ & $0$ \\
\hline
$\mathcal{T}$-Symmetric, $M_{z}$-Broken Fragile TI & (i-k) & (\ref{eq:my2Dquad}) and~(\ref{eq:breakI}) & $2$ & $1.5$ & $-1.5$ & $0.1$ & $1.3$ & $0.4$ & $0$ \\
\hline
$M_{z}\times\mathcal{T}$-Symmetric Quadrupole & (l-n) & (\ref{eq:my2Dquad}) and~(\ref{eq:quad}) & $2$ & $1.5$ & $-1.5$ & $0.1$ & $1.3$ & $0$ & $0.5$ \\
\hline
\end{tabular}
\caption{Parameters used in Eqs.~(\ref{eq:my2Dquad}),~(\ref{eq:breakI}), and~(\ref{eq:quad}) to realize the 2D insulating phases shown in Fig.~\ref{fig:Phases} and in Fig.~\ref{fig:2Dmain} of the main text.}
\label{tb:2D}
\end{table}

For the 3D semimetallic phases shown in Fig.~\ref{fig:HingeSMmain} of the main text, we used the parameters listed in Table~\ref{tb:3D} in the 3D Hamiltonians:
\begin{eqnarray}
\mathcal{H}_{H1}({\bf k}) &=& \mathcal{H}({\bf k}) + U({\bf k}) + t_{H}\tau^{z}\cos(k_{z}) \label{eq:magHinge} \\
\mathcal{H}_{H2}({\bf k}) &=& \mathcal{H}({\bf k}) + U({\bf k})\sin(k_{z}) + t_{H}\tau^{z}\cos(k_{z}).
\label{eq:hinge}
\end{eqnarray}
Eq.~(\ref{eq:hinge}) is invariant under the symmetries of space group (SG) 123 $P4/mmm1'$, whose generating symmetries are represented by:
\begin{eqnarray}
M_{x}\mathcal{H}(k_{x},k_{y},k_{z})M_{x}^{-1} &=& \sigma^{x}\mathcal{H}(-k_{x},k_{y},k_{z})\sigma^{x},\ M_{z}\mathcal{H}(k_{x},k_{y},k_{z})M_{z}^{-1} = \sigma^{z}\mathcal{H}(k_{x},k_{y},-k_{z})\sigma^{z}, \nonumber \\
C_{4z}\mathcal{H}(k_{x},k_{y},k_{z})C_{4z}^{-1} &=& \tau^{z}\left(\frac{\mathds{1}_{\sigma} - i\sigma^{z}}{\sqrt{2}}\right)\mathcal{H}(k_{y},-k_{x},k_{z})\tau^{z}\left(\frac{\mathds{1}_{\sigma} + i\sigma^{z}}{\sqrt{2}}\right),\ \mathcal{T}\mathcal{H}(k_{x},k_{y},k_{z})\mathcal{T}^{-1} = \sigma^{y}\mathcal{H}^{*}(-k_{x},-k_{y},-k_{z})\sigma^{y}, \nonumber \\
\label{eq:3DnewSymRep}
\end{eqnarray}
whereas Eq.~(\ref{eq:magHinge}) is only invariant under magnetic SG $P4/m'mm$ (123.341 in the BNS notation~\cite{MagneticBook}), whose generating symmetries are $M_{x}$, $C_{4z}$, and $M_{z}\times\mathcal{T}$ as represented in Eq.~(\ref{eq:3DnewSymRep}).  Additionally, in Fig.~\ref{fig:HingeSMmain}(e,h) of the main text, we show a noncentrosymmetric Dirac semimetal in SG 99 $P4mm1'$ that results from breaking $M_{z}$ (and $\mathcal{I}$) symmetry while keeping the other symmetries of SG 123 in $P4/mmm1'$ through the addition of Eq.~(\ref{eq:breakI}) to Eq.~(\ref{eq:hinge}).

\begin{table}[h]
\begin{tabular}{|c|c|c|c|c|c|c|c|c|c|c|c|}
\hline
\multicolumn{12}{|c|}{Parameters for the 3D Models in Fig.~\ref{fig:HingeSMmain} of the Main Text} \\
\hline
Phase & Panels & Equations & $t_{1}$ & $t_{2}$ & $v_{m}$ & $t_{PH}$ & $v_{s}$ & $t_{H}$ & $u\ (\mathcal{H}_{H1})$ & $u\ (\mathcal{H}_{H2})$ & $v_{M_{z}}$ \\
\hline
\hline
$M_{z}\times\mathcal{T}$-Symmetric Magnetic HOFA Dirac & (c,f) & (\ref{eq:magHinge}) & $1$  & $2$ & $-1.5$ & $0.1$ & $1.3$ & $2.8$ & $0.5$ & $0$ & $0$ \\
\hline
$M_{z}$- and $\mathcal{T}$-Symmetric HOFA Dirac & (d,g) & (\ref{eq:hinge}) & $1$ & $2$ & $-1.5$ & $0.1$ & $1.3$ & $2.8$ & $0$ & $0.5$ & $0$ \\
\hline
$\mathcal{T}$-Symmetric HOFA Dirac without 2D Surface States & (e,h) & (\ref{eq:hinge}) and~(\ref{eq:breakI}) & $1.5$ & $0.4$ & $-1.5$ & $0.1$ & $1.3$ & $2.3$ & $0$ & $0.5$ & $0.3$ \\
\hline
\end{tabular}
\caption{Parameters used in Eqs.~(\ref{eq:breakI}), (\ref{eq:quad}),~(\ref{eq:magHinge}), and~(\ref{eq:hinge}) to realize the 3D semimetallic phases shown in Fig.~\ref{fig:HingeSMmain} of the main text and Fig.~\ref{fig:Phases}.}
\label{tb:3D}
\end{table}

All three of the Dirac semimetal phases listed in Table~\ref{tb:3D} exhibit higher-order Fermi arcs (HOFAs) along their 1D hinges.  The $M_{z}\times\mathcal{T}$-symmetric magnetic Dirac semimetal ($\mathcal{H}_{H1}({\bf k})$ in Eq.~(\ref{eq:magHinge})) and the $M_{z}$- and $\mathcal{T}$-symmetric Dirac semimetal ($\mathcal{H}_{H2}({\bf k})$ in Eq.~(\ref{eq:hinge})) both exhibit twofold band degeneracies at generic $k$ points due the presence of the combined antiunitary symmetry $\mathcal{I}\times\mathcal{T}$.  Of these, the $\mathcal{T}$-symmetric semimetal also exhibits a topologically nontrivial plane at $k_{z}=0$, for which the 2D mirror Chern number $C_{M_{z}}=2$.  Finally, by relaxing $M_{z}$ (and $\mathcal{I}$) symmetry while keeping $\mathcal{T}$ by adding Eq.~(\ref{eq:breakI}) to Eq.~(\ref{eq:hinge}), we realize a previously uncharacterized variant of topological Dirac semimetal~\cite{NagaosaDirac} without topological surface states, but with topological hinge states connected to the 0D corner states of a 2D fragile phase (Fig.~\ref{fig:HingeSMmain}(h) of the main text).  Specifically, for the noncentrosymmetric Dirac semimetal phase of Eqs.~(\ref{eq:breakI}) and~(\ref{eq:hinge}), the Hamiltonian of the $k_{z}=0$ plane exhibits fragile topology and Kramers pairs of corner modes, and the $k_{z}$ planes between $k_{z}=0$ and the Dirac points are equivalent to $M_{z}\times\mathcal{T}$-broken QIs in wallpaper group $p4m$ (layer group $p4mm$).  This ``fragile'' topological Dirac semimetal therefore represents a gapless tuning cycle between a 2D fragile phase and a trivial insulator, similar to the gapped pumping cycles between 2D fragile and trivial phases that characterize higher-order TIs (HOTIs)~\cite{WiederAxion}.  Like with the (magnetic) HOTIs analyzed in Ref.~\onlinecite{WiederAxion}, when trivial bands are added to the fragile topological Dirac semimetal phase of Eqs.~(\ref{eq:breakI}) and~(\ref{eq:hinge}), the Hamiltonian of the $k_{z}=0$ plane no longer describes a fragile phase, but instead characterizes an obstructed atomic limit with the same corner modes as the fragile phase from which it originated (Appendix~\ref{sec:fragile} and Refs.~\onlinecite{WiederAxion,BernevigMoTe2}).

\section{Topological Equivalence of the Spinful $s$-$d$ Hybridized Model and Spinless Flux-Threaded Model of Quadrupole Insulators}
\label{sec:equivalence}

Here, we show that the spinful $s-d_{x^{2}-y^{2}}$-hybridized model of a QI introduced in this paper (Eqs.~(\ref{eq:my2DquadMain}) and~(\ref{eq:quadMain}) of the main text, reproduced in Eqs.~(\ref{eq:my2Dquad}) and~(\ref{eq:quad}), respectively) is topologically equivalent to the spinless QI model with threaded flux introduced in Ref.~\onlinecite{multipole}.  We begin with the $\mathcal{T}$-broken Hamiltonian formed by adding Eqs.~(\ref{eq:my2Dquad}) and (\ref{eq:quad}):
\begin{equation}
\mathcal{H}_{M}({\bf k})=\mathcal{H}({\bf k}) + U({\bf k}).
\label{eq:tbreakSD}
\end{equation}
Eq.~(\ref{eq:tbreakSD}) is invariant under the action of the type-III magnetic layer group $p4/m'mm$, which is generated by:
\begin{equation}
\{C_{4z}|00\},\ \{M_{x}|00\},\ \{M_{z}\times\mathcal{T}|00\},
\end{equation}
as well as the 2D lattice translations $T_{x,y}$.  Throughout this work, we have employed the magnetic group labeling convention of Refs.~\onlinecite{BigBook,MagneticBook}, for which the combined operation of mirror and time-reversal (for this layer group, the operation $M_{z}\times\mathcal{T}$) is denoted as $m'$.  The dispersion relation of Eq.~(\ref{eq:tbreakSD}) can be calculated explicitly and compared to that of the quadrupole model in Ref.~\onlinecite{multipole} (Eq.~(6) in Ref.~\onlinecite{multipole}).  Because the QI model introduced in Ref.~\onlinecite{multipole} is particle-hole symmetric, we begin by tuning the quadrupole model introduced in this work (Eqs.~(\ref{eq:my2Dquad}) and~(\ref{eq:quad})) to the particle-hole symmetric limit in which $t_{PH}=v_{s}=0$, while keeping all of the other symmetries of $p4/m'mm$.  In this limit, the spectrum of Eqs.~(\ref{eq:my2Dquad}) and~(\ref{eq:quad}) is given by:
\begin{eqnarray}
E^{2}({\bf k}) &=& t_{1}^{2}\left[\cos^{2}(k_{x}) + \cos^{2}(k_{y}) + 2\cos(k_{x})\cos(k_{y})\right] + v_{m}^{2} + t_{2}^{2}\left[\cos^{2}(k_{x}) + \cos^{2}(k_{y}) - 2\cos(k_{x})\cos(k_{y})\right] \nonumber \\
&+& 2t_{1}v_{m}\left[\cos(k_{x}) + \cos(k_{y})\right] + u^{2}\left[\sin^{2}(k_{x}) + \sin^{2}(k_{y})\right],
\label{eq:temptempPHDispersion}
\end{eqnarray}
where bands are doubly degenerate due to the presence of the combined antiunitary symmetry $\mathcal{I}\times\mathcal{T}=M_{x}M_{y}(M_{z}\times\mathcal{T})$.  In the limit that:
\begin{equation}
t_{1}=t_{2}=\frac{1}{\sqrt{2}}u=t,
\label{eq:sub1}
\end{equation}
Eq.~(\ref{eq:temptempPHDispersion}) reduces to:
\begin{equation}
E^{2}({\bf k}) = 4t^{2} + v_{m}^{2} + 2tv_{m}\left[\cos(k_{x}) + \cos(k_{y})\right] .
\label{eq:temptempPHDispersion2}
\end{equation}
We can perform the substitution:
\begin{equation}
t=\frac{\lambda}{\sqrt{2}},\ v_{m}=\gamma\sqrt{2},
\label{eq:sub2}
\end{equation}
under which the dispersion relation of this spinful $s-d$-hybridized model (Eq.~(\ref{eq:temptempPHDispersion2})) becomes exactly equal to that of the spinless, flux-threaded model of a QI in Ref.~\onlinecite{multipole} (Eq.~(6) in Ref.~\onlinecite{multipole}).  As with the earlier QI model in Eq.~(6) in Ref.~\onlinecite{multipole}, our model displays gap closures at $\Gamma$ for $\gamma/\lambda = -1$ and at $M$ for $\gamma/\lambda = 1$.  Calculating the nested Wilson loop of the lower two bands of Eq.~(\ref{eq:tbreakSD}), we confirm that they exhibit a nested Berry phase of $0$ for $|\gamma/\lambda|>1$ and a nested Berry phase of $\pi$ for $|\gamma/\lambda|<1$.  This nested Berry phase is also symmetry-indicated by the eigenvalues of $C_{4z}$ of the occupied corepresentations at $\Gamma$ and $M$~\cite{WladTheory}.  In both models, the representations of the mirror symmetries $M_{x}$ and $M_{y}$ anticommute at all four TRIM points, such that states at the TRIM points are twofold degenerate despite the absence of time-reversal symmetry.  In the trivial insulating phases of both models, the occupied bands at both $C_{4z}$-invariant TRIM points have the same $C_{4z}$ eigenvalues: $\lambda_{C_{4z}}(\Gamma)=\lambda_{C_{4z}}(M) = (1\pm i)/\sqrt{2}$ or $\lambda_{C_{4z}}(\Gamma)=\lambda_{C_{4z}}(M) = -(1\pm i)/\sqrt{2}$.  In the QI phases of both models, either $\Gamma$ or $M$ has an occupied pair of states with $C_{4z}$ eigenvalues $\lambda_{C_{4z}} = (1\pm i)/\sqrt{2}$, whereas the other $C_{4z}$-invariant TRIM point has an occupied pair with $\lambda_{C_{4z}} = -(1\pm i)/\sqrt{2}$.  This can also be understood from the elementary band representations (EBRs) induced from the site-symmetry representations of spinful $s$ and $d$ orbitals~\cite{ZakBandrep1,ZakBandrep2,Bandrep3,QuantumChemistry,Bandrep1,Bandrep2,JenFragile1,AdrianFragile,JenFragile2,ZhidaBLG,BarryFragile,WladTheory,HermeleSymmetry,AshvinIndicators,ChenTCI} (see Appendix~\ref{sec:bandrep}). 

Furthermore, we can show that within each of the two phases of Eq.~(\ref{eq:tbreakSD}), there exist points in parameter space at which both models are unitarily related through a $k$-independent transformation.  As the QI phase is distinguished from a trivial insulator with Wannier orbitals located on the unit-cell atoms (\emph{i.e.} an unobstructed atomic limit~\cite{QuantumChemistry}) by a $\mathbb{Z}_{2}$ topological invariant~\cite{multipole,WladTheory}, it follows that all points in parameter space with the same symmetry are in topologically equivalent phases as long as one can tune between them without closing a gap~\cite{TKNN,AndreiTI,CharlieTI,SSH,RiceMele,ThoulessWannier,QHZ,AndreiXiZ2,AlexeyVDBWannier,AlexeyVDBTI,QuantumChemistry,ZakPhase,VDBpolarization}.  Therefore, to demonstrate the overall topological equivalence between the two models, we only need to demonstrate the topological equivalence between them at one point in parameter space within each of the two gapped phases of the different models.

We begin in the limit that both models have the same dispersion relations.  The model of a QI introduced in Eq.~(6) of Ref.~\onlinecite{multipole} can be written as:
\begin{eqnarray}
\mathcal{H}'({\bf k})&=&(\gamma + \lambda\cos(k_{x}))\Gamma_{1} + (\gamma + \lambda\cos(k_{y}))\Gamma_{2} + \lambda\sin(k_{x})\Gamma_{3} + \lambda\sin(k_{y})\Gamma_{4},
\label{eq:oldQI}
\end{eqnarray}
whereas the $\mathcal{T}$-broken $s-d$-hybridized model in Eq.~(\ref{eq:tbreakSD}) with the substitutions in Eqs.~(\ref{eq:sub1}) and~(\ref{eq:sub2}) can be written as 
\begin{equation}
\mathcal{H}_{M}({\bf k})=\left(\gamma\sqrt{2} + \frac{\lambda}{\sqrt{2}}[\cos(k_{x})+\cos(k_{y})]\right)\tilde{\Gamma}_{1} + \frac{\lambda}{\sqrt{2}}\bigg(\cos(k_{x}) -\cos(k_{y})\bigg)\tilde{\Gamma}_{2} + \lambda\sin(k_{x})\tilde{\Gamma}_{3} + \lambda\sin(k_{y})\tilde{\Gamma}_{4},
\label{eq:newQI}
\end{equation}
where $\{\Gamma_{i}\}$ and $\{\tilde{\Gamma}_{i}\}$ are sets of four-component Dirac matrices that each separately form a Clifford algebra.  Without loss of generality, we choose the matrix representations:
\begin{equation}
\Gamma_{i}=\tilde{\Gamma}_{i},\ \text{for }i=1\text{ to }5,
\end{equation}
and develop for each phase an exact transformation matrix at one point in parameter space.  We note that, because the coefficients of $\Gamma_{3,4}$ and $\tilde{\Gamma}_{3,4}$ in Eqs.~(\ref{eq:oldQI}) and~(\ref{eq:newQI}), respectively, are proportional to the same functions of $k$, we expect that the $k$-dependence of the unitary transformation that maps $\mathcal{H}'({\bf k})$ to $\mathcal{H}_{M}({\bf k})$ will lie in the subspace of $\Gamma_{1,2}$  for generic values of $k$.  

We first consider relating the two models' trivial (normal) insulating (NI) phases, defined as exhibiting a nested Berry phase of $0$.  The quadrupole moments of the insulating phases of both Hamiltonians are entirely determined by a single parameter: the ratio $|\gamma/\lambda|$.  Both models are gapless only when $|\gamma|=|\lambda|$, and both are topologically trivial for $|\gamma|>|\lambda|$.  We first consider the limit $\lambda\rightarrow 0$, which defines:
\begin{eqnarray}
\mathcal{H}'_{NI}({\bf k})&=&(\gamma)\Gamma_{1} + (\gamma)\Gamma_{2},
\end{eqnarray}
and
\begin{equation}
\mathcal{H}_{M,NI}({\bf k})=(\gamma\sqrt{2})\Gamma_{1}.
\end{equation}
Expressing the Hamiltonians as vectors of coefficients of the Dirac matrices:
\begin{equation}
\mathcal{H}_{M}({\bf k})={\bf v}({\bf k})\cdot{\bf \Gamma},\ \mathcal{H}'({\bf k})={\bf v}'({\bf k})\cdot{\bf \Gamma},
\label{eq:vdot}
\end{equation}
it is trivial to show that for $v' = \bar{V}v$,
\begin{equation}
\bar{V} = \left(\begin{array}{cc}
R\left(\frac{\pi}{4}\right) & \mathbf{0} \\
\mathbf{0} & \mathds{1}\end{array}\right), 
\end{equation}
where $R\left(\frac{\pi}{4}\right)$ is an orthogonal rotation matrix in the $2\times 2$ $\Gamma_{1,2}$ subspace:  
\begin{equation}
R\left(\frac{\pi}{4}\right) = \left(\begin{array}{cc}
1/\sqrt{2} & -1/\sqrt{2} \\
1/\sqrt{2} & 1/\sqrt{2} \end{array}\right).
\end{equation}
As $\bar{V}$ is $k$-independent, both models' trivial phases are topologically equivalent.  

We now relate the QI phases of both models.  A characteristic point within this phase can be obtained by choosing $\gamma/\lambda=0$, which defines:
\begin{equation}
\mathcal{H}'_{QI}({\bf k})=\lambda\cos(k_{x})\Gamma_{1} + \lambda\cos(k_{y})\Gamma_{2} + \lambda\sin(k_{x})\Gamma_{3} + \lambda\sin(k_{y})\Gamma_{4}
\end{equation}
and
\begin{equation}
\mathcal{H}_{M,QI}({\bf k})=\frac{\lambda}{\sqrt{2}}\left(\cos(k_{x})+\cos(k_{y})\right)\Gamma_{1} + \frac{\lambda}{\sqrt{2}}\left(\cos(k_{x}) -\cos(k_{y})\right)\Gamma_{2} + \lambda\sin(k_{x})\Gamma_{3} + \lambda\sin(k_{y})\Gamma_{4}.
\end{equation}
Again expressing the transformation matrix $\bar{V}$ in terms of the vectors of the coefficients of the Dirac matrices (Eq.~(\ref{eq:vdot})) for each Hamiltonian, the transformation $v'=\bar{V}v$ is satisfied by:
\begin{equation}
\bar{V} = \left(\begin{array}{cc}
S\left(\frac{\pi}{4}\right) & \mathbf{0} \\
\mathbf{0} & \mathds{1}\end{array}\right), 
\end{equation}
where $S\left(\frac{\pi}{4}\right)$ is an orthogonal rotoinversion matrix in the $2\times 2$ $\Gamma_{1,2}$ subspace:  
\begin{equation}
S\left(\frac{\pi}{4}\right) = \left(\begin{array}{cc}
1/\sqrt{2} & 1/\sqrt{2} \\
1/\sqrt{2} & -1/\sqrt{2} \end{array}\right).
\end{equation}
As $\bar{V}$ is here also $k$-independent, both models' QI phases are topologically equivalent. 

\section{Quadrupole Insulators in Spinful Magnetic Wallpaper Group $p4m$ as Obstructed Atomic Limits}
\label{sec:bandrep}

We analyze the EBRs~\cite{ZakBandrep1,ZakBandrep2,QuantumChemistry,Bandrep1,Bandrep2,Bandrep3,JenFragile1} of the four-band, $\mathcal{T}$-broken model of a $s-d_{x^{2}-y^{2}}$-hybridized quadrupole insulator (QI) introduced in this paper (Eqs.~(\ref{eq:my2Dquad}) and~(\ref{eq:quad})).  In this section, we show that the QI phase of this model, and consequently the QI phase of the topologically equivalent model introduced in Ref.~\onlinecite{multipole} (Appendix~\ref{sec:equivalence}), is an obstructed atomic limit~\cite{QuantumChemistry} of a 2D magnetic wallpaper (or layer) group.  Our model consists of two spin-$1/2$ $s$ and two spin-$1/2$ $d_{x^{2}-y^{2}}$ orbitals at the $1a$ Wyckoff position of type-I magnetic wallpaper group $p4m$, which is generated by $C_{4z}$ and $M_{x}$:
\begin{equation}
\{C_{4z}|00\},\ \{M_{x}|00\},
\end{equation}
as well as the 2D lattice translations $T_{x,y}$.  Our analysis in this section will also apply to the type-III magnetic layer group~\cite{MagneticBook} $p4/m'mm$, which has an additional $\mathcal{I}\times\mathcal{T}$ symmetry:
\begin{equation}
\{\mathcal{I}\mathcal{T}|00\},
\end{equation}
such that:
\begin{equation}
p4/m'mm = (E)p4m \cup (\mathcal{I}\times\mathcal{T})p4m,
\end{equation}
where $E$ is the identity operation.  By examining all momentum space (co)representations of these two groups~\cite{BigBook,QuantumChemistry,Bandrep1,Bandrep2,Bandrep3}, we find that the additional generator that separates them, $\mathcal{I}\times\mathcal{T}$, does not increase the connectivity of any of the double valued EBRs of either 2D magnetic group.  More specifically, because the little co-groups of the $\Gamma$ and $M$ points are isomorphic to $4mm$, and the little co-group of the $X$ point is isomorphic to $mm2$ (Fig.~\ref{fig:2Dmain}(b) of the main text), the double-valued (co)representations at all of the TRIM points are already twofold degenerate whether or not $\mathcal{I}\times\mathcal{T}$ is present; the additional antiunitary symmetry simply serves in $p4/m'mm$ to enforce twofold band degeneracies at all of the other points in the BZ (and to make the Wilson loop particle-hole symmetric at each $k$ point~\cite{ArisInversion,Alexandradinata14a,Cohomological,HourglassInsulator,DiracInsulator}).  Here, and throughout this work, the little co-group is defined as the little group modulo lattice translations~\cite{BigBook,QuantumChemistry}.  We also note that, in position space, the addition to $p4m$ of $\{\mathcal{I}\mathcal{T}|00\}$ does not change the locations, multiplicities, or group-subgroup relations of the site-symmetry groups of any of the Wyckoff positions.  Therefore, we can perform the more general analysis here of the Wannier description of band representations of wallpaper group $p4m$, and conclude that the same analysis applies to its supergroup $p4/m'mm$.  As the original model of a QI is topologically equivalent to a spinful model in $p4/m'mm$ (Appendix~\ref{sec:equivalence}), and as the models introduced in this work (Appendix~\ref{sec:parameters}) describe QIs in both $p4/m'mm$ (Eqs.~(\ref{eq:my2Dquad}) and~(\ref{eq:quad})) and $p4m$ ((Eqs.~(\ref{eq:my2Dquad}),~(\ref{eq:breakI}), and~(\ref{eq:quad})) our more general analysis in this section of (obstructed) atomic limits in $p4m$ applies to both the original spinless QI model introduced in Ref.~\onlinecite{multipole} as well as to all of the QI models introduced in this work.

To understand the band representations that correspond to the QI phase, we first examine the $C_{4z}$-invariant, multiplicity-one maximal Wyckoff positions of the type-I magnetic wallpaper group $p4m$ and the irreducible representations of their site-symmetry groups.  The site-symmetry groups of the $1a$ and $1b$ Wyckoff positions of $p4m$ are isomorphic to the point group $4mm$, and the site-symmetry group of the $2c$ position is isomorphic to the point group $mm2$ (Fig.~\ref{fig:2Dmain}(a) of the main text).  Consulting the point group tables on the Bilbao Crystallographic Server (BCS)~\cite{QuantumChemistry,BilbaoPoint}, we find that a spinless $s$ orbital transforms as the 1D single-valued irreducible representation $A_{1}$, whereas a spinless $d_{x^{2}-y^{2}}$ orbital, whose wavefunction is odd under diagonal mirror $M_{x\pm y}$ and $C_{4z}$, transforms as the irreducible representation $B_{1}$.  To add spin, we consult the direct product tables of Altmann and Herzig (Table 52.9 on page 491 of Ref.~\onlinecite{PointGroupTables}), 
\begin{equation}
A_{1}\otimes\bar{E}_{1/2} = \bar{E}_{1/2} \equiv \bar{E}_{1},\ B_{1}\otimes\bar{E}_{1/2}=\bar{E}_{3/2}\equiv\bar{E}_{2},
\label{eq:ereps}
\end{equation}
where for convenience, we provide double-valued irreducible representations in both the notation of Altmann and Herzig~\cite{PointGroupTables} ($\bar{E}_{1/2,3/2}$) and that of the BCS~\cite{Bandrep1,Bandrep2,Bandrep3,BilbaoPoint} ($\bar{E}_{1,2}$).  The notation of Altmann and Herzig in particular provides some physical intuition for these representations: $\bar{E}_{1/2}$ corresponds to the spin-$1/2$ representation of a spinful $s$ orbital, and $\bar{E}_{3/2}$ corresponds to the $m_{j}=\pm 3/2$ piece of the $J=5/2$ representation of spinful $d$ orbitals~\cite{PointGroupTables,NewFermions,RhSiArc,CoSiArc,KramersWeyl}.  Both double-valued irreducible representations $\bar{E}_{1,2}$ are two-dimensional, and their dimension does not double when $\mathcal{T}$ (or $\mathcal{I}\times\mathcal{T}$) is introduced.  These representations correspond to doubly-degenerate pairs of spinful atomic orbitals with different complex-conjugate pairs of $C_{4z}$ eigenvalues:
\begin{equation}
\chi_{\bar{E}_{1}}(C_{4z})=\frac{1+i}{\sqrt{2}} + \frac{1-i}{\sqrt{2}}=\sqrt{2},\ \chi_{\bar{E}_{2}}(C_{4z})=\frac{-1+i}{\sqrt{2}} + \frac{-1-i}{\sqrt{2}}=-\sqrt{2},
\label{eq:c4evalsE}
\end{equation}
where $\chi_{\rho}(h)$ is the character of the unitary symmetry $h$ in the irreducible representation $\rho$, and is equal to the sum of the eigenvalues of $h$ in $\rho$.  The twofold degeneracy of states characterized by $\bar{E}_{1,2}$ is here enforced by $\{M_{x},M_{y}\}=0$, instead of $\mathcal{T}$, which is absent for the calculations in this section (we will later also enforce $\mathcal{T}$ symmetry in Appendix~\ref{sec:fragile}).

Finally, before detailing the EBRs of $p4m$, we establish the set of possible little co-group representations at the high-symmetry points in the BZ (Fig.~\ref{fig:2Dmain}(b) of the main text).  Because wallpaper group $p4m$ is symmorphic, it is sufficient to examine the little co-group at each high-symmetry point.  The little co-groups of $\Gamma$ and $M$ (Fig.~\ref{fig:2Dmain}(b) of the main text) are isomorphic to $4mm$, which has two double-valued representations, denoted $\bar{\rho}_{6,7}$.  When we are referring to the little co-group representation at $\Gamma$, we will write:
\begin{equation}
\bar{\Gamma}_{7}\equiv\bar{E}_{1}\equiv\bar{\rho}_{7},\ \bar{\Gamma}_{6}\equiv\bar{E}_{2}\equiv\bar{\rho}_{6},
\label{eq:krepsGamma}
\end{equation}
while at $M$:
\begin{equation}
\bar{M}_{7}\equiv\bar{E}_{1}\equiv\bar{\rho}_{7},\ \bar{M}_{6}\equiv\bar{E}_{2}\equiv\bar{\rho}_{6},
\label{eq:krepsM}
\end{equation}
such, that, incorporating Eq.~(\ref{eq:c4evalsE}), the characters of the little co-group representations $\bar{\rho}_{6,7}$ are:
\begin{equation}
\chi_{\bar{\rho}_{6}}(C_{4z})=-\sqrt{2},\ \chi_{\bar{\rho}_{7}}(C_{4z})=\sqrt{2}.
\label{eq:c4evals}
\end{equation}
At the $X$ and $X'$ points, the little co-groups are isomorphic to the double-valued point group $mm2$, which has only a single double-valued irreducible representation $\bar{E}$, which is two-dimensional and corresponds to a pair of states with $M_{x,y}$ eigenvalues $\pm i$.  This twofold degeneracy does not require time-reversal symmetry; instead, it is enforced by the anticommutation relation $\{M_{x},M_{y}\}=0$.  

In the four-band quadrupole Hamiltonian (Eqs.~(\ref{eq:my2Dquad}) and~(\ref{eq:quad})), the $\mathcal{T}$-symmetric spin-orbit coupling term $v_{s}\tau^{y}\sigma^{z}\sin(k_{x})\sin(k_{y})$ and quadrupolar magnetic term $u[\tau^{y}\sigma^{y}\sin(k_{x}) + \tau^{y}\sigma^{x}\sin(k_{y})]$ vanish at the $\Gamma$ ($k_{x}=k_{y}=0$) and $M$ ($k_{x}=k_{y}=\pi$) points, and the Hamiltonian at these points takes the simplified form: 
\begin{equation}
\mathcal{H}(\Gamma) = (2t_{1}  + v_{m})\tau^{z} + 2t_{PH}\mathds{1}_{\tau\sigma},\ \mathcal{H}(M) = (-2t_{1}  + v_{m})\tau^{z} - 2t_{PH}\mathds{1}_{\tau\sigma},
\label{eq:spinlessTau}
\end{equation}
where eigenstates can be indexed by their \emph{spinless} fourfold rotation eigenvalues.  Most precisely, because none of the terms in Eq.~(\ref{eq:spinlessTau}) contain $\sigma$ matrices, Eq.~(\ref{eq:spinlessTau}) exhibits $SU(2)$ spin rotation symmetry, in addition to the spinful $C_{4z}$ symmetry inherited from Eqs.~(\ref{eq:my2Dquad}) and~(\ref{eq:quad}):
\begin{equation}
C_{4z} = \tau^{z}\left(\frac{\mathds{1}-i\sigma^{z}}{\sqrt{2}}\right).
\end{equation}
The combination of $SU(2)$ spin rotation symmetry and spinful $C_{4z}$ results in an additional spinless fourfold rotation symmetry, represented at the $\Gamma$ and $M$ points by:
\begin{equation}
\tilde{C}_{4z} = \tau^{z}.
\end{equation}
The states that transform as $s$ ($d$) orbitals correspond to its positive (negative) eigenvalues.  We find that the trivial and QI phases of the $s-d$-hybridized model introduced in this work can be distinguished by the irreducible representations of their occupied bands at $\Gamma$ and at $M$, which are determined by the numerical prefactors of $\tau^{z}$ in Eq.~(\ref{eq:spinlessTau}) using the values in Table~\ref{tb:2D}.  The trivial phase displays either $\bar{\rho}_{7}$ or $\bar{\rho}_{6}$ at both $\Gamma$ and at $M$; the QI phase displays $\bar{\rho}_{6}$ at $\Gamma$ and $\bar{\rho}_{7}$ at $M$, or $\bar{\rho}_{7}$ at $\Gamma$ and $\bar{\rho}_{6}$ at $M$, in agreement with the results of Ref.~\onlinecite{WladTheory}.

To relate these momentum-space representations to EBRs, we induce the site-symmetry irreducible representations of spinful $s$ and $d_{x^{2}-y^{2}}$ orbitals ($\bar{E}_{1,2}$, respectively) at the $1a$ and $1b$ Wyckoff positions into $G=p4m$ and then subduce onto $\Gamma$ and $M$.  As we have previously shown that the only symmetry indicators in $p4m$ are the pairs of $C_{4z}$ eigenvalues within each occupied irreducible representation at $\Gamma$ and at $M$, we can deduce all of the relevant information by calculating the character of $C_{4z}$ in the subduced representations at the $\Gamma$ and $M$ points following the procedure in Ref.~\onlinecite{Bandrep3}.  Since the Wyckoff positions are multiplicity-one, the formula simplifies to:
\begin{equation}
\chi^{{\bf k}}_{G}(h) \equiv e^{-i(R{\bf k})\cdot{\bf t}}\tilde{\chi}[\rho(\{E|-{\bf t}\}h)],
\label{eq:character}
\end{equation}
where:
\begin{equation}
\tilde{\chi}[\rho(g)] = \begin{cases} 
      \chi[\rho(g)] & \text{if } g\in G_{{\bf q}} \\
      0 & \text{if } g \not\in G_{{\bf q}},
   \end{cases}
\end{equation}
and:
\begin{equation}
{\bf t} \equiv h{\bf q} - {\bf q}.
\label{eq:deft}
\end{equation}
In Eqs.~(\ref{eq:character}) -- (\ref{eq:deft}), $G_{{\bf q}}$ is the site-symmetry group at ${\bf q}$, $h$ is an element of the little group at ${\bf k}$ and $R$ is the rotational part of $h$, such that $h=\{R|v\}$.  Applying Eq.~(\ref{eq:deft}) to $h= \{C_{4z} | 00\}$, we find that for the $1a$ position, ${\bf t}=h{\bf q}_{1a} - {\bf q}_{1a} = {\bf 0}$, while for the $1b$ position, ${\bf t} = h{\bf q}_{1b} - {\bf q}_{1b} = -{\bf T}_x$, where ${\bf T}_x$ denotes a lattice translation in the $x$ direction.  Thus, from Eq.~(\ref{eq:character}), an EBR induced from the $1a$ position has characters:
\begin{equation}
\chi^{\Gamma}(C_{4z}) = \chi^{M}(C_{4z}) = \tilde{\chi}^{1a}(C_{4z}),
\end{equation}
While for the $1b$ position, 
\begin{equation}
\chi^{\Gamma}(C_{4z}) = -\chi^{M}(C_{4z}) = \tilde{\chi}^{1b}(C_{4z}).
\end{equation}
The two possible EBRs induced from each site are described in Table~\ref{tb:bandrep}; the same information can be found using the~\textsc{BANDREP} tool on the BCS~\cite{QuantumChemistry,Bandrep1,Bandrep2,Bandrep3}.  Each of the EBRs shown in this table is twofold connected, with nondegenerate bands along all lines and the plane of the BZ away from the TRIM points and with twofold degeneracies at the TRIM points where $M_{x,y}$ intersect and anticommute.  In $p4/m'mm$, the layer supergroup of $p4m$, all of the EBRs maintain the same connectivity and symmetry eigenvalues.  In wallpaper group $p4m$, singly degenerate bands along the four high-symmetry mirror lines $M_{x,y}$ and $M_{x\pm y}$ can be labeled with either $\pm i$ mirror eigenvalues and bands at the TRIM points come in doublets with complex conjugate pairs of mirror eigenvalues.  Conversely, in layer group $p4/m'mm$, all of the bands along the mirror lines and at the TRIM points come in doublets with complex conjugate pairs of mirror eigenvalues related by $\mathcal{I}\times\mathcal{T}$ symmetry.  Furthermore, in $p4/m'mm$, bands at generic momenta in the 2D BZ also appear in doublets protected by $\mathcal{I}\times\mathcal{T}$ symmetry.  Therefore, when bands are inverted at $\Gamma$ in our four-band model in $p4/m'mm$ (Eqs.~(\ref{eq:my2Dquad}) and~(\ref{eq:quad})), the model is only gapless at a single point in parameter space at $\Gamma$.  Conversely, for a four-band model in $p4m$ with two occupied bands, depending on the band dispersion, the model either directly transitions from a trivial phase to a QI (as occurs in the fragile topological Dirac semimetal in Fig.~\ref{fig:HingeSMmain}(e,h) of the main text), or instead transitions from a trivial insulator (or QI) into a 2D semimetal with $8n$ twofold degenerate nodal points protected by $M_{x,y}$ and $M_{x\pm y}$ symmetries~\cite{WiederLayers}.  In this work, we specifically restrict consideration to models in $p4m$ that transition between trivial insulators and QIs when bands are inverted at $\Gamma$.

\begin{table}[h]
\begin{tabular}{c|c|c|c|c}
\hline
Spinful Atomic & Site-Symmetry & Wyckoff & Representation  & Representation \\
Orbitals & Representation of $4mm$  & Position  of $p4m$& Subduced at $\Gamma$ & Subduced at $M$ \\
\hline
$s$ & $\bar{E}_{1}$ & $1a$ & $\bar{\rho}_{7}$ & $\bar{\rho}_{7}$ \\ 
\hline
$d_{x^{2}-y^{2}}$ & $\bar{E}_{2}$ & $1a$ & $\bar{\rho}_{6}$ & $\bar{\rho}_{6}$ \\
\hline
$s$ & $\bar{E}_{1}$ & $1b$ & $\bar{\rho}_{7}$ & $\bar{\rho}_{6}$ \\ 
\hline
$d_{x^{2}-y^{2}}$ & $\bar{E}_{2}$ & $1b$ & $\bar{\rho}_{6}$ & $\bar{\rho}_{7}$ \\
\hline
\end{tabular}
\caption{The elementary band representations (EBRs) induced into type-I magnetic wallpaper group $p4m$ from the double-valued site symmetry representations that transform as spinful $s$ and $d_{x^{2}-y^{2}}$ orbitals at the $1a$ and $1b$ Wyckoff positions (Fig.~\ref{fig:2Dmain}(a) of the main text), and the resulting subduced little group representations at $\Gamma$ and $M$ (Fig.~\ref{fig:2Dmain}(b) of the main text), employing the notation in Eqs.~(\ref{eq:krepsGamma}) and~(\ref{eq:krepsM}).  Band inversion at $\Gamma$ or $M$ between irreducible representations $\bar{\rho}_{6}$ and $\bar{\rho}_{7}$ for bands induced from orbitals at the $1a$ position results in a set of occupied bands with the same symmetry eigenvalues as an EBR induced from the $1b$ position.  All representations were obtained using the point group~\cite{BilbaoPoint} and~\textsc{BANDREP}~\cite{QuantumChemistry,Bandrep1,Bandrep2,Bandrep3} tools on the Bilbao Crystallographic Server (BCS).}
\label{tb:bandrep}
\end{table}

\begin{figure}[t]
\centering
\includegraphics[width=0.85\textwidth]{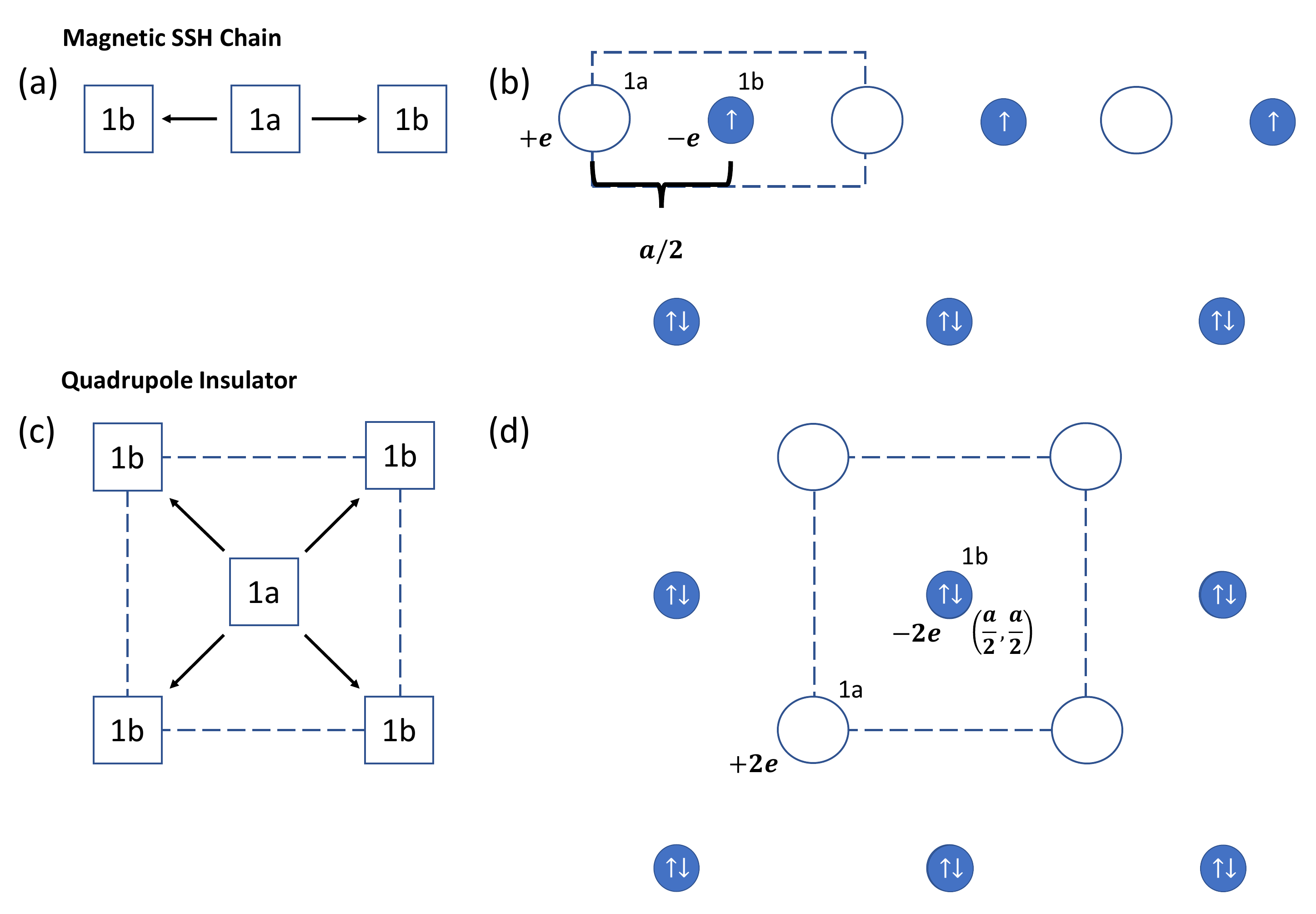}
\caption{Obstructed atomic limits and Wannier center homotopy in 1D and 2D.  (a) The Wannier center homotopy for the spinful, $\mathcal{T}$-broken (magnetic) SSH chain~\cite{SSH,RiceMele} in magnetic rod group $(p\bar{1})_{\text{RG}}$ (Refs.~\onlinecite{subperiodicTables,MagneticBook} and Appendix~\ref{sec:SGs}).  In the trivial phase, one Wannier orbital at the $1a$ position is occupied.  If a gap closes between two bands from spin-up $s$ and $p$ orbitals at $1a$, then there are sufficient Wannier orbitals to ``slide'' them out along the general position $(x)$ to the $1b$ position in an $\mathcal{I}$-symmetric manner; the location $x$ of the two Wannier centers is a gauge-dependent quantity~\cite{QuantumChemistry,MarzariReview}.  (b) When the two orbitals are slid to the $1b$ position and the bulk gap is reopened, a Wannier description of just the lower band is again allowed.  When the lower band is again occupied, the resulting state now exhibits a dipole moment (Eq.~(\ref{eq:dipoleMoment})) of $e/2$ (modulo $e$) per unit cell (dashed rectangle in (b)).  (c) The Wannier center homotopy for a QI in magnetic wallpaper group $p4m$.  In the trivial phase, two Wannier orbitals at the $1a$ position are occupied.  If a gap closes between four bands from pairs of spinful $s$ and $d$ orbitals at the $1a$ position, then there are sufficient Wannier orbitals to ``slide'' them out along along the $4d$ position (Fig.~\ref{fig:2Dmain}(a) of the main text) to the $1b$ position in a $p4m$-invariant manner.  When the gap is closed, the location of the four Wannier centers at $4d$ $(x,x)$ again becomes a gauge-dependent quantity~\cite{QuantumChemistry,MarzariReview}.  (d) When the four orbitals are slid to the $1b$ position and the gap between the lower two and upper two bands is reopened, a Wannier description of just the lower two bands is again allowed.  When these lower two bands are occupied, the resulting state exhibits trivial $x$- and $y$-directed dipole moments (modulo $e$) (Eq.~(\ref{eq:trivialDipoleForQI})), and a nontrivial $xy$ quadrupole moment (Eq.~(\ref{eq:quadrupoleMoment})) of $e/2$ (modulo $e$) per unit cell (dashed square in (d)), in agreement with the value obtained for the spinless, flux-threaded QI introduced in Ref.~\onlinecite{multipole}.}
\label{fig:Wannier}
\end{figure}

We now show that the QI phase that results from inverting bands originating from spinful $s$ and $d$ orbitals at the $1a$ position of $p4m$ is an obstructed atomic limit with Wannier centers at the $1b$ position.  From the contents of Table~\ref{tb:bandrep}, it is clear that when $1a$ bands from $s$ and $d$ orbitals are inverted at $\Gamma$ in a four-band model, the bottom two bands will have the same symmetry eigenvalues as bands from the $1b$ position ($\bar{\rho}_{6}$ at $\Gamma$ and $\bar{\rho}_{7}$ at $M$).  To formally show that the resulting insulator is an obstructed atomic limit, we look at the common Wyckoff position $4d$ (Fig.~\ref{fig:2Dmain}(a) of the main text), which lies at $(\pm x,\pm x)$, such that:
\begin{equation}
G_{1a}\cap G_{1b} = G_{4d},  
\end{equation}
where $G_{4d}$ is isomorphic to point group~\cite{BilbaoPoint,BCS1,BCS2} $m$, as its only symmetry is $M_{x\pm y}$. For both of the one-dimensional irreducible representations $^{1}\bar{E}$ and $^{2}\bar{E}$  of $G_{4d}$ we find that~\cite{BigBook,Bandrep1}:
\begin{equation}
(^{1,2}\bar{E})_{4d}\uparrow G_{1a} = (\bar{E}_{1})_{1a} \oplus (\bar{E}_{2})_{1a},\ (^{1,2}\bar{E})_{4d}\uparrow G_{1b} = (\bar{E}_{1})_{1b} \oplus (\bar{E}_{2})_{1b}, 
\label{eq:EBRsum}
\end{equation}
where $(\bar{\sigma})_{{\bf q}}$ is the EBR induced from the site-symmetry representation $\bar{\sigma}$ of the site-symmetry group of the Wyckoff position at ${\bf q}$.  From this we can conclude the equivalence~\cite{ZakBandrep1,ZakBandrep2,BacryMichelZak,QuantumChemistry,Bandrep1,Bandrep2,Bandrep3,JenFragile1,JenFragile2,ZhidaBLG,AdrianFragile,BarryFragile}:
\begin{align}
[\bar{E}_{1}\oplus\bar{E}_{2}]_{1a}\uparrow G &\equiv [\bar{E}_{1}\oplus\bar{E}_{2}]_{1b}\uparrow G,
\label{eq:sumequiv}
\end{align}
where $G=p4m$.  To define just $(\bar{E}_{2})_{1b}$ in terms of other EBRs, we introduce the symbol $\ominus$, 
\begin{align}
[\bar{E}_{1}\oplus\bar{E}_{2}]_{1a}\uparrow G \ominus [\bar{E}_{1}]_{1b}\uparrow G &\equiv [\bar{E}_{2}]_{1b}\uparrow G,
\label{eq:quadWannier}
\end{align}
where an equivalence formed with $\ominus$ is only defined if a corresponding equivalence with $\oplus$ of the form of Eq.~(\ref{eq:sumequiv}) is also defined.  From this it is clear that we can realize, via induction of four bands from the $1a$ position and band inversion, an occupied pair of bands that is equivalent to an EBR induced from the $1b$ position, or an obstructed atomic limit~\cite{QuantumChemistry,JenFragile1}.  This highlights the direct similarity between the QI and the nontrivial (obstructed atomic limit) phase of the SSH model.  In the SSH chain, which is a 1D chain with inversion centers at the $1a$ and $1b$ positions, the induction of two bands from the $1a$ position and band inversion analogously gives, at half filling, an occupied band that is equivalent to an EBR induced from the $1b$ position~\cite{QuantumChemistry} (Fig.~\ref{fig:Wannier}(a,b)).  Furthermore, more physically, the SSH transition can be expressed as $s-p$ hybridization~\cite{SSH,RiceMele,QuantumChemistry}, whereas in this section we have shown that the QI phase results from $s-d$ hybridization.

Eq.~(\ref{eq:quadWannier}) shows that the four bands corresponding to spinful $s$ and $d$ orbitals at the $1a$ position of $p4m$ can realize a trivial insulator at half-filling; that is, the valence (and conduction) bands possess localized, symmetric Wannier functions centered at the $1a$ position.  When the gap closes and reopens, the valence bands still possess a Wannier description, but one which instead corresponds to $s$ or $d$ orbitals centered at the $1b$ position.  In the intermediate gapless regime, it is not well-defined to compute the Wannier functions of only two bands, but one can compute the Wannier functions of \emph{both} the valence and conduction bands and observe that they correspond to orbitals centered at the $4d$ position $(x,x)$, where $x$ is a gauge-dependent quantity~\cite{MarzariReview,QuantumChemistry}.  We view this entire process as a Wannier center homotopy, where the gap closing and reopening ``slides'' the Wannier orbitals along the $4d$ position in a $p4m$-symmetric manner (Fig.~\ref{fig:Wannier}(c)), ultimately realizing an atomic insulator with Wannier centers lying on a different Wyckoff position ($1b$) than the ionic centers ($1a$) (Fig.~\ref{fig:Wannier}(d)).  This corresponds to an obstructed atomic limit in the language of Refs.~\onlinecite{QuantumChemistry,JenFragile1}.

We emphasize that the transition between a trivial insulator and the QI obstructed atomic limit in $p4m$ can only be realized as a function of a single-parameter in a spinful (double-valued) 2D magnetic group; unlike with the spinful SSH or Rice-Mele chains~\cite{SSH,RiceMele}, there is no analogous $\mathcal{T}$-symmetric limit of the quadrupole as an obstructed atomic limit (as opposed to a bipartite atomic-limit transition) with the symmetries of $p4m$ and single-parameter phase transitions.  Specifically, the singly-degenerate (magnetic spinful or spinless)~\cite{SSH,RiceMele} and $\mathcal{T}$-symmetric (spinful, nonmagnetic)~\cite{TRPolarization} formulations of the SSH chain are both Wannierizable~\cite{QuantumChemistry}, and differ from trivial insulators (unobstructed atomic limits) by phase transitions characterized by a single parameter~\cite{SSH,RiceMele,TRPolarization} (analogous to $\gamma/\lambda$ in the text following Eq.~(\ref{eq:sub2})).  Conversely, as we will show in Appendix~\ref{sec:fragile}, the Wannier description of the QI established in this section is no longer valid when $\mathcal{T}$ symmetry is restored; instead the analogous set of occupied bands exhibits fragile topology.  Since topological phases that are not Wannierizable do not exist in 1D~\cite{ArisBerry,ArisInversion}, this fragile phase represents one of the simplest topological obstructions to forming an obstructed atomic limit (though another commonly cited example occurs in magnetic layer groups with only $\mathcal{I}$ symmetry~\cite{ArisInversion,FragileKoreanInversion,WiederAxion,BernevigMoTe2,HarukiFragile,WiederDefect}).  In fact, the results of Appendix~\ref{sec:fragile} imply that with both $\mathcal{T}$ symmetry and the symmetries of $p4m$, an obstructed atomic limit from $1a$ to $1b$ can only be realized with a minimum of eight total bands (specifically because Wannier orbitals at the intermediate $4d$ ($x,x$) position must be doubly degenerate).  An example of a closely related obstructed atomic limit with four occupied (and four unoccupied) bands in $p41'$ ($p4m1'$ with broken $M_{x,y}$) was introduced in Ref.~\onlinecite{HigherOrderTIChen}, and is implied in that work to differ from a trivial insulator through a phase transition that is a function of more than one parameter (when only physical (space/layer group) symmetries are enforced).  Furthermore, we can show that eight-band phase transitions between $1a$ trivial insulators and $1b$ obstructed atomic limits in $p4m1'$ must also be functions of more than one parameter.  Specifically, because neither wallpaper group $p4m1'$ nor its layer supergroup $p4/mmm1'$ host symmetry-stabilized four-dimensional corepresentations~\cite{WiederLayers,DiracInsulator,BigBook} (though there is a single four-dimensional (including spin) corepresentation when SOC is neglected~\cite{BigBook,Bandrep1}), then all band-inversion transitions in $p4m1'$ and $p4/mmm1'$ involving eight bands must be functions of more than one parameter, and cannot occur simultaneously without fine tuning.  Therefore, instead of being driven by a single band inversion, like the QI~\cite{multipole,WladTheory}, an obstructed atomic limit with four valence and four conduction bands from $1a$ to $1b$ in $p4m1'$ generically represents an example of a 2D corner-mode phase driven by ``double band inversion'', analogous to the $\mathcal{I}$-symmetric corner-mode phases analyzed in Refs.~\onlinecite{YoungkukMonopole,BernevigMoTe2,WiederAxion,KoreanFragile,FragileKoreanInversion}.  

Finally, we analyze the multipole moments of the QI phase and compare them to the dipole moment of the spinful, $\mathcal{T}$-broken (magnetic) SSH chain~\cite{SSH,RiceMele}.  The spinful magnetic SSH chain can be considered a spin-polarized half-filled 1D crystal in magnetic rod group $(p\bar{1})_{\text{RG}}$ (Ref.~\onlinecite{subperiodicTables,MagneticBook} and Appendix~\ref{sec:SGs}), which is generated by:
\begin{equation}
\{\mathcal{I}|0\},\ \{E|1\},
\end{equation}
where $E$ is the identity operation.  We consider this chain to have lattice spacing $a$ and two total bands originating from spin-up $s$ and $p$ orbitals at the $1a$ position ($x=0$) with a gap at half filling; the lower (occupied) and upper (unoccupied) bands each form a Wannier orbital, initially located at the $1a$ position (Fig.~\ref{fig:Wannier}(a)).  When a gap is closed between the occupied and unoccupied bands at a 1D TRIM point, a Wannier description of the valence and conduction bands taken together is still admitted, and the two Wannier orbitals of these bands may then ``slide'' along the general position $2c$ ($x$) in an $\mathcal{I}$-symmetric manner.  More precisely, the location $\pm x$ of these two Wannier orbitals becomes a gauge-dependent quantity~\cite{QuantumChemistry,MarzariReview}.  If the two Wannier orbitals remain at the $1a$ position or slide all the way to the $1b$ position ($x=a/2$), a Wannier description of only the lower band in energy is again permitted after a symmetry-preserving gap is reopened, and specifically characterizes an obstructed atomic limit if the Wannier orbital of the occupied band occupies the $1b$ position~\cite{QuantumChemistry}.  In this obstructed atomic limit, the inversion eigenvalues of the occupied band indicate whether it is induced from a spinful $s$ or $p$ orbital at the $1b$ position~\cite{QuantumChemistry}.  To calculate the dipole moment of each unit cell of the spinful magnetic SSH obstructed atomic limit, we use the charge counting depicted in Fig.~\ref{fig:Wannier}(b) in which an uncovered atom (charge $+e$) lies at $x=0$ ($1a$) and an electron occupying a Wannier orbital (charge $-e$) lies at $x=a/2$ ($1b$):
\begin{equation}
P^{SSH} = \sum_{i}q_{i}x_{i} = -(e) (a/2) = - ae/2. 
\label{eq:dipoleMoment}
\end{equation}  
We confirm that $P^{SSH}$ is nontrivial by calculating:
\begin{equation}
\left(\frac{1}{a}\right)P^{SSH}\text{ mod e} = e/2,
\label{eq:modulo2}
\end{equation}
which coincides with the established value for the polarization of the singly-degenerate (magnetic) SSH chain~\cite{SSH,RiceMele}. 

For the QI phase, we now exploit the Wannier center homotopy to perform the analogous analysis.  As established in this section and depicted in Fig.~\ref{fig:Wannier}(d), the QI is an obstructed atomic limit in magnetic wallpaper group $p4m$ formed from ions (charge $+2e$) lying at $(x,y)=(0,0)$ ($1a$) and two occupied Wannier orbitals (charge $-2e$) lying at $(a/2,a/2)$ ($1b$).  First, using Eq.~(\ref{eq:dipoleMoment}), we calculate the $x$- and $y$-directed dipole moments of the QI unit cell:
\begin{equation}
P^{QI}_{x} = \sum_{i}q_{i}x_{i} = - 2e(a/2)=-ae,\ P^{QI}_{y} = \sum_{i}q_{i}y_{i} = - 2e(a/2)=-ae.
\end{equation}
As in Eq.~(\ref{eq:modulo2}), to determine whether $P_{x,y}^{QI}$ are nontrivial, we calculate the dipole moments in the units of $a$ (\emph{i.e.} the dipole densities) modulo $e$:
\begin{equation}
\left(\frac{1}{a}\right)P^{QI}_{x}\text{ mod e}= \left(\frac{1}{a}\right)P^{QI}_{y}\text{ mod e}= 0,
\label{eq:trivialDipoleForQI}
\end{equation}
which reveals that both $P^{QI}_{x,y}$ exhibit the same values as the trivial phase of the spinful magnetic SSH chain (Fig.~\ref{fig:Wannier}(a,b)).  We then calculate the bulk quadrupole moment per unit cell using the standard formulation in Ref.~\onlinecite{jacksonEM}:
\begin{equation}
Q^{ab} = \frac{1}{2}\sum_{i} q_{i}(3r^{a}_{i}r^{b}_{i} - |{\bf r}_{i}|^{2}\delta^{ab}),
\end{equation}
finding that $Q^{xy}$, in particular, is nonzero (Fig.~\ref{fig:Wannier}(d)):
\begin{equation}
Q^{xy,QI} = \frac{3}{2}\sum_{i}q_{i}x_{i}y_{i} = -\frac{3}{2}\left(\frac{2ea^{2}}{4}\right) = -\left(\frac{3a^{2}}{2}\right)\frac{e}{2}.
\label{eq:quadrupoleMoment}
\end{equation}
As previously with the dipole moments in Eqs.~(\ref{eq:modulo2}) and~(\ref{eq:trivialDipoleForQI}), $Q^{xy}$ can also be expressed as a multipole (quadrupole) density with the units of charge~\cite{multipole}.  Analogously to the simplest dipole, which is a rod centered at the origin with length $a$ and alternating charges $\pm e$ on its ends, the simplest $xy$-quadrupole is a square centered at the origin with side-length $a$ and alternating charges $\pm e$ on its corners~\cite{jacksonEM}, for which $Q^{xy}=3a^{2}/2$ (Eq.~(\ref{eq:quadrupoleMoment})).  We therefore express $Q^{xy,QI}$ in the reduced units of $3a^{2}/2$ and confirm that it is nontrivial by calculating its value modulo~\cite{multipole} $e$:
\begin{equation}
\left(\frac{2}{3a^{2}}\right)Q^{xy,QI}\text{ mod e} = e/2,
\label{eq:quadrupoleMomen2}
\end{equation}
which agrees with the value obtained for the spinless, flux-threaded QI model in Ref.~\onlinecite{multipole}.

\section{Fragile Topology in Wallpaper Group $p4m1'$}
\label{sec:fragile}

In this section, we explore the consequences of restoring $\mathcal{T}$-symmetry to wallpaper group $p4m$ (Appendix~\ref{sec:bandrep}).  We therefore analyze the EBRs of the $\mathcal{T}$-symmetric wallpaper group $p4m1'$, and apply the group theory of band representations~\cite{ZakBandrep1,ZakBandrep2,QuantumChemistry,Bandrep1,Bandrep2,Bandrep3,JenFragile1,AdrianFragile,JenFragile2,ZhidaBLG,BarryFragile,WladTheory,HermeleSymmetry,AshvinIndicators,ChenTCI} to a $\mathcal{T}$-symmetric, $\mathcal{M}_{z}$-broken limit of the 2D model introduced in this work (Eqs.~(\ref{eq:my2Dquad}) and~(\ref{eq:breakI})).  We show that, despite respecting the same unitary symmetries and exhibiting the same eigenvalues as a QI (Appendix~\ref{sec:bandrep}), this model instead hosts two valence and two conduction bands that exhibit ``fragile'' topology~\cite{AshvinFragile,JenFragile1,JenFragile2,AdrianFragile,BarryFragile,ZhidaBLG,AshvinBLG1,AshvinBLG2,AshvinFragile2,YoungkukMonopole,BernevigMoTe2,HarukiFragile,KoreanFragile,FragileKoreanInversion,ZhidaFragileAffine,KooiPartialNestedBerry,WiederAxion,WiederDefect} as a consequence of the additional presence of $\mathcal{T}$ symmetry. 

We again consider spinful $s$ and $d$ orbitals at the $1a$ position of a wallpaper group with $C_{4z}$ and $M_{x,y}$.  However, we now also impose $\mathcal{T}$ symmetry, such that the symmetry group is instead the wallpaper group:
\begin{equation}
G=p4m1',
\label{eq:TFragileGroup}
\end{equation}
the type-II nonmagnetic~\cite{MagneticBook} ($\mathcal{T}$-symmetric) supergroup of the QI wallpaper group $p4m$ (Appendix~\ref{sec:bandrep}).  The site-symmetry groups of the $1a$ and $1b$ positions of $G$ are isomorphic to $4mm1'$.  The corepresentations of $4mm1'$ have the same symmetry eigenvalues and dimensions as the irreducible representations of its unitary subgroup $4mm$, namely $\bar{E}_{1,2}$, whose $C_{4z}$ eigenvalues are given in Eq.~(\ref{eq:c4evals}).  Therefore, the EBRs induced from these corepresentations display the same symmetry eigenvalues as those in Table~\ref{tb:bandrep}.  

However, we observe an important difference between the $\mathcal{T}$-broken case (Appendix~\ref{sec:bandrep}) and the $\mathcal{T}$-symmetric case.  If we start with a four-band, $\mathcal{T}$-symmetric model whose two valence (conduction) bands are equivalent to EBRs induced from $s$ ($d$) orbitals on the $1a$ position and then invert bands at $\Gamma$, the resulting valence (conduction) bands have the same little group corepresentations at $\Gamma$ and at $M$ as the EBRs induced from $d$ ($s$) orbitals at the $1b$ position.  Therefore, they display the same little group corepresentations as the valence bands of a $\mathcal{T}$-broken QI in $p4m$.  However, when we calculate the $x$-directed Wilson loop of the $\mathcal{T}$-symmetric valence bands (Fig.~\ref{fig:2Dmain}(k) of the main text), it now \emph{winds}.  This winding indicates that the valence bands are not Wannierizable~\cite{ThoulessWannier,AndreiXiZ2,AlexeyVDBTI,QuantumChemistry}, and therefore reveals that they do not characterize an obstructed atomic limit like the QI phase previously analyzed in Appendix~\ref{sec:bandrep}.  Instead, as we will show in this section, the presence of $\mathcal{T}$ symmetry enforces a Kramers doubling at the $4d$ Wyckoff position of $p4m1'$ that obstructs the formation of the Wannier center homotopy of the QI (Fig.~\ref{fig:Wannier}(c)).  This obstruction prevents the two occupied bands from being smoothly deformed into EBRs induced from the $1b$ position.  Specifically, unlike previously in Appendix~\ref{sec:bandrep}, the two occupied bands are no longer homotopically equivalent to a $1b$ atomic limit in the presence of $\mathcal{T}$ symmetry~\cite{Bandrep3,QuantumChemistry}.

We formally reveal the obstruction by again attempting to form a homotopy through the $4d$ position at $(x,x)$, which was previously accomplished successfully in Appendix~\ref{sec:bandrep} for the type-I magnetic wallpaper group $p4m$.  We again look at the $4d$ Wyckoff position, whose site-symmetry group satisfies:
\begin{equation}
G_{1a}\cap G_{1b} = G_{4d}.  
\end{equation}
In the presence of $\mathcal{T}$ symmetry, $G_{4d}$ is isomorphic to $m1'$.  Therefore, unlike in Appendix~\ref{sec:bandrep} where the $4d$ position admitted two one-dimensional irreducible representations $^{1,2}\bar{E}$ distinguished by their mirror eigenvalues $\pm i$, here $\mathcal{T}$ symmetry enforces that $G_{4d}$ has only a single, \emph{two-dimensional} corepresentation~\cite{BigBook,Bandrep1} $^{1}\bar{E}^{2}\bar{E}$.  When induced into its site-symmetry supergroups $G_{1a}$ and $G_{1b}$, $^{1}\bar{E}^{2}\bar{E}$ crucially results in a sum of \emph{twice} as many EBRs as previously (Eq.~(\ref{eq:EBRsum})): 
\begin{equation}
(^{1}\bar{E}^{2}\bar{E})_{4d}\uparrow G_{1a} = (2\bar{E}_{1})_{1a} \oplus (2\bar{E}_{2})_{1a},\ (^{1}\bar{E}^{2}\bar{E})_{4d}\uparrow G_{1b} = (2\bar{E}_{1})_{1b} \oplus (2\bar{E}_{2})_{1b}, 
\end{equation}
where $(\bar{\sigma})_{{\bf q}}$ is the EBR induced from the site-symmetry representation $\bar{\sigma}$ of the Wyckoff position at ${\bf q}$, and where:
\begin{equation}
(2\bar{E}_{i})_{{\bf q}} = (\bar{E}_{i})_{{\bf q}}\oplus(\bar{E}_{i})_{{\bf q}}.
\end{equation}
From this we can conclude the equivalence of band representations:
\begin{align}
[2\bar{E}_{1}\oplus2\bar{E}_{2}]_{1a}\uparrow G &\equiv [2\bar{E}_{1}\oplus2\bar{E}_{2}]_{1b}\uparrow G,
\label{eq:fragileWannier}
\end{align}
Eq.~(\ref{eq:fragileWannier}) shows that the set of eight bands consisting of EBRs induced from two Kramers pairs of spinful $s$ orbitals and two Kramers pairs of spinful $d$ orbitals at the $1a$ position is identical to the eight bands consisting of two Kramers pairs of spinful $s$ orbitals and two Kramers pairs of spinful $d$ orbitals on the $1b$ position.  However, no such statement can be made for four bands induced from just one Kramers pair of $s$ orbitals plus one Kramers pair of $d$ orbitals.  Therefore, in $p4m1'$, if four bands are induced from one Kramers pair of spinful $s$ and one Kramers pair of spinful $d$ orbitals at the $1a$ position, and then a gap is closed and reopened such that the $C_{4z}$ eigenvalues of the lower two bands, which we denote as $F$, are no longer the same at $\Gamma$ and $M$ (which is always generically admitted because there are no connectivity-$4$ representation-enforced semimetals~\cite{QuantumChemistry,Bandrep1,Bandrep2,WPVZ,WiederLayers} in $p4m1'$), neither the two valence nor the two conduction bands will be equivalent to an atomic insulator.  This inequivalence holds even though these bands carry the \emph{same} symmetry eigenvalues as bands induced from Kramers pairs of $s$ or $d$ orbitals at the $1b$ position (Table~\ref{tb:bandrep}).

Formally, this can be understood as a failure of the Wannier center homotopy depicted in Fig.~\ref{fig:Wannier}(c,d) in the presence of $\mathcal{T}$ symmetry.  When $\mathcal{T}$ symmetry is absent, four Wannier orbitals may slide from the $1a$ to the $1b$ position along the $4d$ position in a manner that respects the symmetries of type-I magnetic wallpaper group $p4m$.  However, when $\mathcal{T}$-symmetry is present, four orbitals are insufficient to satisfy the analogous process in $p4m1'$, as Kramers' theorem mandates that spinful Wannier orbitals be twofold degenerate at each of the four sites of the $4d$ Wyckoff position.  As shown in numerous works~\cite{AshvinFragile,JenFragile1,JenFragile2,AdrianFragile,BarryFragile,ZhidaBLG,AshvinBLG1,AshvinBLG2,AshvinFragile2,YoungkukMonopole,BernevigMoTe2,HarukiFragile,KoreanFragile,FragileKoreanInversion,ZhidaFragileAffine,KooiPartialNestedBerry,WiederAxion,WiederDefect}, if a set of bands fails to satisfy a Wannier center homotopy because of an insufficient number of occupied Wannier orbitals, then those bands exhibit ``fragile'' topology.  Therefore, the two bands denoted as $F$ in the previous paragraph represent a fragile topological phase in $p4m1'$ that exhibits the same symmetry eigenvalues as a QI obstructed atomic limit in $p4m$ (Eq.~(\ref{eq:quadWannier})), which itself exhibits the same symmetry eigenvalues as an EBR induced from a Kramers pair of spinful $d$ orbitals the $1b$ position of $p4m1'$ (Table~\ref{tb:bandrep}):
\begin{equation}
F\stackrel{I}{\equiv}[\bar{E}_{1}\oplus\bar{E}_{2}]_{1a}\uparrow G \ominus [\bar{E}_{1}]_{1b}\uparrow G\stackrel{I}{\equiv}[\bar{E}_{2}]_{1b},
\label{eq:fragileIrreps}
\end{equation}
where $G=p4m1'$ (unlike previously in Eq.~(\ref{eq:quadWannier})).  The symbol $\stackrel{I}{\equiv}$ in Eq.~(\ref{eq:fragileIrreps}) denotes an ``irreducible-representation equivalence'' -- a weaker form of equivalence than previously employed in Appendix~\ref{sec:bandrep} in which bands simply exhibit the same set of (co)representations~\cite{JenFragile1,JenFragile2,BarryFragile,AshvinFragile2,WiederAxion,WiederDefect}.  A more detailed discussion of these contrasting notions of equivalence will appear in Ref.~\onlinecite{JenFragile2}.

Crucially, unlike the fragile bands previously analyzed in Refs.~\onlinecite{AshvinFragile,JenFragile1,WiederAxion,ZhidaBLG,AshvinBLG1,AshvinBLG2,FragileKoreanInversion,ZhidaFragileAffine}, $F$ is \emph{already} irreducible-representation-equivalent to a linear combination of EBRs with only positive coefficients (as opposed to the $\ominus$ in Eq.~(\ref{eq:fragileIrreps})), which was previously determined in Ref.~\onlinecite{JenFragile1} to be a necessary (but not sufficient) condition for a set of bands to be Wannierizable.  We are only able to deduce that $F$ is not Wannierizable by attempting (and failing) to derive a four-band Wannier center homotopy between $1a$ and $1b$ in $p4m1'$ (Eqs.~(\ref{eq:TFragileGroup}) through~(\ref{eq:fragileWannier})).  $F$ therefore represents a previously uncharacterized variant of fragile phase that eludes the symmetry-eigenvalue-based diagnosis schemes previously employed in Refs.~\onlinecite{AshvinFragile,JenFragile1,WiederAxion,ZhidaBLG,AshvinBLG1,AshvinBLG2,FragileKoreanInversion,ZhidaFragileAffine}.

Before analyzing $F$ with tight-binding models, we will demonstrate that when an EBR from the $2c$ position of $p4m1'$, which lies at $(1/2,0),\ (0,1/2)$, is added to $F$, the resulting set of bands remains irreducible-representation-equivalent to a sum of EBRs with only positive coefficients, as well as (when the two fragile conduction bands $\tilde{F}$ are taken into account) contains the same $C_{4z}$ symmetry eigenvalues as bands that satisfy the eight-band Wannier center homotopy from $1a$ to $1b$ in $p4m1'$ derived in Eqs.~(\ref{eq:TFragileGroup}) through~(\ref{eq:fragileWannier}).  First, we note that Eq.~(\ref{eq:fragileWannier}) implies that:
\begin{equation}
[\bar{E}_{1}\oplus\bar{E}_{2}]_{1a}\uparrow G \stackrel{I}{\equiv} [\bar{E}_{1}\oplus\bar{E}_{2}]_{1b}\uparrow G.
\label{eq:1a1bIrrepEquiv}
\end{equation}
Therefore, if $F$ characterizes the (fragile) valence bands of a four-band model in $p4m1'$ originating from spinful $s$ and $d$ orbitals at the $1a$ position, then the conduction bands of this model $\tilde{F}$, which are also fragile (Eqs.~(\ref{eq:TFragileGroup}) through~(\ref{eq:fragileWannier})), satisfy:
\begin{equation}
\tilde{F} \stackrel{I}{\equiv} [\bar{E}_{1}\oplus\bar{E}_{2}]_{1a}\uparrow G \ominus F \stackrel{I}{\equiv}[\bar{E}_{1}]_{1b}.
\label{eq:conductionBandIrrepEquiv}
\end{equation}
Next, we form an equivalence between EBRs induced from the $1a$ position of $p4m1'$ and EBRs induced from the $2c$ position.  The $2c$ position is invariant under site-symmetry group $mm21'$ and has a unique two-dimensional corepresentation~\cite{BigBook,Bandrep1} $\bar{E}$ due to the anticommutation relation $\{M_{x},M_{y}\}=0.$  We first examine the $4e$ position at $(x,0),\ (0,x)$, which interpolates between $1a$ and $2c$ (Fig.~\ref{fig:2Dmain}(a) of the main text and Ref.~\onlinecite{BigBook}).  The site-symmetry group $G_{4e}$ of the $4e$ Wyckoff position is isomorphic to point group~\cite{BilbaoPoint} $m1'$, and therefore, like the $4d$ position, has a single corepresentation $^{1}\bar{E}^{2}\bar{E}$.  $G_{4e}$ satisfies:
\begin{equation}
G_{1a}\cap G_{2c} = G_{4e},  
\end{equation}
where:
\begin{equation}
|G_{1a}|/|G_{4e}| = 4,\ |G_{2c}|/|G_{4e}| = 2,
\end{equation}
where $|G|$ is the number of elements in $G$; thus, $|G|/|H|=N$ indicates that $|H|$ is an index-$N$ subgroup of $G$.  We also compute:
\begin{equation}
(^{1}\bar{E}^{2}\bar{E})_{4e} \uparrow G_{1a} = (2\bar{E}_{1})_{1a} \oplus (2\bar{E}_{2})_{1a},\ (^{1}\bar{E}^{2}\bar{E})_{4e} \uparrow G_{2c} = (2\bar{E})_{2c}, 
\label{eq:2cEBRsum}
\end{equation}
which implies that:
\begin{align}
[2\bar{E}_{1}\oplus2\bar{E}_{2}]_{1a}\uparrow G &\equiv [2\bar{E}]_{2c}\uparrow G.
\label{eq:2cWannier}
\end{align}
Eq.~(\ref{eq:2cWannier}) displays a similar issue as in Eq.~(\ref{eq:quadWannier}): we are unable to define an equivalence between bands induced from $2c$ and single copies of bands induced from the $1a$ position.  However, we can still use Eq.~(\ref{eq:2cWannier}) to define an irreducible-representation equivalence:
\begin{align}
[\bar{E}_{1}\oplus\bar{E}_{2}]_{1a}\uparrow G &\stackrel{I}{\equiv} [\bar{E}]_{2c}\uparrow G.
\label{eq:2cIrrepEquiv}
\end{align}
We then combine Eqs.~(\ref{eq:fragileIrreps}), (\ref{eq:1a1bIrrepEquiv}),~(\ref{eq:conductionBandIrrepEquiv}), and~(\ref{eq:2cIrrepEquiv}) to deduce that:
\begin{equation}
F \oplus \tilde{F} \oplus [\bar{E}]_{2c}\uparrow G \stackrel{I}{\equiv}  [2\bar{E}_{1}\oplus 2\bar{E}_{2}]_{1b}\uparrow G.
\label{eq:trivializeMyEBRSum}
\end{equation}
To conclude, Eq.~(\ref{eq:trivializeMyEBRSum}) implies that the fragile valence and conduction bands $F$ and $\tilde{F}$, which originated from spinful $s$ and $d$ orbitals at the $1a$ position of $p4m1'$, along with bands induced from spinful $s$ orbitals at the $2c$ position of $p4m1'$, exhibit the same symmetry eigenvalues as eight bands induced from the $1b$ position.  As shown in Eq.~(\ref{eq:fragileWannier}), this implies that if those eight bands could also be formed into maximally localized, symmetric Wannier functions at $1b$, then they could also be freely ``slid'' to $1a$ while respecting the symmetries of $p4m1'$.

\begin{figure}[t]
\centering
\includegraphics[width=0.85\textwidth]{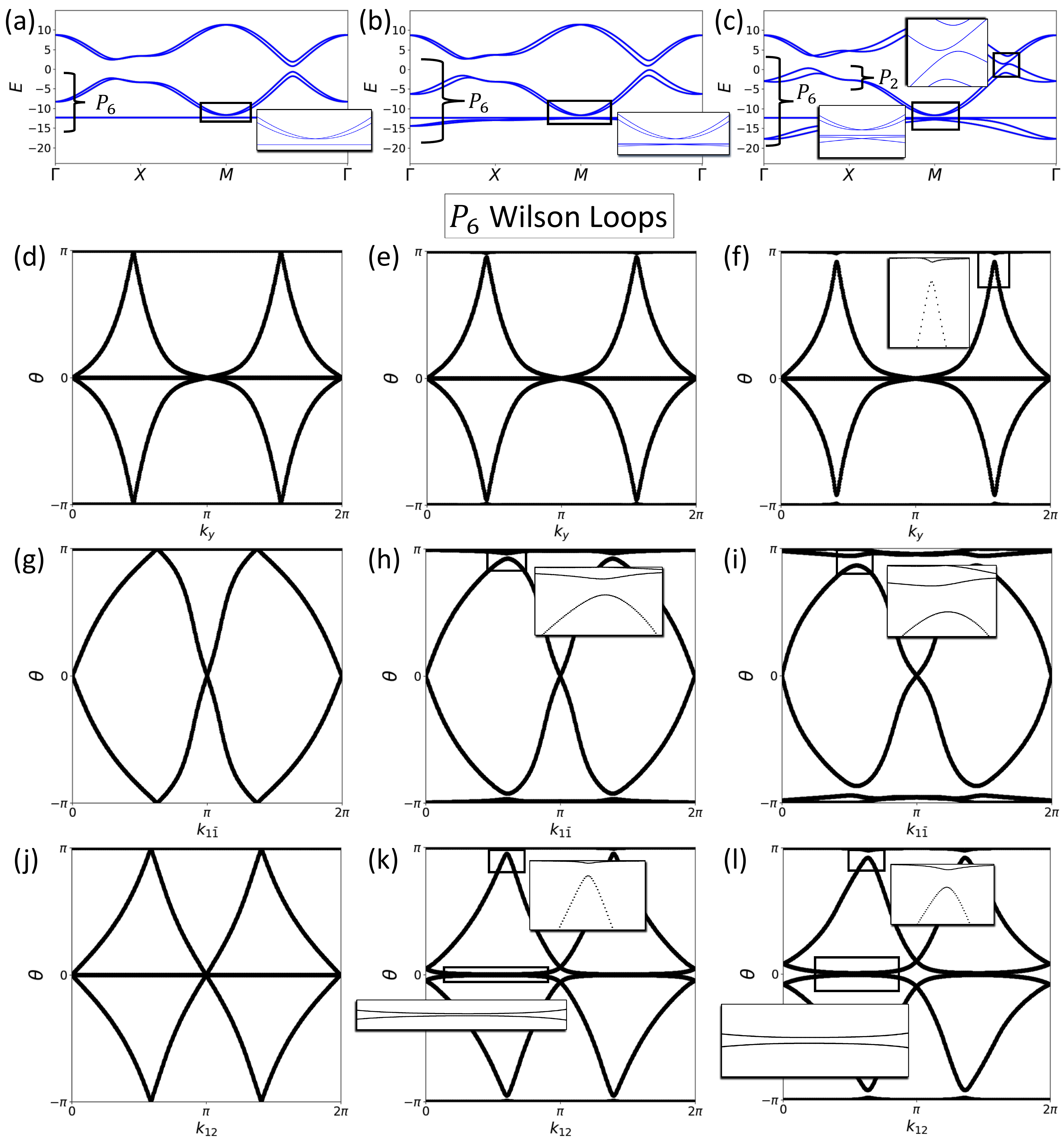}
\caption{Trivialized fragile-phase Wilson loops.  (a-c) The bulk bands and (d-f) $x$-directed Wilson loop, (g-i) $(x+y)$-directed Wilson loop, and (j-l) $(2x+y)$-directed Wilson loop over the lower six bands ($P_{6}$) of $\mathcal{H}_{F}({\bf k})$ in Eq.~(\ref{eq:myFragile}) with additional Kramers pairs of $s$ orbitals placed at the $2c$ position of wallpaper group $p4m1'$ (Fig.~\ref{fig:2Dmain}(a) of the main text) and coupled through Eq.~(\ref{eq:2cCoupling}).  Figures are plotted for $v_{C} = 0$, $2.5$, and $5$ in Eq.~(\ref{eq:2cCoupling}), respectively.  In the inset panels, we show narrowly avoided crossings in the energy and Wilson spectra.  These three Wilson loops span the set of symmetry-inequivalent Wilson loops (where only one Wilson loop without edge-projecting symmetries~\cite{DiracInsulator}, the $(2x+y)$-directed Wilson loop, is plotted for simplicity).  We observe that the addition of $2c$ orbitals trivializes all of the six-band Wilson loops of this model, even though the bands from the $2c$ orbitals remain separated from the original two occupied bands ($P_{2}$ in (c)) by an energy gap, which is explicitly shown along high-symmetry lines in (b) and (c), and was additionally numerically confirmed throughout the BZ interior.}
\label{fig:fragile}
\end{figure}

Though Eq.~(\ref{eq:trivializeMyEBRSum}) does not on its own additionally imply that $F \oplus [\bar{E}]_{2c}\uparrow G$ is Wannierizable, we will also numerically demonstrate in this section that a tight-binding model in $p4m1'$ with the occupied bands $F \oplus [\bar{E}]_{2c}\uparrow G$ exhibits trivial winding in all symmetry-inequivalent Wilson loops.  Here, we define symmetry-inequivalent Wilson loops as sets of Wilson loops that that are not related to each other by bulk crystal symmetries, and where in this section, for convenience, we only sample one Wilson loop (the $(2x+y)$-directed Wilson loop) that does not preserve \emph{any} edge-projecting crystal symmetry~\cite{DiracInsulator} (aside from perpendicular lattice translations).

To numerically demonstrate the removal of Wilson loop winding, which we here refer to as the ``trivialization'' of the Wilson loop, we begin by reproducing the tight-binding model that realizes the two sets of bands with fragile topology shown in Fig.~\ref{fig:2Dmain}(i-k) of the main text:
\begin{eqnarray}
\mathcal{H}_{F}({\bf k})&=&t_{1}\tau^{z}[\cos(k_{x}) + \cos(k_{y})] + t_{2}\tau^{x}[\cos(k_{x}) - \cos(k_{y})] + v_{m}\tau^{z} \nonumber \\
&+& t_{PH}\mathds{1}_{\tau\sigma}[\cos(k_{x}) + \cos(k_{y})]  + v_{s}\tau^{y}\sigma^{z}\sin(k_{x})\sin(k_{y}) \nonumber \\
&+& v_{M_{z}}\left[\tau^{z}\sigma^{y}\sin(k_{x}) - \tau^{z}\sigma^{x}\sin(k_{y})\right].
\label{eq:myFragile}
\end{eqnarray}
$\mathcal{H}_{F}({\bf k})$ is invariant under wallpaper group $p4m1'$.  To tune Eq.~(\ref{eq:myFragile}) into a fragile phase with the occupied bands $F$ (Eq.~(\ref{eq:fragileIrreps})), we choose the parameters $t_{1} =5,\ t_{2} = 1.5,\ v_{m} = -1.5,\ t_{PH} = 0.1, v_{s} = 1.3,$ and $v_{M_{z}} = 0.4$.  In its fragile phase, the spectrum of $\mathcal{H}_{F}({\bf k})$ exhibits two sets of bands separated by an energy gap at half filling.  When placed on a ribbon geometry, the fragile phase of Eq.~(\ref{eq:myFragile}) exhibits no edge modes (Fig.~\ref{fig:2Dmain}(j) of the main text), but when placed on a finite-sized square, it exhibits four Kramers pairs of corner modes (Appendix~\ref{sec:TCIBoundary}) that, depending on the energetics (here, the value of $t_{2}$), can ``float'' into the gap, as seen at $\bar{\Gamma}$ in Fig.~\ref{fig:HingeSMmain}(h) of the main text.  As shown in Appendix~\ref{sec:TCIBoundary}, even when the corner modes lie within the bulk manifolds, the fragile phase can still be distinguished from a trivial insulator by counting the number of states above and below the gap in the energy spectrum calculated with square (open) boundary conditions.

\begin{figure}[h]
\centering
\includegraphics[width=0.4\textwidth]{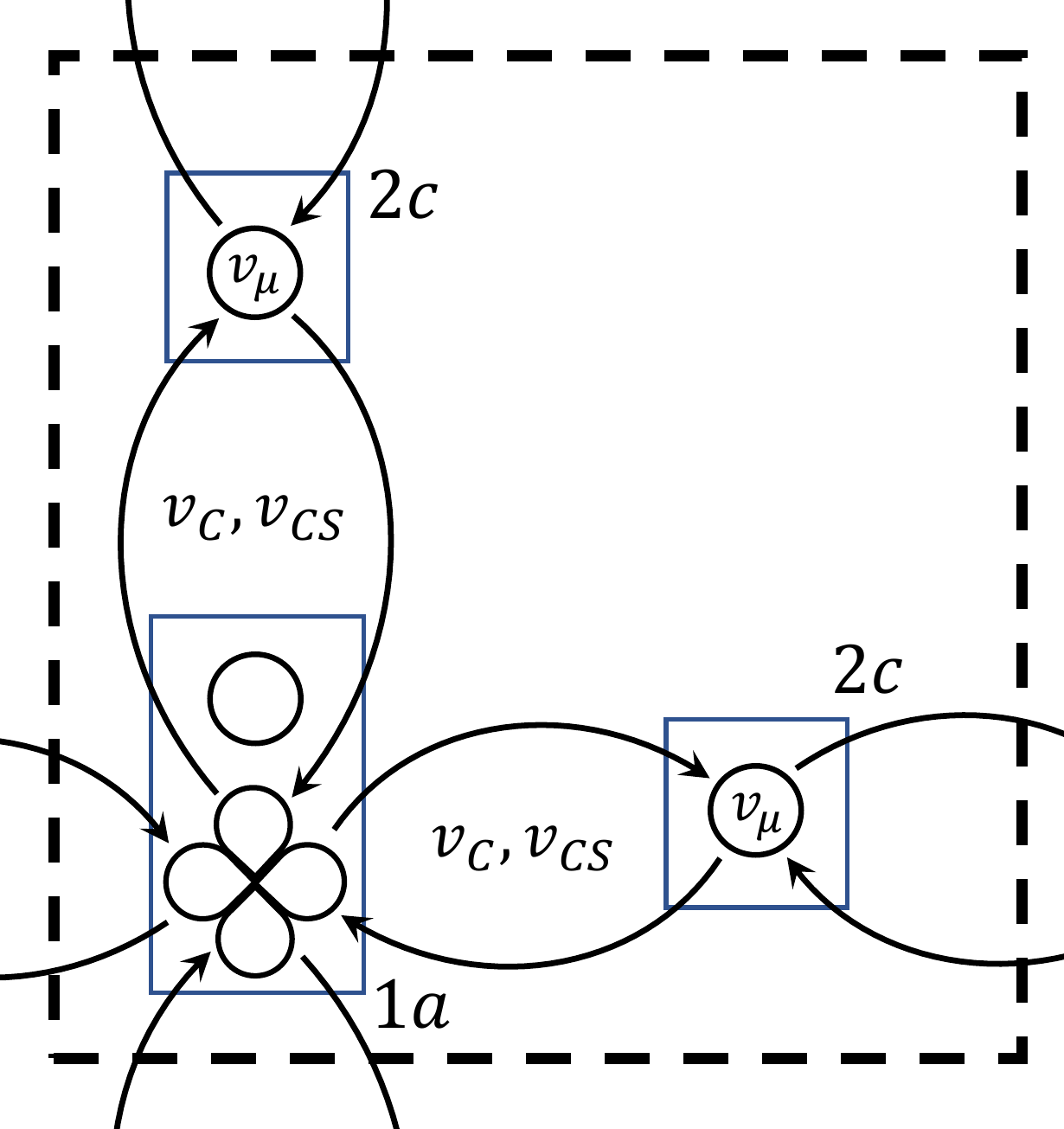}
\caption{Schematic of the hopping terms added in $V_{C}({\bf k})$ (Eq.~(\ref{eq:2cCoupling})) to trivialize the fragile valence bands $F$ of Eq.~(\ref{eq:myFragile}) (Fig.~\ref{fig:fragile}).  The symbols at each Wyckoff position of $\mathcal{T}$-symmetric wallpaper group $p4m1'$ represent Kramers pairs of spinful orbitals; there are two $s$ and two $d$ orbitals at $1a$, and two $s$ orbitals at each site of the $2c$ position.  The hopping terms $v_{C}$ and $v_{CS}$ in Eq.~(\ref{eq:2cCoupling}) couple the $d$ orbitals at $1a$ to the $s$ orbitals at $2c$; there are no terms that couple the $s$ orbitals at $1a$ to the $s$ orbitals at $2c$.}
\label{fig:fragileHop}
\end{figure}

The $x$-directed Wilson loop over the lower two bands (or alternatively over the upper two bands) of the fragile phase of Eq.~(\ref{eq:myFragile}) exhibits the same winding as would a TCI with mirror Chern number $C_{M_{z}}=2$, even though $M_{z}$ is broken when $v_{M_{z}}\neq 0$ (conversely, when $v_{M_{z}}$ is tuned to zero without closing a bulk gap, an edge gap closes and the lower two bands actually do characterize a $C_{M_{z}}=2$ TCI (Fig.~\ref{fig:2Dmain}(f-h) of the main text)).  Unlike in a 2D TCI phase, the $x$-directed Wilson loop over the lower two bands of the fragile phase of Eq.~(\ref{eq:myFragile}) (Fig.~\ref{fig:2Dmain}(k)) exhibits crossings that are not protected by the corepresentations of the $(10)$-edge line group~\cite{subperiodicTables,MagneticBook,DiracInsulator} $pm1'$.  Instead, the Wilson crossings in Fig.~\ref{fig:2Dmain}(k) are only protected by \emph{bulk} symmetries (specifically $C_{2z}\times\mathcal{T}$); more specific details regarding $C_{2z}\times\mathcal{T}$-protected Wilson crossings are discussed in Refs.~\onlinecite{BarryFragile,ZhidaBLG}.  In this work, we demonstrate, through the addition of spinful $s$ orbitals at the $2c$ position, that the winding Wilson spectrum of the valence bands of Eq.~(\ref{eq:myFragile}) becomes trivialized, a hallmark of ``fragile'' topology~\cite{AshvinFragile,JenFragile1,JenFragile2,AdrianFragile,BarryFragile,ZhidaBLG,AshvinBLG1,AshvinBLG2,AshvinFragile2,YoungkukMonopole,BernevigMoTe2,HarukiFragile,KoreanFragile,FragileKoreanInversion,ZhidaFragileAffine,KooiPartialNestedBerry,WiederAxion,WiederDefect}.  We choose to place the additional orbitals at the $2c$ position, rather than at the $1a$ position, which we have already demonstrated can relieve the obstruction to forming an obstructed atomic limit (Eq.~(\ref{eq:fragileWannier})), because the particular couplings that trivialize the set of Wilson spectra in this model are considerably easier to deduce using bands from $2c$, and because bands from $2c$ can be irreducible-representation-equivalent to bands from $1a$ (Eq.~(\ref{eq:2cIrrepEquiv})).

\begin{figure}[H]
\centering
\includegraphics[width=0.9\textwidth]{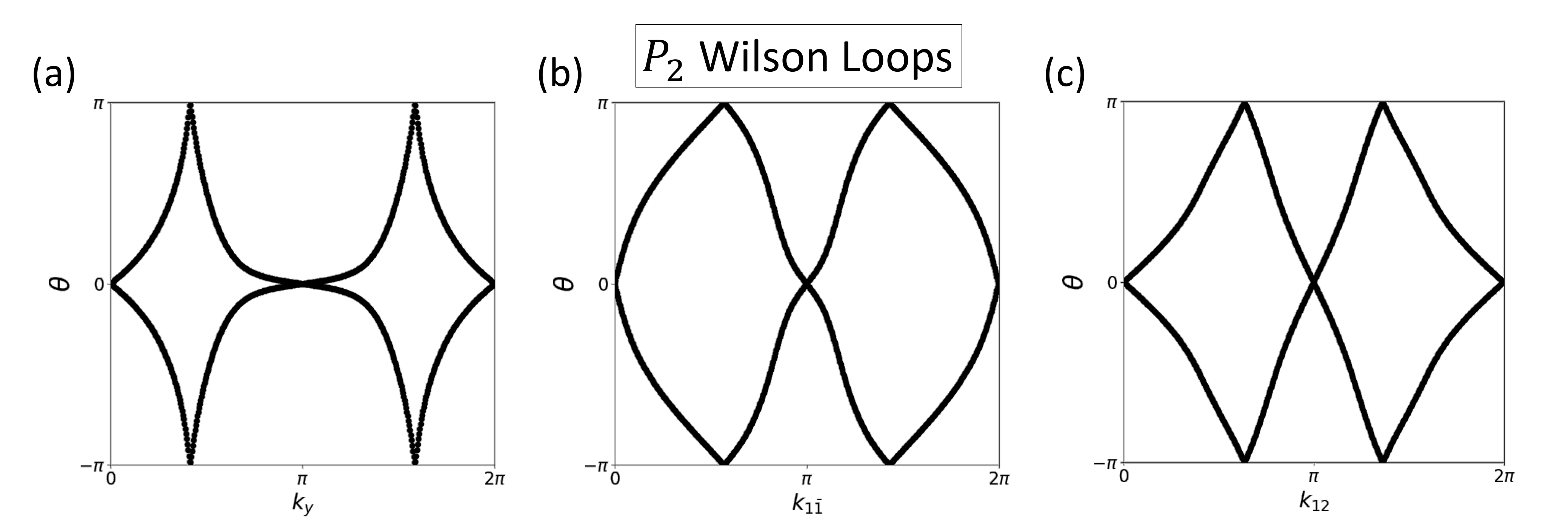}
\caption{Wilson loops of isolated fragile bands.  (a) The $x$-directed, (b) $(x+y)$-directed, and (c) $(2x+y)$-directed Wilson loops over the fragile valence bands of Eq.~(\ref{eq:myFragile}) ($P_{2}$ in Fig.~\ref{fig:fragile}(c)), which remain separated from additional bands from the $2c$ position by an energy gap (this is shown along high-symmetry lines in Fig.~\ref{fig:fragile}(c), and was additionally numerically confirmed throughout the BZ interior).  Figures are plotted using the same parameters as Fig.~\ref{fig:fragile}(c).  All three Wilson loops (a-c) still wind~\cite{BarryFragile}, even though the Wilson loops over the lower $6$ bands ($P_{6}$) have become trivialized by the additional trivial bands from the $2c$ position (Fig.~\ref{fig:fragile}(f,i,l)).}
\label{fig:fragile2}
\end{figure}

To demonstrate how the winding of the Wilson spectrum of the lower two bands of Eq.~(\ref{eq:myFragile}) is fragile, we introduce spinful $s$ orbitals at the $2c$ position (Fig.~\ref{fig:fragileHop}) through the addition of the terms:
\begin{equation}
V_{C}({\bf k})= v_{\mu}P_{\mu^{s}} + v_{C}\left[\mu^{x}_{1}\cos\left(\frac{k_{x}}{2}\right) + \mu^{x}_{2}\cos\left(\frac{k_{y}}{2}\right)\right] - v_{CS}\left[\mu^{x}_{1}\sigma^{y}\sin\left(\frac{k_{x}}{2}\right) - \mu^{x}_{2}\sigma^{x}\sin\left(\frac{k_{y}}{2}\right)\right],
\label{eq:2cCoupling}
\end{equation} 
where $\mu^{x}_{1}$ ($\mu^{x}_{2}$) represents sublattice hopping between the $d_{x^{2}-y^{2}}$ orbitals at the $1a$ position and the $s$ orbitals at the $x=1/2,\ y=0$ ($x=0,\ y=1/2$) site of the $2c$ position, and where $P_{\mu^{s}}$ is a projection matrix into the $2c$ $s$-orbital subspace of the eight-band Hamiltonian $\mathcal{H}_{F}({\bf k})+V_{C}({\bf k})$.  As shown in Fig.~\ref{fig:fragileHop}, $v_{C}$ ($v_{CS}$) represents (spin-) orbital coupling between Kramers pairs of spinful $s$ orbitals at the $2c$ position and spinful $d_{x^{2}-y^{2}}$ orbitals at the $1a$ position, and $v_{\mu}$ represents a chemical potential on the $2c$ $s$ orbitals in $\mathcal{H}_{F}({\bf k})+V_{C}({\bf k})$ (Eqs.~(\ref{eq:myFragile}) and~(\ref{eq:2cCoupling})); there is no hopping between the spinful $s$ orbitals at $1a$ and the $s$ orbitals at $2c$.  We then calculate the bulk band structure and $x$-, $(x+y)$-, and $(2x+y)$-directed Wilson loops over the lower six bands of the eight-band Hamiltonian using the parameters $v_{\mu} = 8.25v_{m}$; $v_{C}=0$, $2.5$, and $5$; and $v_{CS} = 0.45 v_{C}$ (Fig.~\ref{fig:fragile}).  As the bulk BZ (Fig.~\ref{fig:2Dmain}(b) of the main text) has independent mirror lines along $k_{x,y}=0,\pi$ and $k_{x}=\pm k_{y}$, these three Wilson loops comprise the set of symmetry-inequivalent Wilson loops that pass through the $\Gamma$ point.  Aside from the $(2x+y)$-directed Wilson loop, there are of course other, low-symmetry Wilson loops, but they will exhibit spectra adiabatically related to that of the $(2x+y)$-directed loop.  As can be observed in Fig.~\ref{fig:fragile}(f,i,l), all possible Wilson spectra over the lower six bands of this model ($P_{6}$ in Fig.~\ref{fig:fragile}(a-c)) become trivialized when the two fragile bands are coupled to four bands from spinful $s$ orbitals at the $2c$ position, \emph{even though the bands from these $s$ orbitals are separated from the bands with fragile topology by an energy gap} (which is explicitly shown along high-symmetry lines in Fig.~\ref{fig:fragile}(b,c), and was additionally numerically confirmed throughout the BZ interior).  However, if the Wilson projector for the same system is chosen over just the two fragile valence bands ($P_{2}$ in Fig.~\ref{fig:fragile}(c)), then the Wilson loop still winds (Fig.~\ref{fig:fragile2}), \emph{despite the presence of the additional bands}.  This is further explored in Refs.~\onlinecite{BarryFragile,WiederAxion}.

\section{Alternative Realizations of $\mathcal{T}$-Symmetric HOFA Semimetals}
\label{sec:altHOFAsem}

In the main text, we presented two realizations of $\mathcal{T}$-symmetric Dirac semimetals with HOFAs: in Fig.~\ref{fig:HingeSMmain}(d,g) of the main text, we showed a tetragonal Dirac semimetal characterized by a nontrivial plane with mirror Chern number $C_{M_{z}}=2$ at $k_{z}=0$, and in Fig.~\ref{fig:HingeSMmain}(e,h) of the main text, we showed a tetragonal Dirac semimetal with broken $M_{z}$ and $\mathcal{I}$ symmetries, for which the $k_{z}=0$ plane is in the fragile topological phase examined in Appendix~\ref{sec:fragile}.  Here, we present two additional realizations of $\mathcal{T}$-symmetric Dirac semimetals with HOFA states.  There are likely many more semimetallic systems with other variants of HOFAs, including both those with $4mm$-symmetric Dirac points representative of the quantum critical point of the QI phase~\cite{multipole}, and also additional HOFA semimetals with nodal features equivalent to other 2D atomic limit or fragile transitions~\cite{HigherOrderTIChen,HermeleSymmetry,BernevigMoTe2}.  For simplicity, we will focus in this section on HOFA semimetals within the same space group (SG 123 $P4/mmm1'$) that vary by the locations and multiplicities of their Dirac points and by their bulk symmetry (parity) eigenvalues, and leave for future works the enumeration of all possible 2D one-parameter atomic-limit and fragile-phase transitions (Appendix~\ref{sec:fragile}), and their corresponding 3D HOFA semimetals.

\subsection{HOFA Semimetals with Band Inversion on a Line}
\label{sec:double}

One of the simplest realizations of a $\mathcal{T}$-symmetric Dirac semimetal for which the only boundary modes are hinge-localized Fermi arcs occurs when there are multiple inversions between bands with distinct $C_{4z}$ eigenvalues along $4mm$-invariant lines in a HOFA-supporting SG (Appendix~\ref{sec:SGs}).  Though we do not present material candidates in this work with this band inversion structure, we do formulate in this section tight-binding models with multiple band inversions, and show that they exhibit HOFAs as their only boundary (surface and hinge) modes.

\begin{figure}[h]
\centering
\includegraphics[width=0.95\textwidth]{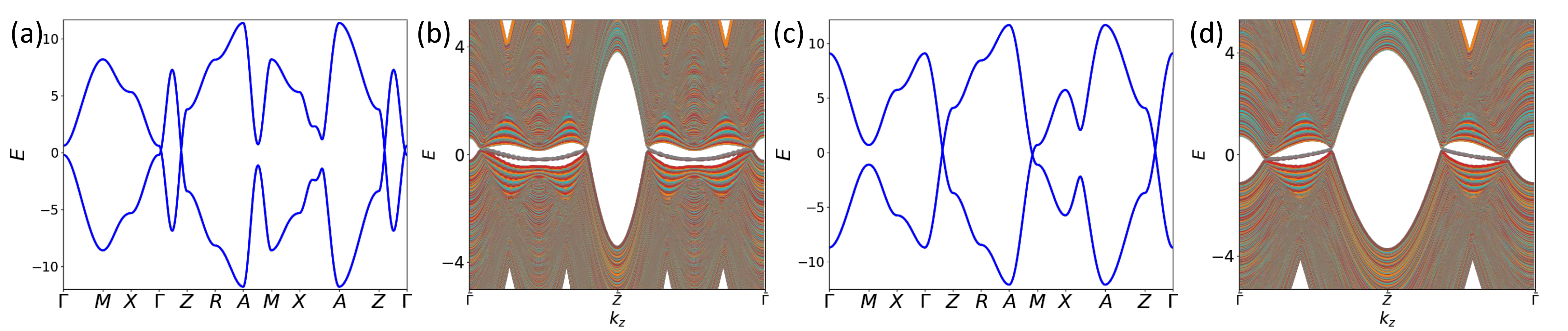}
\caption{Time-reversal-symmetric Dirac semimetals in tetragonal space group 123 $P4/mmm1'$ with four Dirac points.  The transition between 2D trivial insulating and QI phases can be driven by band inversion at either $k_{x}=k_{y}=0$ or at $k_{x}=k_{y}=\pi$ (Appendix~\ref{sec:bandrep} and Refs.~\onlinecite{multipole,WladTheory}).  Therefore, in a 3D system with BZ planes invariant under type-I magnetic wallpaper group $p4m$ or its supergroup $p4/m'mm$, such as a crystal in SG 123 $P4/mmm1'$, 3D Dirac points along \emph{both} $\Gamma Z$ and ${MA}$ represent quadrupole transitions.  Therefore, a $z$-directed rod (Fig.~\ref{fig:pumpMain}(a) of the main text) of either (a) a Dirac semimetal with two Dirac points along $\Gamma Z$ (and two related by $\mathcal{T}$ symmetry along $Z \Gamma$) or of (c) a Dirac semimetal with one Dirac point along $\Gamma Z$ and one along $M A$ (in addition to their time-reversal partners), will exhibit (b,d) HOFAs spanning the projections of the 3D bulk Dirac points as its only boundary modes.}
\label{fig:double}
\end{figure}

To begin, consider a Dirac semimetal in SG 123 ($P4/mmm1'$) formed from orbitals at the $1a$ Wyckoff position ($\{x,y,z\}=\{0,0,0\}$) that exhibits four total Dirac points along the $k_{x}=k_{y}=0$ line (Fig~\ref{fig:double}(a)).  Each one of these Dirac points is formed by an inversion of the irreducible corepresentations $\bar{\rho}_{6,7}$ of the little group of this BZ line (Appendix~\ref{sec:bandrep}).  The Hamiltonian of each 2D plane of the BZ indexed by $k_{z}$ ($\mathcal{H}(k_{z})$) is therefore characterized by a $\mathbb{Z}_{2}$ quantity dictating whether it is equivalent to a 2D trivial insulator or to a QI with corner modes.  This $\mathbb{Z}_{2}$ number is determined at each $k_{z}$ by the $C_{4z}$ eigenvalues of the occupied corepresentations along the two BZ lines with $4mm$ symmetry: $\Gamma Z$ ($k_{x}=k_{y}=0$) and $MA$ ($k_{x}=k_{y}=\pi$).  Specifically, using the results of Appendix~\ref{sec:bandrep}, along $\Gamma Z$ and $M A$, the occupied bands at a fixed value of $k_{z}$ exhibit the following set of corepresentations:
\begin{equation}
\{\bar{\rho}\}_{0,\pi}(k_{z}) = a(k_{z})_{0,\pi}\bar{\rho}_{7} \oplus b(k_{z})_{0,\pi}\bar{\rho}_{6},
\end{equation}
where $a(k_{z})_{0,\pi}$ and $b(k_{z})_{0,\pi}$ are integers indicating the multiplicities of corepresentations $\bar{\rho}_{7,6}$ in the set $\{\bar{\rho}\}$ and where:
\begin{equation}
a_{0,\pi}(k_{z})+b_{0,\pi}(k_{z})=\nu/2,
\label{eq:FillingFixA}
\end{equation}
where $\nu/2$ is equal to half the number of occupied bands, as $\dim\bar{\rho}_{6}=\dim\bar{\rho}_{7}=2$, and where $0,\pi$ indicate BZ lines at $k_{x}=k_{y}=0,\pi$, respectively.  When $a_{0}(k_{z})=a_{\pi}(k_{z})$ (and hence $b_{0}(k_{z})=b_{\pi}(k_{z})$), Table~\ref{tb:bandrep} implies that the symmetry eigenvalues of this BZ plane indexed by $k_{z}$ match those of a 2D spinful magnetic atomic insulator with $p4m$ symmetry and $\nu/2$ pairs of spinful orbitals at the $1a$ position (though again here, as in Appendix~\ref{sec:bandrep}, the spinful orbitals are paired by $M_{x,y}$, and not by $\mathcal{T}$).  Conversely if $a_{0}(k_{z})\neq a_{\pi}(k_{z})$ (and hence $b_{0}(k_{z})\neq b_{\pi}(k_{z})$), the symmetry eigenvalues of the bands in the BZ plane at $k_{z}$ match those of a 2D spinful magnetic atomic insulator with $\nu/2 - |a_{0}(k_{z}) - a_{\pi}(k_{z})|$ pairs of spinful atomic orbitals at $1a$ and $|a_{0}(k_{z}) - a_{\pi}(k_{z})|$ pairs at $1b$.

Continuing, we consider advancing $k_{z}$ until we encounter a gapless (Dirac) point along either $\Gamma Z$ or $M A$.  A Dirac point along $\Gamma Z$ changes $a_{0}(k_{z})$ and $b_{0}(k_{z})$ by one; similarly a Dirac point along $M A$ changes $a_{\pi}(k_{z})$ and $b_{\pi}(k_{z})$ by one.  Thus, the symmetry eigenvalues of the occupied bands in the 2D planes above or below a Dirac point are the same as those of 2D atomic insulators which differ by one in their number of Wannier centers at the $1a$ position.  Consequently, if there is a Dirac point along $\Gamma Z$ that lies above $k_{z1}$ and below $k_{z2}$, then:
\begin{equation}
a_{0}(k_{z2}) = a_{0}(k_{z1}) \pm 1,\ b_{0}(k_{z2}) = b_{0}(k_{z1}) \mp 1,\ a_{\pi}(k_{z2}) = a_{\pi}(k_{z1}),\ b_{\pi}(k_{z2}) = b_{\pi}(k_{z1}),
\end{equation}
and thus $\mathcal{H}(k_{z2})$ is equivalent to a 2D insulator with one fewer pair of Wannier orbitals at $1a$ and one more pair at $1b$.  Because the values of $a_{0,\pi}(k_{z})$ and $b_{0,\pi}(k_{z})$ are not independent (Eq.~(\ref{eq:FillingFixA})), we can therefore restrict our discussion to changes in $a_{0,\pi}(k_{z})$, which combined with Eq.~(\ref{eq:FillingFixA}), is sufficient information to determine the (polarization) topology of $k_{z}$-indexed planes in this semimetal.  If we now assume that $a_{0}(k_{z1})=a_{\pi}(k_{z1})$, then $\mathcal{H}(k_{z2})$ will therefore be nontrivial and exhibit QI corner states.  If there is then a second Dirac point along $\Gamma Z$ lying above $k_{z2}$ and below $k_{z3}$, then:
\begin{equation}
a_{0}(k_{z3}) = a_{0}(k_{z2}) \pm 1,\   a_{\pi}(k_{z3}) = a_{\pi}(k_{z2}),
\end{equation}
leading to an overall relation: 
\begin{equation}
a_{0}(k_{z3}) = a_{0}(k_{z1})\text{ or } a_{0}(k_{z3}) =a_{0}(k_{z1})\pm 2. 
\label{eq:OverallSyms}
\end{equation}
Therefore, $\mathcal{H}(k_{z3}$) is equivalent to a 2D insulator with \emph{either} the same number of Wannier orbitals at $1a$ ($a_{0}(k_{z3}) =a_{0}(k_{z1})= a_{\pi}(k_{z3})$) as $\mathcal{H}(k_{z1}$), or to one with two pairs of Wannier orbitals at $1b$ ($a_{0}(k_{z3}) =  a_{0}(k_{z1})\pm 2 = a_{\pi}(k_{z3})\pm 2$).  However, keeping only the symmetries of magnetic wallpaper group $p4m$ or its layer supergroup $p4/m'mm$ (Appendix~\ref{sec:bandrep}), a set of bands containing only trivial bands and two copies of the QI exhibits a net-trivial quadrupole moment~\cite{WladTheory} of $e\mod e=0$.  Therefore, both of the possibilities in Eq.~(\ref{eq:OverallSyms}) imply the absence of protected QI corner states at $k_{z3}$.  We can realize this semimetal by tuning parameters to place two, time-reversed pairs of QI-nontrivial regions at $k_{z}$ values away from the TRIM points (Fig.~\ref{fig:double}(b)).  If the symmetry indicators at the $k_{z}=0,\pi$ planes are otherwise trivial (\emph{e.g.} the symmetry eigenvalues do not indicate 2D quantum spin Hall or mirror Chern $C_{M_{z}}=2$ phases~\cite{FuKaneInversion,AndreiTI,AndreiInversion,NagaosaDirac}), and if all the non-symmetry-indicated Wilson loops are trivial (ruling out the case of $C_{M_{z}}=4n$ where n is a nonzero integer~\cite{ChenBernevigTCI}), then this semimetal will not exhibit any topological surface states on its 2D faces~\cite{NagaosaDirac}, and will only exhibit HOFAs connecting the hinge projections of its four bulk 3D Dirac points.  We realize a Dirac semimetal in this phase (Fig.~\ref{fig:double}(a,b)) by introducing the term:
\begin{equation}
V_{H2}({\bf k}) = t_{H2}\cos(2k_{z}),
\label{eq:H2}
\end{equation}
to $\mathcal{H}_{H2}({\bf k})$ in Eq.~(\ref{eq:hinge}), and choosing the parameters:
\begin{equation}
t_{1}=2,\ t_{2} = 1.5,\ v_{m}=-1.5,\ t_{PH}=0.1,\ v_{s} = 1.3,\ v_{M_{z}}=0,\ t_{H}=1.6,\ u (\mathcal{H}_{H2})=0.5, t_{H2} = -4.5.
\end{equation}

A similar configuration of HOFAs can also be achieved in a Dirac semimetal in SG 123 ($P4/mmm1'$) that is formed from orbitals at the $1a$ Wyckoff position and exhibits a time-reversed pair of Dirac points along $k_{x,y}=0$ and a time-reversed pair along $k_{x,y}=\pi$ (Fig.~\ref{fig:double}(c)).  To explain this, we follow the same logic as previously and place a Dirac point along $\Gamma Z$ at a value of $k_{z}$ between $k_{z1,2}$ where $k_{z2}>k_{z1}>0$.  The Dirac point between $k_{z1}$ and $k_{z2}$ has a time-reversal partner with $k_{z}<0$, to which the following arguments also apply.  At $k_{z2}$, 
\begin{equation}
a_{0}(k_{z2}) = a_{0}(k_{z1}) \pm 1,\   a_{\pi}(k_{z2}) = a_{\pi}(k_{z1}),
\end{equation}
again indicating that $\mathcal{H}(k_{z2})$ is equivalent to a 2D insulator with one fewer pair of Wannier orbitals at $1a$ and one pair at $1b$, and specifically is equivalent to a QI if $a_{0}(k_{z1}) = a_{\pi}(k_{z1})$.  We then place a second Dirac crossing instead \emph{along $M A$} ($k_{x}=k_{y}=\pi$) above $k_{z2}$ and below a third momentum $k_{z3}$ for which $k_{z3}>k_{z2}$, such that:
\begin{equation}
a_{0}(k_{z3}) = a_{0}(k_{z2}),\   a_{\pi}(k_{z3}) = a_{\pi}(k_{z2})\pm 1,
\end{equation}
which indicates that $\mathcal{H}(k_{z3})$ is equivalent to a 2D insulator with \emph{either} the same number of Wannier orbitals at $1a$  ($a_{0}(k_{z3}) = a_{\pi}(k_{z3})$) as $\mathcal{H}(k_{z1})$, or to one with two pairs of Wannier orbitals at $1b$ ($a_{0}(k_{z3}) = a_{\pi}(k_{z3}) \pm 2$), both of which correspondingly carry a trivial quadrupole moment of $e\mod e=0$ and do not exhibit corner states.  The Dirac point between $k_{z2}$ and $k_{z3}$ also has a time-reversal partner with $k_{z}<0$, to which these arguments also apply.

We realize a semimetal in SG 123 $P4/mmm1'$ with this configuration of Dirac points (Fig.~\ref{fig:double}(c)) by choosing in Eqs.~(\ref{eq:hinge}) and~(\ref{eq:H2}) the parameters:
\begin{equation}
t_{1}=2,\ t_{2} = 1.5,\ v_{m}=-1.5,\ t_{PH}=0.1,\ v_{s} = 1.3,\ v_{M_{z}}=0,\ t_{H}=6.4,\ u (\mathcal{H}_{H2})=0.5, t_{H2} = 0.  
\label{eq:twoLineDiracs}
\end{equation}
Plotting the bands of Eqs.~(\ref{eq:hinge}) and~(\ref{eq:H2}) with the parameters in Eq.~(\ref{eq:twoLineDiracs}) on a $z$-directed rod (Fig.~\ref{fig:double}(d)), it is clear that HOFA states connect the projections of the bulk Dirac points along ${\Gamma Z}$ to those along $M A$, and that there are no other boundary modes.  Specifically, at $k_{z}=0,\pi$, the $C_{4z}$ eigenvalues of the occupied bands are the same at $k_{x}=k_{y}=0,\pi$, and match those of either a trivial insulator or a TCI with~\cite{ChenBernevigTCI} $C_{M_{z}}\mod 4 = 0$.  Through calculations of the $x$-directed Wilson loop and the $(100)$-surface states (the results of which are implicit in the rod bands shown in Fig.~\ref{fig:double}(d)), we confirm that the Hamiltonians of the $k_{z}=0,\pi$ planes are topologically trivial ($C_{M_{z}}=0$).

Finally, for completeness, we note that depending on the values of $a_{0,\pi}(k_{z})$ and $b_{0,\pi}(k_{z})$, sets of four occupied bands can exhibit the same symmetry eigenvalues as \emph{either} superposed atomic limits at the $1a$ and $1b$ position of $p4m$, or a single, four-band atomic limit at the $2c$ position (Eqs.~(\ref{eq:1a1bIrrepEquiv}) and~(\ref{eq:2cIrrepEquiv})).  Specifically, because placing pairs of spinful $s$ orbitals at the $2c$ position of $p4m$ (Fig.~\ref{fig:2Dmain}(a) of the main text) induces four bands with the combined symmetry eigenvalues of bands from $s$ orbitals at the $1a$ and $1b$ positions (Appendix~\ref{sec:bandrep} and Refs.~\onlinecite{QuantumChemistry,Bandrep1,Bandrep2,Bandrep3}), then, for generic numbers of occupied bands with different complex-conjugate pairs of $C_{4z}$ eigenvalues, a change of $a_{0,\pi}(k_{z})$ indicates \emph{either} a shift of two occupied (and two unoccupied) Wannier orbitals from $1a$ to $1b$ or a shift of four occupied (and four unoccupied) Wannier orbitals from $1a$ and $1b$ to $2c$.  We emphasize this point because the irreducible-representation equivalence between EBRs from $1a$ plus $1b$ and EBRs from $2c$ in $p4m$ is key to understanding atomic-limit transitions in \emph{nonsymmorphic} 2D wallpaper groups~\cite{MagneticBook,DiracInsulator}, such as $p4g$, which, while largely beyond the scope of this work, can also exhibit corner modes~\cite{BrickLattice} that are equivalent to the HOFA states of $4mm$-symmetric \emph{nonsymmorphic} 3D Dirac semimetals (Appendix~\ref{sec:pinnedHOFAs}).

\subsection{$p_{z}-d_{x^{2}-y^{2}}$-Hybridized HOFA Dirac Semimetals}
\label{sec:pd}

We can also realize a $\mathcal{T}$-symmetric semimetal with HOFA states in SG 123 $P4/mmm1'$ by substituting the $s$ orbitals used to form $\mathcal{H}_{H2}({\bf k})$ in Eq.~(\ref{eq:hinge}) with $p_{z}$ orbitals, still at the $1a$ position.  We first discuss how the representations of crystalline symmetries are different under this orbital substitution than in the $s-d$-hybridized cases highlighted in the main text, and then demonstrate how this difference affects the topology in $k_{z}$-indexed BZ planes.

\begin{figure}[h]
\centering
\includegraphics[width=0.8\textwidth]{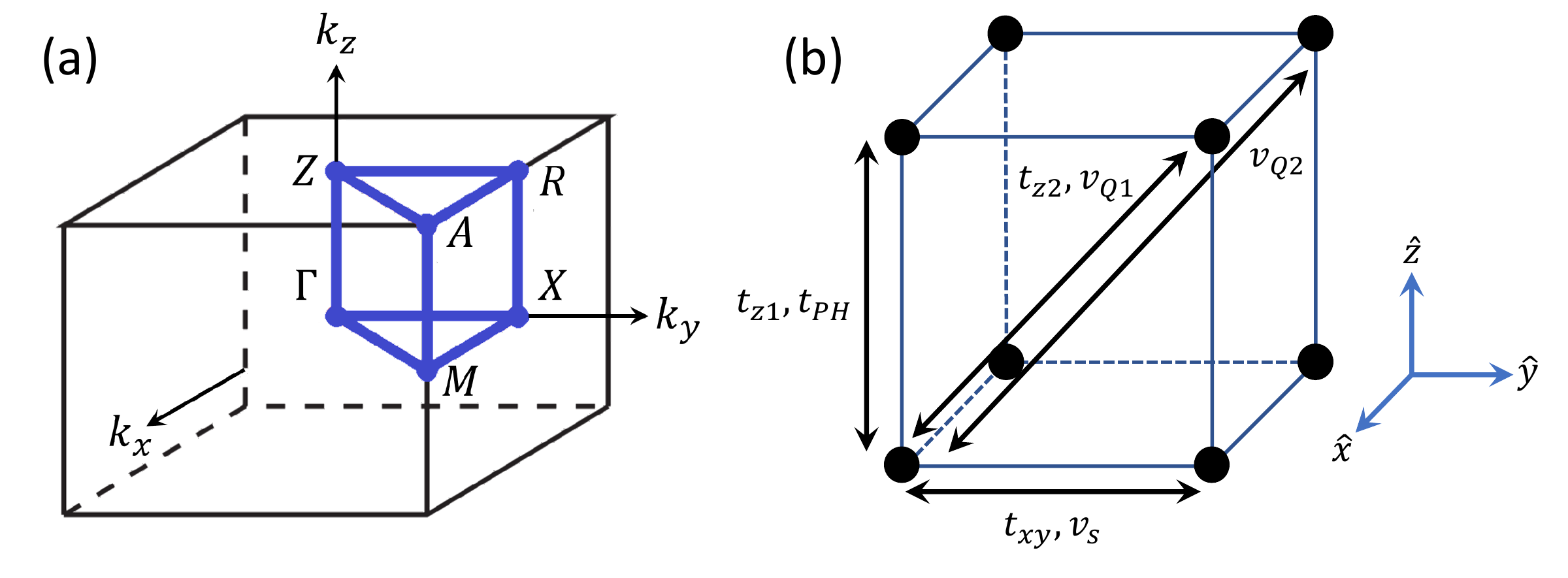}
\caption{BZ and hopping schematic for Eq.~(\ref{eq:pdTI}).  (a) The BZ~\cite{BCTBZ,BCS1,BCS2} of SG 123 $P4/mmm1'$.  (b) Schematic of the hoppings of the model of a $p-d$-hybridized HOFA semimetal in Eq.~(\ref{eq:pdTI}).  Only one set of hoppings is shown per term in Eq.~(\ref{eq:pdTI}); the full set of hoppings is generated by transforming the arrows in (b) under the symmetries of SG 123 $P4/mmm1'$.}
\label{fig:pdHop}
\end{figure}

The site-symmetry group of the $1a$ position of SG 123 is $4/mmm1'$ ($D_{4h}$), which is an index-2 supergroup of $4mm1'$:
\begin{equation}
4/mmm1' = (E)4mm1' \cup (\mathcal{I})4mm1'.
\end{equation}
As $\mathcal{I}h=h\mathcal{I}$ for all $h\in 4mm1'$, then there are simply twice as many corepresentations of $4/mmm1'$ as there are of $4mm1'$, each of which takes the form $\bar{E}_{1,2/g,u}$ where $g$ ($u$) indicates a two-dimensional corepresentation with positive (negative) inversion eigenvalues~\cite{BigBook,BilbaoPoint}.  Specifically, the $p_{z}$ orbitals transform as $\bar{E}_{1,u}$ and the $d_{x^{2}-y^{2}}$ orbitals transform as $\bar{E}_{2,g}$.  When these corepresentations are induced into the space group $G=P4/mmm1'$ and then subduced onto the TRIM points at $k_{x}=k_{y}=0,\pi$ (Fig.~\ref{fig:pdHop}(a)), the resulting little co-group corepresentations are~\cite{QuantumChemistry,Bandrep1,Bandrep2,Bandrep3}:
\begin{eqnarray}
(\bar{E}_{1u}\uparrow G)\downarrow \Gamma &\equiv& (\bar{E}_{1u}\uparrow G)\downarrow Z \equiv (\bar{E}_{1u}\uparrow G)\downarrow M \equiv (\bar{E}_{1u}\uparrow G)\downarrow A \equiv\bar{\rho}_{7}^{-}, \nonumber \\
(\bar{E}_{2g}\uparrow G)\downarrow \Gamma &\equiv& (\bar{E}_{2g}\uparrow G)\downarrow Z \equiv (\bar{E}_{2g}\uparrow G)\downarrow M \equiv (\bar{E}_{2g}\uparrow G)\downarrow A \equiv\bar{\rho}_{6}^{+}, \nonumber \\
\label{eq:kreps}
\end{eqnarray}
where placing $\mathcal{I}$-odd ($\bar{E}_{1,2u}$) (-even ($\bar{E}_{1,2g}$)) orbitals at $1a$ results in all little co-group corepresentations having negative (positive) inversion eigenvalues, whose sum is given by the characters:
\begin{equation}
\chi_{\bar{\rho}_{6,7}^{+}}(\mathcal{I}) = +2,\ \chi_{\bar{\rho}_{6,7}^{-}}(\mathcal{I}) = -2.
\label{eq:inversionEigs}
\end{equation}
The little co-group corepresentations $\bar{\rho}_{6,7}^{\pm}$ also inherit the $C_{4z}$ characters of the corepresentations $\bar{\rho}_{6,7}$ of their $\mathcal{I}$-broken subgroups $4mm$ and $4/m'mm$ (Eq.~(\ref{eq:c4evals})):
\begin{equation}
\chi_{\bar{\rho}^{\pm}_{6}}(C_{4z})=-\sqrt{2},\ \chi_{\bar{\rho}^{\pm}_{7}}(C_{4z})=\sqrt{2}.
\label{eq:c4evalsWithI}
\end{equation}
Crucially, BZ planes indexed by $k_{z}\neq 0,\pi$ are only invariant under magnetic layer group $p4/m'mm$, and thus, as discussed in Appendices~\ref{sec:bandrep} and~\ref{sec:double}, their Hamiltonians are equivalent to 2D trivial insulators or QIs as indicated only by the number of occupied bands labeled by $\bar{\rho}_{6}$ at $k_{x}=k_{y}=0,\pi$ (discussed in the text following Eq.~(\ref{eq:OverallSyms})).  Therefore, whether all of the valence bands of a Dirac semimetal in SG 123 $P4/mmm1'$ have the same inversion eigenvalues at the $C_{4z}$-invariant TRIM points ($k_{x}=k_{y}=0,\pi$,\ $k_{z}=0,\pi$), or if some of the inversion eigenvalues are different, the semimetal will still exhibit HOFA states, because the (obstructed-atomic-limit, specifically QI) topology of the $k_{z}$-indexed planes away from the TRIM points is unaffected by the topology in high-symmetry planes elsewhere in the BZ.  Consequently, a Dirac semimetal in SG 123 $P4/mmm1'$ should exhibit HOFA states whether it is formed from hybridized $s$ and $d_{x^{2}-y^{2}}$ orbitals or from $p_{z}$ and $d_{x^{2}-y^{2}}$ orbitals. 

\begin{figure}[t]
\centering
\includegraphics[width=1.0\textwidth]{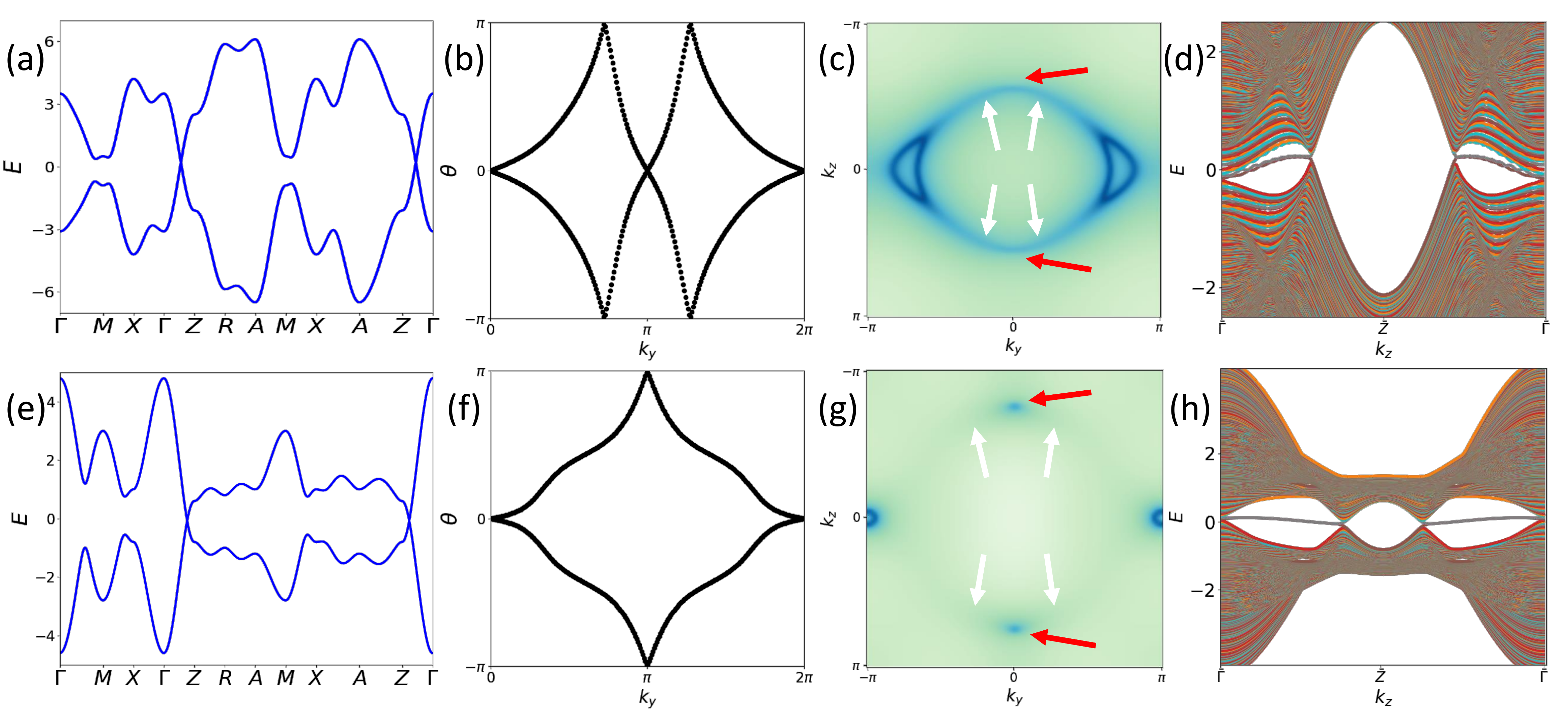}
\caption{Bulk, surface, and hinge states in $s-d$- and $p-d$-hybridized HOFA Dirac semimetals in SG 123 $P4/mmm1'$.  (a) Bulk bands of the $s-d_{x^{2}-y^{2}}$-hybridized, $\mathcal{I}$-, $M_{z}$-, and $\mathcal{T}$-symmetric Dirac semimetal described by $\mathcal{H}_{H2}({\bf k})$ in Eq.~(\ref{eq:hinge}) and highlighted in the main text, plotted using the parameters in Table~\ref{tb:3D}.  Here, all of the occupied bands at $k_{z}=0$ have the same inversion eigenvalues, and this plane is QSH-trivial.  However, because spin-1/2 corepresentations and spin-3/2 corepresentations with the same inversion eigenvalues and opposite pairs of $C_{4z}$ eigenvalues ($\bar{\rho}_{6,7}^{+}$ in Eqs.~(\ref{eq:inversionEigs}) and~(\ref{eq:c4evalsWithI})) are inverted at $\Gamma$ relative to those at $M$, the Hamiltonian of the $k_{z}=0$ plane is equivalent to a mirror TCI with mirror Chern number~\cite{NagaosaDirac,ChenBernevigTCI} $C_{M_{z}}=2$ (Table~\ref{tb:corepssd}), as indicated by (b) the $x$-directed Wilson spectrum and (c) $(100)$-surface states.  In the surface spectrum in (c), the only surface-localized states are a time-reversed pair of TCI cones at $k_{z}=0$; the remains of four Fermi arcs (and their time-reversal partners), which appear in four tightly grouped pairs (white arrows), can be seen connecting the TCI cones at $k_{z}=0$ to the projections of the bulk Dirac points (red arrows).  As there is no topological invariant that requires these surface Fermi arcs to cross the Fermi energy~\cite{KargarianDiracArc1,KargarianDiracArc2,ThomaleArc,KaminskiDiracArc}, they can be gapped out; here that is accomplished by the bulk quadrupolar SOC term $U({\bf k})\sin(k_{z})$ in Eq.~(\ref{eq:hinge}).  (d) $z$-directed rod bands of the $s-d$-hybridized semimetal; HOFA states are clearly visible connecting the hinge projections of the bulk Dirac points to the projections of the topological surface cones.  (e) Bulk bands of the $p_{z}-d_{x^{2}-y^{2}}$-hybridized, $\mathcal{I}$-, $M_{z}$-, and $\mathcal{T}$-symmetric Dirac semimetal described by Eq.~(\ref{eq:pdTI}).  Here, states with opposite inversion eigenvalues ($\bar{\rho}_{7}^{-}$ and $\bar{\rho}_{6}^{+}$ in Eqs.~(\ref{eq:inversionEigs}) and~(\ref{eq:c4evalsWithI})) are inverted at $\Gamma$ relative to those at $M$ (Eq.~(\ref{eq:pdcoreps})), and therefore by the Fu-Kane parity criterion~\cite{FuKaneInversion}, the Hamiltonian of the  $k_{z}=0$ plane is equivalent to a 2D TI (Table~\ref{tb:corepspd}), as indicated by its (f) $x$-directed Wilson spectrum and (g) $(100)$-surface states.  In the surface spectrum in (g), the only surface-localized state is a TI cone at $k_{z}=0,\ k_{y} = \pi$; the (extremely) faint remains of two Fermi arcs (and their time-reversal partners) (white arrows) can be seen connecting the TI cones to the projections of the bulk Dirac points (red arrows).  These surface Fermi arcs have been almost completely gapped out by bulk quadrupolar SOC ($v_{Q1,2}$ in Eq.~(\ref{eq:pdTI})).  (h) $z$-directed rod bands of the $p-d$-hybridized semimetal; HOFA states are clearly visible connecting the hinge projections of the bulk Dirac points to the projections of the topological surface cones, as they were in (d).}
\label{fig:pd}
\end{figure}

To demonstrate this, we form a tight-binding model with $p_{z}$ and $d_{x^{2}-y^{2}}$ orbitals placed at the $1a$ position of SG 123 $P4/mmm1'$:
\begin{eqnarray}
\mathcal{H}({\bf k})&=&\left[t_{z1}\cos(k_{z}) + t_{z2}\cos(k_{z})\times\left(\cos(k_{x})+\cos(k_{y})\right)\right]\tau^{z}+ t_{xy}\left[\cos(k_{x}) + \cos(k_{y})\right]\tau^{z} \nonumber \\
&+&  t_{PH}\cos(k_{z})\mathds{1}_{\tau\sigma} +  v_{s}\left[\sin(k_{x})\tau^{x}\sigma^{y} + \sin(k_{y})\tau^{x}\sigma^{x}\right] +  v_{Q1}\tau^{y}\sin(k_{z})\left[\cos(k_{x})-\cos(k_{y})\right] \nonumber \\
&+& v_{Q2}\tau^{x}\sigma^{z}\sin(k_{x})\sin(k_{y})\sin(k_{z}),
\label{eq:pdTI}
\end{eqnarray}
where $\tau$ indicates the orbital degree of freedom, $\sigma$ indicates the $\mathcal{T}$-odd spin degree of freedom, and where $\mathds{1}_{\tau\sigma}$ is the $4\times 4$ identity.  We note that after the submission of this work, it was recognized that Eq.~(\ref{eq:pdTI}) contains the same terms as the simplified Dirac semimetal model employed in Ref.~\onlinecite{BednikToyDirac} (as well as additional terms that respect the symmetries of SG 123 $P4/mmm1'$).  In Eq.~(\ref{eq:pdTI}), all of the terms correspond to nearest-neighbor hopping, except for the quadrupolar SOC terms $v_{Q1}$ and $v_{Q2}$, which correspond to second-neighbor hopping in the $xy$-plane and third-neighbor hopping along the $xyz$ diagonal, respectively (Fig.~\ref{fig:pdHop}(b)).  These terms, like $U({\bf k})\sin(k_{z})$ in Eqs.~(\ref{eq:quad}) and~(\ref{eq:hinge}) for the $s-d$-hybridized Dirac semimetal, are equivalent to the term employed in Ref.~\onlinecite{KargarianDiracArc1} to disconnect the surface Fermi arcs of Dirac semimetals.  Specifically, $v_{Q1,2}$ enforce the breaking of $\mathcal{T}$- and $M_{z}$ symmetries in $k_{z}$-indexed BZ planes away from $k_{z}=0,\pi$, and gap the surface states at generic values of $k_{z}\neq 0,\pi$.  When $v_{Q1,2}$ in Eq.~(\ref{eq:pdTI}) (and $u$ in Eqs.~(\ref{eq:quad}) and~(\ref{eq:hinge})) are strong, this causes the surface states to close off into rings with a narrow width in $k_{z}$, as opposed to connecting all the way to the surface projections of the bulk Dirac points (Fig.~\ref{fig:pd}(c,g)).  This is allowed because unlike in Weyl semimetals, the surface Fermi arcs in Dirac semimetals are not protected by a robust bulk topological invariant, and are rather just a consequence of the continuity of surface states from topological surface TI and TCI cones to trivial surface states in the vicinities of the projections of bulk Dirac points that can be pushed away from the Fermi energy by bulk SOC~\cite{KargarianDiracArc1,KargarianDiracArc2,ThomaleArc,KaminskiDiracArc}.  This causes the surface states of Dirac semimetals to either appear in closed loops disconnected from the projections of the bulk Dirac points (Fig.~\ref{fig:pd}(c,g) and Refs.~\onlinecite{KargarianDiracArc1,KargarianDiracArc2,KaminskiDiracArc}), or allows them to be completely absent, depending on the topology in high-symmetry BZ planes~\cite{ThomaleArc,NagaosaDirac}.  

The generating symmetries of the $p-d$-hybridized model in Eq.~(\ref{eq:pdTI}) are represented at the eight TRIM points by:
\begin{equation}
\mathcal{T}=i\sigma^{y}K,\ \mathcal{I} = \tau^{z},\ M_{x,y} = -i\sigma^{x,y},
\label{eq:pdTRIMInversion}
\end{equation}
and by an additional $C_{4z}$ symmetry at the four $C_{4z}$-invariant TRIM points:
\begin{equation}
C_{4z} = \tau^{z}\left(\frac{\mathds{1}-i\sigma^{z}}{\sqrt{2}}\right),
\label{eq:pdC4}
\end{equation}
where the prefactor of $\tau^{z}$ in Eq.~(\ref{eq:pdC4}) reflects that $p_{z}$ ($d_{x^{2}-y^{2}}$) orbitals are even (odd) under $C_{4z}$.  Eqs.~(\ref{eq:pdTRIMInversion}) and~(\ref{eq:pdC4}) also imply the presence of $C_{2z}$ and $M_{z}$ symmetries, which are represented at each TRIM point by:
\begin{equation}
C_{2z} = (C_{4z})^{2} = -i\sigma^{z},\ M_{z} = \mathcal{I}C_{2z} = -i\tau^{z}\sigma^{z}.
\end{equation}
To form a semimetal with broken particle-hole symmetry and Dirac points along $\Gamma Z$, we choose in Eq.~(\ref{eq:pdTI}) the parameters:
\begin{equation}
t_{z1}=0.9,\ t_{z2}=0.9,\ t_{xy}=1,\ t_{PH}=0.1,\ v_{s}=0.8,\ v_{Q1} = 0.6,\ v_{Q2} = 0.25.  
\label{eq:pdParams}
\end{equation}
In Fig.~\ref{fig:pd}, we plot the bulk bands, $x$-directed Wilson bands at $k_{z}=0$, $(100)$-surface states, and $z$-directed rod bands of (a-d) the $s-d$-hybridized Dirac semimetal highlighted in the main text (Eq.~(\ref{eq:hinge})) and of (e-h) the $p-d$-hybridized Dirac semimetal described by Eq.~(\ref{eq:pdTI}).  Along the four $C_{2z}$-invariant lines of the BZ at $k_{x,y}=0,\pi$, Eq.~(\ref{eq:pdTI}) takes the simplified form:
\begin{eqnarray}
\mathcal{H}(0,0,k_z) &=& [(t_{z1}+2t_{z2})\cos(k_{z}) + 2t_{xy}]\tau_{z} + a\cos(k_{z})\mathds{1}_{\tau\sigma}, \nonumber \\
\mathcal{H}(\pi,\pi,k_z) &=& [(t_{z1}-2t_{z2})\cos(k_{z}) - 2t_{xy}]\tau_{z} + a\cos(k_{z})\mathds{1}_{\tau\sigma}, \nonumber \\
\mathcal{H}(0,\pi,k_z) &=& \mathcal{H}(\pi,0,k_z) = t_{z1}\cos(k_{z})\tau^{z} + a\cos(k_{z})\mathds{1}_{\tau\sigma},
\label{eq:prefactors}
\end{eqnarray}
which, as,
\begin{equation}
[\tau^{z},\mathcal{H}(0,0,k_{z})]=[\tau^{z},\mathcal{H}(\pi,\pi,k_{z})]=[\tau^{z},\mathcal{H}(0,\pi,k_{z})]=0,
\end{equation}
indicates that the inversion and $C_{4z}$ eigenvalues of the occupied states at the TRIM points at half filling in Eq.~(\ref{eq:pdTI}) in its Dirac semimetallic phase are entirely determined by the signs of the prefactors of $\tau^{z}$ in Eq.~(\ref{eq:prefactors}).   Defining $\bar{\sigma}({\bf k})$ as the corepresentation of the two occupied bands at ${\bf k}$, we find that:
\begin{equation}
\bar{\sigma}(\Gamma) = \bar{\rho}_{7}^{-},\ \bar{\sigma}(Z)=\bar{\sigma}(M) = \bar{\sigma}(A) = \bar{\rho}_{6}^{+},\ \bar{\sigma}(X) =\bar{\sigma}(X') = \bar{\varrho}_{5}^{-},\ \bar{\sigma}(R) = \bar{\sigma}(R') = \bar{\varrho}_{5}^{+},
\label{eq:pdcoreps}
\end{equation}
where the $C_{4z}$-symmetric points $\Gamma,\ Z,\ M,$ and $A$ have little co-groups isomorphic to $4/mmm1'$ and therefore the corepresentations in Eq.~(\ref{eq:kreps}).  Conversely, the $X$ and $R$ (and $X'$ and $R'$) points are not $C_{4z}$ symmetric, and therefore they have little co-groups isomorphic to $mmm1'$, which has two, two-dimensional corepresentations~\cite{BigBook,BilbaoPoint,Bandrep1,Bandrep2,Bandrep3}:
\begin{equation}
(\bar{E}_{1,2/g,u}\uparrow G)\downarrow X \equiv (\bar{E}_{1,2/g,u}\uparrow G)\downarrow R\equiv\bar{\varrho}_{5}^{+,-},
\end{equation}
with inversion characters:
\begin{equation}
\chi_{\bar{\varrho}_{5}^{\pm}}(\mathcal{I}) = \pm 2.
\label{eq:inversionEigsNoC4}
\end{equation}
We observe that the only TRIM points with negative inversion eigenvalues are $\Gamma$, $X$, and $X'$, which all lie in the $k_{z}=0$ plane.  Using the results of Refs.~\onlinecite{ChenBernevigTCI,ChenTCI}, we can express the mirror Chern numbers $C_{M_{z}}(0,\pi)$ of the Hamiltonians of the $k_{z}=0,\pi$ planes in terms of the occupied corepresentations:
\begin{eqnarray}
C_{M_{z}}(k_{z}=0,\pi)\text{ mod }4 &=& \bigg(\left[n_{0,0,k_{z}}(\bar{\rho}_{7}^{-}) + n_{\pi,\pi,k_{z}}(\bar{\rho}_{7}^{-})\right] + 2\left[n_{0,0,k_{z}}(\bar{\rho}_{6}^{+}) + n_{\pi,\pi,k_{z}}(\bar{\rho}_{6}^{+})\right] \nonumber \\
&+& 3\left[n_{0,0,k_{z}}(\bar{\rho}_{6}^{-}) + n_{\pi,\pi,k_{z}}(\bar{\rho}_{6}^{-})\right] + 2n_{\pi,0,k_{z}}(\bar{\varrho}_{5}^{-})\bigg)\text{ mod }4,
\label{eq:MirrorChernIndicator}
\end{eqnarray}
where $n_{k_{x},k_{y},k_{z}}(\bar{\sigma})$ is equal to the number of copies of the corepresentation $\bar{\sigma}$ that appear in the valence manifold~\cite{ChenBernevigTCI,SlagerKanePrx,AshvinIndicators,ChenTCI,AshvinTCI} at ${\bf k}=(k_{x},k_{y},k_{z})$, where we restrict $k_{z}=0,\pi$, and where $C_{M_{z}}\text{ mod }2$ is equivalent to the Fu-Kane parity index $z_{2}$ for a 2D TI~\cite{FuKaneInversion}.  For the $p-d$-hybridized semimetal described by Eq.~(\ref{eq:pdTI}) with the parameters in Eq.~(\ref{eq:pdParams}), Eq.~(\ref{eq:pdcoreps}) implies the values of $n_{k_{x},k_{y},k_{z}}(\bar{\sigma})$ shown in Table~\ref{tb:corepspd}.  As shown in Table~\ref{tb:corepspd}, for the $p-d$-hybridized HOFA Dirac semimetal in Eq.~(\ref{eq:pdTI}), the Hamiltonian of the $k_{z}=0$ plane is equivalent to a 2D TI, whereas the Hamiltonian of the $k_{z}=\pi$ plane exhibits the same symmetry eigenvalues as a trivial insulator~\cite{ChenBernevigTCI,SlagerKanePrx,AshvinIndicators,ChenTCI,AshvinTCI}.  This is the same high-symmetry-plane topology that occurs in many previously identified centrosymmetric 3D Dirac semimetals~\cite{NagaosaDirac,KargarianDiracArc1}, such as Na$_3$Bi~\cite{NaDirac,SchnyderDirac,ZJDirac2} and two of the candidate HOFA-semimetals highlighted in this work (Fig.~\ref{fig:DFTmain} of the main text and Appendix~\ref{sec:DFT}): KMgBi~\cite{KMgBi1,KMgBi2,KMgBi3} and Cd$_3$As$_2$~\cite{ZJDirac,ZJSurface}.  In Fig.~\ref{fig:pd}(f,g), we show the $x$-directed Wilson loop at $k_{z}=0$ and the $(100)$-surface states of Eq.~(\ref{eq:pdTI}), respectively.  The Wilson loop exhibits the characteristic winding of a 2D TI, and the surface spectrum correspondingly consists of a single twofold-degenerate cone at $k_{z}=0$,\ $k_{y}=\pi$.

\begin{table}[H]
\centering
\begin{tabular}{|c|c|c|c|c|c|c|c|}
\hline
\multicolumn{8}{|c|}{Number of Each Occupied Corepresentation in Eq.~(\ref{eq:MirrorChernIndicator}) for} \\
\multicolumn{8}{|c|}{the $p-d$-Hybridized HOFA Dirac Semimetal in Eq.~(\ref{eq:pdTI})} \\
\hline
$k_{z}=0$ & $n_{0,0,0}(\bar{\rho}_{7}^{-})$ & $n_{0,0,0}(\bar{\rho}_{6}^{+})$ & $n_{0,0,0}(\bar{\rho}_{6}^{-})$ & $n_{\pi,\pi,0}(\bar{\rho}_{7}^{-})$ & $n_{\pi,\pi,0}(\bar{\rho}_{6}^{+})$ & $n_{\pi,\pi,0}(\bar{\rho}_{6}^{-})$ & $n_{\pi,0,0}(\bar{\varrho}_{5}^{-})$ \\
\hline
Number & $1$ & $0$ & $0$ & $0$ & $1$ & $0$ & $1$ \\
\hline 
\multicolumn{8}{|r|}{$C_{M_{z}}(0)\text{ mod }4 = 1$,\ $z_{2}(0)=1$} \\
\hline
\hline
$k_{z}=\pi$ & $n_{0,0,\pi}(\bar{\rho}_{7}^{-})$ & $n_{0,0,\pi}(\bar{\rho}_{6}^{+})$ & $n_{0,0,\pi}(\bar{\rho}_{6}^{-})$ & $n_{\pi,\pi,\pi}(\bar{\rho}_{7}^{-})$ & $n_{\pi,\pi,\pi}(\bar{\rho}_{6}^{+})$ & $n_{\pi,\pi,\pi}(\bar{\rho}_{6}^{-})$ & $n_{\pi,0,\pi}(\bar{\varrho}_{5}^{-})$ \\
\hline
Number & $0$ & $1$ & $0$ & $0$ & $1$ & $0$ & $0$ \\
\hline
\multicolumn{8}{|r|}{$C_{M_{z}}(\pi)\text{ mod }4 = 0$,\ $z_{2}(\pi)=0$} \\
\hline
\end{tabular}
\caption{The number of occupied corepresentations, mirror Chern numbers $C_{M_{z}}(k_{z})$ (modulo $4$)~\cite{ChenBernevigTCI,ChenTCI} (Eq.~(\ref{eq:MirrorChernIndicator})), and 2D TI indices~\cite{FuKaneInversion} $z_{2}(k_{z})$ ($C_{M_{z}}\text{ mod }2$) for the $p-d$-hybridized HOFA Dirac semimetal in Eq.~(\ref{eq:pdTI}) with the parameters in Eq.~(\ref{eq:pdParams}).  The Hamiltonian of the $k_{z}=0$ plane is equivalent to a 2D TI, whereas the Hamiltonian of the $k_{z}=\pi$ plane exhibits the same symmetry eigenvalues as a trivial insulator~\cite{ChenBernevigTCI,SlagerKanePrx,AshvinIndicators,ChenTCI,AshvinTCI}.}
\label{tb:corepspd}
\end{table}

\begin{table}[H]
\centering
\begin{tabular}{|c|c|c|c|c|c|c|c|}
\hline
\multicolumn{8}{|c|}{Number of Each Occupied Corepresentation in Eq.~(\ref{eq:MirrorChernIndicator}) for} \\
\multicolumn{8}{|c|}{the $s-d$-Hybridized HOFA Dirac Semimetal in the Main Text (Eq.~(\ref{eq:hinge}))} \\
\hline
$k_{z}=0$ & $n_{0,0,0}(\bar{\rho}_{7}^{-})$ & $n_{0,0,0}(\bar{\rho}_{6}^{+})$ & $n_{0,0,0}(\bar{\rho}_{6}^{-})$ & $n_{\pi,\pi,0}(\bar{\rho}_{7}^{-})$ & $n_{\pi,\pi,0}(\bar{\rho}_{6}^{+})$ & $n_{\pi,\pi,0}(\bar{\rho}_{6}^{-})$ & $n_{\pi,0,0}(\bar{\varrho}_{5}^{-})$ \\
\hline
Number & $0$ & $1$ & $0$ & $0$ & $0$ & $0$ & $0$ \\
\hline 
\multicolumn{8}{|r|}{$C_{M_{z}}(0)\text{ mod }4 =2 $,\ $z_{2}(0)=0$} \\
\hline
\hline
$k_{z}=\pi$ & $n_{0,0,\pi}(\bar{\rho}_{7}^{-})$ & $n_{0,0,\pi}(\bar{\rho}_{6}^{+})$ & $n_{0,0,\pi}(\bar{\rho}_{6}^{-})$ & $n_{\pi,\pi,\pi}(\bar{\rho}_{7}^{-})$ & $n_{\pi,\pi,\pi}(\bar{\rho}_{6}^{+})$ & $n_{\pi,\pi,\pi}(\bar{\rho}_{6}^{-})$ & $n_{\pi,0,\pi}(\bar{\varrho}_{5}^{-})$ \\
\hline
Number & $0$ & $0$ & $0$ & $0$ & $0$ & $0$ & $0$ \\
\hline
\multicolumn{8}{|r|}{$C_{M_{z}}(\pi)\text{ mod }4 = 0$,\ $z_{2}(\pi)=0$} \\
\hline
\end{tabular}
\caption{The number of occupied corepresentations, mirror Chern numbers $C_{M_{z}}(k_{z})$ (modulo $4$)~\cite{ChenBernevigTCI,ChenTCI} (Eq.~(\ref{eq:MirrorChernIndicator})), and 2D TI indices~\cite{FuKaneInversion} $z_{2}(k_{z})$ ($C_{M_{z}}\text{ mod }2$) for the $s-d$-hybridized HOFA Dirac semimetal highlighted in the main text (Eq.~(\ref{eq:hinge}) with the parameters in Table~\ref{tb:3D}).  The Hamiltonian of the $k_{z}=0$ plane is equivalent to a 2D TCI with $C_{M_{z}}\text{ mod 4}=2$, whereas the Hamiltonian of the $k_{z}=\pi$ plane exhibits the same symmetry eigenvalues as a trivial insulator~\cite{ChenBernevigTCI,SlagerKanePrx,AshvinIndicators,ChenTCI,AshvinTCI}.}
\label{tb:corepssd}
\end{table}

Conversely, in the $s-d$-hybridized HOFA Dirac semimetal highlighted in the main text (Eq.~(\ref{eq:hinge})), neither of the high-symmetry BZ planes indexed by $k_{z}$ is topologically equivalent to a 2D TI.  Instead, as shown in Table~\ref{tb:corepssd}, in the $s-d$-hybridized HOFA Dirac semimetal in Eq.~(\ref{eq:hinge}) with the parameters in Table~\ref{tb:3D}, the Hamiltonian of the $k_{z}=0$ plane is equivalent to a 2D TCI~\cite{NagaosaDirac} ($C_{M_{z}}\text{ mod }4=2$).  This occurs because, unlike in the $p-d$-hybridized semimetal (Eq.~(\ref{eq:pdcoreps})), all of the bulk corepresentations $\bar{\sigma}$ in the $s-d$-hybridized semimetal have positive inversion characters $\chi_{\bar{\sigma}}(\mathcal{I})$.  In the surface states of both the $s-d$-hybridized semimetal (Fig.~\ref{fig:pd}(c)) and Eq.~(\ref{eq:hinge})) and the $p-d$-hybridized semimetal (Fig.~\ref{fig:pd}(g) and Eq.~(\ref{eq:pdTI})), only the gapped remnants of surface Fermi arcs (white arrows) are visible connecting the TI cones at $k_{z}=0$ (TCI cones of the $s-d$-hybridized model in (c)) to the projections of the bulk Dirac points (red arrows).  In both models, bulk quadrupolar SOC ($U({\bf k})\sin(k_{z})$ in Eq.~(\ref{eq:hinge}) and the $v_{Q1,2}$ terms in Eq.~(\ref{eq:pdTI})) has gapped the surface Fermi arc states and pushed them away from the Fermi energy, as described in Refs.~\onlinecite{KargarianDiracArc1,KargarianDiracArc2,ThomaleArc,KaminskiDiracArc}.  Calculating the bands of a $z$-directed rod of the $p-d$-hybridized semimetal described by Eq.~(\ref{eq:pdTI}) (Fig.~\ref{fig:pd}(h)), we observe four HOFA states connecting the projections of the bulk 3D Dirac points to those of the 2D face cones at $k_{z}=0$, the same number observed in the rod bands of the $s-d$-hybridized semimetal (Fig.~\ref{fig:pd}(d)).

\section{Evolution of the 1D Edge States of 2D TIs to the Corner Modes of QIs and Fragile Topological Phases}
\label{sec:boundary}

In this section, we use low-energy field theory to track the evolution of the 1D edge modes of a 2D TI and TCI in the presence of symmetry-breaking potentials.  Specifically, we show that the presence of quadrupolar magnetism (\emph{i.e.}, magnetism that preserves type-I wallpaper group~\cite{MagneticBook} $p4m$) can gap the edge states of a 2D TI and TCI and leave behind zero-dimensional corner modes.  In a 3D crystal, the Hamiltonians of 2D planes of the BZ indexed by $k_{z}$ can characterize 2D TIs and TCIs~\cite{NagaosaDirac} at the $\mathcal{T}$- (or $M_{z}$-) invariant values of $k_{z}$.  Thus, in a space group with BZ planes that preserve the crystal symmetries of the QI phase (Appendix~\ref{sec:bandrep}), one can consider the Hamiltonian of a BZ plane indexed by $k_{z}\neq 0,\pi$ as equivalent to a 2D $\mathcal{T}$-broken insulator with bulk quadrupolar magnetism.  In this description, the low-energy theory derived here tracks the evolution in $k_{z}$ of the 2D TI and TCI surface states of a 3D topological Dirac semimetal into the HOFA states on its 1D hinges as the strength of the effective magnetism in each 2D plane grows with increasing $k_{z}$.  During the final stages of preparing this complete work, the low-energy $k\cdot p$ theory of a related QI was also analyzed in Ref.~\onlinecite{DiracKP}, though that work did not relate their $k\cdot p$ theory to TIs, TCIs, HOFA states, and fragile topology, as we do in this section. 

However, it is important to note that \emph{not every} 2D TI and TCI can be gapped into a QI; only topological (crystalline) insulators with the same symmetries and occupied bulk $C_{4z}$ eigenvalues as those of the QI obstructed atomic limit (Appendices~\ref{sec:bandrep} and~\ref{sec:double}) can transition into QIs when gapped with quadrupolar magnetism, although other symmetries can also realize 2D insulators with topological corner states~\cite{HigherOrderTIChen,EmilCorner,WladPhotonCorner,WladCorners,FrankCorners,TeoWladTCI,EzawaCorner,ZeroBerry,FulgaAnon,HermeleSymmetry,BernevigMoTe2,YoungkukBLG}.  We show in Appendix~\ref{sec:TIboundary} that a 2D TI formed of hybridized $p_{z}$ and $d_{x^{2}-y^{2}}$ orbitals at the $1a$ position of wallpaper group $p4m1'$ or layer group $p4/mmm1'$ (Appendix~\ref{sec:bandrep}) gaps into a QI under $p4m$-preserving magnetism.  We then show in Appendix~\ref{sec:TItoTrivial} that a very similar 2D TI, formed instead of $s$ and $p_{z}$ orbitals, gaps into a trivial insulator under $p4m$-preserving magnetism.  Then, in Appendix~\ref{sec:TCIBoundary}, we show that the edge modes of the 2D TCI phase of Eq.~(\ref{eq:my2Dquad}) can evolve into two different kinds of 0D corner states, depending on the symmetries that are broken.  Specifically, depending on whether $\mathcal{T}$ or $M_{z}$ are broken while preserving $p4m$, the resulting 2D phase is either a QI or a fragile TI with the same corner charges (modulo $e$) and symmetry eigenvalues (Eq.~(\ref{eq:fragileIrreps})) as a QI.  If we break $\mathcal{T}$, the resulting phase is a QI, whether or not $M_{z}\times\mathcal{T}$ is preserved (Appendix~\ref{sec:bandrep}), whereas if we keep $\mathcal{T}$ and break $M_{z}$, the resulting phase is fragile (Appendix~\ref{sec:fragile}).  We first show that if $M_{z}$- and $\mathcal{T}$- symmetries are broken while $p4m$ is preserved, the edge states of this 2D TCI evolve into the singly degenerate spinful corner states of a $M_{z}\times\mathcal{T}$-broken 2D QI in $p4m$ (Appendix~\ref{sec:parameters}).  This $M_{z}\times\mathcal{T}$-broken QI phase appears in the $k_{z}\neq 0$ planes with HOFAs in the 3D Dirac semimetal in Fig.~\ref{fig:HingeSMmain}(e,h) of the main text.  Returning to the 2D TCI phase of Eq.~(\ref{eq:my2Dquad}), we then show that if just $M_{z}$ symmetry is broken while preserving $p4m1'$ (and without closing the bulk gap), the 1D TCI edge states evolve into the quarter-filled Kramers pairs of corner modes~\cite{ChenConvo,HigherOrderTIChen} of the fragile topological phase~\cite{AshvinFragile,JenFragile1,JenFragile2,AdrianFragile,BarryFragile,ZhidaBLG,AshvinBLG1,AshvinBLG2,AshvinFragile2,YoungkukMonopole,BernevigMoTe2,HarukiFragile,KoreanFragile,ZhidaFragileAffine,FragileKoreanInversion,KooiPartialNestedBerry,WiederAxion,WiederDefect} discussed in Appendix~\ref{sec:fragile} (equivalent to the Hamiltonian of the $k_{z}=0$ plane of the fragile topological Dirac semimetal in Fig.~\ref{fig:HingeSMmain}(e,h) of the main text).  Finally, in Appendix~\ref{sec:noMirror}, we show that the topological corner modes of a $4mm$-symmetric QI or fragile phase remain anomalous when the system is cut into in a geometry that breaks $M_{x,y}$ while preserving $C_{4z}$.  This is a necessary intermediate step in demonstrating that topological HOFA states are still present in nonsymmorphic Dirac semimetals (such as the archetypal Dirac semimetal $\alpha$-Cd$_3$As$_2$ in SG 142 ($I4_{1}/acd1'$)~\cite{ZJDirac,NagaosaDirac,CavaDirac2,StableCadmium}) whose glide reflections are formed from the combination of $M_{x,y}$ and lattice translations in the $xy$-plane, which cannot be preserved in $z$- (fourfold-axis-) directed nanorod geometries (Appendix~\ref{sec:unpinnedHOFAs}).

We also numerically calculate in Appendix~\ref{sec:boundaryNumbers} the position-space localization of the bulk and surface states of a HOFA Dirac semimetal terminated in a slab geometry.  We find that 2D BZ planes with Hamiltonians equivalent to QIs bind gapped edge Fermi arc-like states~\cite{BoundaryGreen}, and that 2D planes equivalent to trivial insulators generically have no edge states.  Taking planes at successive values of $k_{z}$ passing through one of the bulk Dirac points, we find that the localization lengths of all surface (and hinge) states diverge exactly at the Dirac point.  We therefore find no evidence supporting the presence of the additional surface cones bound to the 2D face projections of the bulk Dirac points.

\subsection{Gapping the Edge Modes of a 2D Topological Insulator with Quadrupolar Magnetism}
\label{sec:TIboundary}

We will first show that a 2D TI in layer group $p4/mmm1'$ (equivalent to the Hamiltonian of the $k_{z}=0$ plane of Eq.~(\ref{eq:pdTI})), formed of $p_{z}$ and $d_{x^{2}-y^{2}}$ orbitals at the $1a$ position (Fig.~\ref{fig:2Dmain}(a) of the main text), gaps into a 2D QI in the presence of magnetism that preserves the symmetries of 2D point group $4mm$.  We begin with this $p-d$-hybridized 2D TI, instead of the $s-d$-hybridized 2D TCI (Fig.~\ref{fig:2Dmain}(f-h) of the main text and Eq.~(\ref{eq:my2Dquad})) that is more closely related to the original formulation of the QI~\cite{multipole}, because the bulk $k\cdot p$ theory of a $p-d$-hybridized TI is simpler to analyze due to its linear dispersion.  In Appendix~\ref{sec:TCIBoundary}, we perform the analogous analysis of the quadratically dispersing $s-d$-hybridized TCI.  In four-band models of both $p_{z}-d_{x^{2}-y^{2}}$-hybridized 2D TIs and $s-d_{x^{2}-y^{2}}$-hybridized 2D TCIs, the bands of the two models exhibit the same $C_{4z}$ eigenvalues (Appendix~\ref{sec:pd}).  Instead, as shown in Tables~\ref{tb:corepspd} and~\ref{tb:corepssd}, the valence and conduction bands of the two insulators are distinguished by their parity eigenvalues.  As we will see in this section and in Appendix~\ref{sec:TCIBoundary}, when $\mathcal{I}$, $M_{z}$, and $\mathcal{T}$ symmetries are broken while $p4m$ is preserved, both $p_{z}-d_{x^{2}-y^{2}}$-hybridized 2D TIs and $s-d_{x^{2}-y^{2}}$-hybridized 2D TCIs (formed from orbitals at the $1a$ position of $p4/mmm1'$ as shown in Appendices~\ref{sec:bandrep} and~\ref{sec:pd}) evolve into QIs.

To demonstrate the presence of the 0D boundary modes in the QI phase that results from gapping a $p-d$-hybridized 2D TI in layer group $p4/mmm1'$, we will first derive the low-energy continuum $k\cdot p$ theory for the bulk of a $p-d$-hybridized 2D TI.  We will then use this $k\cdot p$ theory to solve a nested pair of Jackiw-Rebbi domain wall problems~\cite{JackiwRebbi}, \emph{i.e.} one for the (gapped) 1D edge states and another for the 0D corner states.

To form the bulk $k\cdot p$ Hamiltonian of a 2D $p-d$-hybridized TI~\cite{AndreiTI}, we begin by expanding Eq.~(\ref{eq:pdTI}) about the $\Gamma$ point to linear order and fixing $k_{z}=0$:
\begin{equation}
\mathcal{H}_{\Gamma}({\bf k}) = m\tau^{z} + vk_{x}\tau^{x}\sigma^{y} + vk_{y}\tau^{x}\sigma^{x},
\label{eq:TIGamma}
\end{equation}
where we have taken $t_{PH}\rightarrow 0$, have combined all of the terms proportional to $\tau^{z}$ into a single coefficient $m$, and have relabeled $v_{s}\rightarrow v$.  We also note that in Eq.~(\ref{eq:pdTI}), we have suppressed factors of the lattice constants $a_{x,y}=a$; it will be useful for future approximations to highlight that most precisely, $v=v_{s}a$, where we have specialized to units where $a=1$.  Eq.~(\ref{eq:TIGamma}) has the symmetries of the little co-group of the $\Gamma$ point, which is isomorphic to point group $4/mmm1'$. The Hamiltonian transforms for each symmetry $g\in 4/mmm1'$ under:
\begin{equation}
\mathcal{H}_{\Gamma}(k_{x},k_{y}) \rightarrow g\mathcal{H}_{\Gamma}(gk_{x}g^{-1},gk_{y}g^{-1})g^{-1},
\end{equation}
given in the notation of Refs.~\onlinecite{bernevigBook,ArisBerry,ArisInversion,HourglassInsulator,Cohomological,DiracInsulator,VDBSheet}.  We summarize this transformation as:
\begin{equation}
g:\ g\mathcal{H}_{\Gamma}(gk_{x}g^{-1},gk_{y}g^{-1})g^{-1}.
\label{eq:transformNotation}
\end{equation}
In the notation of Eq.~(\ref{eq:transformNotation}), $\mathcal{H}_{\Gamma}(k_{x},k_{y})$ transforms in the symmetry representation given by:
\begin{align}
&\mathcal{T}:\ \sigma^{y}\mathcal{H}^{*}_{\Gamma}(-k_{x},-k_{y})\sigma^{y},\ M_{z}:\ \tau^{z}\sigma^{z}\mathcal{H}_{\Gamma}(k_{x},k_{y})\tau^{z}\sigma^{z},\ M_{x}:\ \sigma^{x}\mathcal{H}_{\Gamma}(-k_{x},k_{y})\sigma^{x}, \nonumber \\ 
&M_{y}:\ \sigma^{y}\mathcal{H}_{\Gamma}(k_{x},-k_{y})\sigma^{y},\ C_{4z}:\ \tau^{z}\left(\frac{\mathds{1}_{\sigma} - i\sigma^{z}}{\sqrt{2}}\right)\mathcal{H}_{\Gamma}(k_{y},-k_{x})\tau^{z}\left(\frac{\mathds{1}_{\sigma} + i\sigma^{z}}{\sqrt{2}}\right),
\label{eq:pdKspaceSyms}
\end{align}
where $\mathds{1}_{\sigma}$ is the $2\times 2$ identity in $\sigma$ space.  Eq.~(\ref{eq:pdKspaceSyms}) implies an inversion symmetry $\mathcal{I}=M_{x}M_{y}M_{z}$ that transforms $\mathcal{H}_{\Gamma}(k_{x},k_{y})$ under the representation:
\begin{equation}
\mathcal{I}:\ \tau^{z}\mathcal{H}_{\Gamma}(-k_{x},-k_{y})\tau^{z}.
\label{eq:pdISym}
\end{equation}
The $\tau^{z}$ contributions to $M_{z},\ \mathcal{I}$, and $C_{4z}$ in Eqs.~(\ref{eq:pdKspaceSyms}) and~(\ref{eq:pdISym}) reflect that $\mathcal{H}_{\Gamma}({\bf k})$ describes a 2D TI formed of hybridized $p_{z}$ and $d_{x^{2}-y^{2}}$ orbitals, as $p_{z}$ ($d_{x^{2}-y^{2}}$) orbitals are odd (even) under $M_{z}$ and $\mathcal{I}$ and even (odd) under $C_{4z}$.  $\mathcal{H}_{\Gamma}({\bf k})$ also exhibits a unitary particle-hole symmetry:
\begin{equation}
\{\mathcal{H}_{\Gamma}({\bf k}),\Pi\} = 0,\ \Pi = \tau^{x}\sigma^{z},
\label{eq:posPH}
\end{equation}
which we will relax in future steps in this calculation.  Eq.~(\ref{eq:TIGamma}) also respects a second unitary particle-hole symmetry:
\begin{equation}
\{\mathcal{H}_{\Gamma}({\bf k}),\tilde{\Pi}\} = 0,\ \tilde{\Pi} = \tau^{y}.
\label{eq:posPH2}
\end{equation}

We then Fourier transform $\mathcal{H}_{\Gamma}({\bf k})$ such that $k_{x,y}\rightarrow -i\partial_{x,y}$, and take $m$ to have a spatial dependence $m\rightarrow m(x,y)$.  Specifically, we choose the bulk gap $m(x,y)$ to be strongly negative within a region bounded by a circle of radius $R\gg a$, and strongly positive for values outside of this circle; we also take $m(x,y)$ to be isotropic in $\theta = \tan^{-1}(y/x)$.  This distribution of $m$ suggests that the position-space Hamiltonian is more naturally described in polar coordinates, and therefore we transform:
\begin{eqnarray}
\partial_{x} &=& \cos(\theta)\partial_{r} -\frac{1}{r}\sin(\theta)\partial_{\theta}, \nonumber \\
\partial_{y} &=& \sin(\theta)\partial_{r} + \frac{1}{r}\cos(\theta)\partial_{\theta},
\label{eq:polarCoordTrans}
\end{eqnarray}
such that the Hamiltonian in Eq.~(\ref{eq:TIGamma}) now takes the form:
\begin{equation}
\mathcal{H}_{\Gamma}(r,\theta) = m(r)\tau^{z} - iv\tau^{x}\left[\sigma^{1}(\theta)\partial_{r} + \frac{1}{r}\sigma^{2}(\theta)\partial_{\theta}\right],
\label{eq:polarTI}
\end{equation}
where, through a canonical transformation,
\begin{eqnarray}
\sigma^{1}(\theta) &=& \sin(\theta)\sigma^{x} + \cos(\theta)\sigma^{y}=\left(\begin{array}{cc}
0 & -ie^{i\theta} \\
ie^{-i\theta} & 0 \end{array}\right),\nonumber \\
\sigma^{2}(\theta) &=& \cos(\theta)\sigma^{x} -\sin(\theta)\sigma^{y} = \left(\begin{array}{cc}
0 & e^{i\theta} \\
e^{-i\theta} & 0 \end{array}\right), \nonumber \\
\{\sigma^{1}(\theta),\sigma^{2}(\theta)\}&=&0,\ \sigma^{1}(\theta)\sigma^{2}(\theta) = -i\sigma^{z}. 
\label{eq:pdSigmaDef}
\end{eqnarray}
In this section, we employ a similar notation for the position-space, polar-coordinate forms of the symmetries of $\mathcal{H}_{\Gamma}(r,\theta)$ as we previously employed in Eq.~(\ref{eq:transformNotation}); for each symmetry $g$, the Hamiltonian transforms under:
\begin{equation}
\mathcal{H}_{\Gamma}(r,\theta) \rightarrow g\mathcal{H}_{\Gamma}(grg^{-1},g\theta g^{-1})g^{-1}.
\end{equation}
We summarize this transformation as:
\begin{equation}
g:\ g\mathcal{H}_{\Gamma}(grg^{-1},g\theta g^{-1})g^{-1}.
\label{eq:polarNotation}
\end{equation}
In the notation of Eq.~(\ref{eq:polarNotation}), $\mathcal{H}_{\Gamma}(r,\theta)$ transforms in the symmetry representation given by:
\begin{align}
&\mathcal{T}:\ \sigma^{y}\mathcal{H}^{*}_{\Gamma}(r,\theta)\sigma^{y},\ M_{z}:\ \tau^{z}\sigma^{z}\mathcal{H}_{\Gamma}(r,\theta)\tau^{z}\sigma^{z},\ M_{x}:\ \sigma^{x}\mathcal{H}_{\Gamma}(r,\pi-\theta)\sigma^{x}, \nonumber \\ 
&M_{y}:\ \sigma^{y}\mathcal{H}_{\Gamma}(r,-\theta)\sigma^{y},\ C_{4z}:\ \tau^{z}\left(\frac{\mathds{1}_{\sigma} - i\sigma^{z}}{\sqrt{2}}\right)\mathcal{H}_{\Gamma}(r,\theta+\pi/2)\tau^{z}\left(\frac{\mathds{1}_{\sigma} + i\sigma^{z}}{\sqrt{2}}\right), \nonumber \\
&\mathcal{I}:\ \tau^{z}\mathcal{H}_{\Gamma}(r,\theta+\pi)\tau^{z},
\label{eq:pdPolarSyms}
\end{align}
and both particle-hole symmetries remain in the same form as previously in Eqs.~(\ref{eq:posPH}) and~(\ref{eq:posPH2}):
\begin{equation}
\{\mathcal{H}_{\Gamma}(r,\theta),\Pi\}=0,\ \Pi = \tau^{x}\sigma^{z},
\label{eq:PolarPH}
\end{equation}
and:
\begin{equation}
\{\mathcal{H}_{\Gamma}(r,\theta),\tilde{\Pi}\}=0,\ \tilde{\Pi} = \tau^{y}.
\label{eq:PolarPH2}
\end{equation}
We search for zero-energy bound states of Eq.~(\ref{eq:polarTI}) on a disc geometry:
\begin{equation} 
\mathcal{H}_{\Gamma}(r,\theta)|\psi(r,\theta)\rangle = 0.  
\label{eq:polarTIdoubley}
\end{equation}
To solve Eq.~(\ref{eq:polarTIdoubley}), we separate variables by left-multiplying by $\tau^{x}$; after canceling a factor of $-i$:
\begin{equation}
\left[m(r)\tau^{y} + v\sigma^{1}(\theta)\partial_{r}\right]|\psi(r,\theta)\rangle = -\frac{v}{r}\sigma^{2}(\theta)\partial_{\theta}|\psi(r,\theta)\rangle.
\label{eq:separated}
\end{equation}
Because we are only interested in solving Eq.~(\ref{eq:separated}) in the (linear) $k\cdot p$ regime, we will use a series of approximations to find a diagonal solution; though these approximations are not strictly necessary, they provide considerable convenience in the early stages of this calculation while ultimately not affecting the final (topological) result.  First, we recognize that the bound state $|\psi(r,\theta)\rangle$ is almost entirely localized at $r\approx R$, where $m(r)\rightarrow 0$.  As $R\gg a$, where $a$ is the lattice spacing and $v\propto a$, the right-hand side of Eq.~(\ref{eq:separated}) vanishes to leading order:
\begin{equation}
\left[m(r)\tau^{y} + v\sigma^{1}(\theta)\partial_{r}\right]|\psi(r,\theta)\rangle \approx 0.
\label{eq:ApproxEqSep}
\end{equation}
We note that Eq.~(\ref{eq:separated}) can still be exactly solved without exploiting this approximation (it is the $k\cdot p$ differential equation for the edge states of a circular 2D TI, whose exact solution is a 1D Dirac fermion subject to the effects of curvature~\cite{MirlinCircleTI,FranzWormhole}).  Nevertheless, for the purpose of the explicit proofs in this section, Eq.~(\ref{eq:ApproxEqSep}) is advantageous in that it can be diagonalized simply by left-multiplying by $\tau^{y}$ and integrating, which allows us to circumvent at this stage of the calculation some of the complications that arise from the circular geometry (\emph{e.g.} the fact that $\partial_{\theta}$ acts on $\sigma^{1,2}(\theta)$ and the $1/r$ dependence of the right-hand side of Eq.~(\ref{eq:separated})):
\begin{equation}
|\psi_{1,2}(r,\theta)\rangle = \frac{1}{\sqrt{N}}e^{-\frac{1}{v}\int_{R}^{r}m(r')dr'}|\tau^{y}_{\pm}\sigma^{1}_{\pm}(\theta)\rangle = \mathcal{R}(r)|\tau^{y}_{\pm}\sigma^{1}_{\pm}(\theta)\rangle,
\label{eq:boundbound}
\end{equation}
where the normalization constant $N$ has the units of length squared, as the radial part of the measure in polar coordinates is $rdr$:
\begin{equation}
\int_{0}^{\infty}rdr|\mathcal{R}(r)|^{2} = 1. 
\end{equation}
In Eq.~(\ref{eq:boundbound}):
\begin{equation}
|\tau^{y}_{\pm}\sigma^{1}_{\pm}(\theta)\rangle = |\tau^{y}_{\pm}\rangle\otimes|\sigma^{1}_{\pm}(\theta)\rangle, 
\end{equation}
where $|\tau^{i}_{\pm},\sigma^{j}_{\pm}\rangle$ are the eigenstates with eigenvalues $\pm1$ of the $2\times2$ Pauli matrices $\tau^{i}$ and $\sigma^{j}$.  To leading order, Eq.~(\ref{eq:boundbound}) indicates that there are two nondispersing zero modes localized on the boundary of this circle with radius $R$, or close to the region where $m(r)=0$.  We note that, in the limit that the mass $m(r')$ in Eq.~(\ref{eq:boundbound}) is rapidly changing in the vicinity of $r\approx R$ (\emph{i.e.}, that $\left|\frac{dm(r')}{dr'}\right| \gg \frac{v}{R^{2}}$), the radial component $\mathcal{R}(r)$ simplifies:
\begin{equation}
|\mathcal{R}(r)|^{2} \rightarrow \frac{1}{r}\delta(r-R).
\label{eq:deltaFunction}
\end{equation}
However, more generally, like with the SSH chain~\cite{SSH,RiceMele}, the presence or absence of zero modes of the form of Eq.~(\ref{eq:boundbound}) that are localized in the vicinity of $r\approx R$ does not depend on the form of $m(r)$ -- it only depends on whether $m(r)$ changes sign at~\cite{CharlieChapter} $r=R$.

For subsequent calculations, we will find that the symmetries of the edge Hamiltonian appear in a more familiar form in the rotated basis:
\begin{eqnarray}
|\phi_{1}(r,\theta)\rangle &=& \frac{1}{\sqrt{2}}\left(|\psi_{1}(r,\theta)\rangle + |\psi_{2}(r,\theta)\rangle\right) = \frac{\mathcal{R}(r)}{\sqrt{2}}\left(\begin{array}{c}
-e^{i\theta} \\
0 \\
0 \\
1\end{array}\right)=R(r)|\xi_{1}(\theta)\rangle \nonumber \\
|\phi_{2}(r,\theta)\rangle &=& -\frac{ie^{-i\theta}}{\sqrt{2}}\left(|\psi_{1}(r,\theta)\rangle - |\psi_{2}(r,\theta)\rangle\right) = \frac{\mathcal{R}(r)}{\sqrt{2}}\left(\begin{array}{c} 
0 \\
e^{-i\theta} \\
1 \\
0\end{array}\right)=R(r)|\xi_{2}(\theta)\rangle,
\label{eq:defPhi}
\end{eqnarray}
a transformation that we are free to make because $|\phi_{1,2}(\theta)\rangle$ are degenerate (zero modes) at all values of $\theta$ at this stage of the calculation.

We now perturbatively restore the angular velocity term from Eq.~(\ref{eq:polarTI}) by projecting it into the basis of the edge states of $|\phi_{1,2}(r,\theta)\rangle$, integrating out $r$, and exploiting Eq.~(\ref{eq:deltaFunction}): 
\begin{eqnarray}
\mathcal{H}_{edge,ij}^{TI}(\theta) &=& -iv\langle\phi_{i}(r,\theta)|\frac{1}{r}\tau^{x}\sigma^{2}(\theta)\partial_{\theta}|\phi_{j}(r,\theta)\rangle \nonumber \\
 &=& -iv\int_{0}^{\infty}rdr\left(\frac{|\mathcal{R}(r)|^{2}}{r}\right)\left[\langle\langle\xi_{i}(\theta)|\tau^{x}\sigma^{2}(\theta)|\partial_{\theta}\xi_{j}(\theta)\rangle\rangle + \langle\langle\xi_{i}(\theta)|\tau^{x}\sigma^{2}(\theta)|\xi_{j}(\theta)\rangle\rangle\partial_{\theta}\right] \nonumber \\
\mathcal{H}_{edge}^{TI}(\theta) &=& \frac{v}{R}\left(\frac{1}{2}\mathds{1}_{s} + is^{z}\partial_{\theta}\right),
\label{eq:projecty}
\end{eqnarray}
where $s^{z}$ is a Pauli matrix and $\mathds{1}_{s}$ is the identity matrix in the $2\times 2$ basis of $|\phi_{1,2}(r,\theta)\rangle$, the $\langle\langle$ and $\rangle\rangle$ symbols in the second line indicate $\theta$-independent contractions over $4\times 4$ matrices.  The constant term $(v/2R)\mathds{1}_{s}$ arises due to the action of $\partial_{\theta}$ on $|\phi_{1,2}(r,\theta)\rangle$ in Eq.~(\ref{eq:projecty}).  The form of this term depends on the choice of gauge in Eq.~(\ref{eq:defPhi}); it will be useful for future calculations to note that the constant term disappears under the anti- ($4\pi$-) periodic gauge transformation:
\begin{equation}
|\xi_{1}(r,\theta)\rangle \rightarrow e^{-i\theta/2}|\xi_{1}(r,\theta)\rangle = |\tilde{\xi}_{1}(r,\theta)\rangle,\ |\xi_{2}(r,\theta)\rangle \rightarrow e^{i\theta/2}|\xi_{2}(r,\theta)\rangle = |\tilde{\xi}_{2}(r,\theta)\rangle,
\label{eq:firstAntiperiod}
\end{equation}
where $|\partial_{\theta}\tilde{\xi}_{1,2}\rangle$ are the positive and negative eigenstates of $\tau^{x}\sigma^{2}(\theta)$:
\begin{eqnarray}
|\tilde{\xi}_{1}(\theta)\rangle &=& \frac{1}{\sqrt{2}}\left(\begin{array}{c}
-e^{\frac{i\theta}{2}} \\
0 \\
0 \\
e^{-\frac{i\theta}{2}}\end{array}\right),\ |\partial_{\theta}\tilde{\xi}_{1}(\theta)\rangle = \frac{i}{2\sqrt{2}}\left(\begin{array}{c}
-e^{\frac{i\theta}{2}} \\
0 \\
0 \\
-e^{-\frac{i\theta}{2}}\end{array}\right),\ \tau^{x}\sigma^{2}(\theta)|\partial_{\theta}\tilde{\xi}_{1}(\theta)\rangle = \frac{i}{2\sqrt{2}}\left(\begin{array}{c}
-e^{\frac{i\theta}{2}} \\
0 \\
0 \\
-e^{-\frac{i\theta}{2}}\end{array}\right) \nonumber \\
|\tilde{\xi}_{2}(\theta)\rangle &=&  \frac{1}{\sqrt{2}}\left(\begin{array}{c}
0 \\
e^{-\frac{i\theta}{2}} \\
e^{\frac{i\theta}{2}} \\
0\end{array}\right),\ |\partial_{\theta}\tilde{\xi}_{2}(\theta)\rangle = \frac{i}{2\sqrt{2}}\left(\begin{array}{c}
0 \\
-e^{-\frac{i\theta}{2}} \\
e^{\frac{i\theta}{2}} \\
0\end{array}\right),\ \tau^{x}\sigma^{2}(\theta)|\partial_{\theta}\tilde{\xi}_{2}(\theta)\rangle = \frac{i}{2\sqrt{2}}\left(\begin{array}{c}
0 \\
e^{-\frac{i\theta}{2}} \\
-e^{\frac{i\theta}{2}} \\
0\end{array}\right),
\end{eqnarray}
such that in Eq.~(\ref{eq:projecty}):
\begin{equation}
\langle\langle\tilde{\xi}_{i}(\theta)|\tau^{x}\sigma^{2}(\theta)|\tilde{\xi}_{j}(\theta)\rangle\rangle=0 \text{ for } i,j=1,2.  
\end{equation}

In the basis of $|\phi_{1,2}(r,\theta)\rangle$, the symmetries from Eq.~(\ref{eq:pdPolarSyms}) transform $\mathcal{H}_{edge}^{TI}(\theta)$ under the representation:
\begin{align}
&\mathcal{T}:\ s^{y}(\mathcal{H}_{edge}^{TI}(\theta))^{*}s^{y},\ M_{z}:\ s^{z}\mathcal{H}_{edge}^{TI}(\theta)s^{z},\ M_{x}:\ s^{x}\mathcal{H}_{edge}^{TI}(\pi-\theta)s^{x}, \nonumber \\ 
&M_{y}:\ s^{y}\mathcal{H}_{edge}^{TI}(-\theta)s^{y},\ C_{4z}:\ \left(\frac{\mathds{1}_{s} - is^{z}}{\sqrt{2}}\right)\mathcal{H}_{edge}^{TI}(\theta+\pi/2)\left(\frac{\mathds{1}_{s} + is^{z}}{\sqrt{2}}\right), \nonumber \\
&\mathcal{I}:\ \mathcal{H}_{edge}^{TI}(\theta+\pi),
\label{eq:sBasis}
\end{align}
Crucially, in the basis of $|\phi_{1,2}(r,\theta)\rangle$, the particle-hole symmetry from Eq.~(\ref{eq:PolarPH}) takes a $\theta$-dependent form:
\begin{equation}
\Pi(\theta) = s^{1}(\theta),
\label{eq:edgePH}
\end{equation}
where:
\begin{eqnarray}
s^{1}(\theta) &=& \cos(\theta)s^{x} - \sin(\theta)s^{y}=\left(\begin{array}{cc}
0 & e^{i\theta} \\
e^{-i\theta} & 0 \end{array}\right),\nonumber \\
s^{2}(\theta) &=& \sin(\theta)s^{x} + \cos(\theta)s^{y}=\left(\begin{array}{cc}
0 & -ie^{i\theta} \\
ie^{-i\theta} & 0 \end{array}\right),\nonumber \\
\{s^{1}(\theta),s^{2}(\theta)\}&=&0,\ s^{1}(\theta)s^{2}(\theta) = is^{z}. 
\label{eq:defSBasis}
\end{eqnarray}
The $\theta$ dependence of particle-hole symmetry in Eq.~(\ref{eq:edgePH}) and the presence of the constant term $(v/2R)\mathds{1}_{s}$ in Eq.~(\ref{eq:projecty}) reflect the extrinsic curvature of the circular boundary.  Though $(v/2R)\mathds{1}_{s}$ moves the center of the spectrum away from $E=0$, the $\theta$-dependent particle-hole symmetry is still preserved, as the Hamiltonian acts on $\theta$:
\begin{equation}
\{\mathcal{H}_{edge}^{TI}(\theta),\Pi(\theta)\} = 0,
\label{eq:manifestPH1}
\end{equation}
where specifically:
\begin{equation}
\partial_{\theta}s^{1}(\theta) = -s^{2}(\theta) + s^{1}(\theta)\partial_{\theta}.
\label{eq:manifestPH2}
\end{equation}
As in Eq.~(\ref{eq:PolarPH2}), Eq.~(\ref{eq:projecty}) similarly also respects a second theta-dependent particle-hole symmetry of the form:
\begin{equation}
\tilde{\Pi}(\theta) = s^{2}(\theta),
\label{eq:edgePH2}
\end{equation}
where:
\begin{equation}
\partial_{\theta}s^{2}(\theta) = s^{1}(\theta) + s^{2}(\theta)\partial_{\theta}.
\label{eq:manifestPH3}
\end{equation}
Though many previous works have demonstrated the presence of localized 0D modes in systems with sharp corners~\cite{multipole,WladTheory,ZeroBerry,FulgaAnon,HigherOrderTIChen,EmilCorner,WladPhotonCorner,WladCorners,FrankCorners,TeoWladTCI,EzawaCorner,YoungkukBLG}, for which the curvature is zero on the edges and singular on the corners, our explicit calculation of the QI boundary states in a geometry with constant curvature (\emph{i.e.}, on a disc) will allow us to separate the extrinsic effects of sharp corners from the intrinsic (higher-order) topological bulk-boundary (-corner) correspondence of QIs.  It is also important to note that, as $\mathcal{H}_{edge}^{TI}(\theta)$ describes the edge Hamiltonian at $r\sim R$, it is invariant under fewer symmetry restrictions than a Hamiltonian localized in a region containing $r=0$ (\emph{i.e.}, the origin of the symmetry operations of the point group $4/mmm1'$).  At generic values of $\theta$, $\mathcal{H}_{edge}^{TI}(\theta)$ (Eq.~(\ref{eq:projecty})) is only invariant under the $\theta$-preserving action of $\mathcal{T}$ and $M_{z}$ in Eq.~(\ref{eq:sBasis}); the other symmetries of $4/mmm1'$ act at generic angles $\theta$ to relate $\mathcal{H}_{edge}^{TI}(\theta)$ to its value at another, symmetry-related generic angle $\theta'$.  Point group $4/mmm1'$ also has four mirror lines in the $xy$-plane~\cite{BilbaoPoint} (Fig.~\ref{fig:2Dmain}(a) of the main text), $M_{x,y}$ and $M_{x\pm y}$ that fix the angles:
\begin{equation}
\tilde{\theta}_{n} = n\pi/4,\ n\in \mathbb{Z}, 
\label{eq:mirrorFix}
\end{equation}
such that one of $M_{x,y}$ or $M_{x\pm y}$ is a symmetry of $\mathcal{H}_{edge}^{TI}(\theta)$ at each $\tilde{\theta}_{n}$:
\begin{equation}
M_{x,y}\mathcal{H}_{edge}^{TI}(\tilde{\theta}_{n})M_{x,y}^{-1}\text{ or }M_{x\pm y}\mathcal{H}_{edge}^{TI}(\tilde{\theta}_{n})M_{x \pm y}^{-1}.
\end{equation}
When $M_{z}$ and $\mathcal{T}$ symmetries are broken to gap the TI edge states, we will see that QI-nontrivial 0D states become bound to the $\tilde{\theta}_{n}$, \emph{i.e.}, the fixed points (angles) of point group $4mm$.

We now gap $\mathcal{H}_{edge}^{TI}(\theta)$ (Eq.~(\ref{eq:projecty})) by introducing quadrupolar ($p4m$-preserving) magnetism.  We begin by proposing the most general $r$-independent bulk potential to add to $\mathcal{H}_{\Gamma}(r,\theta)$ (Eq.~(\ref{eq:polarTI})):
\begin{equation}
U(\theta) = \sum_{L_{z}=0}^{\infty}\sum_{\mu=\pm}m_{L_{z}}^{\mu}\Gamma^{L_{z},\mu}f^{\mu}_{L_{z}}(\theta),
\label{eq:GeneralU}
\end{equation}
where $\Gamma^{L_{z},\mu}$ is a $4\times 4$ matrix in the basis of $\tau \otimes \sigma$ and $f^{\pm}_{L_{z}}(\theta)$ is a real circular harmonic~\cite{jacksonEM,mcQuarriePchem,harmonics1,harmonics2}:
\begin{equation}
f^{+}_{L_{z}}(\theta) = \cos(L_{z}\theta),\ f^{-}_{L_{z}}(\theta) = \sin(L_{z}\theta),
\label{eq:defCircularHarmonic}
\end{equation}
with angular momentum $L_{z}$.  The sum in Eq.~(\ref{eq:GeneralU}) is taken over all possible products of $4\times 4$ matrices and $f^{\pm}_{L_{z}}(\theta)$ that respect the symmetries of point group $4mm$ (the point group of $p4m$).  In terms of the more familiar spherical harmonics, the functions $f^{\pm}_{L_{z}}(\theta)$ derive from the set of ``cubic harmonics''~\cite{harmonics2}, \emph{i.e.}, the real-valued linear combinations of the spherical harmonics that define the angular dependence of the wavefunctions of the atomic orbitals~\cite{harmonics1,harmonics2}.  Specifically, choosing the $z$-axis to be the plane normal, the real circular harmonics are obtained by taking $z\rightarrow 0$ in the subset of cubic harmonics (atomic orbitals) for which the total angular momentum $L$ equals the magnitude of the $z$-component of the angular momentum $L_{z}$, which we refer to as the angular momentum of the circular harmonic~\cite{jacksonEM,mcQuarriePchem,harmonics1,harmonics2} (Eq.~(\ref{eq:defCircularHarmonic})).

We next explicitly expand Eq.~(\ref{eq:GeneralU}) by choosing all possible mass terms that respect the symmetries of $4mm$ (Eq.~(\ref{eq:pdPolarSyms})) while containing $4\times 4$ matrices $\Gamma^{L_{z},\mu}$ that anticommute with the Dirac matrix coefficients of the bulk mass and angular velocity terms in $\mathcal{H}_{\Gamma}(r,\theta)$ (Eq.~(\ref{eq:polarTI})): $m(r)\tau^{z}$ and $-i(v/r)\tau^{x}\sigma^{2}(\theta)\partial_{\theta}$, respectively.  This guarantees that the terms in $U(\theta)$, when individually added to $\mathcal{H}_{\Gamma}(r,\theta)$, strictly enlarge the bulk gap and open an edge gap~\cite{HOTIBismuth}.  Expressing $U(\theta)$ as a sum of terms that respect the symmetries of $4mm$ ($C_{4z}$ and $M_{x,y}$ in Eq.~(\ref{eq:pdPolarSyms})):
\begin{align}
U(\theta) &= \tau^{x}\sigma^{z}\left[m^{-}_{2}\sin(2\theta) + m^{-}_{6}\sin(6\theta) + m^{-}_{10}\sin(10\theta) + \ldots\ \right] \nonumber \\
&+ \tau^{y}\left[m^{+}_{2}\cos(2\theta) + m^{+}_{6}\cos(6\theta) + m^{+}_{10}\cos(10\theta) + \ldots\ \right] \nonumber \\
&+ \tau^{x}\sigma^{1}(\theta)\left[m_{0}^{+} + m^{+}_{4}\cos(4\theta) + m^{+}_{8}\cos(8\theta) + \ldots\ \right] \nonumber \\
&+ \tau^{y}\sigma^{2}(\theta)\left[m^{-}_{4}\sin(4\theta) + m^{-}_{8}\sin(8\theta) + m^{-}_{12}\sin(12\theta) + \ldots\ \right].
\label{eq:pdMasses}
\end{align}
We observe that the terms in $U(\theta)$ group into circular harmonics of increasing $L_{z}$ multiplied by one of four $4\times 4$ matrices.  We then project $U(\theta)$ into the basis of the edge modes $|\phi_{1,2}(r,\theta)\rangle$, following the procedure in Eq.~(\ref{eq:projecty}) and Ref.~\onlinecite{HOTIBismuth}:
\begin{eqnarray}
U_{edge,ij}(\theta) &=& \langle\phi_{i}(r,\theta)|U(\theta)|\phi_{j}(r,\theta)\rangle \nonumber \\
&=& \sum_{L_{z},\mu}m_{L_{z}}^{\mu}\langle\phi_{i}(r,\theta)|\Gamma^{L_{z},\mu}|\phi_{j}(r,\theta)\rangle f^{\mu}_{L_{z}}(\theta), \\
U_{edge}(\theta) &=&  s^{1}(\theta)\left[m^{-}_{2}\sin(2\theta) + m^{-}_{6}\sin(6\theta) + m^{-}_{10}\sin(10\theta) + \ldots\ \right] \nonumber \\
&+& s^{2}(\theta)\left[m^{+}_{2}\cos(2\theta) + m^{+}_{6}\cos(6\theta) + m^{+}_{10}\cos(10\theta) + \ldots\ \right].
\label{eq:UedgePD}
\end{eqnarray}
We observe that the terms in $U(\theta)$ that commute with $\tau^{y}\sigma^{1}(\theta)$ ($m^{\pm}_{2+4a}$) have nonzero edge projections, whereas the terms that anticommute with $\tau^{y}\sigma^{1}(\theta)$ ($m^{\pm}_{4a}$) project to zero in $U_{edge}(\theta)$ and hence do not open an edge gap.  The nonzero terms in $U_{edge}(\theta)$ break $\mathcal{I},\ M_{z}$, and $\mathcal{T}$ symmetries in the bulk and on the edge, while respecting the combined magnetic symmetries $\mathcal{I}\times\mathcal{T}$ and $M_{z}\times\mathcal{T}$ (Eq.~(\ref{eq:sBasis})).  To understand this result, we form the expression for the projector into the positive eigenspace of $\tau^{y}\sigma^{1}(\theta)$, \emph{i.e.} the space of eigenvectors with eigenvalues $\lambda_{\tau^{y}\sigma^{1}(\theta)}=1$:
\begin{equation}
P_{\lambda_{\tau^{y}\sigma^{1}(\theta)}=1}=\frac{\mathds{1} + \tau^{y}\sigma^{1}(\theta)}{2}.
\label{eq:defProj}
\end{equation}
In order for a generic $4\times 4$ matrix $\Gamma^{L_{z},\mu}$ to have a nonzero projection into the basis of $|\phi_{1,2}(r,\theta)\rangle$, it must satisfy:
\begin{equation}
P_{\lambda_{\tau^{y}\sigma^{1}(\theta)}=1}\Gamma^{L_{z},\mu}P_{\lambda_{\tau^{y}\sigma^{1}(\theta)}=1} \neq 0.
\label{eq:NonzeroP}
\end{equation}
Eq.~(\ref{eq:NonzeroP}) can only be satisfied if:
\begin{equation}
\{\tau^{y}\sigma^{1}(\theta),\Gamma^{L_{z},\mu}\}\neq 0.
\label{eq:antiantiprojector}
\end{equation}
As the basis of edge states $|\phi_{1,2}(r,\theta)\rangle$ is formed from linear combinations of the positive eigenstates of $\tau^{y}\sigma^{1}(\theta)$ (Eq.~(\ref{eq:defPhi})), then Eqs.~(\ref{eq:defProj}),~(\ref{eq:NonzeroP}), and~(\ref{eq:antiantiprojector}) imply that the $m^{\pm}_{4a}$ terms in $U(\theta)$ (Eq.~(\ref{eq:pdMasses})) project to zero in $U_{edge}(\theta)$ (Eq.~(\ref{eq:UedgePD})).  This indicates that, for the $p-d$-hybridized TI in this section, bulk $p4m$-preserving magnetism can only open an edge gap with:
\begin{equation}
L_{z}^{QI} = 2 + 4a,\ a\in\mathbb{Z},
\label{eq:nontrivialLz}
\end{equation}
where the $m^{\pm}_{2}$ terms in Eq.~(\ref{eq:UedgePD}), in particular, are proportional to the circular harmonics of $d_{x^{2}-y^{2}}$ and $d_{xy}$ orbitals, respectively~\cite{jacksonEM,mcQuarriePchem,harmonics1,harmonics2}.

We next confirm that $U_{edge}(\theta)$ is proportional to the representations of $M_{x,y}$ and $M_{x\pm y}=C_{4z}^{\pm 1}M_{x}$ (Eq.~(\ref{eq:sBasis})) at the first four mirror-invariant points in $\theta$ (Eq.~(\ref{eq:mirrorFix})):
\begin{equation}
U\bigg(0\bigg) \propto s^{y},\ U\bigg(\frac{\pi}{4}\bigg) \propto (s^{x}-s^{y}),\ U\bigg(\frac{\pi}{2}\bigg) \propto -s^{x},\ U\bigg(\frac{3\pi}{4}\bigg) \propto (s^{x}+s^{y}),
\label{eq:mirrorRepos}
\end{equation}
and thus verify that $U_{edge}(\theta)$ respects the mirror symmetries at those points.  We define the edge Hamiltonian of the QI to be:
\begin{equation}
\mathcal{H}_{edge}^{QI}(\theta) = \mathcal{H}_{edge}^{TI}(\theta) + U_{edge}(\theta).
\label{eq:fullTIedge}
\end{equation}
If we truncate $U_{edge}(\theta)$ (Eq.~(\ref{eq:UedgePD})) to its leading two $L_{z}=2$ terms, $\mathcal{H}_{edge}^{QI}(\theta)$ exhibits a gap in the long-wavelength limit of:
\begin{equation}
\Delta(\theta) = 2\sqrt{(m^{-}_{2})^{2}\sin^{2}(2\theta) + (m^{+}_{2})^{2}\cos^{2}(2\theta)}.
\label{eq:simpleGapPD}
\end{equation}

With this formality established, we now show that Eq.~(\ref{eq:fullTIedge}) exhibits a quantized quadrupole moment.  We will first demonstrate this in a particle-hole-symmetric limit, after which we will show that the quadrupole moment of Eq.~(\ref{eq:fullTIedge}) remains quantized when particle-hole symmetry is relaxed.  Particle-hole symmetry as represented in Eq.~(\ref{eq:edgePH}) ($\Pi(\theta)=s^{1}(\theta)$) is also a symmetry of all of the $m^{+}_{L_{z}^{QI}}$ mass terms (but not the $m^{-}_{L_{z}^{QI}}$ terms) in Eq.~(\ref{eq:UedgePD}) (most generally, as indicated in Eq.~(\ref{eq:edgePH2}), there is a also second particle-hole symmetry $\tilde{\Pi}(\theta)=s^{2}(\theta)$ that is also a symmetry of $\mathcal{H}_{edge}^{TI}$ and only the $m^{-}_{L_{z}^{QI}}$ mass terms in Eq.~(\ref{eq:UedgePD})).  In the specific particle-hole symmetric limit of Eq.~(\ref{eq:fullTIedge}) in which the only particle-hole symmetry is $\Pi(\theta)=s^{1}(\theta)$, we can first choose $m^{+}_{2}$ to be the only nonzero mass term in $U_{edge}(\theta)$.  In this limit, $\mathcal{H}_{edge}^{QI}(\theta)$ is gapless at $\theta=\theta_{n}$, where the first four independent values of $\theta_{n}$ are:
\begin{equation}
\theta_{n} = \pi/4 + n\pi/2,\ n\in\{0,1,2,3\}.
\label{eq:thetaNPD}
\end{equation}
We then solve for the zero modes bound at $\theta_{n}$, or at the values of $\theta$ at which $\cos(2\theta)$ changes sign, by formulating a Jackiw-Rebbi problem:  
\begin{equation}
\left[\frac{v}{R}\left(\frac{1}{2}\mathds{1}_{s} + is^{z}\partial_{\theta}\right) + m^{+}_{2}\cos(2\theta)s^{2}(\theta)\right]|\tilde{\Theta}(\theta)\rangle =0.
\label{eq:fullThetaPD}
\end{equation}
We will solve for the bound states of Eq.~(\ref{eq:fullThetaPD}) in two steps: first we will remove the constant curvature term $(v/2R)\mathds{1}_{s}$ in Eq.~(\ref{eq:fullThetaPD}) by transforming $|\tilde{\Theta}(\theta)\rangle$ into the antiperiodic gauge in Eq.~(\ref{eq:firstAntiperiod}), which will allow Eq.~(\ref{eq:fullThetaPD}) to be solved using the same method that we used for Eq.~(\ref{eq:ApproxEqSep}).  We will then Taylor expand Eq.~(\ref{eq:fullThetaPD}) around $\theta_{n}$ (Eq.~(\ref{eq:thetaNPD})) to solve for 0D bound states at each $\theta_{n}$.

We first explicitly demonstrate that transforming $|\tilde{\Theta}(\theta)\rangle$ into a wavefunction $|\Theta(\theta)\rangle$ in an antiperiodic gauge removes the constant curvature term in Eq.~(\ref{eq:fullThetaPD}).  We express the transformation between $|\tilde{\Theta}(\theta)\rangle$ and $|\Theta(\theta)\rangle$ as:
\begin{equation}
|\tilde{\Theta}(\theta)\rangle = U(\theta)|\Theta(\theta)\rangle.
\label{eq:UThetaTransformAnti}
\end{equation}
In order for the constant term to vanish, $|\tilde{\Theta}(\theta)\rangle$ must satisfy:
\begin{equation}
\left(\frac{1}{2}\mathds{1}_{s} + is^{z}\partial_{\theta}\right)|\tilde{\Theta}(\theta)\rangle = is^{z}\partial_{\theta}|\Theta(\theta)\rangle,
\label{eq:NoMoreCurves}
\end{equation}
such that $|\Theta(\theta)\rangle$ behaves as if it is a $\theta$-independent eigenstate of a linear Hamiltonian without curvature~\cite{GriffithsBook}.  Eqs.~(\ref{eq:UThetaTransformAnti}) and~(\ref{eq:NoMoreCurves}) imply that:
\begin{equation}
is^{z}\partial_{\theta}U(\theta) = -\frac{1}{2}U(\theta),
\end{equation}
which is satisfied by:
\begin{equation}
U(\theta) = \left(\begin{array}{cc}
e^{i\theta/2} & 0 \\
0 & e^{-i\theta/2} \end{array}\right).
\label{eq:antiperiod2}
\end{equation}
The antiperiodicity of Eq.~(\ref{eq:antiperiod2}) reflects that curvature in a circular (cylindrical) geometry acts as an effective $\pi$ flux~\cite{FranzWormhole}.  While one might be concerned by the antiperiodic boundary conditions of $|\Theta(\theta)\rangle$ in Eqs.~(\ref{eq:UThetaTransformAnti}) and (\ref{eq:antiperiod2}), we note that $|\Theta(\theta)\rangle$ will only be used here as the wavefunction of a single 0D bound state that is exponentially localized within a small vicinity of one of the angles $\theta_{n}$ (Eq.~(\ref{eq:thetaNPD})).  We postulate that, for the wavefunction of each 0D bound state at $\theta=\theta_{n}$, $2\pi$ periodicity can be restored, by adding a $\theta$-dependent local gauge transformation into the gapped region far away from $\theta_{n}$.  For sufficiently large circular boundaries in the thermodynamic limit, these smooth, but highly localized, gauge transformations should have a negligible effect on the (gauge-independent) spectrum~\cite{AshvinTCI}, because, for each 0D bound state, the $2\pi$-periodicity-restoring gauge transformation can be placed in a region where the bound state wavefunction is nearly zero (\emph{i.e.}, at $\theta_{n} + \pi$ for each bound state at $\theta_{n}$).  By substituting Eq.~(\ref{eq:UThetaTransformAnti}) into Eq.~(\ref{eq:fullThetaPD}), we remove the constant curvature term:
\begin{equation}
\left[i\frac{v}{R}s^{z}\partial_{\theta} + m^{+}_{2}\cos(2\theta)s^{2}(\theta)\right]|\Theta(\theta)\rangle =0.
\label{eq:ApproxThetaPD}
\end{equation}

We next expand Eq.~(\ref{eq:ApproxThetaPD}) around $\theta=\theta_{n}+\epsilon$ (Eq.~(\ref{eq:thetaNPD})), where $\epsilon$ is a small parameter, to form an angular Jackiw-Rebbi problem for the zero-energy normalizable bound state at each $\theta_{n}$:
\begin{equation}
\left[i\frac{v}{R}s^{z}\partial_{\epsilon} -(-1)^{n}m_{\theta}\epsilon s^{2}(\theta_{n})\right]|\Theta(\theta_{n},\epsilon)\rangle=0,
\label{eq:MoreApproxThetaPD}
\end{equation}
where $\theta_{n}$ is given in Eq.~(\ref{eq:thetaNPD}) such that $s^{2}(\theta_{n})$ is the matrix $s^{2}(\theta)$ in Eq.~(\ref{eq:defSBasis}) evaluated at $\theta_{n}$, and where:
\begin{equation}
m^{+}_{2}\cos[2(\theta_{n} + \epsilon)]\rightarrow -2m^{+}_{2}\epsilon\sgn[\sin(2\theta_{n})] = -(-1)^{n}m_{\theta}\epsilon,
\label{eq:ThetaNMass}
\end{equation}
where $m_{\theta} =2m^{+}_{2}$.  The factor of $-(-1)^{n}$ in Eq.~(\ref{eq:MoreApproxThetaPD}) enforces that, for increasing $\epsilon$, the domain-wall mass $m^{+}_{2}\cos(2\theta)\rightarrow-(-1)^{n}m_{\theta}\epsilon$ in Eq.~(\ref{eq:ThetaNMass}) exhibits a derivative with the respective signs $\{-,+,-,+\}$ at $\theta_{n} = \{\pi/4,3\pi/4,5\pi/4,7\pi/4\}$.  Next, we solve Eq.~(\ref{eq:ThetaNMass}) for all values of $\theta_{n}$ by left-multiplying by $s^{2}(\theta_{n})$ (exploiting that Eq.~(\ref{eq:defSBasis}) can be rearranged to obtain $s^{z}s^{2}(\theta_{n}) = -is^{1}(\theta_{n})$), and then integrating (exploiting that $\epsilon^{2}/2 = \int_{0}^{\epsilon}\ \epsilon' d\epsilon'$).  We find that, over the circumference of the circle, there are four bound states of the form:
\begin{eqnarray}
|\Theta(\theta_{n},\epsilon)\rangle &=& \frac{1}{\sqrt{N}}e^{-\lambda(\theta_{n})(-1)^{n}\frac{m_{\theta}R}{2v}\epsilon^{2}}|s^{1}(\theta_{n})\rangle_{\lambda(\theta_{n})}, \nonumber \\
&=& \frac{1}{\sqrt{N}}e^{-\frac{m_{\theta}R}{2v}\epsilon^{2}}|s^{1}(\theta_{n})\rangle_{\lambda(\theta_{n})}, 
\label{eq:pioverAll}
\end{eqnarray}
where we have simplified by exploiting that $|s^{1}(\theta_{n})\rangle_{\lambda(\theta_{n})}$ is the eigenstate of $s^{1}(\theta_{n})$ with eigenvalue:
\begin{equation}
\lambda(\theta_{n})=(-1)^{n}.
\label{eq:lambdathetasign}
\end{equation}
We therefore find that when $m_{2}^{+}$ is the only nonzero mass term in Eq.~(\ref{eq:UedgePD}), there are Jackiw-Rebbi zero modes~\cite{JackiwRebbi} localized to the zeroes $\theta_{n}$ of $\cos(2\theta)$:
\begin{equation}
|\Theta(\pi/4,\epsilon)\rangle\propto |s^{1}(\pi/4)\rangle_{+},\ |\Theta(3\pi/4,\epsilon)\rangle\propto |s^{1}(3\pi/4)\rangle_{-},\ |\Theta(5\pi/4,\epsilon)\rangle\propto |s^{1}(5\pi/4)\rangle_{+},\ |\Theta(7\pi/4,\epsilon)\rangle\propto |s^{1}(7\pi/4)\rangle_{-}, 
\label{eq:s1zeromodes}
\end{equation}
where all of the coefficients in the proportionalities are real and of the form of the Gaussian exponential in Eq.~(\ref{eq:pioverAll}).  As $\{C_{4z},s^{1}(\theta)\} =0$ ($\{M_{x,y},s^{1}(\theta)\}=0$), acting with $C_{4z}$ ($M_{x,y}$) on a positive eigenstate of $s^{1}(\theta_{n})$ transforms it to a \emph{negative} eigenstate of $s^{1}(C_{4z}\theta_{n}C_{4z}^{-1}$) ($s^{1}(M_{x,y}\theta_{n}M_{x,y}^{-1}$)) (Eq.~(\ref{eq:polarNotation})).  The set of four states in Eq.~(\ref{eq:s1zeromodes}) is left invariant under $C_{4z}$ and $M_{x,y}$, and thus the four zero modes as a set respect the symmetries of point group $4mm$.

\begin{figure}[h]
\centering
\includegraphics[width=0.7\textwidth]{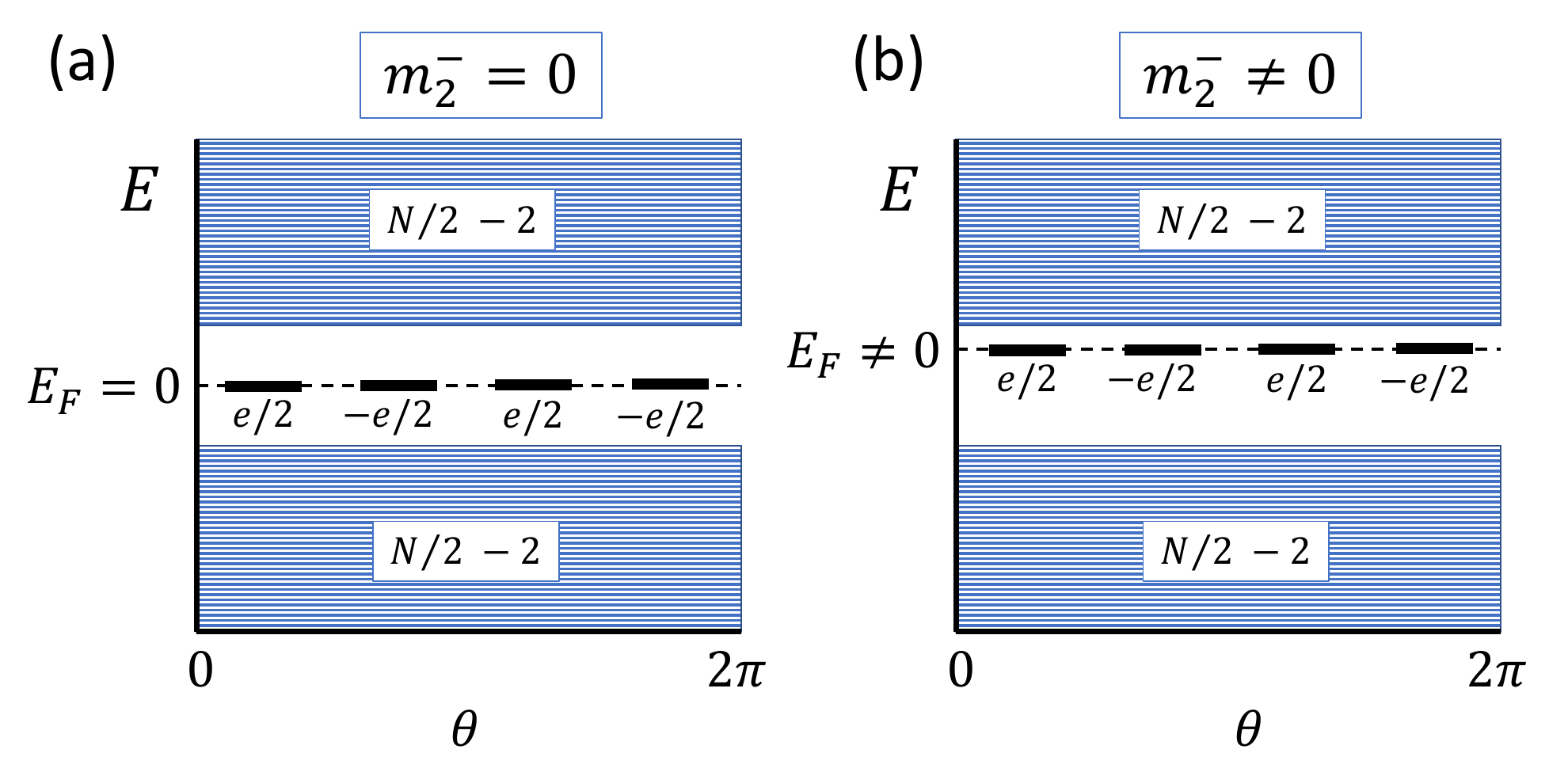}
\caption{Schematic energy spectra of QIs with and without particle-hole symmetry and open boundary conditions (OBC).  (a) The corner modes of the QI can be considered two Jackiw-Rebbi solitons and two antisolitions in the particle-hole-symmetric limit that $m_{2}^{+}$ is the only nonzero mass in Eq.~(\ref{eq:UedgePD}).  In this limit, there are therefore $(N/2)-2$ states in the valence and conduction manifolds, where $N$ is the total number of states.  When the four zero modes are half-filled, they exhibit a charge distribution (Eq.~(\ref{eq:chargePerN})) with an $e/2$ quadrupole moment about $r=0$ (Eq.~(\ref{eq:eOver2ForAll})).  (b) When one of the $m^{-}_{2+4a}$ terms in Eq.~(\ref{eq:UedgePD}) is perturbatively introduced, particle-hole symmetry $\Pi(\theta)$ (Eq.~(\ref{eq:edgePH})) is broken and all four 0D modes are uniformly raised in energy (Eq.~(\ref{eq:offsetE})); this is the same effect as introducing an identity term in the basis of the four 0D modes.  However, if this pattern of boundary modes remains half-filled throughout the breaking of particle-hole symmetry, then the four modes still exhibit an $e/2$ quadrupole moment; only the chemical potential has shifted such that $E_{F}\neq 0$.  The QI phase can thus be identified in a $\mathcal{T}$-broken, spinful, $p4m$-symmetric insulator by drawing a line across a gap in the energy spectrum calculated with OBC and counting the number of states below the gap~\cite{HigherOrderTIChen}, and then by comparing that number to the number of states below the same gap calculated with periodic boundary conditions (PBC).  If the difference in the number of states below the gap is $2 + 4a$, where $a\in\mathbb{Z}$, then the system is a QI (Eq.~(\ref{eq:eOver2ForAll})).  In (a) and (b), we depict schematic OBC spectra of a QI; in both cases, the number of states below the four corner modes is $(N/2)-2$, whereas the number of states below the gap in the PBC spectrum is implied to be $N/2$ in both (a) and (b).  The difference of $2$, combined with the presence of spinful $4mm$ symmetry, indicates that the bulk is a QI.}
\label{fig:cornerModes}
\end{figure}

\begin{figure}[h]
\centering
\includegraphics[width=0.4\textwidth]{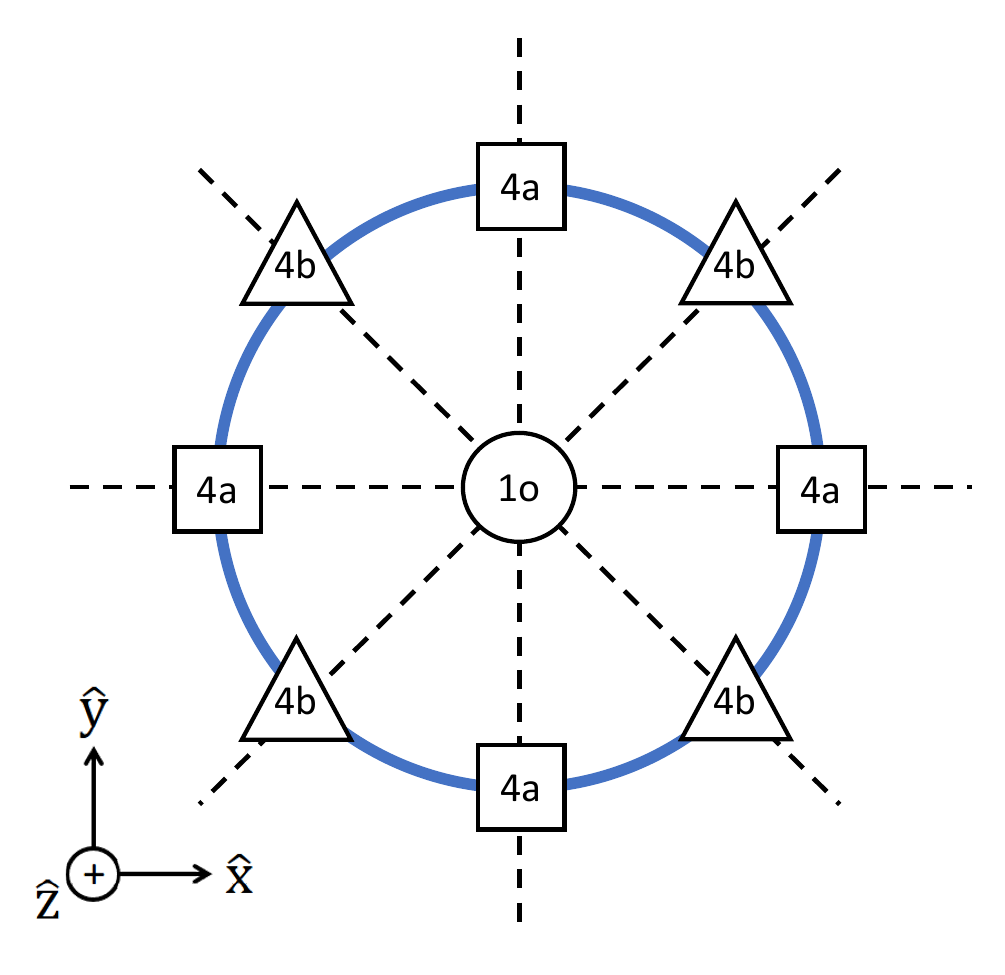}
\caption{The Wyckoff positions of point group~\cite{BilbaoPoint} $4mm$, generated by $M_{x,y}$ and $C_{4z}$.  Mirrors are indicated by dashed lines.  In point groups, the first Wyckoff position is labeled $1o$, to highlight that the origin is a special fixed point in polar (cylindrical) and spherical coordinates.  The general position, $8c$, is not pictured.  In $4mm$, there is only one maximal Wyckoff position~\cite{QuantumChemistry}: the $1o$ position (site-symmetry group $4mm$); the $4a$ and $4b$ positions (site-symmetry group $m$) are non-maximal, because $m$ is a subgroup of $4mm$.  This is different than in wallpaper group~\cite{MagneticBook,DiracInsulator} $p4m$, which is generated by adding 2D translations to $4mm$, in which there are two maximal Wyckoff positions with site-symmetry group $4mm$ ($1a$ and $1b$ in Fig.~\ref{fig:2Dmain}(a)).}
\label{fig:4mm}
\end{figure}

By observing the profile of $m^{+}_{2}\cos(2\theta)$ and choosing the convention in which the $n=0$ corner mode is positively charged, we determine that the zero modes in Eq.~(\ref{eq:s1zeromodes}) are (anti-) solitons at $\theta = \pi /4,5\pi/4$ ($3\pi/4,7\pi/4$) which acquire a charge~\cite{Shockley,SSH,RiceMele,NiemiSemenoff} $+e/2$ ($-e/2$) when $C_{4z}$ is ``softly'' broken~\cite{multipole} to $C_{2z}$ (Fig.~\ref{fig:cornerModes}(a)).  Specifically, in this work, we use the convention in which valence states, when occupied, carry a charge $e$, and conduction states, when occupied, carry a charge $-e$.  The four zero modes in Eq.~(\ref{eq:s1zeromodes}) are formed from a $4mm$-symmetry related set of two valence and two conduction states, each of which is half filled on the average, and can therefore be expressed as fully filled or empty linear combinations of valence (electron) and conduction (hole) states (\emph{i.e.}, solitons and antisolitons) with fractional charge~\cite{NiemiSemenoff,SSH,RiceMele}.  Each 0D bound state then individually carries a charge $\pm e/2$, depending on whether its wavefunction is an even (soliton) or an odd (antisoliton) linear combination of a valence and a conduction state.  However, whereas the energy spectrum of the disc still respects the full point group $4mm$, the charge assignment of the bound states (corner modes) only respects $C_{2z}$; like in the SSH chain~\cite{NiemiSemenoff}, we take this symmetry breaking to be ``soft'' in the sense that the number of electrons (\emph{i.e.}, the filling of the bound states) does not affect the energy spectrum itself.  This charge distribution can be summarized as:
\begin{equation}
q(\theta_{n}) = q_{n} =  \frac{e}{2}(-1)^{n}.
\label{eq:chargePerN}
\end{equation}
The zero modes occupy the $4b$ Wyckoff positions $(\pm x,\pm x)$ of point group~\cite{BilbaoPoint} $4mm$ (Fig.~\ref{fig:4mm}).  As $s^{1}(\theta_{n})\propto s^{x} - (+)\ s^{y}$ for even (odd) $n$ (Eq.~(\ref{eq:defSBasis})), these four zero modes are eigenstates of the diagonal mirrors $M_{x\mp y}$ (Eq.~(\ref{eq:mirrorRepos})).  In every direction, this distribution of charge (Eq.~(\ref{eq:chargePerN})) has a zero dipole moment (Eq.~(\ref{eq:dipoleMoment})).  To determine the $xy$-quadrupole moment, we reexpress Eq.~(\ref{eq:quadrupoleMoment}) as the sum of contributions from $n$ charged particles confined to a ring of radius $R$:
\begin{equation}
Q^{xy} = \frac{3}{2}\sum_{n}q_{n}x_{n}y_{n} = \frac{3R^{2}}{4}\sum_{n}q_{n}\sin(2\theta_{n}),
\label{eq:circularQuadrupole}
\end{equation}
where $q_{n}$ is defined in Eq.~(\ref{eq:chargePerN}).  For the four 0D modes in our calculation, which have fractional charges given by Eq.~(\ref{eq:chargePerN}) and lie at the zeroes of $\cos(2\theta)$, Eq.~(\ref{eq:circularQuadrupole}) indicates that $Q^{xy} = (3R^{2})e/2$.  To compare this with the value obtained through the Wannier description of the QI in Appendix~\ref{sec:bandrep} of $Q^{xy}=e/2$ per square unit cell in the units of $3a^{2}/2$, where $a$ is the lattice spacing, we can imagine that the four 0D states in Fig.~\ref{fig:cornerModes} occupy the corners of a 2D square crystal with $N$ unit cells and an overall diagonal length of $2R$.  For this square, the quadrupole moment of the corner modes is $Q^{xy}=e/2$ in the units of $3a^{2}N/2$, or $3a^{2}/2$ per unit cell, in agreement with the bulk quadrupole moment obtained in Appendix~\ref{sec:bandrep}.

To see that this $e/2$ quadrupole moment is a general bulk property, and not merely a unique feature of the $m^{+}_{2}$ term in Eq.~(\ref{eq:UedgePD}), we perform two more analyses: we first consider setting $m^{+}_{2}\rightarrow 0$ in Eq.~(\ref{eq:UedgePD}) and instead tuning one of the other mass terms away from zero, and we then demonstrate that the quadrupole moment survives under the relaxation of particle-hole symmetry via the introduction of multiple nonzero mass terms in Eq.~(\ref{eq:UedgePD}).

Before breaking particle-hole symmetry, we first consider how the previous analysis in Eqs.~(\ref{eq:thetaNPD}) to~(\ref{eq:circularQuadrupole}) is modified by instead taking $m^{+}_{6}\cos(6\theta)s^{2}(\theta)$ to be the only nonzero term in Eq.~(\ref{eq:UedgePD}).  The $L_{z}=6$ circular harmonic $\cos(6\theta)$ has 12 zeroes on a circle, which occur at $\pi/12 + n\pi/6$ where $n$ is an integer between $0$ and $11$.  More generally, we can state that for a general circular harmonic $\cos(L_{z}\theta)$, it will have $2L_{z}$ zeroes located at:
\begin{equation}
\theta_{n} = \pi/(2L_{z}) + n\pi/L_{z}, n\in \{0,1,\ldots,2L_{z}-1\}.
\label{eq:GeneralZeroes}
\end{equation}
For the specific case of $L_{z}=L_{z}^{QI}=2+4a$ (Eq.~(\ref{eq:nontrivialLz})) for the $m^{\pm}_{L_{z}^{QI}}$ mass terms in Eq.~(\ref{eq:UedgePD}), Eq.~(\ref{eq:GeneralZeroes}) implies that each mass term will individually contribute $2L_{z}=4+8a$ 0D zero modes.  As the first derivative of \emph{any} circular harmonic is at an extremum at a zero of that harmonic and alternates in sign at each zero in increasing $\theta$ (Eq.~(\ref{eq:defCircularHarmonic})), then Eqs.~(\ref{eq:MoreApproxThetaPD}) and~(\ref{eq:chargePerN}) also apply here, \emph{without further modification}.  We can therefore conclude that introducing $m^{+}_{6}$ to the $p-d$-hybridized TI in this section as the only mass term results in a circular boundary with 6 solitons and 6 antisolitions, \emph{i.e.} 0D modes with alternating charge $\pm e/2$ localized at the $\theta_{n}$ in Eq.~(\ref{eq:GeneralZeroes}) for $L_{z}=6$.  This charge distribution is still characterized by  Eq.~(\ref{eq:chargePerN}), with $n$ taken most generally over the range $0$ to $2L_{z}-1$, and thus here specifically from $n=0$ to $11$.  Four of these charges, those with:
\begin{equation}
 n_{4b}=\{0,3,6,9\} 
\end{equation}
 lie at the same locations, \emph{i.e.}, the $4b$ position of $4mm$ (Fig.~\ref{fig:4mm}), as did the four charges with $Q^{xy}=e/2$ for the $L_{z}=2$ case in Fig.~\ref{fig:cornerModes}.  The remaining 8 charges occupy the general position ($8c$) (Fig.~\ref{fig:4mm}).  We use Eq.~(\ref{eq:circularQuadrupole}) to express the quadrupole moment of these 12 charges (Eq.~(\ref{eq:chargePerN})) as a sum over the contributions from those at $4b$ and those at $8c$:
\begin{equation}
Q^{xy} = \frac{3R^{2}}{4}\left(\sum_{n\in n_{4b}}q_{n}\sin(2\theta_{n}) + \sum_{n\not\in n_{4b}}q_{n}\sin(2\theta_{n})\right)
\label{eq:eOver2ForAll}
\end{equation}
where by direct computation we confirm that the charges occupying $4b$ contribute again contribute a $Q^{xy}$ of $e/2$ in the units of $3R^{2}$, as they did previously for the $m_{2}^{+}$ mass term (see the text following Eq.~(\ref{eq:circularQuadrupole})), and the charges at $8c$ (\emph{i.e.} those not at $4b$) contribute a \emph{net-zero} $Q^{xy}$.  This is because, within each quadrant, there are two sites in the $8c$ position with opposite charges $\pm e/2$ but with the \emph{same} value for $\sin(2\theta_{n})$, because the charges are related by $M_{x\pm y}$, which takes $\theta \rightarrow \pm \pi/2 - \theta$.  This implies that \emph{any} set of 0D modes of alternating charge placed at the $8c$ position of $4mm$ will have a net-zero quadrupole moment.  We therefore conclude that the $m^{+}_{6}$ term also only provides an $e/2$ quadrupole moment if it is the only nonzero term in $U_{edge}(\theta)$.  Furthermore, as all of the terms proportional to $s^{2}(\theta)$ in Eq.~(\ref{eq:UedgePD}) are circular harmonics of the form $\cos(L_{z}\theta)$ where $L_{z} = L_{z}^{QI}$ (Eq.~(\ref{eq:nontrivialLz})), then we can conclude from Eqs.~(\ref{eq:GeneralZeroes}) and~(\ref{eq:eOver2ForAll}) that introducing \emph{any} term proportional to $s^{2}(\theta)$ in $U_{edge}(\theta)$ results in a collection of 0D bound states that decomposes into four modes at the $4b$ position and a remaining multiple of $8$ modes at the general position, for which the overall quadrupole moment $Q^{xy}=e/2$.

We also note the previous analysis of the quadrupole moments of the $m_{2,6}^{+}$ mass terms can also, with minor modifications, be applied to any of the terms in $U_{edge}(\theta)$ proportional to $m_{L_{z}^{QI}}^{-}$.  Using the transformation $s^{1}(\theta)\leftrightarrow s^{2}(\theta)$ in Eqs.~(\ref{eq:edgePH}) to Eq.~(\ref{eq:MoreApproxThetaPD}), we can conclude that if the only nonzero mass term in $U_{edge}(\theta)$ is proportional to $s^{1}(\theta)$, it will be proportional to a circular harmonic of the form $\sin(L_{z}\theta)$ where $L_{z}=L_{z}^{QI}$ (Eq.~(\ref{eq:nontrivialLz})), and will therefore exhibit a configuration of $2L_{z}$ 0D states of alternating charges $\pm e/2$ localized on the $r=R$ boundary at:
\begin{equation}
\theta_{n}^{s} = n\pi/L_{z}, n\in \{0,1,\ldots,2L_{z}-1\}.
\label{eq:GeneralZeroesSine}
\end{equation}
These states will instead decompose into 4 modes of alternating charges occupying the $4a$ position $(\pm x, 0);\ (0,\pm x)$ of $4mm$~\cite{BilbaoPoint} (Fig.~\ref{fig:4mm}) and multiples of 8 states of alternating charge occupying the general position.  Plugging the coordinates and charges (Eq.~(\ref{eq:chargePerN})) of the (anti)solitions at $4a$, as well as those at $8c$, into Eq.~(\ref{eq:circularQuadrupole}):
\begin{equation}
Q^{xy} = \frac{3R^{2}}{4}\left(\sum_{n\in n_{4a}}q_{n}\sin(2\theta_{n}) + \sum_{n\not\in n_{4a}}q_{n}\sin(2\theta_{n})\right) = 3R^{2}\left(0 + 0\right).
\label{eq:eOver2ForAllSine}
\end{equation}
However, we can also define an $x^{2}-y^{2}$ quadrupole moment~\cite{jacksonEM} by rotating Eq.~(\ref{eq:circularQuadrupole}) by $\pi/4$:
\begin{equation}
Q^{x^{2}-y^{2}} = \frac{3R^{2}}{4}\sum_{n}q_{n}\cos(2\theta_{n}),
\label{eq:circularQuadrupoleSine}
\end{equation}
for which, using the (anti)soliton locations $\theta^{s}_{n}$ in Eq.~(\ref{eq:GeneralZeroesSine}) and charge assignments in Eq.~(\ref{eq:chargePerN}):
\begin{equation}
Q^{x^{2}-y^{2}} = \frac{3R^{2}}{4}\left(\sum_{n\in n_{4a}}q_{n}\cos(2\theta_{n}) + \sum_{n\not\in n_{4a}}q_{n}\cos(2\theta_{n})\right) = 3R^{2}\left(e/2 + 0\right).
\label{eq:eOver2ForAllSineNontrivial}
\end{equation}
Therefore, as long as there exists a particle-hole symmetry of the form of \emph{either} $s^{1}(\theta)$ or $s^{2}(\theta)$, the introduction to a 2D TI of a term in $U_{edge}(\theta)$ proportional to a circular harmonic with $L_{z}=L_{z}^{QI}=2+4a$ will result in a distribution of boundary zero modes with an $e/2$ quadrupole moment, be it $Q^{xy}=e/2$ or $Q^{x^{2}-y^{2}}=e/2$. 

Finally, to demonstrate the robustness of this result, we show that the corner-mode quadrupole moment persists in the presence of multiple nonzero mass terms (both $m_{2+4a}^{\pm}$) in $U_{edge}(\theta)$, which generically results in the relaxation of all particle-hole symmetries.  To accomplish this, we begin by keeping $m^{+}_{2}\neq 0$ and $m_{2+4a}^{-}=0$ and perturbatively introduce one of the other $m_{2+4a}^{+}$ terms in $U_{edge}(\theta)$ (Eq.~(\ref{eq:UedgePD})).  The added term, which is proportional to $m^{+}_{2+4a}\cos[(2+4a)\theta]s^{2}(\theta)$, is equal to zero at the four values of $\theta_{n}$ in Eq.~(\ref{eq:thetaNPD}) where $\cos(2\theta)$ is zero (\emph{i.e.}, the $4b$ Wyckoff position in Fig.~\ref{fig:4mm}), as well as at $8a$ additional values of $\theta$ that correspond to the $8c$ Wyckoff position in $4mm$.  As shown in Eqs.~(\ref{eq:eOver2ForAll}) and~(\ref{eq:eOver2ForAllSineNontrivial}), respectively, any set of alternating solitons and antisolitons occupying the $8c$ position of $4mm$ contribute net-zero $Q^{xy}$ and $Q^{x^{2}-y^{2}}$ quadrupole moments.  Therefore, the additional $m^{+}_{2+4a}$ mass term does not affect the existing $e/2$ $xy$ quadrupole moment from the $m_{2}^{+}$ term.

We next, setting all of the $m_{2+4a}^{+}$ mass terms in Eq.~(\ref{eq:UedgePD}) back to zero except for $m_{2}^{+}$, consider the effect of adding one of the terms proportional to $m^{-}_{2+4a}\sin[(2+4a)\theta]s^{1}(\theta)$.  Adding one of these $m^{-}_{2+4a}$ terms explicitly breaks particle-hole symmetry (Eq.~(\ref{eq:edgePH})).  At each $\theta_{n}$ in Eq.~(\ref{eq:thetaNPD}), we calculate the first-order energy correction:
\begin{eqnarray}
\Delta_{E}(\theta_{n}) &=& m^{-}_{2+4a}\sin[(2+4a)\theta_{n}]\langle s^{1}(\theta_{n})|_{\lambda(\theta_{n})}s^{1}(\theta_{n})|s^{1}(\theta_{n})\rangle_{\lambda(\theta_{n})} \nonumber \\
 &=& m^{-}_{2+4a}\lambda(\theta_{n})\sin[(2+4a)\theta_{n}] \nonumber \\
 &=& m^{-}_{2+4a}(-1)^{n}(-1)^{n}(-1)^{a} \nonumber \\
 &=& m^{-}_{2+4a}(-1)^{a},
\label{eq:offsetE}
\end{eqnarray}
where we have exploited that $\lambda(\theta_{n})=(-1)^{n}$ (Eq.~(\ref{eq:lambdathetasign})) and that:
\begin{equation}
\sin[(2+4a)\theta_{n}]= (-1)^{n}(-1)^{a},
\end{equation} 
because $\sin(L^{QI}_{z}\theta)$ (Eq.~(\ref{eq:nontrivialLz})) is always at an extremum at a zero $\theta_{n}$ of $\cos{2\theta}$, which are coincident with the zeroes of $\cos(L^{QI}_{z}\theta)$.  Eq.~(\ref{eq:offsetE}) indicates that, beginning with $m_{2}^{+}$ as the only nonzero mass term in Eq.~(\ref{eq:UedgePD}), adding any of the $m^{-}_{2+4a}$ mass terms causes all four (anti)solitons at $\theta_{n}$ to shift together in energy in a $4mm$-symmetric manner, such that the spectrum no longer exhibits zero modes (Fig.~\ref{fig:cornerModes}(b)).  Nevertheless, if the system is half-filled, then within a perturbative range in $m^{-}_{2+4a}$, half of the 0D modes will still be occupied under softly broken $C_{4z}$ symmetry, resulting in a charge distribution with an $e/2$ quadrupole moment (Eq.~(\ref{eq:circularQuadrupole})).  We therefore conclude that the $4mm$-symmetric $p-d$-hybridized TI highlighted in this section can gap into a QI because it admits the presence of a bulk magnetic term with a nonzero edge projection and which is proportional to a circular harmonic with $L_{z} = L_{z}^{QI}$ (Eq.~(\ref{eq:nontrivialLz})).  The preservation of the quantized quadrupole moment away from the particle-hole-symmetric limit is in agreement with the conclusions of Wilczek in Ref.~\onlinecite{WilczekAxion}, in which it is stated that when the masses at Jackiw-Rebbi domain walls are made complex (\emph{i.e.} more than one Pauli matrix is present), there will no longer generically be zero-energy bound states, but the overall distribution of bound charge will remain preserved and reflect the accumulated phase in the complex mass over the domain wall.  This can also be understood from a field-theory perspective by considering the Goldstone-Wilczek formulation~\cite{GoldstoneWilczek,NiemiSemenoff}.

\subsection{Gapping the Edge Modes of a 2D TI into an Magnetic Insulator with Zero Quadrupole Moment}
\label{sec:TItoTrivial}

We will now show that a 2D TI in layer group $p4/mmm1'$, formed of $s$ and $p_{z}$ orbitals at the $1a$ position (Fig.~\ref{fig:2Dmain}(a) of the main text), gaps trivially in the presence of magnetism that preserves the symmetries of wallpaper group $p4m$, unlike the previous $p-d$-hybridized TI in Appendix~\ref{sec:TIboundary}.  We will see that this difference arises because, unlike in the previous $p-d$-hybridized TI (Eq.~(\ref{eq:TIGamma})), the bulk $C_{4z}$ eigenvalues of the occupied bands of an $s-p$-hybridized TI do not match those of a QI (Appendices~\ref{sec:bandrep} and~\ref{sec:double}).  The $k\cdot p$ Hamiltonian of a 2D $s-p$-hybridized TI is: 
\begin{equation}
\mathcal{H}_{\Gamma}({\bf k}) = m\tau^{z} + vk_{x}\tau^{x}\sigma^{y} - vk_{y}\tau^{x}\sigma^{x},
\label{eq:TIGammaSP}
\end{equation}
where we have again suppressed factors of the lattice constants $a_{x,y}=a$.  We note that Eq.~(\ref{eq:TIGamma}) is almost identical to the previous $k\cdot p$ Hamiltonian of a $p-d$-hybridized TI (Eq.~(\ref{eq:TIGamma})); the only difference between the two equations is the minus sign on the $k_{y}$ velocity term.  In the notation of Eq.~(\ref{eq:transformNotation}), $\mathcal{H}_{\Gamma}(k_{x},k_{y})$ (Eq.~(\ref{eq:TIGammaSP})) transforms in the symmetry representation given by:
\begin{align}
&\mathcal{T}:\ \sigma^{y}\mathcal{H}^{*}_{\Gamma}(-k_{x},-k_{y})\sigma^{y},\ M_{z}:\ \tau^{z}\sigma^{z}\mathcal{H}_{\Gamma}(k_{x},k_{y})\tau^{z}\sigma^{z},\ M_{x}:\ \sigma^{x}\mathcal{H}_{\Gamma}(-k_{x},k_{y})\sigma^{x}, \nonumber \\ 
&M_{y}:\ \sigma^{y}\mathcal{H}_{\Gamma}(k_{x},-k_{y})\sigma^{y},\ C_{4z}:\ \left(\frac{\mathds{1}_{\sigma} - i\sigma^{z}}{\sqrt{2}}\right)\mathcal{H}_{\Gamma}(k_{y},-k_{x})\left(\frac{\mathds{1}_{\sigma} + i\sigma^{z}}{\sqrt{2}}\right),
\label{eq:spKspaceSyms}
\end{align}
where $\mathds{1}_{\sigma}$ is the $2\times 2$ identity in $\sigma$ space.  Eq.~(\ref{eq:spKspaceSyms}) implies an inversion symmetry $\mathcal{I}=M_{x}M_{y}M_{z}$ that transforms $\mathcal{H}_{\Gamma}(k_{x},k_{y})$ under the representation:
\begin{equation}
\mathcal{I}:\ \tau^{z}\mathcal{H}_{\Gamma}(-k_{x},-k_{y})\tau^{z}.
\label{eq:spISym}
\end{equation}
The presence of $\tau^{z}$ in $M_{z}$ and $\mathcal{I}$, and the absence of $\tau^{z}$ in $C_{4z}$ in Eqs.~(\ref{eq:spKspaceSyms}) and~(\ref{eq:spISym}) reflect that $\mathcal{H}_{\Gamma}({\bf k})$ describes a 2D TI formed of hybridized $s$ and $p_{z}$ orbitals, as $s$ ($p_{z}$) orbitals are even (odd) under $M_{z}$ and $\mathcal{I}$ and even under $C_{4z}$.  Specifically, unlike previously in the symmetry representation of the $p-d$-hybridized TI in Eq.~(\ref{eq:pdKspaceSyms}) (and like in the symmetry representation of the $k\cdot p$ theory of the $s-d$-hybridized TCI phase of Eq.~(\ref{eq:my2Dquad}) that will be analyzed in Appendix~\ref{sec:TCIBoundary}), there is no prefactor of $\tau^{z}$ in the representation of $C_{4z}$ in Eq.~(\ref{eq:spKspaceSyms}), because the valence and conduction bands of an $s-p_{z}$-hybridized TI have the same $C_{4z}$ eigenvalues (though they have different parity eigenvalues).  $\mathcal{H}_{\Gamma}({\bf k})$ also exhibits the same pair of unitary particle-hole symmetries as previously in Eqs.~(\ref{eq:posPH}) and~(\ref{eq:posPH2}):
\begin{equation}
\{\mathcal{H}_{\Gamma}({\bf k}),\Pi\} = 0,\ \Pi = \tau^{y},
\label{eq:posPHsp}
\end{equation}
and:
\begin{equation}
\{\mathcal{H}_{\Gamma}({\bf k}),\tilde{\Pi}\} = 0,\ \tilde{\Pi} = \tau^{x}\sigma^{z},
\label{eq:posPHsp2}
\end{equation}
which we will again relax in future steps in this calculation, as we did previously in in Appendix~\ref{sec:TIboundary}.  We note that the choice of which particle-hole representation to label with a tilde is arbitrary; the choice of $\Pi=\tau^{y}$ in Eq.~(\ref{eq:posPHsp}) is only distinct from the previous choice in Eq.~(\ref{eq:posPH}) to simplify notation in expressions that will arise later in this section.

As mentioned previously, Eq.~(\ref{eq:TIGammaSP}), is nearly identical to the $k\cdot p$ Hamiltonian of a $p_{z}-d_{x^{2}-y^{2}}$-hybridized TI analyzed in Appendix~\ref{sec:TIboundary} (it only differs from Eq.~(\ref{eq:TIGamma}) by the minus sign of $vk_{y}\tau^{x}\sigma^{x}$).  Nevertheless, the two Hamiltonians cannot be transformed into each other by a unitary transformation that preserves the handedness of the spin and momentum sectors of rotations about the $z$ axis.  Specifically, if we require that $C_{4z}$ is defined as the transformation:
\begin{equation}
k_{x}\rightarrow k_{y},\ k_{y}\rightarrow -k_{x},\ \sigma^{x}\rightarrow \sigma^{y},\ \sigma^{y}\rightarrow -\sigma^{x},
\label{eq:momentumRotation}
\end{equation}
then the unitary transformation $U=\sigma^{y}$ that converts Eq.~(\ref{eq:TIGammaSP}) into Eq.~(\ref{eq:TIGamma}) also changes the sign of $\sigma^{z}$ in $C_{4z}$ in Eq.~(\ref{eq:spKspaceSyms}).  Under this transformation, $C_{4z}$ would continue to rotate momentum counterclockwise as specified in Eq.~(\ref{eq:momentumRotation}), however it would now rotate the spins \emph{clockwise} about $\sigma^{z}$.  Therefore we cannot simultaneously transform Eq.~(\ref{eq:TIGammaSP}) into Eq.~(\ref{eq:TIGamma}) while preserving a physical definition of $C_{4z}$.  This can also be understood by recognizing that in Eq.~(\ref{eq:TIGammaSP}), unlike in Eq.~(\ref{eq:TIGamma}), both the valence and conduction bands have the same complex-conjugate pairs of $C_{4z}$ eigenvalues (they can still be inverted because they possess different inversion eigenvalues~\cite{bernevigBook}).  We will show in this section that, unlike previously in Appendix~\ref{sec:TIboundary}, because the valence and conduction bands of the $s-p_{z}$-hybridized TI in Eq.~(\ref{eq:TIGammaSP}) exhibit the same $C_{4z}$ eigenvalues, then, when Eq.~(\ref{eq:TIGammaSP}) is terminated in a disc geometry and its edge states are gapped with $p4m$-symmetric magnetism, the resulting 0D bound states exhibit a topologically trivial quadrupole moment of $Q^{xy}\text{ mod }e = Q^{x^{2}-y^{2}}\text{ mod }e = 0$.

We again Fourier transform $k_{x,y}\rightarrow -i\partial_{x,y}$ and search for Jackiw-Rebbi bound states on the boundary of a circular region of a large radius $R\gg a$ for which $\sgn[m(r)] = \sgn(r-R)$.  We note that, because Eqs.~(\ref{eq:TIGamma}) and~(\ref{eq:TIGammaSP}) both describe 2D TIs, then the first (bulk-to-edge) Jackiw-Rebbi calculation performed to obtain the edge zero modes will be identical in this section to the previous calculation performed Appendix~\ref{sec:TIboundary}; spectral differences between the two TIs will only begin to manifest when we gap the edge spectrum with $p4m$-symmetric magnetism.

The Hamiltonian in Eq.~(\ref{eq:TIGammaSP}), when converted to polar coordinates using Eq.~(\ref{eq:polarCoordTrans}), takes the same form as Eq.~(\ref{eq:polarTI}):
\begin{equation}
\mathcal{H}_{\Gamma}(r,\theta) = m(r)\tau^{z} - iv\tau^{x}\left[\bar{\sigma}^{1}(\theta)\partial_{r} + \frac{1}{r}\bar{\sigma}^{2}(\theta)\partial_{\theta}\right],
\label{eq:polarTISP}
\end{equation}
but through a different the canonical transformation than Eq.~(\ref{eq:pdSigmaDef}):
\begin{eqnarray}
\bar{\sigma}^{1}(\theta) &=& \sin(\theta)\sigma^{x} - \cos(\theta)\sigma^{y}=\left(\begin{array}{cc}
0 & ie^{-i\theta} \\
-ie^{i\theta} & 0 \end{array}\right),\nonumber \\
\bar{\sigma}^{2}(\theta) &=& \cos(\theta)\sigma^{x} + \sin(\theta)\sigma^{y} = \left(\begin{array}{cc}
0 & e^{-i\theta} \\
e^{i\theta} & 0 \end{array}\right), \nonumber \\
\{\bar{\sigma}^{1}(\theta),\bar{\sigma}^{2}(\theta)\}&=&0,\ \bar{\sigma}^{1}(\theta)\bar{\sigma}^{2}(\theta) = i\sigma^{z}. 
\label{eq:spSigmaDef}
\end{eqnarray}
In the notation of Eq.~(\ref{eq:polarNotation}), $\mathcal{H}_{\Gamma}(r,\theta)$ transforms in the symmetry representation given by:
\begin{align}
&\mathcal{T}:\ \sigma^{y}\mathcal{H}^{*}_{\Gamma}(r,\theta)\sigma^{y},\ M_{z}:\ \tau^{z}\sigma^{z}\mathcal{H}_{\Gamma}(r,\theta)\tau^{z}\sigma^{z},\ M_{x}:\ \sigma^{x}\mathcal{H}_{\Gamma}(r,\pi-\theta)\sigma^{x}, \nonumber \\ 
&M_{y}:\ \sigma^{y}\mathcal{H}_{\Gamma}(r,-\theta)\sigma^{y},\ C_{4z}:\ \left(\frac{\mathds{1}_{\sigma} - i\sigma^{z}}{\sqrt{2}}\right)\mathcal{H}_{\Gamma}(r,\theta+\pi/2)\left(\frac{\mathds{1}_{\sigma} + i\sigma^{z}}{\sqrt{2}}\right), \nonumber \\
&\mathcal{I}:\ \tau^{z}\mathcal{H}_{\Gamma}(r,\theta+\pi)\tau^{z},
\label{eq:spPolarSyms}
\end{align}
and the particle-hole symmetries remain the same as previously in Eqs.~(\ref{eq:posPHsp}) and~(\ref{eq:posPHsp2}):
\begin{equation}
\{\mathcal{H}_{\Gamma}(r,\theta),\Pi\}=0,\ \Pi = \tau^{y}.
\label{eq:PolarPHsp}
\end{equation}
and:
\begin{equation}
\{\mathcal{H}_{\Gamma}(r,\theta),\tilde{\Pi}\}=0,\ \tilde{\Pi} = \tau^{x}\sigma^{z}.
\label{eq:PolarPHsp2}
\end{equation}
Once again, we form a Jackiw-Rebbi problem for the bound states of Eq.~(\ref{eq:polarTISP}):
\begin{equation} 
\mathcal{H}_{\Gamma}(r,\theta)|\psi(r,\theta)\rangle = 0,  
\label{eq:polarTIdoubleySP}
\end{equation}
which, following the procedure in Eqs.~(\ref{eq:polarTIdoubley}),~(\ref{eq:separated}), and~(\ref{eq:ApproxEqSep}), we simplify to:
\begin{equation}
\left[m(r)\tau^{y} +v\bar{\sigma}^{1}(\theta)\partial_{r}\right]|\psi(r,\theta)\rangle \approx 0,
\label{eq:ApproxEqSepSP}
\end{equation}
by left-multiplying by $\tau^{x}$, canceling a factor of $-i$, and recognizing that $(1/r)\sim(1/R)\rightarrow 0$ for a bound state localized at the radius $R$ of a large circle with $R \gg a$ where $a$ is the lattice spacing.  We solve Eq.~(\ref{eq:ApproxEqSepSP}) by left-multiplying by $\tau^{y}$ and integrating:
\begin{equation}
|\psi_{1,2}(r,\theta)\rangle = \frac{1}{\sqrt{N}}e^{-\frac{1}{v}\int_{R}^{r}m(r')dr'}|\tau^{y}_{\pm}\bar{\sigma}^{1}_{\pm}(\theta)\rangle= \mathcal{R}(r)|\tau^{y}_{\pm}\bar{\sigma}^{1}_{\pm}(\theta)\rangle,
\label{eq:spTempBound}
\end{equation}
where:
\begin{equation}
|\tau^{y}_{\pm}\bar{\sigma}^{1}_{\pm}(\theta)\rangle = |\tau^{y}_{\pm}\rangle\otimes|\bar{\sigma}^{1}_{\pm}(\theta)\rangle, 
\end{equation}
where $|\tau^{i}_{\pm},\sigma^{j}_{\pm}\rangle$ are the eigenstates with eigenvalues $\pm1$ of the $2\times2$ Pauli matrices $\tau^{i}$ and $\sigma^{j}$.

As in Appendix~\ref{sec:TIboundary}, we will find that the symmetries of the edge Hamiltonian appear in a more familiar form in the rotated basis:
\begin{eqnarray}
|\bar{\phi}_{1}(r,\theta)\rangle &=& \frac{-ie^{i\theta}}{\sqrt{2}}\left(|\psi_{1}(r,\theta)\rangle - |\psi_{2}(r,\theta)\rangle\right) = \frac{\mathcal{R}(r)}{\sqrt{2}}\left(\begin{array}{c}
0 \\
-e^{i\theta} \\
1 \\
0\end{array}\right) \nonumber \\
|\bar{\phi}_{2}(r,\theta)\rangle &=& \frac{1}{\sqrt{2}}\left(|\psi_{1}(r,\theta)\rangle + |\psi_{2}(r,\theta)\rangle\right) = \frac{\mathcal{R}(r)}{\sqrt{2}}\left(\begin{array}{c} 
e^{-i\theta} \\
0 \\
0 \\
1\end{array}\right),
\label{eq:defPhiSP}
\end{eqnarray}
where, as previously in Eq.~(\ref{eq:defPhi}), we are free to make this transformation because $|\bar{\phi}_{1,2}(\theta)\rangle$ are degenerate (zero modes) at all values of $\theta$ at this stage of the calculation.

To leading order, Eq.~(\ref{eq:spTempBound}) indicates that there are two nondispersing zero modes localized on the boundary of the circular domain wall with radius $R$, or close to the region where $m(r)=0$.  Again, perturbatively restoring the angular velocity term $-i(v/r)\tau^{x}\bar{\sigma}^{2}(\theta)\partial_{\theta}$ and projecting into the basis of the edge states $|\bar{\phi}_{1,2}(r,\theta)\rangle$ using Eq.~(\ref{eq:projecty}), we realize the edge Hamiltonian:
\begin{equation}
\mathcal{H}_{edge}^{TI}(\theta) = -\frac{v}{R}\left(\frac{1}{2}\mathds{1}_{s} + is^{z}\partial_{\theta}\right),
\label{eq:projectySP}
\end{equation}
where $s^{z}$ is a Pauli matrix and $\mathds{1}_{s}$ is the identity matrix in the $2\times 2$ basis of $|\bar{\phi}_{1,2}(r,\theta)\rangle$, and where, as previously in Eq.~(\ref{eq:projecty}), the action of $\partial_{\theta}$ on the edge states $|\bar{\phi}_{1,2}(r,\theta)\rangle$ results in the presence of a gauge-dependent constant term, which is $-(v/2R)\mathds{1}_{s}$ for the choice of gauge in Eq.~(\ref{eq:defPhiSP}).  Eq.~(\ref{eq:projectySP}), the edge Hamiltonian of a $s-p$-hybridized TI, is identical to Eq.~(\ref{eq:projecty}), the previous edge Hamiltonian of a $p-d$-hybridized TI (up to an overall minus sign).

In the basis of $|\bar{\phi}_{1,2}(r,\theta)\rangle$, the symmetries from Eq.~(\ref{eq:spPolarSyms}) transform $\mathcal{H}_{edge}^{TI}(\theta)$ under the representation:
\begin{align}
&\mathcal{T}:\ s^{y}(\mathcal{H}_{edge}^{TI}(\theta))^{*}s^{y},\ M_{z}:\ s^{z}\mathcal{H}_{edge}^{TI}(\theta)s^{z},\ M_{x}:\ s^{x}\mathcal{H}_{edge}^{TI}(\pi-\theta)s^{x}, \nonumber \\ 
&M_{y}:\ s^{y}\mathcal{H}_{edge}^{TI}(-\theta)s^{y},\ C_{4z}:\ \left(\frac{\mathds{1}_{s} - is^{z}}{\sqrt{2}}\right)\mathcal{H}_{edge}^{TI}(\theta+\pi/2)\left(\frac{\mathds{1}_{s} + is^{z}}{\sqrt{2}}\right), \nonumber \\
&\mathcal{I}:\ \mathcal{H}_{edge}^{TI}(\theta+\pi),
\label{eq:sBasisSP}
\end{align}
which is \emph{identical} to the previous edge symmetry representation in Eq.~(\ref{eq:sBasis}).  However, because the bulk representations of $C_{4z}$ are different in Eqs.~(\ref{eq:pdPolarSyms}) and~(\ref{eq:spPolarSyms}), we will see that when $p4m$-symmetric magnetic masses are added to gap the edge states (Eq.~(\ref{eq:projectySP})), a different configuration of (quadrupole-trivial) zero modes emerges in the $s-p$-hybridized case than previously appeared for the $p-d$-hybridized 2D TI in Appendix~\ref{sec:TIboundary}.  Specifically, while the \emph{bulk} symmetry representations of $C_{4z}$ are different in the $p-d$-hybridized 2D TI in Eq.~(\ref{eq:pdPolarSyms}) and in the $s-p$-hybridized 2D TI in Eq.~(\ref{eq:spPolarSyms}), that difference does not carry over into the symmetry representation of the $2\times 2$ \emph{edge} Hamiltonians of the two TIs (Eqs.~(\ref{eq:sBasis}) and~(\ref{eq:sBasisSP}), respectively).  This occurs because the edge Hamiltonian of any 2D TI characterizes a twofold, spin-$1/2$ fermion in 1D~\cite{MirlinCircleTI}, and because there is, up to unitarily equivalent expressions, only one way to represent spinful $C_{4z}$ symmetry in the $2\times 2$ basis of spin-$1/2$ Pauli matrices $\sigma^{i}$ (as opposed to in the $4\times 4$ basis of orbital $\tau^{i}$ and spin $\sigma^{i}$ matrices of the bulk Hamiltonians in Eqs.~(\ref{eq:TIGamma}) and~(\ref{eq:TIGammaSP}), in which there are two inequivalent ways to represent $C_{4z}$).  Nevertheless and crucially, as we will shortly see, the corner spectrum of a 2D TI gapped with $p4m$-symmetric magnetism (\emph{i.e.} the number and quadrupole moment of the 0D modes on its boundary), depends on the \emph{bulk} representation of $C_{4z}$, and not the edge representation, and thus still distinguishes between $p-d$-hybridized TIs and $s-p$-hybridized TIs.

As previously (Eq.~(\ref{eq:edgePH})), in the basis of $|\bar{\phi}_{1,2}(r,\theta)\rangle$, the particle-hole symmetries from Eqs.~(\ref{eq:PolarPHsp}) and~(\ref{eq:PolarPHsp2}) take $\theta$-dependent forms:
\begin{equation}
\Pi(\theta) = \bar{s}^{2}(\theta),\ \tilde{\Pi}(\theta) = \bar{s}^{1}(\theta),
\label{eq:edgePHSP}
\end{equation}
where:
\begin{eqnarray}
\bar{s}^{1}(\theta) &=& \cos(\theta)s^{x} + \sin(\theta)s^{y}=\left(\begin{array}{cc}
0 & e^{-i\theta} \\
e^{i\theta} & 0 \end{array}\right),\nonumber \\
\bar{s}^{2}(\theta) &=& \sin(\theta)s^{x} - \cos(\theta)s^{y}=\left(\begin{array}{cc}
0 & ie^{-i\theta} \\
-ie^{i\theta} & 0 \end{array}\right),\nonumber \\
\{\bar{s}^{1}(\theta),\bar{s}^{2}(\theta)\}&=&0,\ \bar{s}^{1}(\theta)\bar{s}^{2}(\theta) = -is^{z},
\label{eq:defSBasisSP}
\end{eqnarray}
and as previously in Eqs.~(\ref{eq:manifestPH1}) through~(\ref{eq:manifestPH3}), the $\theta$ dependence of $\Pi(\theta)$ and $\tilde{\Pi}(\theta)$ in Eq.~(\ref{eq:edgePHSP}) maintain both representations of particle-hole symmetry through the relations:
\begin{equation}
\partial_{\theta}\bar{s}^{1}(\theta) = -\bar{s}^{2}(\theta) + \bar{s}^{1}(\theta)\partial_{\theta},\ \partial_{\theta}\bar{s}^{2}(\theta) = \bar{s}^{1}(\theta) + \bar{s}^{2}(\theta)\partial_{\theta}.
\label{eq:manifestPHsp2}
\end{equation}

We now add magnetic terms that preserve the symmetries of $p4m$.  Here, there are manifest differences between these terms and the corresponding terms in Appendix~\ref{sec:TIboundary}.  We again work in the long-wavelength limit and propose the most general $r$-independent bulk mass term $U(\theta)$ in the form of Eq.~(\ref{eq:GeneralU}), whose terms individually anticommute with the Dirac matrix coefficients of the mass ($m(r)\tau^{z}$) and angular velocity ($-i(v/r)\tau^{x}\bar{\sigma}^{2}(\theta)\partial_{\theta}$) terms in Eq.~(\ref{eq:polarTISP}) while respecting the bulk representations of $M_{x,y}$ and $C_{4z}$ in Eq.~(\ref{eq:spPolarSyms}).  We express $U(\theta)$ as a sum of terms proportional to circular harmonics of increasing $L_{z}$ (Eq.~(\ref{eq:defCircularHarmonic})):
\begin{align}
U(\theta) &= \tau^{x}\sigma^{z}\left[m^{-}_{4}\sin(4\theta) + m^{-}_{8}\sin(8\theta) + m^{-}_{12}\sin(12\theta) + \ldots\ \right] \nonumber \\
&+ \tau^{y}\left[m^{+}_{0} + m^{+}_{4}\cos(4\theta) + m^{+}_{8}\cos(8\theta) + \ldots\ \right] \nonumber \\
&+ \tau^{x}\bar{\sigma}^{1}(\theta)\left[\tilde{m}^{+}_{0} + \tilde{m}^{+}_{4}\cos(4\theta) + \tilde{m}^{+}_{8}\cos(8\theta) + \ldots\ \right] \nonumber \\
&+ \tau^{y}\bar{\sigma}^{2}(\theta)\left[\tilde{m}^{-}_{4}\sin(4\theta) + \tilde{m}^{-}_{8}\sin(8\theta) + \tilde{m}^{-}_{12}\sin(12\theta) + \ldots\ \right],
\label{eq:spMasses}
\end{align}
and observe that the terms group into circular harmonics with $\Delta L_{z}=4$ multiplied by one of four $4\times 4$ matrices.  We then, as was done to generate Eq.~(\ref{eq:UedgePD}), project $U(\theta)$ into the basis of the edge modes $|\bar{\phi}_{1,2}(r,\theta)\rangle$:
\begin{eqnarray}
U_{edge,ij}(\theta) &=& \langle\bar{\phi}_{i}(r,\theta)|U(\theta)|\bar{\phi}_{j}(r,\theta)\rangle \nonumber \\
&=& \sum_{L_{z},\mu}m^{\mu}_{L_{z}}\langle\bar{\phi}_{i}(r,\theta)|\Gamma^{L_{z},\mu}|\bar{\phi}_{j}(r,\theta)\rangle f^{\mu}_{L_{z}}(\theta), \\
U_{edge}(\theta) &=&  \bar{s}^{1}(\theta)\left[m^{-}_{4}\sin(4\theta) + m^{-}_{8}\sin(8\theta) + m^{-}_{12}\sin(12\theta) + \ldots\ \right] \nonumber \\
&+& \bar{s}^{2}(\theta)\left[m^{+}_{0} + m^{+}_{4}\cos(4\theta) + m^{+}_{8}\cos(8\theta) + \ldots\ \right],
\label{eq:UedgeSP}
\end{eqnarray}
where, in agreement with Eqs.~(\ref{eq:defProj}),~(\ref{eq:NonzeroP}), and~(\ref{eq:antiantiprojector}), the terms in $U(\theta)$ that commute with $\tau^{y}\bar{\sigma}^{1}(\theta)$ ($m^{\pm}_{4a}$) have nonzero edge projections, whereas the terms that anticommute with $\tau^{y}\bar{\sigma}^{1}(\theta)$ ($\tilde{m}^{\pm}_{4a}$) project to zero in $U_{edge}(\theta)$.  The nonzero terms in $U_{edge}(\theta)$ break $\mathcal{I},\ M_{z}$, and $\mathcal{T}$ symmetries, while respecting the combined magnetic symmetries $\mathcal{I}\times\mathcal{T}$ and $M_{z}\times\mathcal{T}$ (Eq.~(\ref{eq:sBasisSP})) (though, as shown in Appendix~\ref{sec:bandrep}, the antiunitary magnetic symmetries $\mathcal{I}\times\mathcal{T}$ and $M_{z}\times\mathcal{T}$, while symmetries of the original QI model in Ref.~\onlinecite{multipole}, are not necessary to protect the QI phase).  As previously in Appendix.~\ref{sec:TIboundary}, half of the mass terms in Eq.~(\ref{eq:UedgeSP}) only respect one of the particle-hole symmetries in Eq.~(\ref{eq:edgePHSP}) (the $m_{L_{z}}^{-}$ terms and $\Pi(\theta) = \bar{s}^{2}(\theta)$) and the other terms only respect the other particle-hole symmetry (the $m_{L_{z}}^{+}$ terms and $\tilde{\Pi}(\theta) = \bar{s}^{1}(\theta)$).  Therefore, when mass terms from both the $m^{\pm}_{L_{z}}$ sets are nonzero, particle-hole symmetry is broken.

Unlike previously in Appendix~\ref{sec:TIboundary}, Eq.~(\ref{eq:UedgeSP}) indicates that, for the $s-p$-hybridized TI in this section, bulk $p4m$-preserving magnetism can only open an edge gap with:
\begin{equation}
L_{z}^{NI} = 4a,\ a\in\mathbb{Z},
\label{eq:trivialLz}
\end{equation}
where the $m^{-}_{4}$ and $m_{0}^{+}$ terms in Eq.~(\ref{eq:UedgeSP}), in particular, are proportional to the circular harmonics of $g_{xy(x^{2}-y^{2})}$ and $s$ orbitals, respectively~\cite{jacksonEM,mcQuarriePchem,harmonics1,harmonics2}.  This result is markedly different than the previous conclusion in Appendix~\ref{sec:TIboundary} that, for a $p-d$-hybridized TI gapped with $p4m$-preserving magnetism, $L_{z}^{QI}=2 + 4a$ (Eq.~(\ref{eq:nontrivialLz})).  We label Eq.~(\ref{eq:trivialLz}) with the typical abbreviation for a normal (trivial) insulator (NI) because, as we will shortly demonstrate, it implies that $p4m$-preserving magnetism can only gap an $s-p_{z}$-hybridized TI into a magnetic insulator with a trivial quadrupole moment.

We define the edge Hamiltonian to be:
\begin{equation}
\mathcal{H}_{edge}^{NI}(\theta) = \mathcal{H}_{edge}^{TI}(\theta) + U_{edge}(\theta).
\label{eq:fullTIedgeSP}
\end{equation}
If we truncate $U_{edge}(\theta)$ (Eq.~(\ref{eq:UedgeSP})) to its leading  $m^{-}_{4}$ and $m_{0}^{+}$ terms, $\mathcal{H}_{edge}^{NI}(\theta)$ exhibits a gap in the long-wavelength limit of:
\begin{equation}
\Delta(\theta) = 2\sqrt{(m^{-}_{4})^{2}\sin^{2}(4\theta) + (m^{+}_{0})^{2}}.
\end{equation}

\begin{figure}[h]
\centering
\includegraphics[width=0.90\textwidth]{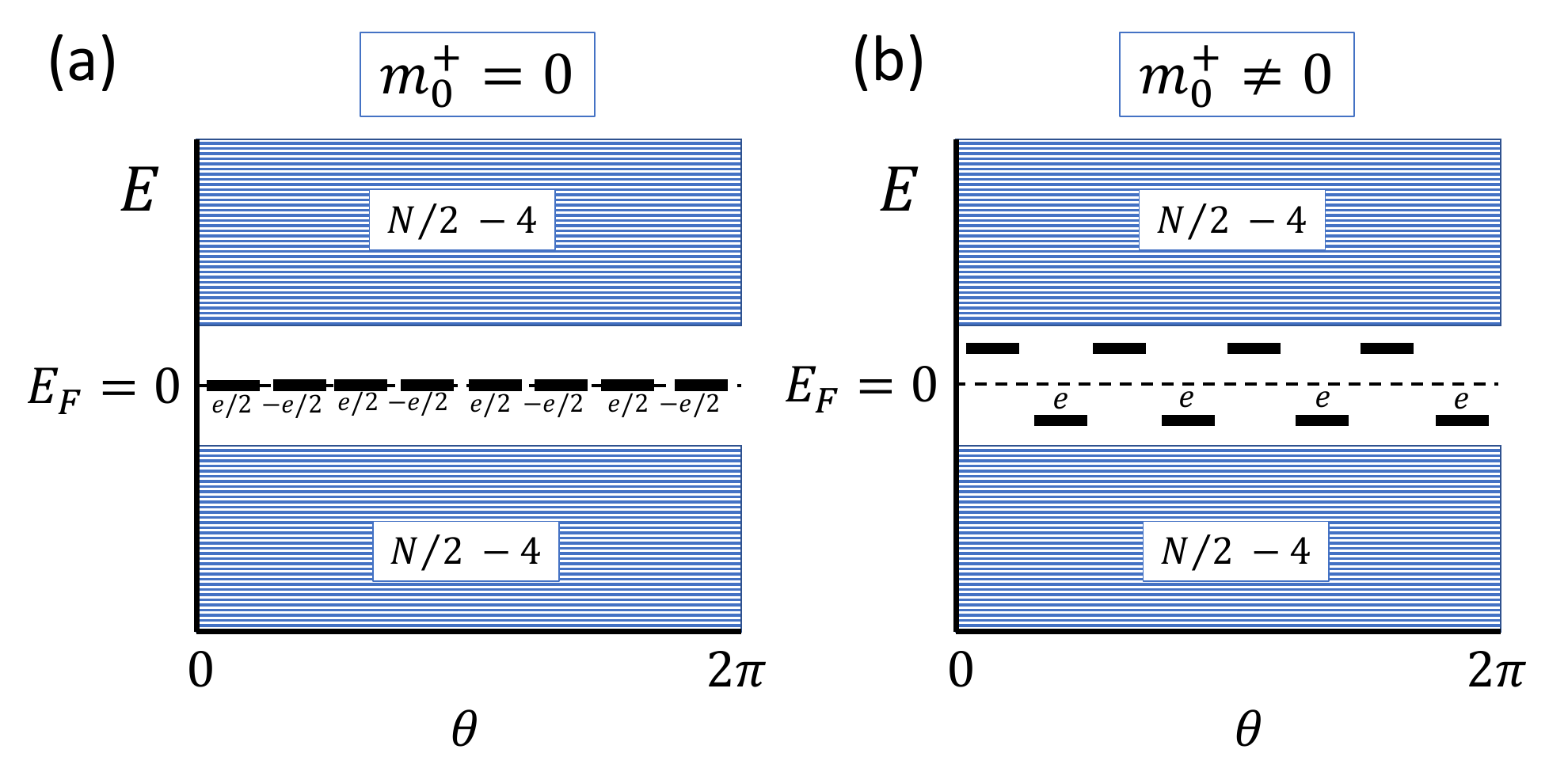}
\caption{Schematic energy spectra of magnetically gapped, $p4m$-symmetric, $s-p_{z}$-hybridized 2D TIs with and without particle-hole symmetry and open boundary conditions (OBC).  (a) For an $s-p_{z}$-hybridized 2D TI with $p4m$ symmetry, quadrupolar magnetism can generically result in the presence of $8a, a\in\mathbb{Z}$ (Eq.~(\ref{eq:trivialLz})) edge solitons (including zero in the case where $m^{+}_{0}$ is the only nonzero mass term in Eq.~(\ref{eq:UedgeSP})).  All of these symmetry-allowed soliton configurations exhibit the same, trivial (net-zero) $xy$ and $x^{2}-y^{2}$ quadrupole moments (Eqs.~(\ref{eq:TrivialQXY}) and~(\ref{eq:TrivialQX2Y2})).  (a) As an example, consider the case in which $m^{-}_{4}$ is, at first, the only nonzero mass term in Eq.~(\ref{eq:UedgeSP}).  This results in \emph{eight} Jackiw-Rebbi 0D bound states at $\theta_{n} = n\pi/4$ (Eq.~(\ref{eq:thetaNSP})), in the presence of particle-hole symmetry (Eq.~(\ref{eq:edgePHSP})).  However, when particle-hole symmetry is relaxed as other symmetry-allowed mass terms are reintroduced, half of the modes can float up and half can float down in energy in a $4mm$-symmetric manner (Eq.~(\ref{eq:offsetESP})).  In both (a) and (b), the occupied 0D modes do not exhibit a topological quadrupole moment (Eqs.~(\ref{eq:TrivialQXY}) and~(\ref{eq:TrivialQX2Y2})).}
\label{fig:TrivialCorners}
\end{figure}

We can understand why $\mathcal{H}_{edge}(\theta)$ does not exhibit a topological quadrupole moment from several perspectives.  First, we can begin in the limit where $m_{0}^{+}$ is the only nonzero mass term in Eq.~(\ref{eq:UedgeSP}).  In this limit, $\mathcal{H}_{edge}(\theta)$ is simply gapped at all values of $\theta$, as opposed to Eq.~(\ref{eq:simpleGapPD}), which binds 0D modes when either $m^{\pm}_{L_{z}}$ is nonzero.  As shown previously in Appendix~\ref{sec:TIboundary}, an edge (and bulk) quadrupole moment is only topological if its value remains fixed (modulo $e$ and up to a choice of orientation, as shown in the text surrounding Eq.~(\ref{eq:circularQuadrupoleSine})) in the presence of any linear combination of symmetry-allowed mass terms.  If there are no edge solitons at all, then $Q^{xy}=Q^{x^{2}-y^{2}}=0$ trivially.

We can further understand the absence of a topological quadrupole moment by next beginning in the limit that $m_{4}^{-}$ is the only nonzero mass in Eq.~(\ref{eq:UedgeSP}), and then perturbatively reintroducing other mass terms.  When $m^{-}_{4}$ is the only nonzero mass, the system is particle-hole symmetric (Eq.~(\ref{eq:edgePHSP})) and the spectrum is gapless at eight values of $\theta$:
\begin{equation}
\theta_{n} = n\pi/4,\text{ for }n\in\{0,2,\ldots,7\}.  
\label{eq:thetaNSP}
\end{equation}
We search for zero-energy bound states at each of the $\theta_{n}$ by forming the Jackiw-Rebbi problem: 
\begin{equation}
\left[-\frac{v}{R}\left(\frac{1}{2}\mathds{1}_{s} + is^{z}\partial_{\theta}\right) + m^{-}_{4}\sin(4\theta)\bar{s}^{1}(\theta)\right]|\tilde{\Theta}(\theta)\rangle =0,
\label{eq:fullThetaSP}
\end{equation}
which we simplify by using the identical procedure in Eqs.~(\ref{eq:fullThetaPD}) to~(\ref{eq:MoreApproxThetaPD}) to remove the constant curvature term $-(v/2R)\mathds{1}_{s}$ and then expanding $\theta$ in a small range $\epsilon$ around each angle $\theta_{n}$ in Eq.~(\ref{eq:thetaNSP}):
\begin{equation}
\left[-i\frac{v}{R}s^{z}\partial_{\epsilon} + (-1)^{n}m_{\theta}\epsilon \bar{s}^{1}(\theta_{n})\right]|\Theta(\theta_{n},\epsilon)\rangle=0,
\label{eq:MoreApproxThetaSP}
\end{equation}
where $m_{\theta}=4m^{-}_{4}$.  We solve Eq.~(\ref{eq:MoreApproxThetaSP}) by left-multiplying by $\bar{s}^{1}(\theta)$ and integrating:
\begin{eqnarray}
|\Theta(\theta_{n},\epsilon)\rangle &=& \frac{1}{\sqrt{N}}e^{-\frac{m_{\theta}R}{2v}\epsilon^{2}}|\bar{s}^{2}(\theta_{n})\rangle_{\lambda(\theta_{n})},
\label{eq:pioverAllSP}
\end{eqnarray}
where we have simplified by exploiting that $|\bar{s}^{2}(\theta_{n})\rangle_{\lambda(\theta_{n})}$ is the eigenstate of $\bar{s}^{2}(\theta_{n})$ with eigenvalue:
\begin{equation}
\lambda(\theta_{n})=(-1)^{n}.
\label{eq:lambdathetasignSP}
\end{equation}
We therefore find that when $m^{-}_{4}$ is the only nonzero mass, there are eight Jackiw-Rebbi zero modes~\cite{JackiwRebbi} bound to zeroes of $\sin(4\theta)$, in agreement with the result obtained in Eq.~(\ref{eq:GeneralZeroesSine}) for a general edge mass term proportional to a circular harmonic $\sin(L_{z}\theta)$.  These eight modes are alternately, in increasing $\theta$, solitions with charge $e/2$ and antisolitions with charge~\cite{Shockley,SSH,RiceMele,NiemiSemenoff} $-e/2$, a charge distribution that can still be summarized using Eq.~(\ref{eq:chargePerN}) (up to an overall sign reflecting an offset in indexing between the (anti)solitons in this and the previous problem (Appendix~\ref{sec:TIboundary})).  The four solitons (antisolitons) therefore occupy the $4a$ ($4b$) Wyckoff position of point group~\cite{BilbaoPoint} $4mm$ (Fig.~\ref{fig:4mm}).  For this arrangement of charges, 
\begin{equation}
n_{4a} = \{0,2,4,6\},\ n_{4b} = \{1,3,5,7\},
\end{equation}
in Eq.~(\ref{eq:thetaNSP}).  Using Eqs.~(\ref{eq:chargePerN}),~(\ref{eq:circularQuadrupole}),~(\ref{eq:circularQuadrupoleSine}) to calculate the quadrupole moments of these (anti)solitions, both:
\begin{equation}
Q^{xy} = \frac{3R^{2}e}{8}\left(\sum_{n\in n_{4a}}\sin(2\theta_{n}) - \sum_{n\in n_{4b}}\sin(2\theta_{n})\right) = 0,
\label{eq:TrivialQXY}
\end{equation}
and:
\begin{equation}
Q^{x^{2}-y^{2}} = \frac{3R^{2}e}{8}\left(\sum_{n\in n_{4a}}\cos(2\theta_{n}) - \sum_{n\in n_{4b}}\cos(2\theta_{n})\right) = 0.
\label{eq:TrivialQX2Y2}
\end{equation}
The overall quadrupole moment is therefore topologically trivial.  

Furthermore, and crucially, we can also show that any linear combination of nonzero mass terms in Eq.~(\ref{eq:UedgeSP}) contributes a trivial quadrupole moment (modulo $e$).  First, using the results of Eqs.~(\ref{eq:GeneralZeroesSine}) through~(\ref{eq:eOver2ForAllSineNontrivial}), we deduce that if \emph{any single} term in $U_{edge}(\theta)$ is nonzero while the other terms are zero, the resulting configuration of zero modes will consist of (up to the overall sign of the charge) four solitions at $4a$, four antisolitons at $4b$, and multiples of eight charges at the general position $8c$ (Fig.~\ref{fig:4mm}).  As Eqs.~(\ref{eq:eOver2ForAll}) and~(\ref{eq:eOver2ForAllSineNontrivial}) show that any set of charges of alternating sign occupying the general position has $Q^{xy}=Q^{x^{2}-y^{2}}=0$, then this, along with Eqs.~(\ref{eq:TrivialQXY}) and~(\ref{eq:TrivialQX2Y2}) implies, that \emph{all} of the terms in $U_{edge}(\theta)$ (Eq.~(\ref{eq:UedgeSP})) individually lead to a configuration of zero modes with trivial quadrupole moments.

We can now explore the effects of adding other mass terms in $U_{edge}(\theta)$ (Eq.~(\ref{eq:UedgeSP})) while keeping $m^{-}_{4}\neq 0$.  First, we consider turning on one of the other mass terms proportional to $m^{-}_{4a}\sin[(4a)\theta]\bar{s}^{1}(\theta)$.  This term will be zero at all of the $\theta_{n}$ in Eq.~(\ref{eq:thetaNSP}) where $\sin(4\theta)$ is zero, as well as at $8(a-1)$ additional values of $\theta$ that correspond to the $8c$ Wyckoff position in $4mm$ (Fig.~\ref{fig:4mm}).  Unlike in the previous discussion in the text following Eq.~(\ref{eq:eOver2ForAll}), in which the 0D modes at $8c$ had alternating charges, in this case, the $8(a-1)$ modes at $8c$ will either \emph{all} be solitons with charge $e/2$ or antisolitons with charge $-e/2$.  However, by direct computation using Eqs.~(\ref{eq:eOver2ForAll}) and~(\ref{eq:eOver2ForAllSineNontrivial}), we confirm that any set of solitons or antisolitons occupying the $8c$ position of $4mm$ contributes net-zero $Q^{xy}$ and $Q^{x^{2}-y^{2}}$ quadrupole moments, and therefore the overall quadrupole moment remains trivial.

We next consider, beginning in the limit that $m^{-}_{4}$ is the only nonzero mass term in Eq.~(\ref{eq:UedgeSP}), the effect of perturbatively introducing a term proportional to $m^{+}_{4a}\cos[(4a)\theta]\bar{s}^{2}(\theta)$.  Adding one of these $m^{+}_{4a}$ terms explicitly breaks particle-hole symmetry (Eq.~(\ref{eq:edgePHSP})).  At each $\theta_{n}$ in Eq.~(\ref{eq:thetaNSP}), we calculate the first-order energy correction:
\begin{eqnarray}
\Delta_{E}(\theta_{n}) &=& m^{+}_{4a}\cos[(4a)\theta_{n}]\langle \bar{s}^{2}(\theta_{n})|_{\lambda(\theta_{n})}\bar{s}^{2}(\theta_{n})|\bar{s}^{2}(\theta_{n})\rangle_{\lambda(\theta_{n})} \nonumber \\
&=& m^{+}_{4a}\lambda(\theta_{n})\cos[(4a)\theta_{n}] \nonumber \\
&=& m^{+}_{4a}(-1)^{n}(-1)^{na} \nonumber \\
&=& m^{+}_{4a}[(-1)^{n}]^{(a+1)},
\label{eq:offsetESP}
\end{eqnarray}
where we have exploited that $\lambda(\theta_{n})=(-1)^{n}$ (Eq.~(\ref{eq:lambdathetasignSP})) and that:
\begin{equation}
\cos[(4a)\theta_{n}]= (-1)^{na},
\end{equation} 
because $\cos[(4a)\theta]$ is always at an extremum at a zero $\theta_{n}$ of $\sin(4\theta)$, which are coincident with the zeroes of $\sin(L_{z}\theta)$ for $L_{z}=L_{z}^{NI}$ (Eq.~(\ref{eq:trivialLz})).  Eq.~(\ref{eq:offsetESP}) implies that if $a$ is even, then the eight 0D states in Fig.~\ref{fig:TrivialCorners}(a) will split into two sets of four states in a $4mm$-symmetric manner, and that if $a$ is odd, then all eight states will shift together in energy by the same amount.  For example, if $a=0$ in Eq.~(\ref{eq:offsetESP}), then the sign of $\Delta_{E}(\theta_{n})$ will alternate with increasing $\theta$, such that four of the modes move down in energy and four of the modes move up in energy while the Fermi level stays in the center (Fig.~\ref{fig:TrivialCorners}).  As this leads to all four of the lower (upper) 0D modes at either the $4a$ or $4b$ ($4b$ or $4a$) position being occupied (unoccupied) (Fig.~\ref{fig:4mm}), the particle-hole-broken system continues to exhibit a trivial quadrupole moment at half filling (Eqs.~(\ref{eq:TrivialQXY}) and~(\ref{eq:TrivialQX2Y2})).  Because mass terms proportional to $m^{+}_{4a}$ in Eq.~(\ref{eq:UedgeSP}) with both odd and even values of $a$ are allowed by symmetry, then generically, the eight corner states in Fig.~\ref{fig:TrivialCorners}(a) will always appear split into groups of four states with trivial quadrupole moments.

We therefore conclude that the $p4m$-symmetric $s-p_{z}$-hybridized TI highlighted in this section \emph{cannot} gap into a QI because it \emph{does not} admit a bulk magnetic mass term that has a nonzero edge projection and is proportional to a circular harmonic with $L_{z}=L_{z}^{QI}=2+4a$ (Eq.~(\ref{eq:nontrivialLz})).  This occurs precisely because the valence and conduction bands of the $k\cdot p$ Hamiltonian (Eq.~(\ref{eq:TIGammaSP})) have the same complex-conjugate pairs of $C_{4z}$ eigenvalues, which forces all possible $4mm$-symmetric magnetic mass terms (Eq.~(\ref{eq:spMasses})) to be proportional to circular harmonics with $L_{z}=L_{z}^{NI}=4a$ (Eq.~(\ref{eq:trivialLz})), which we have shown to generate (anti)solition configurations with trivial quadrupole moments (Eqs.~(\ref{eq:TrivialQXY}) and~(\ref{eq:TrivialQX2Y2})).

\subsection{Gapping the Edge Modes of a 2D TCI with $C_{M_{z}}=2$ with Quadrupolar Magnetism}
\label{sec:TCIBoundary}

In this section, we show that the specific 2D TCI~\cite{TeoFuKaneTCI,NagaosaDirac} with mirror Chern number $C_{M_{z}}=2$ in layer group $p4/mmm1'$ highlighted in this work (Eq.~(\ref{eq:my2Dquad})), equivalent to the Hamiltonian of the $k_{z}=0$ plane of $\mathcal{H}_{H2}({\bf k})$ in Eq.~(\ref{eq:hinge}), gaps into a QI in the presence of magnetism that preserves wallpaper group $p4m$ while breaking $M_{z}$ and $\mathcal{T}$.  We further schematically and numerically show that this TCI gaps into a fragile phase with fractionally charged, quarter-filled (or quarter-empty) Kramers pairs of corner modes under a $p4m$-preserving potential that breaks $M_{z}$ while preserving $\mathcal{T}$.  It is important to note that \emph{not every} 2D TCI can gap to realize fractionally charged 0D states; only TCIs with occupied bands with the same bulk $C_{4z}$ eigenvalues as those of a QI-nontrivial obstructed atomic limit (Appendices~\ref{sec:bandrep} and~\ref{sec:double}) can gap into a QI or fragile TI that exhibits the same corner charges (modulo $e$) as the QI model introduced in Ref.~\onlinecite{multipole}.  We also again emphasize that, as stated in the text before Eq.~(\ref{eq:polarTISP}), $Q^{xy}$ and $Q^{x^{2}-y^{2}}$ are only strictly calculable through Eqs.~(\ref{eq:circularQuadrupole}) and (\ref{eq:circularQuadrupoleSine}) for a $C_{M_{z}}=2$ TCI when $M_{z}$ symmetry is relaxed.  Therefore, as was done previously for the TIs in Appendices~\ref{sec:TIboundary} and~\ref{sec:TItoTrivial}, we will show in this section that by counting the $C_{4z}$ eigenvalues of the occupied bands of a 2D TCI, we can still determine the quadrupole moment that it will exhibit (modulo $e$) when $M_{z}$ is relaxed.

\begin{figure}[h]
\centering
\includegraphics[width=0.45\textwidth]{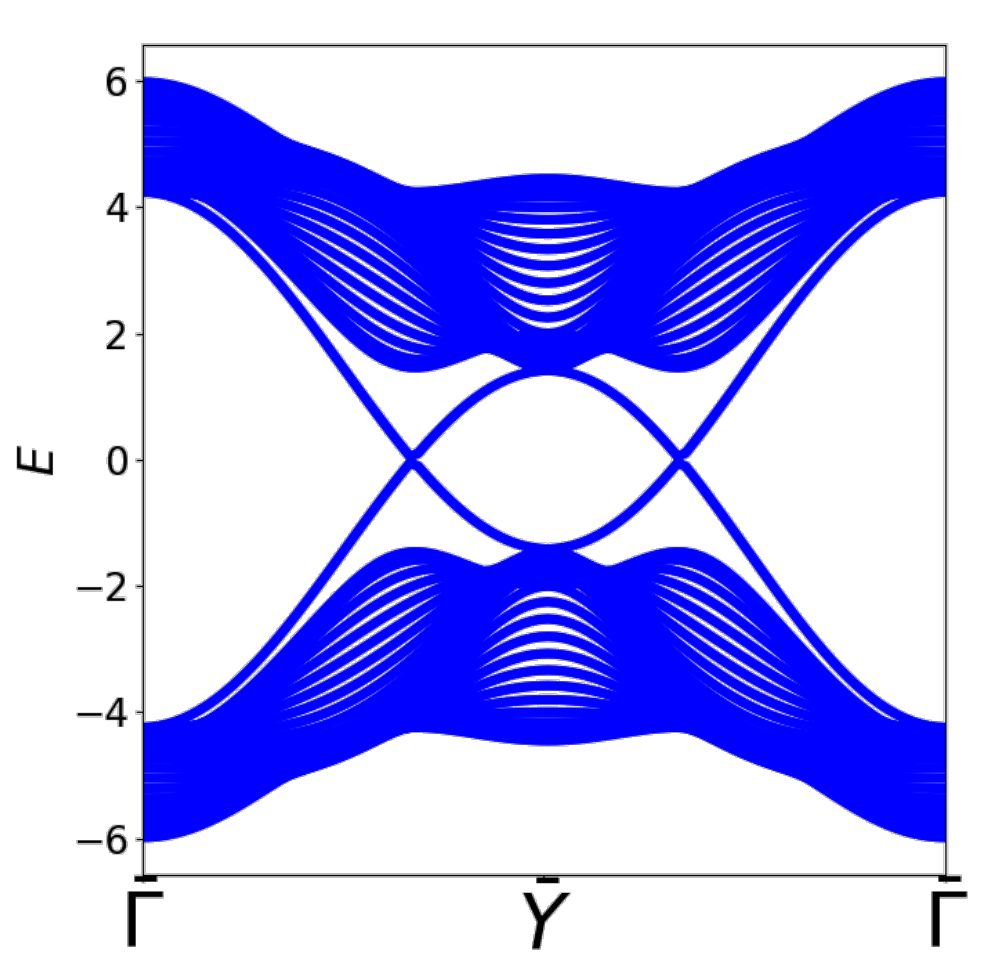}
\caption{The $x$-directed ribbon bands of the $s-d_{x^{2}-y^{2}}$ 2D TCI described by Eq.~(\ref{eq:my2Dquad}), plotted~\cite{PythTB} in the limit that $t_{PH}=0$, $t_{1}= v_{m} = -t_{2} = v_{s}/2 = 2$.  As indicated by the ribbon spectrum, the two occupied bands of this TCI exhibits a mirror Chern number $C_{M_{z}}=2$, which can also be determined by direct computation using Eq.~(\ref{eq:MirrorChernIndicator}).  Unlike in the (diagonal-mirror-symmetric BZ plane of the) experimentally confirmed TCI phase in SnTe~\cite{HsiehTCI}, the TCI phase of Eq.~(\ref{eq:my2Dquad}) is driven by a band inversion between states at $k_{x}=k_{y}=0$ with the same parity eigenvalues and different $C_{4z}$ eigenvalues.  Conversely, in SnTe (both in monolayer form~\cite{LiangTCIMonolayer} and in the diagonal-mirror-symmetric BZ plane of a 3D crystal~\cite{TeoFuKaneTCI,HsiehTCI}), the $C_{M_{z}}=2$ phase is instead driven by band inversion at $(k_{x},k_{y})=(\pi,0)$ and $(0,\pi)$ between bands with opposite parity eigenvalues.  Therefore, as shown in this section, even though the two $C_{M_{z}}=2$ TCI phases are topologically equivalent, they will gap into different corner-mode phases when $M_{z}$ and $\mathcal{I}$ are relaxed, because they exhibit different $C_{4z}$ eigenvalues.}
\label{fig:TCIBound}
\end{figure}

First, we note that the 2D TCI phase highlighted in this work is in this sense distinct from other previous examples of spinful TCIs~\cite{TeoFuKaneTCI,HsiehTCI}.  Whereas the mirror Chern number $C_{M_{z}}=2$ in many previously highlighted TCI phases, such as those in (monolayers~\cite{LiangTCIMonolayer} and mirror-symmetric planes in 3D crystals~\cite{TeoFuKaneTCI,HsiehTCI} of) SnTe~\cite{HsiehTCI}, originated from band-inversion between states with opposite parity eigenvalues at two $C_{4z}$-related TRIM points (\emph{e.g.} $(k_{x},k_{y})=(\pi,0)$ and $(0,\pi$)), in the TCI phase of Eq.~(\ref{eq:my2Dquad}), the bulk topology instead originates from band inversion at a \emph{single}, $C_{4z}$-invariant TRIM point between states with the same parity eigenvalues and different $C_{4z}$ eigenvalues (Eq.~(\ref{eq:MirrorChernIndicator})).  In both TCIs, the band within each $M_{z}$ sector undergoes a direct transition from exhibiting Chern number $C=0$ to $C=\pm 2$.  However, in our model (Eq.~(\ref{eq:my2Dquad})), this change in Chern number $\Delta C = \pm 2$ is driven by a change in the occupied $C_{4z}$ eigenvalues at $\Gamma$ ($(k_{x},k_{y})=(0,0)$), \emph{without a corresponding change in the occupied parity eigenvalues} at $\Gamma$, or at any other TRIM point.  This allows our model to realize a transition from a trivial insulator to a $C_{M_{z}}=2$ TCI while still exhibing the same $C_{4z}$ eigenvalues as a QI (as well as the same $C_{4z}$ eigenvalues as the 2D $p-d$-hybridized TI highlighted in Appendix~\ref{sec:TIboundary}).  Therefore, like the previous $p-d$-hybridized TI, when we break $M_{z}$ and $\mathcal{T}$ symmetries while keeping $p4m$, we will see that the 2D TCI phase of Eq.~(\ref{eq:my2Dquad}) gaps into a QI.  However, as we will also see in this section, unlike for the previous $p-d$-hybridized TI in Appendix~\ref{sec:TIboundary}, we can also realize a corner-mode phase with a nontrivial quadrupole moment \emph{without} breaking $\mathcal{T}$.  Specifically, we can gap the edge states of the TCI phase of Eq.~(\ref{eq:my2Dquad}) by breaking $M_{z}$ while keeping $\mathcal{T}$ symmetry, which we will see results in an insulator (which is fragile, and not Wannierizable like the QI (Appendices~\ref{sec:bandrep} and~\ref{sec:fragile})) that also exhibits corner modes for which $Q^{xy}\text{ mod }e=e/2$ or $Q^{x^{2}-y^{2}}\text{ mod }e=e/2$.

We begin by expanding Eq.~(\ref{eq:my2Dquad}) about the $\Gamma$ point to quadratic order in ${\bf k}$:
\begin{equation}
\mathcal{H}_{\Gamma}(k_{x},k_{y}) = [v_{m} + t(2-\frac{1}{2}(k_{x}^{2} + k_{y}^{2}))]\tau^{z} + C_{1}(k_{x}^{2}-k_{y}^{2})\tau^{x} + 2C_{2}k_{x}k_{y}\tau^{y}\sigma^{z},
\label{eq:TCIkp}
\end{equation}
where we have set $t_{PH}=0$ and:
\begin{equation}
t = t_{1},\ C_{1} = -t_{2}/2,\ C_{2} = v_{s}/2.
\end{equation}
In the notation of Eq.~(\ref{eq:transformNotation}), $\mathcal{H}_{\Gamma}(k_{x},k_{y})$ (Eq.~(\ref{eq:TCIkp})) transforms in the symmetry representation given by:
\begin{align}
&\mathcal{T}:\ \sigma^{y}\mathcal{H}^{*}_{\Gamma}(-k_{x},-k_{y})\sigma^{y},\ M_{z}:\ \sigma^{z}\mathcal{H}_{\Gamma}(k_{x},k_{y})\sigma^{z},\ M_{x}:\ \sigma^{x}\mathcal{H}_{\Gamma}(-k_{x},k_{y})\sigma^{x}, \nonumber \\ 
&M_{y}:\ \sigma^{y}\mathcal{H}_{\Gamma}(k_{x},-k_{y})\sigma^{y},\ C_{4z}:\ \tau^{z}\left(\frac{\mathds{1}_{\sigma} - i\sigma^{z}}{\sqrt{2}}\right)\mathcal{H}_{\Gamma}(k_{y},-k_{x})\tau^{z}\left(\frac{\mathds{1}_{\sigma} + i\sigma^{z}}{\sqrt{2}}\right),
\label{eq:sdKspaceSyms}
\end{align}
where $\mathds{1}_{\sigma}$ is the $2\times 2$ identity in $\sigma$ space.  Eq.~(\ref{eq:sdKspaceSyms}) implies an inversion symmetry $\mathcal{I}=M_{x}M_{y}M_{z}$ that transforms $\mathcal{H}_{\Gamma}(k_{x},k_{y})$ under the representation:
\begin{equation}
\mathcal{I}:\ \mathcal{H}_{\Gamma}(-k_{x},-k_{y}).
\label{eq:sdISym}
\end{equation}
The $\tau^{z}$ contribution to only $C_{4z}$ in Eqs.~(\ref{eq:sdKspaceSyms}) and~(\ref{eq:sdISym}) reflects that Eq.~(\ref{eq:TCIkp}) describes a 2D TCI formed of hybridized $s$ and $d_{x^{2}-y^{2}}$ orbitals, as $s$ ($d_{x^{2}-y^{2}}$) orbitals are even under $M_{x,y,z}$ and $\mathcal{I}$ and even (odd) under $C_{4z}$.

Unlike in Appendices~\ref{sec:TIboundary} and~\ref{sec:TItoTrivial}, Eq.~(\ref{eq:TCIkp}) is quadratic, and thus eludes the (relatively) straightforward Jackiw-Rebbi analysis performed in those sections.  Therefore, to determine the corner modes that result from gapping the TCI edge states, we will employ two distinct approaches.  First, in Appendix~\ref{sec:EBRsforTCI}, we will use EBRs to show that the particular $s-d$-hybridized TCI described by Eq.~(\ref{eq:my2Dquad}) exhibits the combined quadrupole moments (modulo $e$) of the $p-d$- and $s-p$-hybridized 2D TIs analytically examined in Appendices~\ref{sec:TIboundary} and~\ref{sec:TItoTrivial}, respectively.  Next, in Appendix~\ref{sec:fragilePhaseCorners}, to further draw connection between the fragile phase in $p4m1'$ examined in Appendix~\ref{sec:fragile} and the QI, we will use symmetry arguments and the results of numerical calculations, in conjunction with the analytic results derived in Appendices~\ref{sec:TIboundary} and~\ref{sec:TItoTrivial}, to determine the corner spectrum and charges of the particle-hole-\emph{asymmetric} fragile phase from Appendix~\ref{sec:fragile} by smoothly deforming the corner spectrum of the $\Pi$-symmetric QI phase of Eqs.~(\ref{eq:my2Dquad}) and~(\ref{eq:quad}).  We will find in particular that even though this fragile phase is $\mathcal{T}$-symmetric, it still displays Kramers pairs of corner modes, that, because they originate from an imbalanced number of valence and conduction states ($6$ and $2$), which is allowed because $\Pi$ symmetry is absent, still exhibit the same quadrupole moment $Q^{xy}\text{ mod }e=e/2$ or $Q^{x^{2}-y^{2}}\text{ mod }e=e/2$ as the corner modes of a \emph{magnetic} QI in $p4m$ when the overall system is half filled.

\subsubsection{The Quadrupole Moment of an $s-d$-Hybridized TCI from EBRs}
\label{sec:EBRsforTCI}

We begin by introducing the EBRs of layer group~\cite{WiederLayers,DiracInsulator,SteveMagnet,MagneticBook,subperiodicTables}:
\begin{equation}
G=p4/mmm1',
\end{equation}
which we will use to determine the quadrupole moment of the $s-d$-hybridized TCI highlighted in this work (Eq.~(\ref{eq:my2Dquad})) when its edge states are gapped with $M_{z}$-breaking, $p4m$-symmetric magnetism.  Using the~\textsc{BANDREP} tool on the BCS~\cite{QuantumChemistry,Bandrep1,Bandrep2,Bandrep3}, we induce the following EBRs from the $1a$ Wyckoff position of $G$ (Fig.~\ref{fig:2Dmain}(a)):
\begin{eqnarray}
\left(s\right)_{1a}\uparrow G &\equiv& \left(\bar{\rho}_{7}^{+}\right)_{\Gamma} \oplus \left(\bar{\varrho}_{5}^{+}\right)_{X} \oplus \left(\bar{\rho}_{7}^{+}\right)_{M} \nonumber \\
\left(p_{z}\right)_{1a}\uparrow G &\equiv& \left(\bar{\rho}_{7}^{-}\right)_{\Gamma} \oplus \left(\bar{\varrho}_{5}^{-}\right)_{X} \oplus \left(\bar{\rho}_{7}^{-}\right)_{M} \nonumber \\
\left(d_{x^{2}-y^{2}}\right)_{1a}\uparrow G &\equiv& \left(\bar{\rho}_{6}^{+}\right)_{\Gamma} \oplus \left(\bar{\varrho}_{5}^{+}\right)_{X} \oplus \left(\bar{\rho}_{6}^{+}\right)_{M},
\label{eq:EBRsforTCI}
\end{eqnarray}
where we have employed the shorthand of Ref.~\onlinecite{WiederAxion} in which, in the left-hand side of Eq.~(\ref{eq:EBRsforTCI}), $(\bar{\sigma})_{{\bf q}}$ is the EBR induced from the corepresentation of the site-symmetry group of the Wyckoff position at ${\bf q}$ that transforms as the (Kramers pair of) atomic orbitals $\bar{\sigma}$, and where, in the right-hand side of Eq.~(\ref{eq:EBRsforTCI}), $(\bar{\rho})_{{\bf k}}$ is the corepresentation subduced at the TRIM point ${\bf k}$.  In Eq.~(\ref{eq:allChars2}), all of the position- and momentum-space corepresentations are two-dimensional, and we do not show corepresentations at the $X'$ point ($(k_{x},k_{y})=(0,\pi)$), because they are fixed to be the same as the corepresentations at the $X$ point ($(k_{x},k_{y})=(\pi,0)$) by $C_{4z}$ symmetry.  In terms of their $C_{4z}$ and parity ($\mathcal{I}$) eigenvalues, as discussed in Appendix~\ref{sec:pd}:
\begin{equation}
\chi_{\bar{\rho}_{6,7}^{\pm}}(\mathcal{I}) = \pm2,\ \chi_{\bar{\rho}^{\pm}_{6}}(C_{4z})=-\sqrt{2},\ \chi_{\bar{\rho}^{\pm}_{7}}(C_{4z})=\sqrt{2}, \chi_{\bar{\varrho}_{5}^{\pm}}(\mathcal{I}) = \pm 2,
\label{eq:allChars2}
\end{equation}
where $\chi_{\bar{\rho}}(h)$ is the character of the unitary symmetry $h$ in the corepresentation $\bar{\rho}$, and is equal to the sum of the eigenvalues of $h$ in $\bar{\rho}$, and where $\chi_{\bar{\varrho}_{5}^{\pm}}(C_{4z})$ does not appear because the $X$ point in $p4/mmm1'$ is not invariant under $C_{4z}$.   

Next, to form the topological valence bands of the 2D phases analyzed in this section and in Appendices~\ref{sec:TIboundary} and~\ref{sec:TItoTrivial}, we invert bands at the $\Gamma$ point between pairs of EBRs from Eq.~(\ref{eq:EBRsforTCI}):
\begin{eqnarray}
\left[\text{2D TI}\right]_{p_{z}-d_{x^{2}-y^{2}}} \equiv \left(\bar{\rho}_{7}^{-}\right)_{\Gamma} \oplus \left(\bar{\varrho}_{5}^{+}\right)_{X} \oplus \left(\bar{\rho}_{6}^{+}\right)_{M} \nonumber \\
\left[\text{2D TI}\right]_{s-p_{z}} \equiv \left(\bar{\rho}_{7}^{+}\right)_{\Gamma} \oplus \left(\bar{\varrho}_{5}^{-}\right)_{X} \oplus \left(\bar{\rho}_{7}^{-}\right)_{M} \nonumber \\
\left[\text{2D TCI}\right]_{s-d_{x^{2}-y^{2}}} \equiv \left(\bar{\rho}_{7}^{+}\right)_{\Gamma} \oplus \left(\bar{\varrho}_{5}^{+}\right)_{X} \oplus \left(\bar{\rho}_{6}^{+}\right)_{M},
\label{eq:TIsforTCI}
\end{eqnarray}
for which, in the convention of Eq.~(\ref{eq:MirrorChernIndicator}), we calculate the mirror Chern numbers $C_{M_{z}}$ and $\mathbb{Z}_{2}$ TI indices $z_{2}$ to be:
\begin{eqnarray}
C_{M_{z}}\left(\left[\text{2D TI}\right]_{p_{z}-d_{x^{2}-y^{2}}}\right)\text{ mod }4 &=& -1,\ C_{M_{z}}\left(\left[\text{2D TI}\right]_{s-p_{z}}\right)\text{ mod }4 = -1,\ C_{M_{z}}\left(\left[\text{2D TCI}\right]_{s-d_{x^{2}-y^{2}}}\right)\text{ mod }4 = 2, \nonumber \\
z_{2}\left(\left[\text{2D TI}\right]_{p_{z}-d_{x^{2}-y^{2}}}\right) &=& 1,\ z_{2}\left(\left[\text{2D TI}\right]_{s-p_{z}}\right) = 1,\ z_{2}\left(\left[\text{2D TCI}\right]_{s-d_{x^{2}-y^{2}}}\right) = 0.
\end{eqnarray}

Crucially, Eqs.~(\ref{eq:EBRsforTCI}) and~(\ref{eq:TIsforTCI}) imply the equivalence:
\begin{equation}
\left[\text{2D TI}\right]_{p_{z}-d_{x^{2}-y^{2}}} \oplus \left[\text{2D TI}\right]_{s-p_{z}} \equiv \left[\text{2D TCI}\right]_{s-d_{x^{2}-y^{2}}} \oplus \left(p_{z}\right)_{1a}\uparrow G.
\label{eq:QIfromEBRs}
\end{equation}
In Appendix~\ref{sec:TIboundary} (\ref{sec:TItoTrivial}) we showed that 2D TIs formed from $p-d$-hybridization ($s-p_{z}$-hybridization) transition into QIs (trivial insulators) upon gapping their edge states with $p4m$-symmetric magnetism.  This implies that when $p4m$-symmetric magnetism is applied to an insulator whose occupied bands are given by the left-hand side of Eq.~(\ref{eq:QIfromEBRs}), then this insulator will transition into a QI (\emph{i.e.} an insulator whose bulk bands and corner modes exhibit $Q^{xy}\text{ mod }e=e/2$ or $Q^{x^{2}-y^{2}}\text{ mod }e=e/2$).  Furthermore, because $\left(p_{z}\right)_{1a}\uparrow G$ is an unobstructed (trivial) atomic limit at the $1a$ position, then it will not exhibit a quadrupole moment under the application of $p4m$-symmetric magnetism (Appendix~\ref{sec:bandrep}).  Therefore, because the left- and right-hand sides of Eq.~(\ref{eq:QIfromEBRs}) must characterize equivalent, QI-nontrivial insulators in the presence of $p4m$-symmetric magnetism, we conclude that the $s-d$-hybridized TCI phase of Eq.~(\ref{eq:my2Dquad}) will transition into a QI under $M_{z}$-breaking, $p4m$-symmetric magnetism.

\subsubsection{Fractionally Charged Corner Modes in a Fragile Phase in $p4m1'$}
\label{sec:fragilePhaseCorners}

\begin{figure}[h]
\centering
\includegraphics[width=0.5\textwidth]{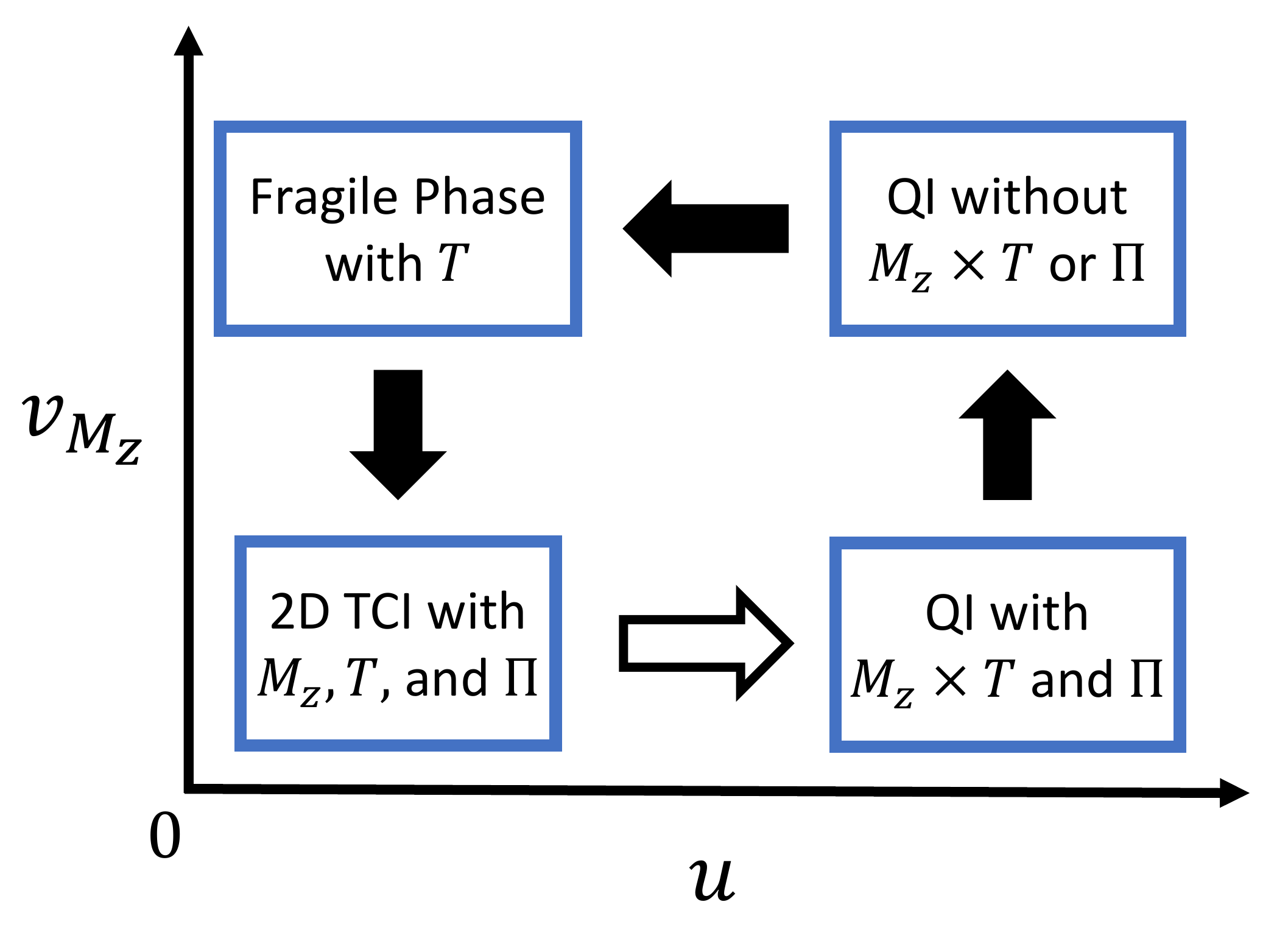}
\caption{Schematic for our analysis of the bulk, edge, and corner spectra of related QI, fragile, and $s-d$-hybridized TCI phases (Appendix~\ref{sec:parameters}).  Even though the 2D $C_{M_{z}}=2$ TCI in $p4/mmm1'$ examined in this section originates from $s-d$ hybridization (Eqs.~(\ref{eq:TCIkp}) through~(\ref{eq:sdISym})), when the symmetries are lowered to $p4/m'mm$, it transitions into a particle-hole- ($\Pi$-) symmetric QI whose linear $k\cdot p$ theory (Eq.~(\ref{eq:finalQuadJackiw})) can be mapped to the $k\cdot p$ theory of a (magnetically gapped) $p_{z}-d_{x^{2}-y^{2}}$ TI (Eq.~(\ref{eq:TIGamma})), which we previously analyzed in Appendix~\ref{sec:TIboundary}.  We can therefore schematically work backwards from this $\Pi$-symmetric QI phase (black arrows) to avoid performing the more complicated analysis of the edge and corner modes of the quadratic $k\cdot p$ theory of a $\Pi$-symmetric $C_{M_{z}}=2$ TCI (Eq.~(\ref{eq:TCIkp}) and Fig.~\ref{fig:TCIBound}).  This allows us to exploit the previous analytic expressions for the boundary (corner) modes and quadrupole moments of a $\Pi$-symmetric QI in Appendix~\ref{sec:TIboundary} to infer the evolution of the edge and corner spectra of our 2D models in Appendix~\ref{sec:parameters} through the direct transition from the $\Pi$-symmetric TCI to the $\Pi$-symmetric QI (white arrow), as well as through the transition from a $\Pi$-broken QI to a $\mathcal{T}$-symmetric fragile TI  in $p4m1'$ with fractionally charged corner modes (Fig.~\ref{fig:TCIcorners}).}
\label{fig:TCIphases}
\end{figure}

\begin{figure}[h]
\centering
\includegraphics[width=0.68\textwidth]{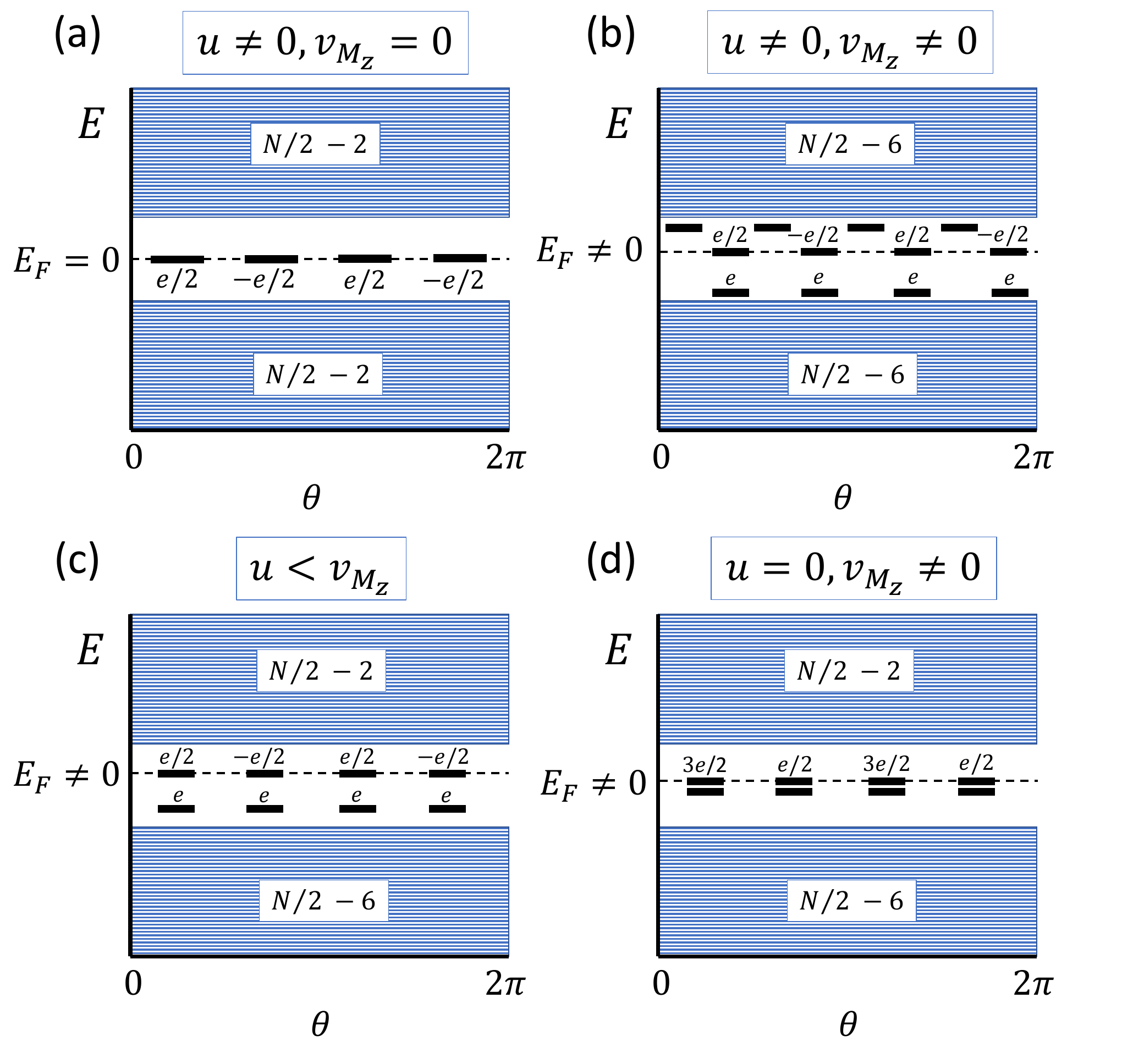}
\caption{Schematic open-boundary-condition (OBC) energy spectra for the transition from a $p4m$- and particle-hole- ($\Pi$-) symmetric QI to a $p4m$- and $\mathcal{T}$-symmetric fragile phase with broken $\Pi$ symmetry.  (a) The half-filled corner modes of a $\Pi$-symmetric QI, characterized by $u\neq 0$ and $v_{M_{z}}=0$ in Eqs.~(\ref{eq:my2Dquad}),~(\ref{eq:breakI}), and~(\ref{eq:quad}) (Fig.~\ref{fig:cornerModes}(a)).  In terms of $M_{z}$ and $\mathcal{T}$ symmetries, both $u$ and $v_{M_{z}}$ break $M_{z}$, whereas $u$ ($v_{M_{z}}$) breaks (respects) $\mathcal{T}$ symmetry; both $u$ and $M_{z}$ respect the symmetries of $p4m$ (Eq.~(\ref{eq:symsParams})).  (b)  Keeping $u\neq 0$, we can introduce eight more corner modes by breaking $\Pi$-symmetry through the introduction of nonzero $v_{M_{z}}$ in Eq.~(\ref{eq:breakI}).  These eight modes are equivalent to the QI-trivial corner states of the 2D $s-p_{z}$-hybridized TI gapped with $p4m$-symmetric magnetism shown in Fig.~\ref{fig:TrivialCorners}(b); as shown in the text surrounding Eq.~(\ref{eq:offsetESP}), the eight trivial modes generically appear in two energetically-split sets of four $4mm$-symmetry-related states.  (c) Further breaking $\Pi$ symmetry and reducing the strength of $u$ in Eq.~(\ref{eq:quad}), we return the four empty, QI-trivial corner modes to the conduction manifold, and push the four occupied trivial corner modes closer to the half-filled (QI-nontrivial) modes in energy.  We numerically observe this process in the rod bands near $k_{z}=0$ in Fig.~\ref{fig:HingeSMmain}(h) of the main text, in which a trivial set of HOFAs from the valence manifold begins to approach the HOFAs at the spectral center.  (d) Turning $u$ completely to zero while keeping $v_{M_{z}}\neq 0$, we keep the bulk and edge gap open and restore $\mathcal{T}$ symmetry, resulting in the fragile phase detailed in Appendix~\ref{sec:fragile}.  The restoration of $\mathcal{T}$ symmetry forces the corner modes to become doubly degenerate, as they characterize spinful electrons (Fig.~\ref{fig:HingeSMmain}(h)).  However, as we have not closed a bulk or edge gap, the filling of the corner modes persists from the $\Pi$-symmetric QI phase in (a), resulting in quarter-filled (and quarter-empty) corner modes that exhibit the same $xy$ quadrupole moment, taken modulo $e$, as the corner modes of a magnetic QI in $p4m$ (Appendix~\ref{sec:TIboundary}).  This is consistent with our determination in Appendix~\ref{sec:EBRsforTCI} that the specific $s-d$-hybridized TCI phase of Eq.~(\ref{eq:my2Dquad}) exhibits the same nontrivial ($e/2$) quadrupole moment as a $p-d$-hybridized 2D TI when the edge states of the two insulators are respectively gapped.  The eight corner modes in (d) modes are not pinned to $E=0$, as $\Pi$ is strongly broken.  Nevertheless, as there are eight corner modes and the valence and conduction manifolds differ by four states, acting on the modes with a chemical potential that pushes them into one of the bulk manifolds while preserving the bulk and edge gaps will not resolve the (anomalous~\cite{ChenConvo,WiederAxion}) mismatch between the number of states in the valence and conduction manifolds.}
\label{fig:TCIcorners}
\end{figure}

In this section, we will use symmetry arguments bolstered by numerical observations to show that the QI phase of Eqs.~(\ref{eq:my2Dquad}) and~(\ref{eq:quad}) can transition into a fragile TI in $p4m1'$ (Appendix~\ref{sec:fragile}) that exhibits the same corner charges as a QI (taken modulo $e$).  Because we previously showed in Appendix~\ref{sec:EBRsforTCI} that the specific $s-d$-hybridized TCI in Eq.~(\ref{eq:TCIkp}) can transition into a QI under breaking $M_{z}$ and $\mathcal{T}$, then this will allow us to conclude that the same TCI can transition into a fragile phase with corner charges when $M_{z}$ is broken while preserving $\mathcal{T}$ (Fig.~\ref{fig:TCIphases}).  We begin by expanding Eqs.~(\ref{eq:my2Dquad}) and~(\ref{eq:quad}) to linear order about the $\Gamma$ point:
\begin{equation}
\mathcal{H}_{Q}({\bf k}) = m\tau^{z} + u\tau^{y}(\sigma^{y}k_{x} + \sigma^{x}k_{y}),
\label{eq:finalQuadJackiw}
\end{equation}
in the limit that $t_{PH}=0$ and where $m=2t_{1}+v_{m}$.  Eq.~(\ref{eq:finalQuadJackiw}) is the bulk $k\cdot p$ theory of a QI with $M_{z}\times\mathcal{T}$ and $\Pi$ symmetries (Appendix~\ref{sec:parameters}); it is specifically invariant under $C_{4z}$, $M_{x,y}$, $M_{z}\times\mathcal{T}$, and $\mathcal{I}\times\mathcal{T}$ in the symmetry representation in Eqs.~(\ref{eq:sdKspaceSyms}) and~(\ref{eq:sdISym}).  We note that, because Eq.~(\ref{eq:finalQuadJackiw}) originates from Eqs.~(\ref{eq:my2Dquad}) and~(\ref{eq:quad}), then, in the limit that $u\rightarrow 0$, Eq.~(\ref{eq:finalQuadJackiw}) reduces to the linear $k\cdot p$ Hamiltonian of the 2D TCI phase of Eq.~(\ref{eq:my2Dquad}), which is just given by the $m\tau^{z}$ terms in Eqs.~(\ref{eq:TCIkp}) and~(\ref{eq:finalQuadJackiw}).  As $\mathcal{H}_{Q}({\bf k})$ does not contain terms proportional to $\tau^{x}$ or $\tau^{y}\sigma^{z}$, which only arise in a quadratic-order expansion of Eq.~(\ref{eq:my2Dquad}), then it is slightly underconstrained, and exhibits additional, artificial symmetries:
\begin{equation}
\tilde{\mathcal{T}}:\ \tau^{z}\sigma^{y}\mathcal{H}^{*}_{Q}(-{\bf k})\tau^{z}\sigma^{y},\ \tilde{\mathcal{I}}:\ \tau^{z}\mathcal{H}_{Q}(-{\bf k})\tau^{z},
\label{eq:fakesyms} 
\end{equation}
in the notation of Eq.~(\ref{eq:transformNotation}).  Notably, under a unitary transformation:
\begin{equation}
U=\frac{1}{\sqrt{2}}\left(\mathds{1}_{\tau\sigma} + i\tau^{z}\right),
\label{eq:fakeUnitaryTransform}
\end{equation}
which rotates $\tau^{x,y}$ while preserving $\tau^{z}$ and all of the spin matrices $\sigma^{i}$:
\begin{equation}
\tau^{y}\rightarrow\tau^{x},\ \tau^{x}\rightarrow -\tau^{y},\ \tau^{z}\rightarrow\tau^{z},
\end{equation}
Eq.~(\ref{eq:finalQuadJackiw}) transforms into:
\begin{equation}
\tilde{\mathcal{H}}_{Q}({\bf k}) = U\tilde{\mathcal{H}}_{Q}({\bf k})U^{\dag} = m\tau^{z} + u\tau^{x}(\sigma^{y}k_{x} + \sigma^{x}k_{y}),
\label{eq:finalFakeTI}
\end{equation}
and the symmetries as represented in Eqs.~(\ref{eq:sdKspaceSyms}) and~(\ref{eq:fakesyms}) transform into:
\begin{align}
&\tilde{\mathcal{T}}:\ \sigma^{y}\tilde{\mathcal{H}}^{*}_{Q}(-k_{x},-k_{y})\sigma^{y},\ \tilde{\mathcal{I}}:\ \tau^{z}\tilde{\mathcal{H}}_{Q}(-k_{x},-k_{y})\tau^{z},\ M_{x}:\ \sigma^{x}\tilde{\mathcal{H}}_{Q}(-k_{x},k_{y})\sigma^{x}, \nonumber \\ 
&M_{y}:\ \sigma^{y}\tilde{\mathcal{H}}_{Q}(k_{x},-k_{y})\sigma^{y},\ C_{4z}:\ \tau^{z}\left(\frac{\mathds{1}_{\sigma} - i\sigma^{z}}{\sqrt{2}}\right)\tilde{\mathcal{H}}_{Q}(k_{y},-k_{x})\tau^{z}\left(\frac{\mathds{1}_{\sigma} + i\sigma^{z}}{\sqrt{2}}\right), \nonumber \\
&\tilde{M}_{z}:\ \tau^{z}\sigma^{z}\tilde{\mathcal{H}}_{Q}(k_{x},k_{y})\tau^{z}\sigma^{z},
\label{eq:fakePDTISyms}
\end{align}
where $\tilde{M}_{z} = \tilde{\mathcal{I}}M_{x}^{-1}M_{y}^{-1}$.  Eqs.~(\ref{eq:finalFakeTI}) and~(\ref{eq:fakePDTISyms}) are exactly equal to, respectively, the $k\cdot p$ Hamiltonian (Eq.~(\ref{eq:TIGamma})) and the symmetry representation (Eq.~(\ref{eq:pdKspaceSyms})) of the 2D $p_{z}-d_{x^{2}-y^{2}}$-hybridized TI previously analyzed in Appendix~\ref{sec:TIboundary}.  Therefore, we conclude that when Fourier transformed and placed on a disc geometry, Eq.~(\ref{eq:finalFakeTI}) will only admit bulk, edge-projecting mass terms with $L_{z}^{QI}=2+4a$ (Eq.~(\ref{eq:nontrivialLz})), and will therefore, exhibit a configuration of 0D (anti)solitons on its boundary with a nontrivial quadrupole moment $Q^{xy}\text{ mod }e=e/2$ (Eq.~(\ref{eq:circularQuadrupole})) or $Q^{x^{2}-y^{2}}\text{ mod }e=e/2$ (Eq.~(\ref{eq:eOver2ForAllSineNontrivial})), as depicted in Fig.~\ref{fig:TCIcorners}(a).  Because Eq.~(\ref{eq:finalQuadJackiw}) (and hence Eq.~(\ref{eq:finalFakeTI})) originated from adding Eq.~(\ref{eq:quad}) to Eq.~(\ref{eq:my2Dquad}) to gap its TCI edge states without closing a bulk gap, then this implies that the 2D $s-d$-hybridized TCI phase of Eq.~(\ref{eq:my2Dquad}) gaps into \emph{the same} $\Pi$-symmetric QI as does a $p-d$-hybridized 2D TI (Appendix~\ref{sec:TIboundary}) when their edge states are gapped by breaking $M_{z}$ and $\mathcal{T}$ while preserving $p4m$, in agreement with the result obtained from EBRs in Appendix~\ref{sec:EBRsforTCI}.

Next, we will exploit this result to demonstrate that a $\mathcal{T}$-symmetric fragile phase in $p4m1'$ that is connected to a ($\mathcal{T}$-broken) $p4m$-symmetric QI without closing a bulk or edge gap also exhibits fractionally charged corner modes.  We will first use symmetry arguments (bolstered by explicit numerical calculations) to track the effects of introducing Eq.~(\ref{eq:breakI}) and tuning $v_{M_{z}}$ to be nonzero.  The QI phase of Eq.~(\ref{eq:my2Dquad}) and~(\ref{eq:quad}) is invariant under $p4m$ and $M_{z}\times\mathcal{T}$ symmetries (magnetic layer group~\cite{MagneticBook,subperiodicTables} $p4/m'mm$), as well as $\Pi$ symmetry, where specifically the terms proportional to $u$ in Eq.~(\ref{eq:quad}) break $\mathcal{T}$ and $M_{z}$ symmetries while preserving their product $M_{z}\times\mathcal{T}$.  We begin by assuming that $u$ is large, and that this corresponds to the dominant mass term in Eq.~(\ref{eq:pdMasses}) being $m^{+}_{2}\cos(2\theta)$, giving the $L_{z}=2$ distribution of corner charges in Fig.~\ref{fig:TCIcorners}(a), which we observe in our numerical calculations (\emph{e.g.}, the hinge states in Fig.~\ref{fig:HingeSMmain}(f-h) of the main text).  We next break $M_{z}\times\mathcal{T}$ and particle-hole symmetries by introducing nonzero $v_{M_{z}}$ in Eq.~(\ref{eq:breakI}); particle-hole symmetry is specifically broken because, when $v_{M_{z}}$ is nonzero, there are no $4\times 4$ matrices that anticommute with the combination of Eqs.~(\ref{eq:my2Dquad}),~(\ref{eq:quad}), and~(\ref{eq:breakI}).  In the bulk spectrum, because $M_{x,y}$ are still enforced, and $\mathcal{I}=M_{x}M_{y}M_{z}$, $v_{M_{z}}$ also breaks $\mathcal{I}\times\mathcal{T}$ in the bulk, causing bands to become singly degenerate (observable at $k_{z}\neq 0,\pi$ in the bulk band structure of the fragile topological Dirac semimetal in Fig.~\ref{fig:HingeSMmain}(e) of the main text).  Tuning $v_{M_{z}}$ away from zero does not change the filling of the four corner modes, but does allow additional, QI-trivial corner modes to float into the gap (Fig.~\ref{fig:TCIcorners}(b)).  We depict this process in Fig.~\ref{fig:TCIcorners}(b) as the donation from the valence and conduction manifolds of eight additional corner charges (four from each manifold), representative of the zero modes of a QI-trivial $L_{z}=4$ mass term (Eq.~(\ref{eq:trivialLz})) split with a $p4m$-symmetric, $\Pi$-breaking potential (Fig.~\ref{fig:TrivialCorners}(b)).  More generally, as shown in Eq.~(\ref{eq:UedgeSP}), QI-trivial 0D states (\emph{i.e.} those that do not carry a $Q^{xy}$ or $Q^{x^{2}-y^{2}}$ modulo $e$ of $e/2$ (Eqs.~(\ref{eq:eOver2ForAll}) and~(\ref{eq:eOver2ForAllSineNontrivial}), respectively)) can appear in any arrangement consistent with a linear combination of circular harmonics with $L_{z}=L_{z}^{NI}=4a$ (Eq.~(\ref{eq:trivialLz})) (energetically split in a manner respecting the symmetries of $p4m$ (Fig.~\ref{fig:TrivialCorners})); we are in this section only choosing the $L_{z}=4$ harmonic to be associated with nonzero $v_{M_{z}}$ because it is the harmonic with the smallest $L_{z}$ that is consistent with the trivial corner (hinge) states that we observe in our numerics (Fig.~\ref{fig:HingeSMmain}(h) of the main text).  As with all of the $L_{z}\neq 0$ mass terms in Eq.~(\ref{eq:UedgeSP}), the $L_{z}=4$ mass term creates four (occupied) modes of the same charge (which we here take to be $e$) at the same angles as the (QI-nontrivial) $m^{+}_{2}\cos(2\theta)$ term in Eq.~(\ref{eq:UedgePD}), as well as four (unoccupied) modes with a different charge (here $0$) at four angles that are offset from the first four angles by $\pi/4$; the eight total 0D boundary modes originating from the $L_{z}=4$ mass term are equivalent to the QI-trivial bound states of an $s-p_{z}$-hybridized 2D TI gapped with $p4m$-symmetric magnetism (Fig.~\ref{fig:TrivialCorners}(b) and Appendix~\ref{sec:EBRsforTCI}).  We depict this $\Pi$-breaking arrangement of QI-trivial and non-trivial corner modes in Fig.~\ref{fig:TCIcorners}(b).

Next, we begin to restore $\mathcal{T}$ symmetry by tuning $u$ towards zero in Eq.~(\ref{eq:quad}), while keeping $v_{M_{z}}\neq 0$ in Eq.~(\ref{eq:breakI}); in our numerical calculations using the parameters in Table~\ref{tb:2D}, this causes the four unoccupied QI-trivial corner modes to return to the conduction manifold and the fully occupied trivial corner modes to approach the nontrivial 0D states in energy (Fig.~\ref{fig:TCIcorners}(c)).  This process can specifically be observed in the rod bands near $k_{z}=0$ of the noncentrosymmetric fragile topological Dirac semimetal shown in Fig.~\ref{fig:HingeSMmain}(h) of the main text, in which a QI-trivial set of HOFAs from the valence manifold begins to approach the QI-nontrivial HOFAs at the spectral center.  Specifically, in Fig.~\ref{fig:HingeSMmain}(h) of the main text, near $k_{z}=0$, $k_{z}$ acts exactly like $u$ to break $M_{z}$ and $\mathcal{T}$, because the SOC term in Eq.~(\ref{eq:hinge}) is proportional to $u\sin(k_{z})$.  Next, we set $u=0$ while keeping $v_{M_{z}}\neq 0$, resulting in the preservation of the bulk and edge gaps and the restoration of $\mathcal{T}$ symmetry (Fig.~\ref{fig:TCIphases}).  As shown in Appendix~\ref{sec:fragile}, the resulting insulator has the bulk symmetries of $\mathcal{T}$-symmetric wallpaper group $p4m1'$, and exhibits fragile topology.  To satisfy Kramers' theorem in the fragile phase, the trivial corner modes join with the half-filled, nontrivial corner modes to form $3/4$-filled Kramers pairs of corner states, which carry alternating charges $3e/2$ and $e/2$ (Fig.~\ref{fig:TCIcorners}(d) and $k_{z}=0$ in Fig.~\ref{fig:HingeSMmain}(h) of the main text).  Taken modulo $e$, this is the same charge per corner as the original $\Pi$-symmetric QI in Fig.~\ref{fig:TCIcorners}(a), indicating that this fragile phase represents a previously unrecognized example of an insulator without maximally localized symmetric Wannier functions~\cite{ThoulessWannier,AndreiXiZ2,AlexeyVDBTI,QuantumChemistry,AshvinFragile,JenFragile1,AdrianFragile,JenFragile2,ZhidaBLG,BarryFragile,YoungkukMonopole,KoreanFragile,WiederAxion}, but with a nontrivial multipole moment~\cite{multipole,WladTheory}.  This is consistent with our determination in Appendix~\ref{sec:EBRsforTCI} that the specific $s-d$-hybridized TCI phase of Eq.~(\ref{eq:my2Dquad}) exhibits the same nontrivial ($e/2$) quadrupole moment as a $p-d$-hybridized 2D TI when their edge states are respectively gapped.  Because $\Pi$-symmetry can be strongly broken in this fragile phase, then these Kramers pairs of corner states may not necessarily appear as midgap modes.  Nevertheless, because they are accompanied by an anomalous mismatch in the number of states in the valence and conduction manifolds, their presence in the spectrum can still be detected by counting the number of states above and below the gap in the energy spectrum of a 2D, $p4m1'$-symmetric insulator.  Specifically, the corner modes of this fragile phase (or of the obstructed atomic limit that results from adding trivial bands (Appendix~\ref{sec:fragile})) can be diagnosed by calculating the energy spectrum with open boundary conditions (OBC) and periodic boundary conditions (PBC) and comparing the number of states below an energy gap; if the number of states below the gap in the OBC and PBC spectra differs by $6 + 8\mathbb{Z}$ (or $2 + 8\mathbb{Z}$) then an anomalous number of fragile-phase corner modes are present in the spectrum\cite{ChenConvo,WiederAxion} (though, depending on energetics, the states themselves may lie within the bulk manifolds) (Fig.~\ref{fig:TCIcorners}(d)).  Finally, we tune $v_{M_{z}}\rightarrow 0$, which closes the edge gap, restores $\mathcal{T}$, $\Pi$, and $M_{z}$ symmetries, and induces a $C_{M_{z}}=2$ TCI phase (Fig.~\ref{fig:2Dmain}(f-h) of the main text).  Because the topology of our model is indifferent to the order in which parameters are tuned as long as the bulk gap is not closed and the symmetries of $p4m$ are not broken, then this process must be equivalent to the reverse of the direct transition from the $\Pi$-symmetric TCI to the $\Pi$-symmetric QI with four 0D corner states (Fig.~\ref{fig:TCIphases}, white arrow).

The $3/4$-filled corner modes that appear in the $p4m1'$-symmetric fragile phase (Fig.~\ref{fig:TCIcorners}(d)) can also be understood from the perspective of the two pairs of helical edge states of the parent mirror TCI phase (Fig.~\ref{fig:TCIphases}).  Naively, one might expect that for all $C_{M_{z}}=2$ TCIs with $p4m$ symmetry, each pair of edge states gaps under a $p4m$-symmetric, $M_{z}$-breaking potential to give four $\pm e/2$-charged corner modes, resulting in an overall trivial quadrupole moment (taken modulo $e$).  However, because the $C_{4z}$ eigenvalues of the specific 2D TCI in Eq.~(\ref{eq:my2Dquad}) match those of a QI (as well as those of a $p-d$-hybridized 2D TI), as shown in Eq.~(\ref{eq:QIfromEBRs}), then the we instead observe that the two pairs of helical modes gap differently when $M_{z}$ is broken.  Specifically, one pair of helical modes gaps to give an anomalous configuration of corner charges with:
\begin{equation}
L_{z,1} = L_{z}^{QI} = 2+ 4a,\ a\in \mathbb{Z},
\end{equation}
where $L_{z}^{QI}$ is derived in the text surrounding Eq.~(\ref{eq:nontrivialLz}), and the other pair gaps to give a non-anomalous configuration of corner charges with:
\begin{equation}
L_{z,2} = L_{z}^{NI}=4a,\ a\in \mathbb{Z},
\end{equation}
where $L_{z}^{NI}$ is derived in the text surrounding Eq.~(\ref{eq:trivialLz}).  When the TCI gaps directly into the QI through the introduction of $u$, which breaks $M_{z}$ and $\mathcal{T}$ while preserving $p4m$, then $L_{z,1}$ and $L_{z,2}$ are free to take their minimum values of $2$ and $0$, respectively.  However, when the TCI gaps into a $p4m1'$-symmetric fragile phase, then $\mathcal{T}$ symmetry requires $L_{z,2}$ to be a nonzero multiple of $4$ (Eq.~(\ref{eq:trivialLz})), so that there are fully-filled (or empty) modes at the $4b$ position of $4mm$ (Fig.~\ref{fig:4mm}) to pair with the existing QI corner states.  These quarter-filled (or quarter-empty) Kramers pairs of fragile-phase corner modes are depicted in Fig.~\ref{fig:TCIcorners}(d) and appear in our numerical calculations ($k_{z}=0$ in Fig.~\ref{fig:HingeSMmain}(h) of the main text).  Our discovery that 2D TCIs with the same number of edge states, depending on their bulk symmetry eigenvalues, can gap to give different corner-mode phases is reminiscent of a similar phenomenon that occurs in 3D TCIs and higher-order topological insulators (HOTIs)~\cite{AshvinIndicators,AshvinTCI,ChenTCI}.  Specifically, in a 3D TCI, each 2D surface is characterized by pairs of twofold Dirac cones from bulk double band inversion, which can be deformed into unstable fourfold Dirac cones~\cite{DiracInsulator,BernevigMoTe2,Steve2D}, and can therefore be gapped by breaking crystal symmetries while preserving $\mathcal{T}$~\cite{HigherOrderTIBernevig,BernevigMoTe2,AshvinIndicators,AshvinTCI,ChenTCI}.  Depending on the bulk symmetry eigenvalues, the pairs of twofold Dirac cones gap to provide either an integer (non-anomalous) or a half-integer (anomalous) contribution to the surface quantum spin Hall effect~\cite{DiracInsulator,WiederDefect}, which taken over all Dirac cones and surfaces, determines the presence or absence of intrinsic helical hinge modes, and thus whether the 3D TCI has transitioned into a HOTI or a trivial insulator.  Because HOTIs can be expressed as pumping cycles of QIs and other 2D corner-mode phases~\cite{WladTheory,HigherOrderTIBernevig,HigherOrderTIChen,WiederAxion} there is likely a more explicit link between the 2D and 3D cases.

Finally, we note that the observation that six (or two) of the eight fragile-phase corner modes (four Kramers pairs) are filled is subtly in agreement with the results of Appendices~\ref{sec:bandrep} and~\ref{sec:fragile}.  To see this, we first briefly summarize the details of the Wannier center homotopy for a half-filled QI, which we will then adapt to analyze the analogous homotopy of the three-quarters-filled trivialized fragile phase in $p4m1'$ examined in Appendix~\ref{sec:fragile}.  In Appendix~\ref{sec:bandrep}, we developed a Wannier center homotopy of the QI obstructed atomic limit (Fig.~\ref{fig:Wannier}(c,d)).  In this homotopy, four Wannier centers ``slide'' from the $1a$ to the $1b$ Wyckoff position of $p4m$ (Fig.~\ref{fig:2Dmain}(a) of the main text), representing the transition between a trivial insulator with two valence and two conduction bands both originating from orbitals at the $1a$ position, and an obstructed atomic limit with two valence and two conduction bands, which can each be formed into maximally localized, symmetric Wannier orbitals at the $1b$ position (Fig.~\ref{fig:Wannier}(c,d)).  When this pattern of Wannier center ``sliding'' is terminated on the boundary of a finite-sized QI, four of the bulk Wannier orbitals (one per corner) become the 0D corner states of the QI.  As there are four Wannier orbitals in the Wannier center homotopy, and two orbitals originate from valence bands if the system remains half-filled, then we conclude that each of the four corner states is half-filled, in agreement with the results of Appendix~\ref{sec:TIboundary}.

A similar analysis reveals that the $3/4$-filled corner modes of the fragile phase analyzed in this section can \emph{also} originate from a Wannier center homotopy, even though they can appear on the corners of an insulator that is not Wannierizable.  When our four-band fragile model in $p4m1'$ (Eqs.~(\ref{eq:my2Dquad}) and~(\ref{eq:breakI})) is half-filled, the two valence and two conduction bands each exhibit fragile topology.  In Appendix~\ref{sec:fragile}, we proposed through an irreducible-representation equivalence (Eq.~(\ref{eq:trivializeMyEBRSum})) and demonstrated numerically (Fig.~\ref{fig:fragile}) that this obstruction to forming a Wannier description can be lifted by introducing four (trivial) valence bands.  Taken over the now six total valence bands (two fragile and four trivial) and two fragile conduction bands, a homotopic description of sliding eight Wannier centers from the $1a$ to the $1b$ position of $p4m1'$ can be formed (Eq.~(\ref{eq:fragileWannier})).  When this eight-band pattern of sliding Wannier orbitals is terminated on the boundary of a finite-sized system, it results in eight corner states (one Kramers pair per corner).  Counting the occupancy of the eight sliding Wannier centers, as stated previously, six originate from valence bands and two originate from conduction bands, yielding an overall $3/4$-filled, eight-band obstructed atomic limit that carries the same corner degeneracy and charges as the original four-band fragile insulator at half filling (Fig.~\ref{fig:TCIcorners}(d)).  Specifically, because unobstructed (trivial) atomic limits do not exhibit corner charges along a boundary with a trivial insulator with atoms at the same Wyckoff positions (Appendix~\ref{sec:bandrep}), then we conclude that the $3/4$-filled Kramers pairs of corner modes must be a consequence of the (fragile) topology of the two valence bands (and the two conduction bands) of our original four-band model (Eqs.~(\ref{eq:my2Dquad}) and~(\ref{eq:breakI})).  This equivalence between the corner charges of a fragile phase (the four-band fragile TI in $p4m1'$ characterized by Eqs.~(\ref{eq:my2Dquad}) and~(\ref{eq:breakI})) at one filling (here $1/2$) and those of an obstructed atomic limit at a (sometimes) different filling (here $3/4$) is also explored in Ref.~\onlinecite{WiederAxion}, and can be inferred from the results of Ref.~\onlinecite{WladCorners}.

\subsection{Relaxation of $M_{x,y}$ Symmetry in QIs and Related Fragile Phases}
\label{sec:noMirror}

In this section, we discuss the consequences of relaxing $M_{x,y}$ symmetry while preserving $C_{4z}$ symmetry on the corner spectra of the previous QI and fragile phases (Appendices~\ref{sec:TIboundary} and~\ref{sec:TCIBoundary}, respectively).  First, we begin with the $\mathcal{T}$-broken QI phase in $p4m$, and then later, we subsequently reintroduce $\mathcal{T}$ symmetry to analyze the fragile phase in $p4m1'$ discussed in Appendices~\ref{sec:fragile} and~\ref{sec:TCIBoundary}.  

\begin{figure}[h]
\centering
\includegraphics[width=\textwidth]{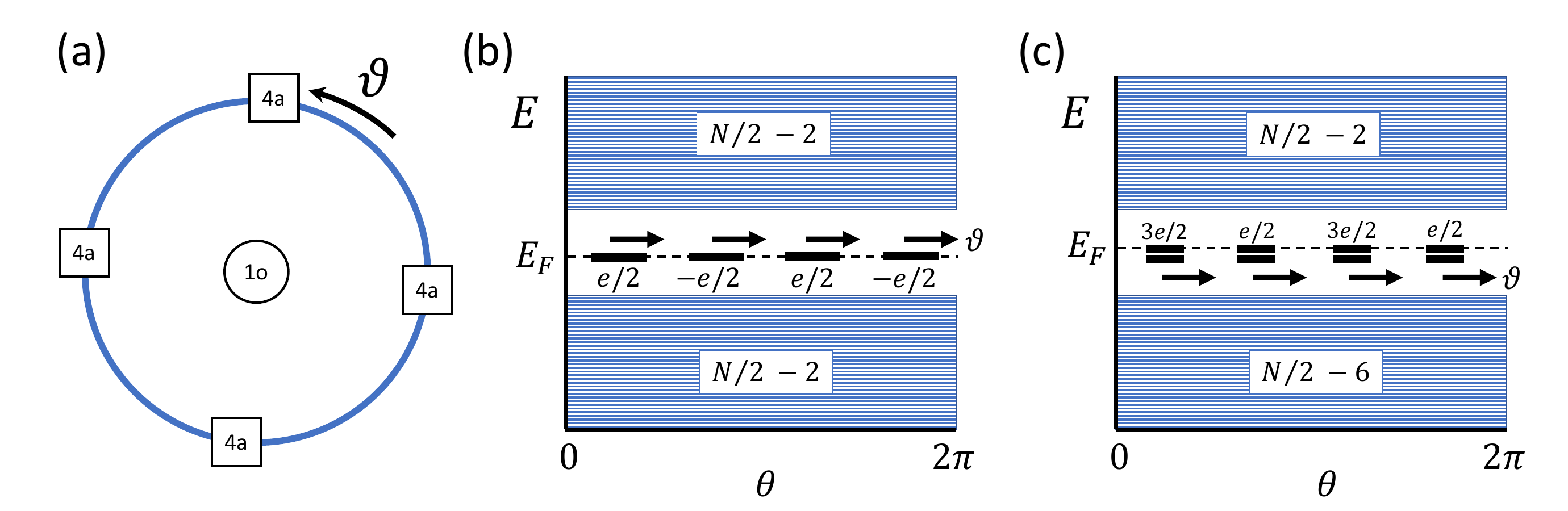}
\caption{(a) The Wyckoff positions of point group~\cite{BilbaoPoint} $4$, generated by $C_{4z}$.  The $4a$ position characterizes four points at $C_{4z}$-related angles with an overall free angular parameter $\vartheta$.  (b)  Breaking $M_{x,y}$ while preserving $C_{4z}$ for the QI (Fig.~\ref{fig:cornerModes}), allows the four corner modes to rotate freely as a set; however, as long as the bulk and edge gaps remain open and $C_{4z}$ is preserved, there will remain an anomalous absence of $L_{z}^{QI}=2+4n$, $n\in\mathbb{Z}$ (Eq.~(\ref{eq:nontrivialLz})) states from the valence manifold of the open-boundary-condition (OBC) spectrum relative to the spectrum calculated with periodic boundary conditions (PBC).  (c)  Breaking $M_{x,y}$ while preserving $C_{4z}$ and $\mathcal{T}$ for the fragile phase in $p4m1'$ described in Appendix~\ref{sec:TCIBoundary} similarly allows the four, three-quarters-filled (or quarter-filled) Kramers pairs of corner states to freely rotate as a set; however, if a bulk or edge gap is not closed and $C_{4z}$ and $\mathcal{T}$ are preserved, there will remain an anomalous absence of $6 + 8n$ (or $2 + 8 n$) states from the valence manifold of the OBC spectrum relative to the PBC spectrum.}
\label{fig:4only}
\end{figure}

In Appendix~\ref{sec:TIboundary}, we showed that the QI with $M_{x,y}$ and $C_{4z}$ symmetries can be diagnosed by observing that $L_{z}^{QI}=2+4n$, $n\in\mathbb{Z}$ (Eq.~(\ref{eq:nontrivialLz})) states are missing from the valence (and conduction) manifolds of the spectrum of a $p4m$-symmetric insulator calculated with $4mm$-symmetric open boundary conditions (OBC), relative to the spectrum calculated with periodic boundary conditions (PBC) (Fig.~\ref{fig:cornerModes}).  When the $L_{z}^{QI}$ missing states from the valence manifold appear in the bulk gap, they, along with $L_{z}^{QI}$ states from the conduction manifold, represent four corner modes localized to either the $4a$ or the $4b$ Wyckoff position of point group $4mm$ (as well as 8n trivial states localized at the general position $8c$ (Fig.~\ref{fig:4mm})).  Specifically, because of the in-plane mirrors $M_{x,y}$, if only four states are present at the same energy in the OBC spectrum, then they \emph{must} appear in a finite-sized QI with $4mm$ at the fixed angles of either $M_{x,y}$ ($4a$):
\begin{equation}
\theta_{4a} = n\pi/2,\ n\in \mathbb{Z},
\label{eq:4aAngles}
\end{equation}
or at the fixed angles of $M_{x\pm y}$ ($4b$):
\begin{equation}
\theta_{4b} = n\pi/2 + \pi/4,\ n\in\mathbb{Z}.
\label{eq:4bAngles}
\end{equation}

Next, we consider introducing a perturbation that breaks $M_{x,y}$ symmetries while preserving $C_{4z}$ and does not close a bulk gap.  Using the point group tables on the BCS~\cite{BilbaoPoint}, we determine that this perturbation transforms as the irreducible representation~\cite{DDP,DiracInsulator,CracknellMagneticTrans,LongerCracknellSymmetry} $A_{2}$ of $4mm$, because:
\begin{equation}
\chi_{A_{2}}(C_{4z}) = \chi_{A_{2}}(C_{2z})=1,\ \chi_{A_{2}}(M_{x,y})=-1,
\end{equation}
where $\chi_{\rho}(h)$ is the character of the symmetry $h$ in the irreducible representation~\cite{PointGroupTables,BilbaoPoint}.  Applying an $A_{2}$ perturbation reduces the point group of the finite-sized QI from $4mm$ to $4$.  As shown in Fig.~\ref{fig:4only}(a), point group $4$ does not distinguish between any points on a circle; instead it only hosts a $4a$ position that labels four $C_{4z}$-related angles:
\begin{equation}
\tilde{\theta}_{4a} = \theta_{4b} + \vartheta,
\end{equation}
where $\theta_{4b}$ refers to the coordinates of the $4b$ sites of $4mm$ (Eq.~(\ref{eq:4bAngles})) and $\vartheta$ is a free angle.  Here, we have chosen $\vartheta$ with respect to the $4b$ position of $4mm$, rather than the $4a$ position, because the corner modes of previous QI models~\cite{multipole,WladTheory} with $4mm$ appeared at $4b$.  In terms of the circular harmonics $f^{\pm}_{L_{z}}(\theta)$ whose zeroes determine the locations of the QI corner modes (Eqs.~(\ref{eq:GeneralU}) through~(\ref{eq:s1zeromodes})), the reduction from $4mm$ to $4$ removes the distinction between $f^{+}_{L_{z}}(\theta)$ and $f^{-}_{L_{z}}(\theta)$, resulting in a new set of circular harmonics given by:
\begin{equation}
f_{L_{z}}^{\vartheta}(\theta) = \cos\left[L_{z}(\theta-\vartheta)\right].
\label{eq:newCircular4}
\end{equation}
Eq.~(\ref{eq:newCircular4}) implies that perturbatively relaxing $M_{x,y}$ unpins the four corner modes from the fixed angles of $4mm$ (Fig.~\ref{fig:4only}(b)), permitting them to be rotated by a symmetry-allowed boundary term that does not change the bulk topology.  This allows the quadrupole moment of the corner modes (Eqs.~(\ref{eq:circularQuadrupole}) and~(\ref{eq:circularQuadrupoleSine})) to freely rotate between $Q^{xy}=e/2$, $Q^{x^{2}-y^{2}}=0$; $Q^{xy}=0$, $Q^{x^{2}-y^{2}}=e/2$; and all intermediate values with a total quadrupole moment of $e/2$.  However, and crucially, because $C_{4z}$ symmetry still relates the four corner modes to each other, a bulk-gap-preserving chemical potential can only move the four corner modes together in energy (and rotate them as a whole about the origin ($1o$ in Fig.~\ref{fig:4only}(a))), but it cannot lift the anomalous absence of $L_{z}^{QI}$ states from the valence and conduction manifolds in the OBC spectrum.  Therefore, if a QI is terminated on a boundary that only preserves $C_{4z}$, but not $M_{x,y}$, it will still exhibit four anomalous corner modes with an $e/2$ quadrupole moment whose orientation is a free parameter $\vartheta$.  This is analogous to the $e/2$ dipole moment of the inversion- ($\mathcal{I}$)- and $C_{2z}\times\mathcal{T}$- protected fragile phases introduced in Refs.~\onlinecite{BernevigMoTe2,WiederAxion}.  In the fragile phases in those works, the bulk (fragile or obstructed-atomic-limit) topology guaranteed the presence of $2+4a$ corner modes that exhibit a net $e/2$ dipole moment.  However, unlike the symmetries of $4mm$ (Fig.~\ref{fig:4mm}) neither $\mathcal{I}$ nor $C_{2z}\times\mathcal{T}$ fixes any points on the boundary of a circle (they only relate pairs of points).  Therefore, the direction of the anomalous $e/2$ dipole moment of the corner modes of the fragile phases in Refs.~\onlinecite{BernevigMoTe2,WiederAxion} is a free parameter, analogous to $\vartheta$ in Eq.~(\ref{eq:newCircular4}).

We note that, in this section, we only consider the case in which $M_{x,y}$ is perturbatively broken in the bulk.  This guarantees that the bulk gap does not close when $M_{x,y}$ is broken, and keeps the bulk bands adiabatically connected to those of a QI.  This also allows us to avoid symmetry-allowed intermediate Chern insulating phases between a trivial insulator and a QI in wallpaper group~\cite{WiederLayers,DiracInsulator} $p4$, the $M_{x,y}$-broken subgroup of $p4m$.  Specifically, if the bulk in-plane mirrors of $p4m$ are preserved, then bands at $\Gamma$ and $M$ remain twofold degenerate, and the ($C_{4z}$) eigenvalues of the occupied bands only distinguish between QI and trivial phases (Appendix~\ref{sec:bandrep}).  Conversely, if $M_{x,y}$ are broken, then bands at all TRIM points become singly degenerate.  With only singly-degenerate bands, then transitions between trivial insulating and QI phases can only be facilitated through multiple, independent band inversions.  However, unlike with wallpaper group $p4m$, a system with only $C_{4z}$ symmetry can generically pass through intermediate Chern insulating phases when singly-degenerate bands with different $C_{4z}$ (or $C_{2z}$) eigenvalues are inverted~\cite{ChenBernevigTCI}.  Therefore, because the phase boundaries separating trivial, Chern, and quadrupole insulators in $p4$ are considerably more complicated than the simple boundary separating trivial and QI phases in $p4m$ (Appendix~\ref{sec:bandrep}), we leave the complete analysis of QI transitions in $p4$ for future works.

Finally, we note that the arguments in this section also apply to the corner modes of the fragile phase in $p4m1'$ examined in Appendix~\ref{sec:TCIBoundary} with minimal modification.  As shown in Fig.~\ref{fig:TCIcorners}, the eight three-quarters-filled (or quarter-empty) corner modes of a fragile phase in $p4m1'$ are localized at the $4a$ or $4b$ Wyckoff position of $4mm1'$ (Fig.~\ref{fig:4mm}), the point group generated by adding $\mathcal{T}$ to $4mm$.  Here, the presence of $\mathcal{T}$ symmetry requires that all states are at least twofold degenerate; therefore, the constraints imposed by $C_{4z}$ and $\mathcal{T}$ symmetry require that a bulk-gap-preserving chemical potential moves all eight corner states together in energy.  When $M_{x,y}$ are relaxed without breaking $C_{4z}$ or $\mathcal{T}$, then the point group of the finite-sized system is reduced from $4mm1'$ to $41'$, and the eight corner modes can be rotated as a set by a free angle $\vartheta$ (Fig.~\ref{fig:4only}(c)).  However, as previously with the QI, because a $C_{4z}$- and $\mathcal{T}$-preserving potential that does not close the bulk or edge gap must move all eight corner modes together in energy, then the valence manifold of the OBC spectrum will still display an anomalous absence of $6 + 8n$ (or $2 + 8 n$) states when calculated relative to the PBC spectrum.  Therefore, if this fragile phase is terminated on a boundary that only preserves $C_{4z}$ and $\mathcal{T}$, but not $M_{x,y}$, it will still exhibit four anomalous Kramers pairs of corner modes with an $e/2$ quadrupole moment whose orientation is a free parameter $\vartheta$ (though, due to strongly broken particle-hole symmetry, the corner modes may be buried in the valence or conduction manifolds in the OBC spectrum).

\subsection{Numerical Investigations of the Surface States of HOFA Dirac Points}
\label{sec:boundaryNumbers}

A Dirac point in a HOFA semimetal represents the bulk quantum critical point between QI and trivial phases (Appendices~\ref{sec:bandrep} and~\ref{sec:double}).  In previous works, QI phase transitions have been shown to be accompanied by changes in the edge polarization that correspond to ``edge'' quantum critical points~\cite{multipole,WladTheory,BoundaryGreen,TaylorToy}.  It is therefore natural to ask whether there are additional states in the surface spectrum (besides the projections of the bulk Dirac points) that would be representative of a QI edge gap closure, analogous to the surface states of the higher-order ``surface-only'' semimetals proposed in Ref.~\onlinecite{TaylorToy}, which appeared while this extensive work was in preparation.  To accomplish this, we begin with the $\mathcal{T}$-broken model of a HOFA semimetal in Eq.~(\ref{eq:magHinge}), and place it on a slab geometry that is infinite in the $y$ and $z$ directions and finite with 500 layers in the $x$ direction.   Here, this semimetallic system, while not particle-hole symmetric, still features bulk Dirac crossings at $k_{x}=k_{y}=0, k_{z}=k^{\pm}_{d}$ that lie in the spectral center $N/2$ (Fig.~\ref{fig:HingeSMmain}(c) of the main text), where $N=500\times 4$, with the factor of $4$ coming from the two spin-$1/2$ $s$ and two spin-$1/2$ $d$ orbitals in each unit cell.  This guarantees that, taking $E_{d}$ to be the energy of the bulk Dirac points at $k_{z}=k_{d}$, the spectrum at each $k$ point near $k_{d}$ exhibits $N/2$ states with energy $E>E_{d}$ and $N/2$ states with energy $E\leq E_{d}$. 

\begin{figure}[h]
\centering
\includegraphics[width=1.0\textwidth]{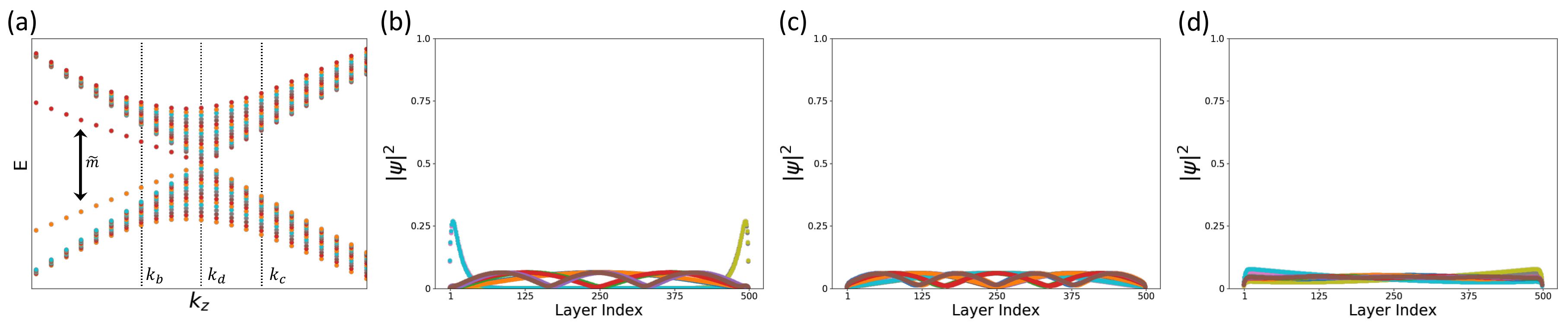}
\caption{Numerical demonstration of the absence of additional surface states in HOFA Dirac semimetals.  (a) Energy spectrum of the $\mathcal{T}$-broken Dirac semimetal in Eq.~(\ref{eq:magHinge}), plotted as a function of $k_{z}$ in the vicinity of $k_{d}$ for a Dirac point located at $k_{x}=k_{y}=0$, $k_{z}=k_{d}$, and restricting to the 64 states closest to the center of the spectrum $E=E_{d}$.  Above and below $E_{d}$, the spectrum at each $k$ point exhibits $N/2$ states, where $N$ is the total number of states in a slab with $N/4$ layers, where the factor of $4$ originates from the two spin-$1/2$ $s$ and two spin-$1/2$ $d$ orbitals in each unit cell.  We take $k_{y}=0$ for all calculations. Two pairs of gapped surface Fermi arc states, whose splitting $\tilde{m}$ in energy (a) scales as $u\sin(k_{d}-k_{z})$ in Eq.~(\ref{eq:magHinge}), can be observed at $k_{z}<k_{d}$.  At fixed values of $k_{z}$ less than $k_{d}$ (but still closer to $k_{d}$ than to $k_{z}=0$), the surface Fermi arc states also exhibit an increasing energy gap with increasing $|k_{y}|$ (not pictured).  (b) Orbital-summed wavefunction magnitude $\sum_{s,d,\sigma}|\psi|^{2}$ of states in (a) at $k_{b}$, plotted as a function of layer index along $x$, the finite direction of the slab; exactly four states can be observed localized on the two $\pm x$-normal boundaries, corresponding to the gapped Fermi arc states in (a).  (c) Orbital-summed wavefunction magnitude of the states at $k_{c}$ in (a); all of these states are localized in the bulk; \emph{i.e.} they are centered about layer index $250$ and decay as layer index approaches the boundary values of $1$ and $500$.  (d) Orbital-summed wavefunction magnitude of the states exactly at the bulk Dirac point at $k_{d}$.  All of the system states are perfectly delocalized; we do not detect additional surface states.}
\label{fig:Numerics}
\end{figure}

We plot in Fig.~\ref{fig:Numerics}(b,c,d) the orbital-summed wavefunction magnitude $\sum_{s,d,\sigma}|\psi|^{2}$ of the 64 energy eigenstates nearest $E=E_{d}$ as a function of $k_{z}$, taking $k_{y}$=0 (Fig.~\ref{fig:Numerics}(a)).  We define $k_{b}$ ($k_{c}$) to be a point at $k_{y}=0$ with $k_{z}< (>)\ k_{d}$ (Fig.~\ref{fig:Numerics}(a)).  We then define a state to be edge localized if more than 75\% of its probability density lies within within the first or the last 100 layers of the slab.  If a state exhibits equal probability density on all layers, we consider it to be delocalized.  If a state is neither edge localized nor delocalized, then we consider it to be bulk localized.  At $k_{y}=0,\ k_{z}=k_{b}$, all of the modes are bulk localized, except for two pairs of gapped surface Fermi arc states (Fig.~\ref{fig:Numerics}(a,b)), whose splitting $\tilde{m}$ in energy scales as $u\sin(k_{d}-k_{z})$ in Eq.~(\ref{eq:magHinge}).  Remaining at $k_{y}=0$ and taking increasing values of $k_{z}$ that pass through the bulk Dirac point at $k_{d}$, all of the states become perfectly delocalized at $k_{z}=k_{d}$ (\emph{i.e.}, they display equal probability density on all layers), and then become bulk localized at $k_{c}>k_{d}$ (Fig.~\ref{fig:Numerics}(a,c,d)).  It is clear that, aside from the four surface Fermi arc states at $k<k_{d}$, which separate into groups of two arcs on each of the two slab surfaces, there are no additional bound states on the 2D faces of this system; at $k_{z}=k_{d}$, all of the modes in the bulk and on the faces and hinges become delocalized.  We therefore do not observe any signatures of additional surface states bound to the projections of the bulk 3D Dirac points.

\section{Space Groups Supporting Dirac Semimetals with Quadrupolar HOFA States}
\label{sec:SGs}

In this section, we deduce the set of 3D space groups (SGs) capable of hosting Dirac semimetal phases with HOFA states.  For the purposes of this work, we restrict consideration to Dirac points equivalent to the critical point between 2D QI and trivial phases, as discussed in Appendices~\ref{sec:bandrep} and~\ref{sec:double}.  Other nodal points equivalent to other 2D critical points occur in other SGs (both with and without SOC~\cite{BernevigMoTe2,WiederAxion}), and if the symmetries that enforce their anomalous corner (hinge) modes can be preserved on a rod, they will also exhibit HOFAs~\cite{BernevigMoTe2}; we leave the complete enumeration of such bulk nodal points and variants of HOFA states for future works.  In Appendix~\ref{sec:symmetry}, we use the symmetries of quasi-one-dimensional rods, known as the crystallographic ``rod groups~\cite{subperiodicTables},'' to derive a set of SGs in which Dirac semimetals exhibit HOFA states derived from QIs.  In real materials, unlike in the models examined in this work, there are generically multiple kinds of atoms, each with different valence atomic orbitals, occupying different Wyckoff positions throughout the unit cell~\cite{ICSD}.  Therefore, instead of searching for candidate HOFA semimetals by restricting to specific cases of atomic-orbital hybridization (\emph{e.g.}, $s-d_{x^{2}-y^{2}}$- or $p_{z}-d_{x^{2}-y^{2}}$-hybridization at the $1a$ position, like the models in Eqs.~(\ref{eq:hinge}) and~(\ref{eq:pdTI}) respectively), we will exploit the analysis in Appendices~\ref{sec:TIboundary} through~\ref{sec:noMirror} to perform the more general search for topological semimetals whose low-energy theories bind and position-space symmetries protect QI-nontrivial corner- (hinge-) states.  We will find that HOFA states are generically present on the hinges of Dirac semimetals whose SGs have point group~\cite{BilbaoPoint} $4mm$ (or higher) when they are cut into rods that preserve a fourfold axis.  Then, in Appendix~\ref{sec:bodycenter}, we will demonstrate that body-centered Dirac semimetals can also exhibit HOFA states, even though fourfold axes do not coincide with crystal lattice vectors in body-centered SGs~\cite{BigBook}.  Specifically, when a body-centered Dirac semimetal is cut into a rod that preserves a fourfold axis, the lattice vectors of the finite-sized rod cannot coincide with the original lattice vectors of the bulk crystal, and so one might be concerned that HOFA states do not appear along the rod, due to the same zone-folding effects that negate the presence of edge states on armchair-terminated graphene~\cite{GrapheneEdge1,GrapheneEdge2,GrapheneReview,GrapheneEdgeMullen,GrapheneEdgeFan}.  However, by explicitly performing the BZ folding from a Dirac semimetal in a body-centered SG to a rod that preserves a fourfold axis, we will show that HOFA states are still generically present on the rod hinges.  We find that this occurs because the 3D HOFA Dirac points examined in this work arise from band inversion (``enforced semimetals'' in nomenclature of Ref.~\onlinecite{AndreiMaterials}), and are thus free to shift in momentum along high-symmetry BZ lines, whereas, conversely, the Dirac points in graphene are pinned by band connectivity to the high-symmetry BZ points~\cite{GrapheneReview,QuantumChemistry} $K$ and $K'$ (``enforced semimetal with Fermi degeneracy'' in the nomenclature of Ref.~\onlinecite{AndreiMaterials}).

In this work, we define a rod as a 1D crystal that is invariant under 3D symmetry operations; the symmetries of these systems are given by the rod groups~\cite{subperiodicTables,MagneticBook}.  For our purposes, we specialize to the crystallographic rod groups, which only contain symmetry elements that are also allowed in the 3D space groups~\cite{subperiodicTables}.  We introduce the subscript $RG$ to distinguish the symbols for rod groups from those for layer groups~\cite{WiederLayers,DiracInsulator,SteveMagnet,MagneticBook}, as there are rod groups and layer groups whose symbols are otherwise indistinguishable~\cite{subperiodicTables,MagneticBook} (\emph{e.g.}, $(p4mm1')_{RG}$ and $p4mm1'$).  Each of the crystallographic rod groups, when in-plane lattice translations $T_{x,y}$ are added to it, is isomorphic to a 3D space group.  For example, the $\mathcal{T}$-symmetric rod group $(p4mm1')_{\text{RG}}$ is generated by $T_{z},\ C_{4z}$, and $M_{x}$, and is related to SG 99 $P4mm1'$ by:
\begin{equation}
P4mm1' \equiv E(p4mm1')_{\text{RG}} \cup T_{x}(p4mm1')_{\text{RG}},
\label{eq:moduloT}
\end{equation}
where $E$ is the identity operation and the other in-plane translation $T_{y}$ is generated by $C_{4z}\times T_{x}$. 

In Appendix~\ref{sec:symmetry}, we will specifically show that when Dirac semimetals whose SGs have point group~\cite{BilbaoPoint,BCS1,BCS2} $4mm$ (or higher) are cut into a rod whose rod group has point group $4$ (or higher), the rod will exhibit quadrupolar HOFA states.  From an experimental perspective, the theoretical process of ``cutting a 3D crystal into a quasi-1D rod with point group $4$'' is equivalent to growing or cutting a sample into a nanowire whose long axis is coincident with a bulk fourfold axis.  It is possible that HOFA states may also be observable through momentum-resolved probes of the interior hinges of the pits of $C_{4}$-symmetric arrangements of step edges, as high-symmetry step edge configurations have been shown in experiment to exhibit the same $d-2$-dimensional hinge states as nanowires~\cite{HOTIBismuth}.

\subsection{Symmetry Conditions for Dirac Points with HOFA States Derived from QIs}
\label{sec:symmetry}

In this section, we derive the symmetry conditions for Dirac semimetals to exhibit anomalous HOFA states derived from QIs.  We divide this process into two steps, based on the analyses performed in Appendices~\ref{sec:TIboundary},~\ref{sec:double}, and~\ref{sec:noMirror}.  First, in Appendix~\ref{sec:pinnedHOFAs}, we use the results of Appendices~\ref{sec:TIboundary} and~\ref{sec:double} to show the more narrow result that Dirac semimetals whose SGs have point group $4mm$ or higher will exhibit HOFA states pinned to the fixed angles of $4mm$ (\emph{i.e.}, the $4a$ or $4b$ position in Fig.~\ref{fig:4mm}) if they can be cut into rods with $4mm$ (or higher) symmetry.  In these Dirac semimetals, the HOFA states are pinned to the same angles (the $4a$ or $4b$ position in Fig.~\ref{fig:4mm}) at all values of $k$ along the rod axis (Fig.~\ref{fig:RodGroups}(b)).  Then, in Appendix~\ref{sec:unpinnedHOFAs}, we use the results of Appendix~\ref{sec:noMirror} to further extend consideration to tetragonal and cubic SGs with bulk $4mm$ symmetry that \emph{cannot} be cut into rods with point group $4mm$, but which nevertheless support Dirac semimetals with HOFA states.  In these Dirac semimetals, the quadrupolar HOFAs generically appear at each rod $k$ point at free angles related by fourfold symmetry (Fig.~\ref{fig:4only}).  The list of SGs supporting HOFA Dirac semimetals obtained in Appendix~\ref{sec:unpinnedHOFAs} consequently includes the list of SGs obtained in Appendix~\ref{sec:pinnedHOFAs}; we therefore reproduce the more general list from Appendix~\ref{sec:unpinnedHOFAs} in Table~\ref{tb:SGsMain} of the main text.

\begin{figure}[h]
\centering
\includegraphics[width=0.65\textwidth]{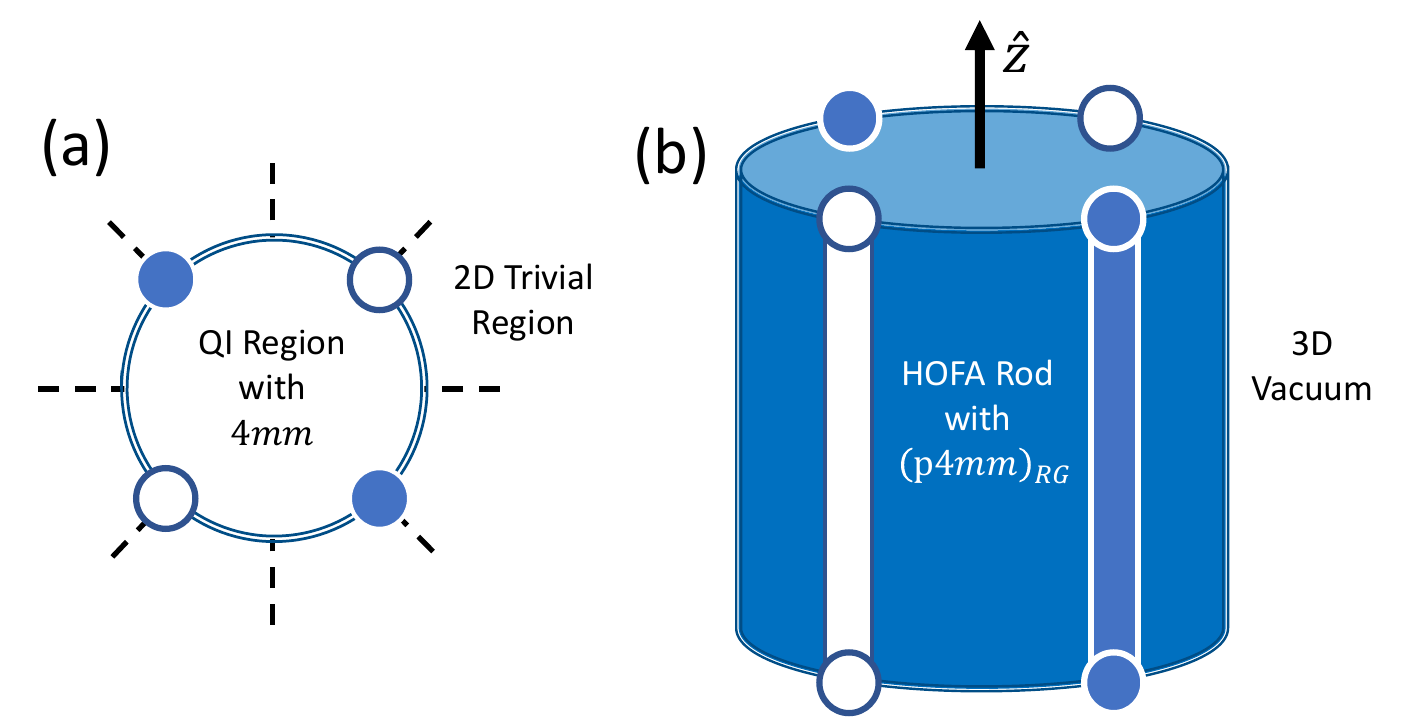}
\caption{The boundary states of finite-sized QI discs and HOFA Dirac semimetal rods.  (a) In 2D, a finite-sized region in position space with the bulk topology of a QI will exhibit corner states bound to reflection-fixed angles (\emph{i.e.} $\theta_{n}^{4a,4b}$ in Eqs.~(\ref{eq:4aForSGs}) and~(\ref{eq:4bForSGs}) and Fig.~\ref{fig:4mm}) if the point group of the QI region has $4mm$ (or higher) symmetry (Appendix~\ref{sec:boundary}).  (b) Extrapolating to 3D, a Dirac semimetal with bulk $4mm$ (or higher) symmetry will exhibit HOFA states that are fixed to $\theta=\theta_{n}^{4a,4b}$ if, in position space, it is cut into a rod that has a fourfold axis (either a rotation or a screw axis) and two perpendicular reflections (either mirrors or glides whose fractional translations are parallel to the fourfold axis).  This implies that the rod is symmetric under the action of a rod group~\cite{subperiodicTables} that is a supergroup of either $(p4mm)_{RG}$, $(p4_{2}cm)_{RG}$, $(p4_{2}mc)_{RG}$, or $(p4cc)_{RG}$.  The $\mathcal{T}$-symmetric rod supergroups of these (magnetic~\cite{MagneticBook}) rod groups are provided in Table~\ref{tb:RodGroups}.  Note that unlike a nanorod of a higher-order topological insulator~\cite{WladTheory,HigherOrderTIBernevig,HigherOrderTIChen,HigherOrderTIPiet,DiracInsulator,ChenRotation,AshvinTCI,HOTIBismuth,BernevigMoTe2}, the HOFA semimetal nanorod in (b) is gapless in its interior (and possibly also on its faces (Appendix~\ref{sec:pd})), because of its bulk (and possible surface) nodal points.  Nevertheless, in the limit that translation in the $z$ direction is still approximately preserved, the HOFA states depicted in (b) may still be detected through momentum-resolved probes along the hinges of the rod.}
\label{fig:RodGroups}
\end{figure}

\subsubsection{Dirac Semimetals with HOFA States Pinned to Fixed Angles}
\label{sec:pinnedHOFAs}

We begin by determining the SGs in which Dirac semimetals, when cut into $4mm$-symmetric rods, exhibit HOFA states that are pinned at each QI-nontrivial rod $k$ point (Appendix~\ref{sec:double}) to high-symmetry angles corresponding to the $4a$ or $4b$ Wyckoff position of point group $4mm$ (Fig.~\ref{fig:4mm}).  For the $4a$ position, these angles are~\cite{BilbaoPoint,BCS1,BCS2}:
\begin{equation}
\theta^{4a}_{n}=\{0,\pi/2,\pi,3\pi/2\},
\label{eq:4aForSGs}
\end{equation}
and for the $4b$ position, the angles are:
\begin{equation}
\theta^{4b}_{n}=\{\pi/4,3\pi/4,5\pi/4,7\pi/4\}.  
\label{eq:4bForSGs}
\end{equation}
We note that, as we will subsequently see in Appendix~\ref{sec:unpinnedHOFAs}, the list of SGs obtained in this section is \emph{not} the most general list of SGs in which Dirac semimetals exhibit quadrupolar HOFA states.  For completeness, however, we will still here complete the more restrictive tabulation of the SGs that support Dirac semimetals with HOFAs pinned to $\theta_{n}^{4a,4b}$ (Eqs.~(\ref{eq:4aForSGs}) and~(\ref{eq:4bForSGs})) as an intermediate step towards the complete list that will appear in Appendix~\ref{sec:unpinnedHOFAs} (and is reproduced in Table~\ref{tb:SGsMain} of the main text) of SGs supporting Dirac semimetals with HOFA states directly derived from the specific QI phase introduced in Ref.~\onlinecite{multipole}.

To obtain the relevant SGs, we first analyze the conditions that allowed the low-energy theory of the 2D QI in Appendix~\ref{sec:boundary} to exhibit corner modes pinned to $\theta_{n}^{4a,4b}$ in Eqs.~(\ref{eq:4aForSGs}) and~(\ref{eq:4bForSGs}).  In Appendix~\ref{sec:boundary}, we demonstrated that the presence of four corner modes localized to $\theta_{n}^{4a,4b}$ in the low-energy theory of a 2D QI occurred under three conditions:
\begin{enumerate}
\item The bulk differed from a trivial (unobstructed) atomic limit through a band inversion at a $k$ point whose little co-group had a subgroup (possibly itself) isomorphic to $4mm$.
\item The finite-sized QI region in position space (\emph{i.e.} in Appendix~\ref{sec:boundary}, the circle whose interior was a QI) was invariant under the action of point group $4mm$ (Fig.~\ref{fig:RodGroups}(a)).
\item The valence and conduction bands of the bulk $k\cdot p$ theory had different complex-conjugate pairs of $C_{4z}$ eigenvalues (\emph{i.e.}, transformed under different (co)representations of $4mm$ (Appendix~\ref{sec:bandrep})).  
\end{enumerate}
These constraints guarantee that the presence of 0D modes, which localize on the corners of a square geometry~\cite{multipole,HigherOrderTIBernevig,DDP,HOTIBismuth}, is a consequence of a topological quadrupole moment $Q^{xy}=e/2$ or $Q^{x^{2}-y^{2}}=e/2$ (Eqs.~(\ref{eq:circularQuadrupole}) and~(\ref{eq:circularQuadrupoleSine})) that cannot be removed without breaking a symmetry or closing a \emph{bulk} gap (Appendix~\ref{sec:boundary}).  If $C_{4z}$ is relaxed, then the boundary mass terms may change locally, allowing for corner modes to be removed by \emph{surface} gap closures~\cite{multipole,WladTheory}.  In this work, we are only concerned with systems that exhibit boundary states (both edge (surface) and corner (hinge)) as a consequence of their bulk topology.  This focus allows us to predict and analyze robust, intrinsic $d-2$-dimensional modes based on bulk topology, as opposed to predicting extrinsic corner (hinge) states whose presence depends on surface physics.  We therefore exclude 2D insulators whose anomalous corner modes may be removed by closing an edge (\emph{i.e.} Wilson) gap without closing a bulk gap, such as the QI phases without fourfold rotation symmetry in Refs.~\onlinecite{multipole,WladTheory}.  When these constraints are extended to 3D semimetals, we therefore also exclude the ``surface-only'' HOFA-semimetal phases introduced in Ref.~\onlinecite{TaylorToy}, in which the presence of HOFA states is \emph{entirely} dependent on the details of surface potentials, and is thus not a consequence of the bulk topology and difficult to predict in real materials through density functional theory.

To identify 3D SGs that support Dirac semimetals with HOFA states pinned to $\theta_{n}^{4a,4b}$ in Eqs.~(\ref{eq:4aForSGs}) and~(\ref{eq:4bForSGs}), we therefore require three conditions: 
\begin{enumerate}
\item  There exist lines in the BZ whose little groups contain $4mm$.
\item  When a crystal in this SG is cut into a rod that is finite in two dimensions and infinite along the direction of the fourfold axis from condition 1, its rod group has a point group that contains $4mm$.  This implies that the finite-sized rod is symmetric under the action of a rod group~\cite{subperiodicTables} that is a (possibly $\mathcal{T}$-symmetric) supergroup of one of the type-I magnetic rod groups~\cite{MagneticBook,ITCA,subperiodicTables} $(p4mm)_{RG}$, $(p4_{2}cm)_{RG}$, $(p4_{2}mc)_{RG}$, or $(p4cc)_{RG}$, as those are the lowest-symmetry rod groups that contain these symmetries.
\item The BZ line from condition 1 must have at least two distinct two-dimensional (co)representations, characterized by different complex-conjugate pairs of $C_{4z}$ eigenvalues, that can cross to form a symmetry-stabilized Dirac point.  As shown in Appendix~\ref{sec:double}, this, along with the reflection symmetries from condition 1, guarantees that this Dirac point is equivalent to the critical point between 2D trivial and QI phases.  This condition excludes, for example, BZ lines in nonsymmorphic SGs along which additional crystal symmetries beyond $4mm$ act to make corepresentations fourfold degenerate~\cite{BigBook,WiederLayers,Bandrep1}.
\end{enumerate}

In order to cut a 3D crystal into a rod with a point group that contains $4mm$, that crystal must have two reflection planes (mirrors or glides) that intersect on a fourfold axis (rotation or screw); for $z$-directed rods, this requirement necessarily excludes glide reflections with translations in the $xy$-plane, which cannot be preserved on a rod.  The rod group of this rod is a subgroup of the 3D space group of the crystal; cutting an infinite 3D crystal into a rod only lowers the overall symmetry.  Therefore, to identify the relevant SGs, we begin by enumerating the rod groups with point groups that contain $4mm$, of which the nonmagnetic examples are given in Table~\ref{tb:RodGroups}.

\begin{table}[H]
\centering
\begin{tabular}{|c|c|c|c|c|c|c|}
\hline
\multicolumn{7}{|c|}{$\mathcal{T}$-Symmetric Rod Groups with Point Group $4mm1'$ or $4/mmm1'$}  \\
\hline
RG Symbol & RG Number & Isomorphic SG Number & \ \ \ \ & RG Symbol & RG Number & Isomorphic SG Number \\
\hline
\hline
$(p4mm1')_{\text{RG}}$ & 34 & 99 & & $(p4/mmm1')_{\text{RG}}$ & 39 & 123 \\
\hline
$(p4_{2}cm1')_{\text{RG}}$ & 35 & 101 & & $(p4_{2}/mmc1')_{\text{RG}}$ & 38 & 131 \\
$(p4_{2}mc1')_{\text{RG}}$ & & 105 & & $(p4_{2}/mcm1')_{\text{RG}}$ &  & 132 \\
\hline
$(p4cc1')_{\text{RG}}$ & 36 & 103 & & $(p4/mcc1')_{\text{RG}}$ & 40 & 124 \\
\hline
\end{tabular}
\caption{Crystallographic rod groups~\cite{subperiodicTables} with $\mathcal{T}$ symmetry and whose point groups contain $4mm$.  We also list the numbers of their isomorphic space groups under the addition of in-plane lattice translations $T_{x}$ and $T_{y}$ (Eq.~(\ref{eq:moduloT})).  Because the rod groups are subperiodic groups that are finite in all directions in the $xy$-plane and infinite in the $z$ ($c$) direction~\cite{subperiodicTables}, they do not distinguish between symmetries such as the glide reflections $g_{x} = \{M_{x}|00\frac{1}{2}\}$ and $g_{x+y}=\{M_{x+y}|00\frac{1}{2}\}$.  For example, rod group 35 can either characterize a rod with $g_{x} = \{M_{x}|00\frac{1}{2}\}$ symmetry (($p4_{2}cm1')_{RG}$) or a rod with $g_{x} = \{M_{x}|00\frac{1}{2}\}$ symmetry (($p4_{2}mc1')_{RG}$), which are listed under different ``settings'' of rod group 35 on the BCS~\cite{BCS1,BCS2}.  However, when in-plane lattice translations are added to a rod group to convert it into a space group, $g_{x}$ and $g_{x+y}$ are no longer related by a translation-preserving unitary transformation.  Therefore, one rod group can become two different space groups depending on the orientation of its in-plane crystal symmetries~\cite{BCS1,BCS2,subperiodicTables} relative to the added in-plane lattice translations.  In all of the nonsymmorphic rod groups in this table (\emph{i.e.} those whose symbols contain the letter $c$), the glide reflections and screw symmetries only contain fractional translations in the $z$ direction (along the rod axis), because translations in the $xy$-plane are not symmetries of $z$-directed rods.}
\label{tb:RodGroups}
\end{table}

Three of the rod groups in Table~\ref{tb:RodGroups} are noncentrosymmetric (34 -- 36) and three are their centrosymmetric supergroups (38--40).  In particular, the three noncentrosymmetric rod groups are isomorphic, under the addition of in-plane lattice translations, to SGs 99, 101, 103, and 105 (both SGs 101 and 105 have the same symmetries as rod group 35 under the addition of in-plane lattice translations (Table~\ref{tb:RodGroups})).  Consequently, \emph{all} $\mathcal{T}$-symmetric space groups that characterize crystals that can be cut into rods with point group $4mm1'$ or its supergroup $4/mmm1'$ are necessarily supergroups of these four space groups.  For a Dirac semimetal in one of these SGs, if the Dirac points lie along an axis of fourfold rotation, then a rod cut from this crystal along the same axis will exhibit HOFA states, as such a Dirac point is necessarily equivalent to the critical point between 2D trivial and QI phases (Appendix~\ref{sec:double}).  We obtain these space groups by using~\textsc{MINSUP} on the BCS~\cite{BCS1,BCS2,BCSSuper} to find all of the supergroups of the space groups listed in Table~\ref{tb:RodGroups}.  We then impose conditions 1 and 3 explicitly by using Refs.~\onlinecite{BigBook,QuantumChemistry,Bandrep1,Bandrep2,Bandrep3} to identify the BZ lines that admit fourfold Dirac points.  Specifically, these lines have little groups with with $4mm$ (or higher) symmetry and have at least two, two-dimensional corepresentations; when bands with different corepresentations cross along these lines, a fourfold Dirac point forms~\cite{ZJDirac}.  This excludes, for example, BZ lines in nonsymmorphic SGs along which additional crystal symmetries combine with $4mm$ to make all corepresentations four-dimensional~\cite{WiederLayers,BigBook,Bandrep1} (\emph{e.g.}, $MA$ in SG 129 $P4/nmm1'$).  The full list of space groups and BZ lines is listed in Table~\ref{tb:SGrod}.

\begin{table}[h]
\begin{tabular}{|c|c|c|c|c|c|c|c|}
\hline
\multicolumn{8}{|c|}{Space Groups Admitting Dirac Points with Reflection-Fixed HOFA States}  \\
\hline
SG Symbol & SG Number & \ \ \ \ & SG Symbol & SG Number & \ \ \ \ & SG Symbol & SG Number \\
\hline
\hline
$P4mm1'$ & 99$^{\dag}$ & & $P4/mcc1'$ & 124$^{\dag}$ & & $I4/mmm1'$ & 139 \\
\hline
$P4_{2}cm1'$ & 101$^{\dag}$ & & $P4/nmm1'$ & 129 & & $I4/mcm1'$ & 140 \\
\hline
$P4cc1'$ & 103$^{\dag}$ & & $P4/ncc1'$ & 130 & & $Pm\bar{3}m1'$ & 221$^{\ddag}$ \\
\hline
$P4_{2}mc1'$ & 105$^{\dag}$ & & $P4_{2}/mmc1'$ & 131$^{\dag}$ & & $Pm\bar{3}n1'$ & 223$^{\ddag}$ \\
\hline
$I4mm1'$ & 107 & & $P4_{2}/mcm1'$ & 132$^{\dag}$ & & $Fm\bar{3}m1'$ & 225 \\
\hline
$I4cm1'$ & 108 & & $P4_{2}/nmc1'$ & 137 & & $Fm\bar{3}c1'$ & 226 \\
\hline
$P4/mmm1'$ & 123$^{\dag}$ & & $P4_{2}/ncm1'$ & 138 & & $Im\bar{3}m1'$ & 229 \\
\hline
\end{tabular}
\caption{Space groups that admit Dirac points with HOFA states derived from the QI introduced in Ref.~\onlinecite{multipole} and pinned to reflection-fixed rod hinges (\emph{i.e} to $\theta_{n}^{4a,4b}$ in Eqs.~(\ref{eq:4aForSGs}) and~(\ref{eq:4bForSGs}) and in Fig.~\ref{fig:4mm}).  These groups comprise the supergroups of the isomorphic space groups listed in Table~\ref{tb:RodGroups}, obtained using~\textsc{MINSUP} on the BCS~\cite{BCS1,BCS2,BCSSuper}.  For all of these SGs, HOFA Dirac points are always allowed to form along $\Gamma Z$ in tetragonal SGs ($\Gamma M$ if body-centered) and $\Gamma X$ in cubic SGs ($\Gamma H$ if face-centered) (Fig.~\ref{fig:BZs}).  In a subset of the primitive tetragonal groups, denoted with $\dag$, HOFA Dirac points may also form along $MA$.  In a subset of the primitive cubic groups, denoted with $\ddag$, HOFA Dirac points may also form along $M R$.  Quadrupolar HOFA states are also supported in Dirac semimetals in additional SGs, if we relax that constraint that the HOFA states at each QI-nontrivial rod $k$ point are bound to $\theta_{n}^{4a,4b}$ (Eqs.~(\ref{eq:4aForSGs}) and~(\ref{eq:4bForSGs})); a complete tabulation of all SGs supporting Dirac semimetals with free-angle HOFA states is provided in Appendix~\ref{sec:unpinnedHOFAs} and is reproduced in Table~\ref{tb:SGsMain} of the main text.}
\label{tb:SGrod}
\end{table}

\begin{figure}[h]
\centering
\includegraphics[width=1.0\textwidth]{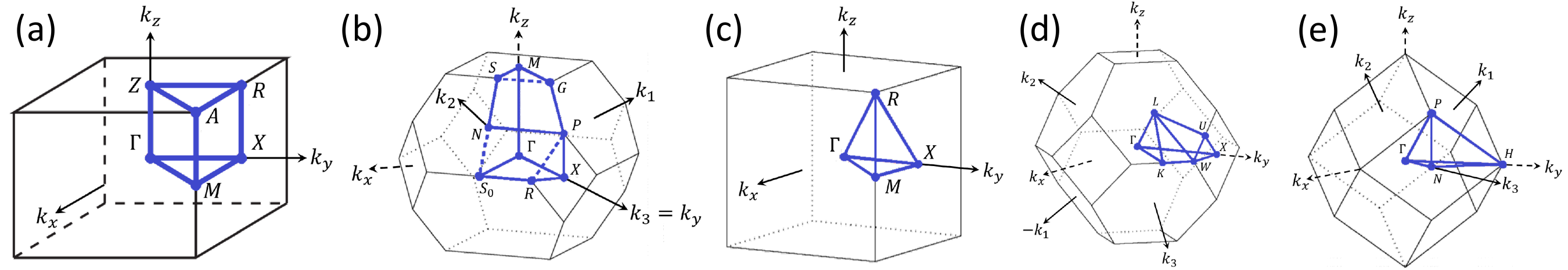}
\caption{The bulk Brillouin zones (BZs), highlighting the $k$-paths from Ref.~\onlinecite{BCTBZ} and labeling according to the BCS convention~\cite{BCS1,BCS2}, of the space groups that support Dirac semimetals with HOFA states (Table~\ref{tb:SGrod}).  The lattices of these space groups are (a) primitive tetragonal, (b) body-centered tetragonal, (c) primitive cubic, (d) face-centered cubic, and (e) body-centered cubic.}
\label{fig:BZs}
\end{figure}

We note that the preceding arguments contain a subtlety when applied to the space supergroups of the rod groups in Table~\ref{tb:RodGroups} with fourfold \emph{screw} axes (rod groups 35 and 38).  Along momentum-space lines with a fourfold screw axis defined by a $b/4$ fractional lattice translation:
\begin{equation}
s_{4_{b}z} = C_{4z}T_{b/4},
\label{eq:screwDef}
\end{equation}
as well as with two orthogonal reflections (\emph{e.g.}, the line $\Gamma Z$ in SG 105 $P4_{2}mc1'$), the symmetries in the basis of the four bands nearest a Dirac point can be represented by~\cite{Steve2D,WiederLayers}:
\begin{equation}
s_{4_{b}z} = \tau^{z}\left(\frac{\mathds{1}_{\sigma} - i\sigma^{z}}{\sqrt{2}}\right)\Lambda(k_{z})=C_{4z}\Lambda(k_{z}),\ M_{x,y} = i\sigma^{x,y},
\label{eq:nonsymmorphic}
\end{equation}
where: 
\begin{equation}
\Lambda(k_{z}) = e^{\frac{i bk_{z}}{4}} = e^{ik_{z} t_{s}}. 
\label{eq:defScrewPhase}
\end{equation}
While $t_{s}=1/2$ in the units of the $z$-direction lattice spacing for the $4_{2}$ screw in SG 105, more generally, the values $t_{s}=1/4,\ 3/4$ are also allowed in other SGs with fourfold screws~\cite{BigBook}.  At first, the symmetry representations in Eq.~(\ref{eq:nonsymmorphic}) appear distinct from those employed in Appendix~\ref{sec:TIboundary} to characterize the corner modes of a QI (Eq.~(\ref{eq:pdKspaceSyms})).  However, we note that, in the absence of antiunitary symmetries, such as $\mathcal{T}$ or $\mathcal{I}\times\mathcal{T}$ (which are present in $\mathcal{T}$- and centro- ($\mathcal{I}$-) symmetric SGs), we are free to rotate the phases of the representations of crystal symmetries without changing their commutation relations~\cite{WiederLayers,SteveMagnet}.  Specifically, here, we are also permitted to employ the symmetry representation:
\begin{equation}
\bar{C}_{4_{b}z} = s_{4_{b}z}\Lambda^{*}(k_{z})=\tau^{z}\left(\frac{\mathds{1}_{\sigma} - i\sigma^{z}}{\sqrt{2}}\right),\ M_{x,y} = i\sigma^{x,y},
\label{eq:rotatedNS}
\end{equation}
where we have labeled $\bar{C}_{4_{b}z}$ with a bar to emphasize that it is not the representation of a real $C_{4z}$ symmetry, but is rather an alternative representation of the fourfold screw symmetry $s_{4_{b}z}$.  Eq.~(\ref{eq:rotatedNS}), is \emph{identical} to the representation used in Appendix~\ref{sec:TIboundary} to predict QI corner modes (Eq.~(\ref{eq:pdKspaceSyms})).  Therefore, a Dirac point with the nonsymmorphic symmetry representation in Eq.~(\ref{eq:nonsymmorphic}) is described by the \emph{same} $k\cdot p$ Hamiltonian as a Dirac point with the symmorphic symmetry representation in Eq.~(\ref{eq:rotatedNS}), in agreement with the methods employed in Ref.~\onlinecite{ZJDirac} to characterize the band-inversion Dirac points in the nonsymmorphic Dirac semimetal Cd$_3$As$_2$.  Furthermore, even though the eigenvalues of screw symmetries, unlike the eigenvalues of rotations, depend on the choice of BZ (\emph{i.e.}, are not $2\pi$-periodic)~\cite{Steve2D,WiederLayers,HourglassInsulator,Cohomological,DiracInsulator}, \emph{within} each BZ, two bands can still be unambiguously labeled with distinct pairs of fourfold screw eigenvalues.  Crucially, because, within each 2D BZ slice (here indexed by $k_{z}$), the presence or absence of QI-nontrivial corner modes (HOFA states) only depends on the \emph{difference} in fourfold rotation eigenvalues between the valence and conduction bands (Appendices~\ref{sec:TIboundary} and~\ref{sec:TItoTrivial}), and because we have already shown that symmorphic Dirac semimetals with $4mm$ (or higher) point-group symmetry exhibit HOFA states when cut into rods with point group $4mm$ (or higher), then Eq.~(\ref{eq:rotatedNS}) allows us to conclude that noncentrosymmetric nonsymmorphic Dirac semimetals with point group $4mm$ \emph{also} exhibit HOFA states.

Furthermore, we can show that Eq.~(\ref{eq:rotatedNS}) also applies in $\mathcal{I}$- (centro-) and $\mathcal{T}$-symmetric nonsymmorphic Dirac semimetals.  Specifically, the HOFA Dirac points analyzed in this work occur away from TRIM points, and therefore, in an $\mathcal{I}$- and $\mathcal{T}$- symmetric Dirac semimetal, the Hamiltonian in the vicinity of each Dirac point only respects the combined magnetic symmetry $\mathcal{I}\times\mathcal{T}$ (in addition to $4mm$).  We can incorporate $\mathcal{I}\times\mathcal{T}$ into the symmetry representation in Eq.~(\ref{eq:rotatedNS}) by choosing a representation for $\mathcal{I}\times\mathcal{T}$ that neither commutes nor anticommutes with~\cite{SteveMagnet} $\bar{C}_{4z}$.  However, because, in the bulk, $\mathcal{I}\times\mathcal{T}$ only serves to make bands doubly degenerate away from~\cite{WiederLayers} $k_{x,y}=0$, and, on the corners, it does not change the anomalous QI state counting in Fig.~\ref{fig:cornerModes}, then we conclude that $\mathcal{I}\times\mathcal{T}$- and fourfold-screw-symmetry-enforced Dirac points also exhibit intrinsic HOFA states.  This argument only breaks down when both $\mathcal{I}$ and $\mathcal{T}$ are individually enforced, which can only occur at TRIM points, and thus does not apply to the band-inversion Dirac points discussed in this work.  While previous works have also introduced nonsymmorphic-symmetry-enforced fourfold Dirac~\cite{SteveDirac,JuliaDirac,Steve2D,WiederLayers,SteveMagnet} (and eightfold double-Dirac~\cite{DDP,NewFermions}) points that are specifically pinned to TRIM points, we will leave a detailed analysis of potential higher-order topology in these \emph{enforced} semimetals for future works, though we do predict that enforced semimetals should also exhibit higher-order topological effects.  Additionally, because our arguments here do not depend on the details of the exact phase $\Lambda(k_{z})$ in Eq.~(\ref{eq:nonsymmorphic}), then they will also apply without further modification to the Dirac semimetals with free-angle HOFA states enforced by $4_{1}$ ($t_{s}=1/4$) and $4_{3}$ ($t_{s}=3/4$) screw symmetries that will be introduced in Appendix~\ref{sec:unpinnedHOFAs}.

Finally, we note that, as implied from the discussion in the main text and $\mathcal{H}_{H1}({\bf k})$ (Eq.~(\ref{eq:magHinge})), HOFA states are also permitted in some magnetic Dirac semimetals.  However, as the number of known magnetic structures is small compared to the number of known materials~\cite{ICSD,BilbaoMagStructures}, it is relatively difficult to identify magnetic materials candidates.  Therefore, we leave the complete enumeration of all magnetic space groups that admit Dirac semimetals with HOFA states for future works.  However, as examples, by applying the procedure used to generate Tables~\ref{tb:RodGroups} and~\ref{tb:SGrod} to the magnetic rod and space groups, we conclude that magnetic Dirac points with HOFAs may form along $\Gamma Z$ in magnetic SGs $P4/m'mm$ (123.341 in the Belov-Nerenova-Smirnova (BNS) notation~\cite{MagneticBook}) and $P_{c}4/ncc$ (130.432 in the BNS notation).  The first group is that of our model, $\mathcal{H}_{H1}({\bf k})$ in Eq.~(\ref{eq:magHinge}), and the second group characterizes the antiferromagnetic (AFM) Dirac semimetal phase of CeSbTe~\cite{SchoopAFM}.  As the Dirac points in the AFM phase of CeSbTe lie along lines with $4mm$ symmetry and its magnetic space group contains $P4cc$, CeSbTe will exhibit reflection-fixed HOFA states when cut into a rod with $(p4cc)_{RG}$ (or higher) symmetry (though the HOFA states may be difficult to separate from the hinge projections of bulk and surface states in CeSbTe).

\subsubsection{Dirac Semimetals with HOFA States at Free Angles}
\label{sec:unpinnedHOFAs}

\vspace{-0.06in}
In this section, building upon the previous discussion in Appendix~\ref{sec:pinnedHOFAs}, we will develop the most general list of SGs in which Dirac semimetals exhibit HOFA states derived from the QI model introduced in Ref.~\onlinecite{multipole}.  Previously, in Appendix~\ref{sec:pinnedHOFAs}, we showed that if both infinite crystals and finite-sized rods of a Dirac semimetal preserve fourfold axes and two in-plane reflection symmetries (\emph{i.e}, have point groups that contain $4mm$), then the semimetal will exhibit HOFA states fixed to the rod hinges at $\theta=\theta_{n}^{4a,4b}$ in Eqs.~(\ref{eq:4aForSGs}) and~(\ref{eq:4bForSGs}) and Fig.~\ref{fig:4mm}.  However, we also previously showed in Appendix~\ref{sec:noMirror} that if a 2D QI is formed from band inversion about a $k$ point whose little co-group contains $4mm$, then its four corner modes remain anomalous when the system is terminated in an $M_{x,y}$-breaking, $C_{4z}$-symmetric geometry.  In this lower-symmetry geometry with point group~\cite{BilbaoPoint} $4$, the corner modes become unpinned from the fixed angles of $4mm$, and their $e/2$ quantized quadrupole moment becomes free to lie at any intermediate angle between $xy$ and $x^{2}-y^{2}$ (Fig.~\ref{fig:4only}).  Exploiting this result, we will determine in this section a list of SGs whose bulk crystals support $4mm$-symmetric HOFA Dirac points, even though they cannot all be cut into rods that simultaneously preserve fourfold axes and in-plane reflections.  Specifically, in some of these SGs, the in-plane reflections (mirrors and glides) also contain in-plane translations relative to the fourfold axes; when they are cut into rods that preserve one of their fourfold axes, their in-plane reflections are necessarily broken.  The list of SGs that we obtain in this section will contain the SGs previously tabulated in Table~\ref{tb:SGrod}, as well as additional SGs with bulk $4mm$ symmetry.

First, we will explain our restriction in this section to SGs with point group $4mm$ or higher, and not to SGs with point group $4$ or higher.  In SGs with point group $4mm$, bands along at least one fourfold axis are generically twofold degenerate, because the double-valued (spinful) in-plane reflections of $4mm$ anticommute (Appendix~\ref{sec:bandrep}).  Therefore, along a line with $4mm$ symmetry, there are two distinct corepresentations with different fourfold rotation eigenvalues, which can cross to form a symmetry-stabilized Dirac point (Appendix~\ref{sec:double}).   However in lower symmetry SGs with only point group $4$, then absent additional symmetries, bands are generically singly degenerate along fourfold axes away from the TRIM points, and can only cross to form Weyl points~\cite{StepanMultiWeyl,KramersWeyl}.  Additionally, while other SGs with other point groups (such as~\cite{BilbaoPoint,BCS1,BCS2} $4/m1'$) also exhibit twofold degenerate corepresentations along fourfold axes that can cross to form symmetry-stabilized Dirac points, we have not explicitly analyzed QI-nontrivial obstructed atomic limits and fragile phases in those SGs.  We will, for now, consider HOFA states in these SGs beyond the scope of the present work due to our focus on Dirac semimetals directly derived from the QI phase introduced in Ref.~\onlinecite{multipole}.  Therefore, we will only focus in this work on Dirac semimetals whose infinite crystals have point group $4mm$, even though our results imply that even more semimetallic phases exist with HOFA states derived from other variants of QIs and 2D corner-mode phases.

To identify 3D SGs that support Dirac semimetals with QI-nontrivial HOFA states at unpinned rod angles, we will find that, unlike previously in Appendix~\ref{sec:pinnedHOFAs}, there is only one independent condition.  To see how the three previous conditions from Appendix~\ref{sec:pinnedHOFAs} reduce in this section to a single independent condition, we focus on the group-subgroup relations of the rod, point, and space groups.  

To begin, we previously established in Appendix~\ref{sec:pinnedHOFAs} that for 2D BZ planes in the vicinity of a bulk Dirac point to exhibit the same bulk topology as a QI, the little group along the high-symmetry BZ line of the Dirac point must have $4mm$ or higher symmetry (condition 1 in Appendix~\ref{sec:pinnedHOFAs}).  Therefore, we require that:
\begin{enumerate}
\item {The SG of the bulk crystal must have point group $4mm$ or higher.} 
\end{enumerate}
For the purposes of identifying SGs that support Dirac semimetals with HOFA states, this requirement subsumes condition 1 in Appendix~\ref{sec:pinnedHOFAs}, because BZ lines cannot have $4mm$ symmetry if the point group of the SG of the crystal does not contain~\cite{BigBook} $4mm$.  Specifically, because we will ultimately obtain in this section a much larger list of SGs than previously obtained in Appendix~\ref{sec:pinnedHOFAs} (Table~\ref{tb:SGrod}), then we will only focus here on whether or not an SG can support \emph{any} Dirac point with HOFA states, and will not further determine the specific BZ lines along which such a Dirac point can form, as we did previously in Table~\ref{tb:SGrod}.

Under this looser restriction, because point groups $4mm$ and $4/m'mm$ both host pairs of two-dimensional (co)representations with different fourfold rotation eigenvalues, then all of the $\Gamma$-point-intersecting BZ lines in SGs with point group $4mm$ or higher also satisfy condition 3 from Appendix~\ref{sec:pinnedHOFAs}.  Specifically, because translations only act as phases in momentum space~\cite{BigBook,Steve2D,WiederLayers,Bandrep1}, then \emph{all} SGs with point group $4mm$ or higher contain at least one fourfold axis in momentum space that intersects the $\Gamma$ point along which $k$ points have little groups that are either isomorphic to $4mm$ or to $4/m'mm$.  For example, taking the fourfold axis to lie along the $z$ direction, whether the $x$ and $y$ (in-plane) reflections of an SG with point group $4mm$ are mirrors or glides, they can still be represented the same way along $k_{x}=k_{y}=0$ (\emph{i.e.} along a line intersecting the $\Gamma$ point (${\bf k}={\bf 0}$)) because in-plane translations are represented as $\exp(ik_{x,y}/b)$ where $b$ depends on the specific SG~\cite{BigBook}, and because translations along the $z$ axis only act as an overall phase that can be removed by a unitary transformation away from the TRIM points (Eq.~(\ref{eq:rotatedNS})).  Using the TRIM point labeling in Fig.~\ref{fig:BZs}, we identify the BZ lines that always support HOFA Dirac points in crystals in these SGs as $\Gamma Z$ and $\Gamma M$ in primitive and body-centered tetragonal crystals, respectively, and $\Gamma X$, $\Gamma X$, and $\Gamma H$ in primitive, face-centered, and body-centered cubic crystals, respectively.  Additionally we note that, as previously discussed in Appendix~\ref{sec:pinnedHOFAs}, some SGs host more than one line that satisfies the previous conditions 1 and 3 from Appendix~\ref{sec:pinnedHOFAs} (such as $MA$ in SG 123 $P4/mmm1'$).  However, we also again note that in many nonsymmorphic SGs, such as SG 100 $P4bm1'$, some BZ lines with $4mm$ symmetry instead exhibit a single four-dimensional corepresentation, due to the commutation relations between the nonsymmorphic crystal symmetries~\cite{BigBook,WiederLayers,DiracInsulator}.  In summary, for the purposes of this section, both conditions 1 and 3 from Appendix~\ref{sec:pinnedHOFAs} are redundant with the requirement that the SG of the bulk crystal must have point group $4mm$ or higher.

Finally, for a Dirac semimetal whose SG contains point group $4mm$ to exhibit HOFA states, we require that when this semimetal is cut into a rod that is finite in two dimensions and infinite along the direction of its fourfold axis, its rod group contains enough symmetries to preserve the nontrivial topology of its intrinsic (anomalous) HOFA states.  As discussed earlier in this section, we previously determined in Appendix~\ref{sec:noMirror} that the minimum symmetry requirement for QI-nontrivial 0D states in 2D is fourfold rotation.  When this constraint is applied to HOFA states in 3D rods, it is promoted to the constraint that the rod respects a rod group with a fourfold (rotation or screw) axis.  This supersedes the more restrictive requirement in condition 2 of Appendix~\ref{sec:pinnedHOFAs} that the rod respects both a fourfold axis as well as perpendicular in-plane reflections.  The requirement that a rod respects a fourfold axis can be reexpressed as the statement that the rod group~\cite{subperiodicTables} has a point group that contains point group $4$.  All of the rod groups with point group $4$ or higher are necessarily supergroups of one of the type-I magnetic rod groups~\cite{MagneticBook,ITCA,subperiodicTables} $(p4)_{RG}$, $(p4_{1})_{RG}$, $(p4_{2})_{RG}$, or $(p4_{3})_{RG}$, as those are the lowest-symmetry rod groups that contain fourfold axes.  Crucially, we find that \emph{all} SGs with point group $4mm$ or higher are supergroups of these four rod groups, and therefore characterize crystals that can be cut into rods with fourfold axes.  Therefore, the requirement that a Dirac semimetal can be cut into a rod with a fourfold axis, which was obtained by relaxing the reflection-symmetry requirement in condition 2 in Appendix~\ref{sec:noMirror}, is \emph{also} redundant with the simple restriction in this section to SGs with point group $4mm$ or higher.  Furthermore, as discussed in the text surrounding Eq.~(\ref{eq:rotatedNS}), because the representation of fourfold screw symmetry in the vicinity of Dirac points in screw-symmetric Dirac semimetals can be rotated into the same form as the representation of fourfold rotation in symmorphic Dirac semimetals, then we predict the presence of HOFA states in both symmorphic and nonsymmorphic Dirac semimetals whose SGs have point group $4mm$ or higher.  Additionally, in Appendix~\ref{sec:bodycenter}, we will show that body-centered Dirac semimetals with $4mm$ symmetry can generically be cut into rods that preserve a fourfold axis, even though they do not preserve the body-centered lattice vectors of the uncut crystal, and that, consequently, they also exhibit HOFA states.  We have therefore shown that that \emph{all} semimetals in SGs with point group $4mm$ or higher hosting Dirac points along BZ lines with little groups that contain point group $4mm$ or $4/m'mm$ can be cut into nanorod geometries that exhibit anomalous (\emph{i.e.} intrinsic) HOFA states.  In Table~\ref{tb:SGrod4Only} we summarize this result and enumerate the SGs with point group $4mm$ or higher that support Dirac semimetals with HOFA states derived from QIs.

\begin{table}[h]
\begin{tabular}{|c|c|c|}
\hline
\multicolumn{3}{|c|}{Space Groups Admitting Dirac Points with HOFA States}  \\
\hline
Point Group Name &  Point Group Symbol & SG Numbers  \\
\hline
\hline
$C_{4v}$ & $4mm1'$ & 99 -- 110 \\
\hline
$D_{4h}$ & $4/mmm1'$ & 123 -- 142 \\
\hline
$O_{h}$ & $m\bar{3}m1'$ & 221 -- 230 \\
\hline
\end{tabular}
\caption{Space groups that admit Dirac points with HOFA states derived from the QI introduced in Ref.~\onlinecite{multipole}.  This list includes all of the SGs listed in Table~\ref{tb:SGrod}, which contains the more restrictive set of SGs for which Dirac semimetals can be cut into rods that exhibit HOFA states pinned to the high-symmetry rod hinges at $\theta=\theta_{n}^{4a,4b}$ in Eqs.~(\ref{eq:4aForSGs}) and~(\ref{eq:4bForSGs}) and Fig.~\ref{fig:4mm}.  In addition to those SGs, this table also includes SGs whose reflections contain in-plane lattice translations, such that they are either glide reflections with in-plane fractional lattice translations (\emph{e.g.}, $g_{x} = \{M_{x}|\frac{1}{2}\frac{1}{2}0\}$ in SG 100 $P4bm1'$) or mirror reflections that do not coincide with the bulk fourfold axes (\emph{e.g.}, $M_{x+y} = \{M_{x+y}|\frac{1}{2}\frac{1}{2}0\}$ and $C_{4z}=\{C_{4z}|000\}$ in SG 100 $P4bm1'$), neither of which can be preserved in a rod geometry that also preserves a fourfold axis.  For all of the SGs in this table, semimetals with Dirac points along lines with $4mm$ or $4/m'mm$ symmetry will exhibit intrinsic HOFA states when cut into nanorods that preserve fourfold axes and are thick compared to the in-plane lattice spacing.  This list is reproduced in Table~\ref{tb:SGsMain} of the main text.}
\label{tb:SGrod4Only}
\end{table}

\vspace{-0.06in}
Though in some of the SGs in Table~\ref{tb:SGrod4Only} (\emph{i.e.} those also listed in Table~\ref{tb:SGrod}), the in-plane reflections coincide with the fourfold axes and do not contain in-plane translations, other SGs (such as SG 100 $P4bm1'$) contain in-plane glide reflections with in-plane lattice translations (\emph{i.e.} $\{M_{x}|\frac{1}{2}\frac{1}{2}0\}$), which cannot be preserved in a fourfold-symmetric rod geometry.  Nevertheless, we can take nanorods in these SGs to be sufficiently thick for the glide symmetries to be approximately preserved in the bulk, such that $4mm$-symmetric Dirac points are only weakly split when in-plane lattice translations are broken in the finite-sized rod geometry.  Specifically, while the angle $\vartheta$ by which the HOFA states are rotated is uncontrollably large (Fig.~\ref{fig:4only}), the weak breaking of glide symmetry still preserves the bulk band ordering, guaranteeing that the HOFA states remain an intrinsic consequence of the bulk Dirac points.

We finally note that, more generally, breaking in-plane reflections in the bulk while preserving a fourfold axis will split a $4mm$-symmetric Dirac point into Weyl points whose arrangement and Chern numbers are restricted by fourfold rotation symmetry~\cite{StepanMultiWeyl} (conversely, the Dirac points can be fully gapped by breaking fourfold rotation symmetry in the bulk).  Most interestingly, this implies that if a fourfold-symmetric Weyl semimetal can be deformed into a Dirac semimetal with HOFA states, then it will also exhibit intrinsic HOFA states at free angles \emph{coexisting} with topological surface Fermi arcs if the Dirac point was split into a pair of Weyl points along the fourfold axis.  Specifically, in this Weyl semimetal, taking $k_{z}$ to lie along the fourfold axis and considering 2D BZ planes at increasing values of $k_{z}$, one will pass first from a trivial insulator into a 2D Chern insulating phase when $k_{z}$ is increased through the first Weyl point.  Then, continuing to increase $k_{z}$, when one passes over the second Weyl point, one will pass from the Chern insulating phase into a 2D corner-mode phases that exhibits a free-angle quadrupole moment (Appendix~\ref{sec:noMirror}).  In the Chern insulating phase, the surface Fermi arcs are equivalent to the chiral edge states of the intermediate Chern insulating phases separating trivial and QI phases in wallpaper group $p4$ discussed in Appendix~\ref{sec:noMirror}.  Because the band connectivity, topology, and Fermi surfaces of this HOFA \emph{Weyl semimetal} are considerably more complicated than those of the HOFA Dirac semimetals introduced in this work, then we leave the analysis and complete enumeration of HOFA Weyl semimetals for future works.

\subsection{HOFA States in Body-Centered Dirac Semimetals}  
\label{sec:bodycenter}

When a body-centered Dirac semimetal is cut into a rod that preserves a fourfold axis, as was prescribed in Appendices~\ref{sec:pinnedHOFAs} and~\ref{sec:unpinnedHOFAs} to observe HOFA states, the lattice vector of the finite-sized rod (\emph{i.e.} the direction and length along which it is periodic) cannot coincide with the original lattice vectors of the bulk crystal, because fourfold axes and lattice vectors do not coincide in body-centered SGs~\cite{BigBook}.  Therefore, one might be concerned that HOFA states do not appear on the rod, due to zone-folding effects similar to those that negate the presence of edge states on armchair-terminated graphene~\cite{GrapheneEdge1,GrapheneEdge2,GrapheneReview,GrapheneEdgeMullen,GrapheneEdgeFan}, which we will review in this section.  We make this analogy to graphene, because the flat-band-like Fermi arc states in graphene can be considered the ``first-order'' analogs of the HOFA states analyzed in this work.  However, we will show in this section that body-centered tetragonal and cubic Dirac semimetals still generically exhibit HOFA states, because their Dirac points are free to shift along high-symmetry lines, whereas the Dirac points in graphene are pinned to high-symmetry BZ points.

\begin{figure}[h]
\centering
\includegraphics[width=0.85\textwidth]{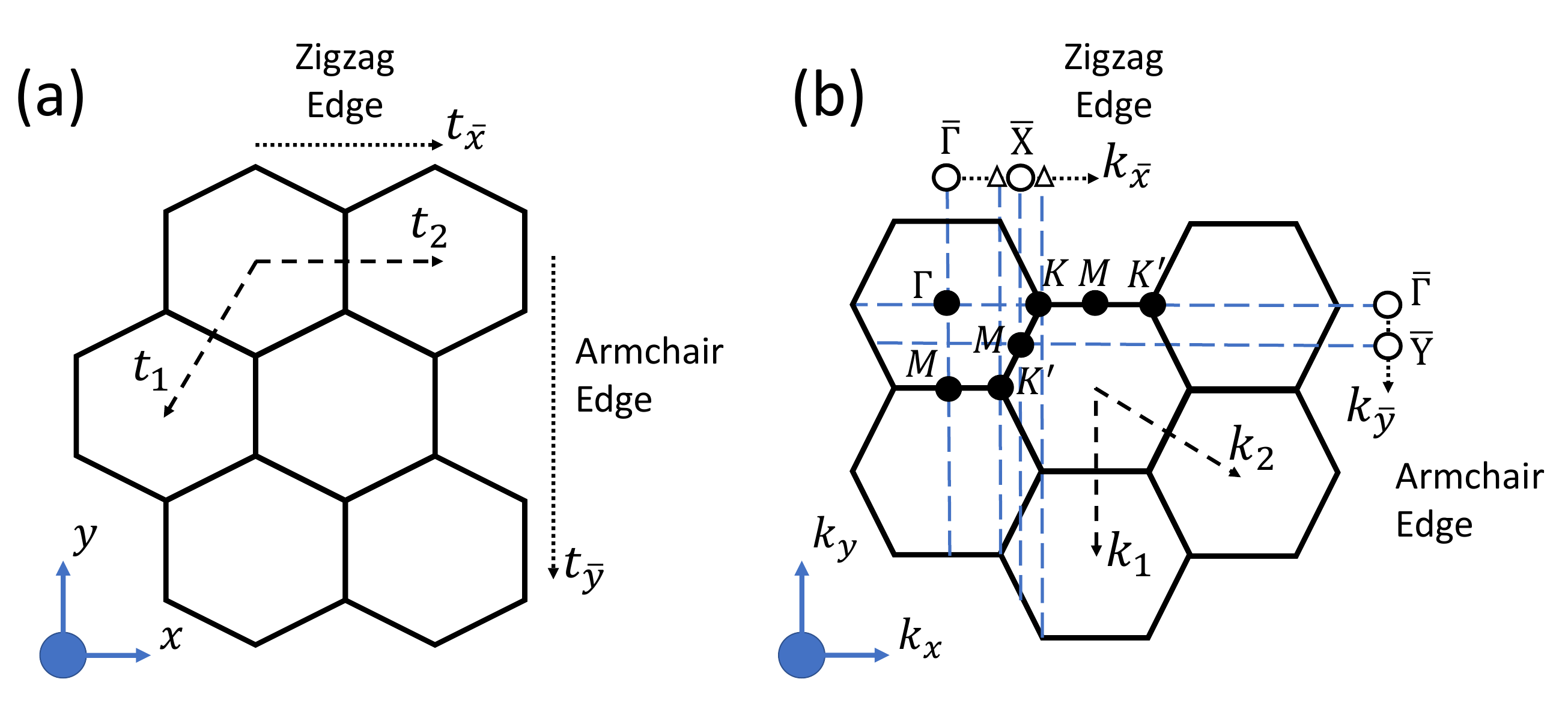}
\caption{Termination dependence of the edge BZ and Fermi arcs in graphene.  (a) The position-space 2D honeycomb lattice of graphene, which can be terminated with either zigzag or armchair edges~\cite{GrapheneReview}.  For the zigzag termination of graphene, the edge lattice vector $t_{\bar{x}}$ is the same as the bulk lattice vector $t_{2}$.  However, for armchair-terminated graphene, the edge lattice vector $t_{\bar{y}}$ is equal to a linear combination of bulk lattice vectors $2t_{1}+t_{2}$.  (b)  The 2D bulk and edge Brillouin Zones (BZs) of graphene.  The termination dependence of the edge lattice vector in graphene is reflected in where the bulk high-symmetry BZ points project in the edge BZ.  For a zigzag-terminated edge, the Dirac points at $K$ and $K'$ project to different points (triangles) on the 1D edge BZ.  These projections are spanned by boundary polarization modes, \emph{i.e.}, arc-like flat-band edge states~\cite{GrapheneEdge1,GrapheneEdge2,GrapheneEdgeMullen,GrapheneEdgeFan}, that are the ``first-order'' analogs of the HOFA states analyzed in this work.  However, on armchair-terminated edges, the $K$ and $K'$ points both project to the same surface TRIM point $\bar{\Gamma}$, and therefore, as there is no region between the projections of the bulk Dirac points in the 1D armchair edge BZ, there are no edge Fermi arcs~\cite{GrapheneEdgeFan}.}  
\label{fig:graphene}
\end{figure}

First, we will review how geometry and periodicity explain the absence of flat-band-like edge Fermi arcs in armchair-terminated graphene.  Then, by explicitly performing the BZ folding from a Dirac semimetal in a body-centered SG to a rod that preserves a fourfold axis, we will show that HOFA states are still generically present on the rod hinges.

We begin by reviewing graphene, a two-dimensional honeycomb carbon lattice~\cite{GrapheneReview,GrapheneEdgeFan}.  The electronic structure of graphene is characterized by two spin-degenerate Dirac cones lying at the high-symmetry BZ points~\cite{GrapheneReview} $K$ and $K'$ (Fig.~\ref{fig:graphene}(b)).  Graphene sheets can be terminated with either zigzag or armchair edges (Fig.~\ref{fig:graphene}(a)).  It has been extensively demonstrated that while zigzag edges exhibit nearly flat surface Fermi arcs, armchair edges do not exhibit low-energy boundary modes~\cite{GrapheneEdge1,GrapheneEdge2,GrapheneEdgeMullen,GrapheneEdgeFan}.  To explain this phenomenon, we employ arguments from Ref.~\onlinecite{GrapheneEdgeFan}, and analyze the relationship between edge and bulk periodicity.  On zigzag edges, the edge lattice vector ($t_{\bar{x}}$) is the same length and direction as the bulk lattice vector $t_{2}$.  Consequently, in the edge BZ (Fig.~\ref{fig:graphene}(b)), the bulk Dirac points at $K$ and $K'$, which have different $k_{x}$ momentum coordiantes, project to different points.  Due to the presence of bulk mirror and inversion symmetries, their projections are spanned by Fermi arcs ``protected'' by the dipole moments of the effective 1D Hamiltonians along the BZ lines that project to them~\cite{GrapheneEdge1,GrapheneEdge2,GrapheneEdgeMullen,GrapheneEdgeFan}.  Conversely, in armchair-terminated graphene, the edge lattice vector $t_{\bar{y}}$ (Fig.~\ref{fig:graphene}(a)) spans multiple unit cells; specifically, $t_{\bar{y}}=2t_{1}+t_{2}$.  Consequently, in the armchair edge BZ (Fig.~\ref{fig:graphene}(b)), the edge reciprocal lattice vector $k_{\bar{y}}$ is shorter than the bulk reciprocal lattice vector $k_{1}$, and the $K$ and $K'$ points project to the \emph{same} point: the edge TRIM point $\bar{\Gamma}$.  Therefore, there is no region in the armchair edge BZ spanning the projections of the bulk Dirac points, and thus armchair terminations do not exhibit edge states at low energies~\cite{GrapheneEdge1,GrapheneEdge2,GrapheneEdgeMullen,GrapheneEdgeFan}.

We can extend the same arguments to 3D body-centered crystals to calculate the hinge projections of bulk Dirac points capable of supporting HOFA states.  In Tables~\ref{tb:SGrod} and~\ref{tb:SGrod4Only}, we show that Dirac points with HOFA states may only form in body-centered crystals along the fourfold axis $\Gamma Z$ ($\Gamma M$) in tetragonal SGs (Fig.~\ref{fig:BZs}(a) and (b), respectively), and along the fourfold axes $\Gamma X$ ($\Gamma H$) in cubic SGs (Fig.~\ref{fig:BZs}(c,d) and (e), respectively).  Therefore, following the arguments in Appendix~\ref{sec:symmetry}, HOFAs may form on the hinges of $z$-directed rods of semimetals in these SGs (and, up to equivalence, on the hinges of rods oriented along the $x$, $y$, and $z$ directions in cubic systems).  However, in body-centered tetragonal and cubic SGs, the $k_{z}$ axis is not parallel to the lattice vectors~\cite{BigBook,BCS1,BCS2}.  Specifically, in tetragonal SGs, the fourfold axis lies along the $z$ direction, whereas the reciprocal lattice vectors are:
\begin{equation}
{\bf k}_{1} = \left(0,\frac{2\pi}{a},\frac{2\pi}{c}\right),\ {\bf k}_{2} = \left(\frac{2\pi}{a},0,\frac{2\pi}{c}\right),\ {\bf k}_{3} = \left(\frac{2\pi}{a},\frac{2\pi}{a},0\right).
\label{eq:BCTLatVec}
\end{equation}
Therefore when a crystal with a body-centered tetragonal or cubic SG is cut into a rod that preserves a fourfold axis and exhibits a lattice periodicity of $c$ in the $z$ direction, distinct points within the bulk BZ will be folded onto the same point in the rod BZ, analogous to armchair-terminated graphene (Fig.~\ref{fig:graphene}(b)).

\begin{figure}[h]
\centering
\includegraphics[width=1.0\textwidth]{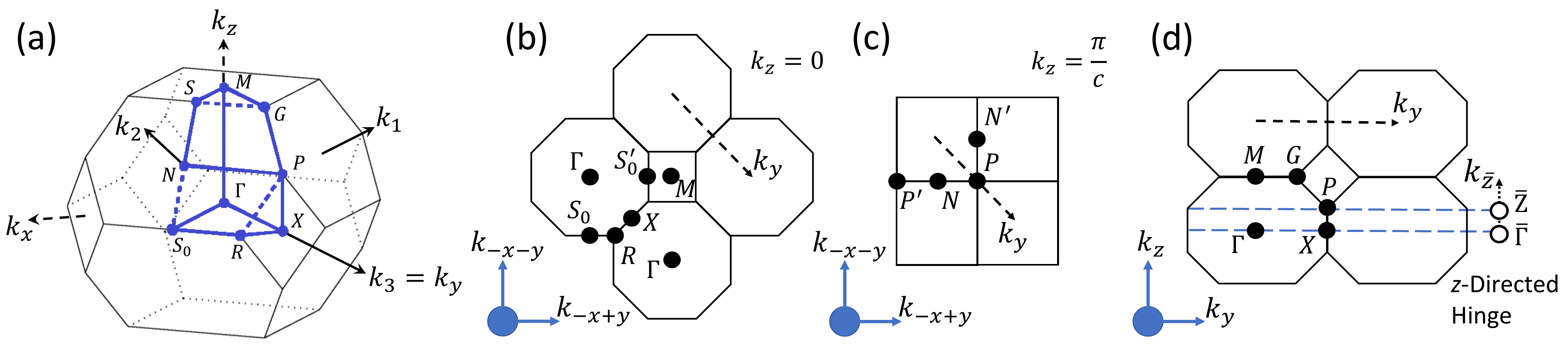}
\caption{Hinge projection of the bulk BZ in body-centered tetragonal crystals.  (a) The full 3D bulk BZ~\cite{BCTBZ,BCS1,BCS2}, (b) the bulk BZ slice at $k_{z}=0$, and (c) the bulk BZ slice at $k_{z} = \pi/c$ of a body-centered tetragonal crystal, where $c$ is the lattice periodicity in the $z$ direction (though, as shown in Eq.~(\ref{eq:BCTLatVec}), $k_{z}=2\pi/c$ is not a reciprocal lattice vector).  (d) In the 1D hinge BZ of a $z$-directed rod, all of the points in the planes at $k_{z}=0,2\pi/c$ project to the hinge TRIM point $\bar{\Gamma}$, and all of the points at $k_{z}=\pm \pi/c$ project to the hinge TRIM point $\bar{Z}$.  As long as Dirac points along a fourfold axis in the BZ do not lie exactly in the $k_{z}=\pm \pi/c$ planes, there will be a finite distance between their projections in the 1D hinge BZ, allowing for the presence of HOFA states under the symmetry conditions in Appendix~\ref{sec:symmetry}.}
\label{fig:BCT}
\end{figure}

We show in Fig.~\ref{fig:BCT}(b,c) the bulk BZ planes that project to hinge TRIM points (Fig.~\ref{fig:BCT}(d)); note that the $\Gamma$ and $M$ points in successive BZs lie in the same planes at $k_{z}=0,2\pi/c$, where $c$ is the lattice spacing in the $c$ direction (though in the bulk crystal, $2\pi/c$ is not a reciprocal lattice vector (Eq.~(\ref{eq:BCTLatVec}))).  Therefore, all of the points in these planes project to the hinge TRIM point $\bar{\Gamma}$ (Fig.~\ref{fig:BCT}(d)).  The $N$ and $P$ points and their time-reversal and fourfold rotation partners $N'$ and $P'$ also lie in the same bulk BZ planes at $k_{z}=\pm \pi/c$ (Fig.~\ref{fig:BCT}(c)), and all project to the same hinge TRIM point $\bar{Z}$.  Therefore, if a Dirac point along $\Gamma M$ lies \emph{exactly} at $\pi/c$, its time-reversal partner will also lie in the same plane in the next BZ, and both Dirac points will project to $\bar{Z}$.  This situation is analogous to the armchair termination of graphene (Fig.~\ref{fig:graphene}(b)): because both Dirac points project to the same point in the hinge BZ, there is no region between their projections for HOFA states to span, and so HOFAs will not be present.  However, as the SGs in Table~\ref{tb:SGrod4Only} do not contain additional symmetries that can force Dirac points to lie exactly at $k_{z}=\pi/c$, it is extremely unlikely for band-inversion-driven~\cite{ZJDirac,WiederLayers} Dirac points in real materials to lie in these planes.  Therefore, HOFAs should still be generic features of fourfold-symmetric rods of body-centered Dirac semimetals with the SGs listed in Table~\ref{tb:SGrod4Only}.

The difference between armchair-terminated graphene and body-centered HOFA Dirac semimetals can be summarized by recognizing that the 3D HOFA Dirac points analyzed in this section arise from band inversion, and are thus free to shift in momentum along high-symmetry BZ lines (they are ``enforced semimetals'' in nomenclature of Ref.~\onlinecite{AndreiMaterials}), whereas, conversely, the Dirac points in graphene are pinned by band connectivity to the high-symmetry BZ points~\cite{GrapheneReview,QuantumChemistry} $K$ and $K'$ (graphene is an ``enforced semimetal with Fermi degeneracy'' in the nomenclature of Ref.~\onlinecite{AndreiMaterials}).  For example, if the Dirac points in a body-centered 3D semimetal were instead hypothetically characterized by four-dimensional corepresentations~\cite{SteveDirac} pinned to the $P$ and $P'$ points (Fig.~\ref{fig:BCT}(a,c)), then they would lie exactly at $k_{z} = \pm \pi/c$, and would not exhibit HOFA states (Fig.~\ref{fig:BCT}(d)).

\section{First-Principles Calculation Details} 
\label{sec:DFT}

\subsection{HOFA States in KMgBi}
\label{sec:kmgbi}

\begin{figure}[t]
\centering
\includegraphics[width=0.65\textwidth]{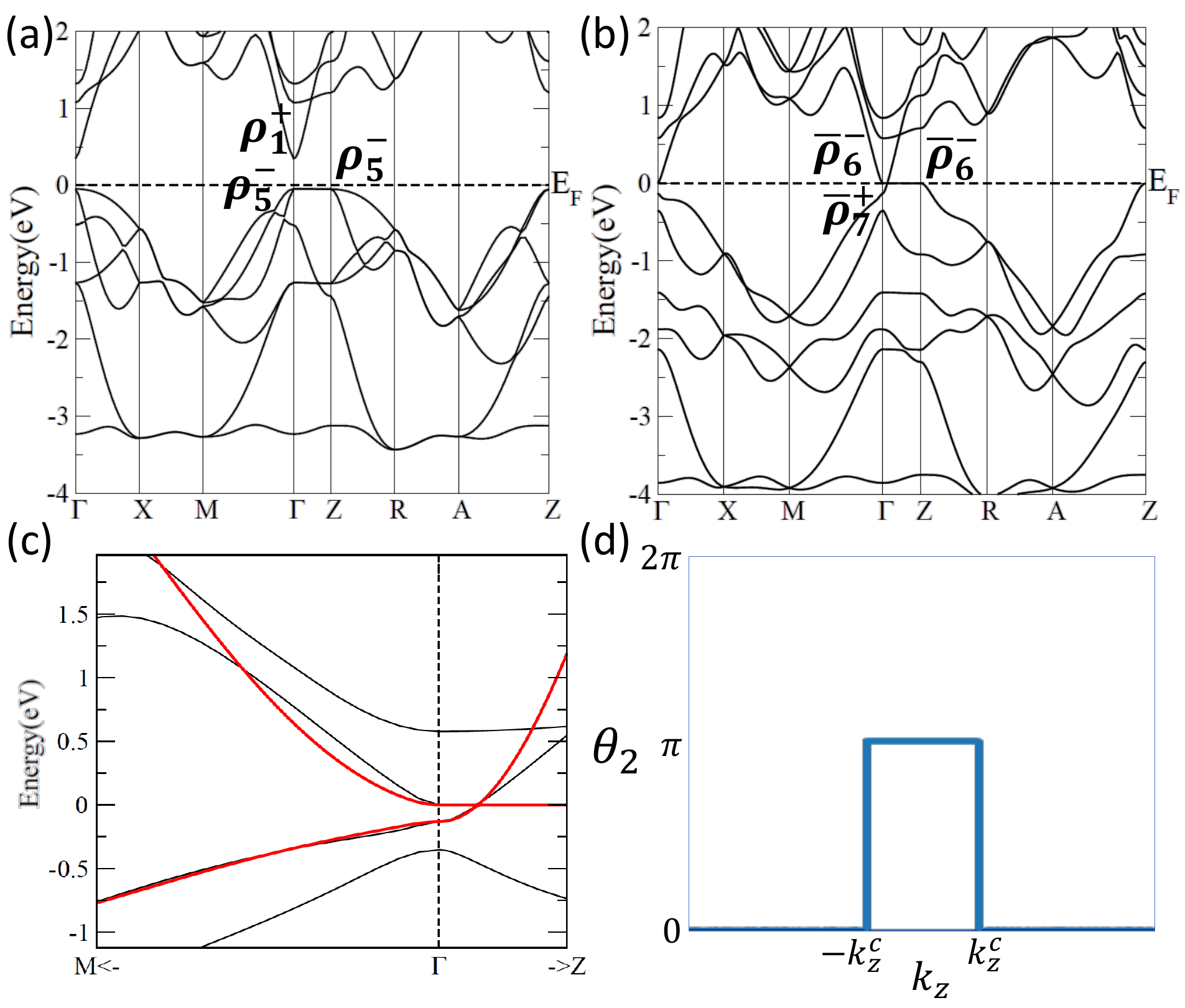}
\caption{The electronic structure of KMgBi in SG 129 $P4/nmm1'$ as calculated from first principles.  Bands (a) without and (b) with spin-orbit coupling (SOC), labeled by the single- and double-valued corepresentations, respectively, of point group $4/mmm1'$ ($D_{4h}$), to which the little co-groups at $\Gamma$ and $Z$ are isomorphic.  In (a), the highest valence band comes from an antibonding combination of Bi $p$ orbitals, and the lowest conduction band comes from a K $s$ orbital.  In (b), the effects of SOC drive a band inversion along $\Gamma Z$, resulting in a time-reversed pair of symmetry-stabilized Dirac points at $k_{x}=k_{y}=0$, $k_{z}=\pm k_{z}^{c}$, where $k_{z}^{c}=0.06085~(2\pi/c)$.  (c) Bands fitted from the $k\cdot p$ Hamiltonian in Eq.~(\ref{eq:ZJ}) (red) to the bands obtained from first-principles calculations (black).  (d) The nested Wilson loop of the $k\cdot p$ Hamiltonian in Eq.~(\ref{eq:ZJ}) with the parameters in Table~\ref{tb:param}.  To obtain the nested Wilson phase $\theta_{2}(k_{z})$ in (d), we first calculate the $x$-directed Wilson loop $W_{1}(k_{y},k_{z})$ over the lower two bulk bands, and then calculate the $y$-directed nested Wilson loop~\cite{multipole,WladTheory,WiederAxion} over the lower Wilson band of $W_{1}(k_{y},k_{z})$.  The resulting nested Wilson loop $W_{2}(k_{z})$ has only a single eigenvalue $\theta_{2}(k_{z})$ at each value of $k_{z}$, and indicates a nontrivial nested Berry phase of $\pi$ for the values of $k_{z}$ with HOFA states in Fig.~\ref{fig:DFTmain}(d) of the main text ($0<|k_{z}|<k_{z}^{c}$).}
\label{fig:DFT}
\end{figure}

Among the previously synthesized materials that fulfill the criteria for HOFAs derived in Appendix~\ref{sec:SGs} (Table~\ref{tb:SGrod4Only}), we identify the candidate HOFA Dirac semimetal KMgBi in SG 129 ($P4/nmm1'$)~\cite{KMgBi1,KMgBi2,KMgBi3} (Inorganic Crystal Structure Database (ICSD)~\cite{ICSD} No. 616748, further details available at~\cite{QuantumChemistry,AndreiMaterials,BCS1,BCS2}~\url{https://topologicalquantumchemistry.org/#/detail/616748}).  We calculate the electronic structure of KMgBi from first principles with the projector augmented wave (PAW)~\cite{paw1} method as implemented in the VASP package~\cite{vasp1,vasp2} (Fig.~\ref{fig:DFT}(a,b) and Fig.~\ref{fig:DFTmain} of the main text).  In KMgBi, the Bi atoms occupy the $2c$ Wyckoff position and each exhibit an oxidation state of $3-$; this implies that all 6 occupied bands near the Fermi energy ($E_{F}$) arise from Bi $p$ orbitals in the limit of vanishing spin-orbit coupling (SOC) (Fig.~\ref{fig:DFT}(a)).  As there are two Bi atoms per unit cell, these $p$-orbitals can form bonding and anti-bonding states.  In SG 129, the $\Gamma$ and $Z$ points have little co-groups isomorphic to point group $4/mmm1' $ ($D_{4h}$), and points along the line $\Gamma Z$ have little groups isomorphic to magnetic point group $4/m'mm$ (Appendix~\ref{sec:bandrep}).  We therefore label bands at $\Gamma$ and $Z$ using the notation employed in Appendix~\ref{sec:bandrep} (Eqs.~(\ref{eq:krepsGamma}) and~(\ref{eq:krepsM})) and Appendix~\ref{sec:pd} (Eq.~(\ref{eq:kreps})) for the irreducible corepresentations of $4/mmm1'$  (Eqs.~(\ref{eq:c4evals}),~(\ref{eq:c4evalsWithI}), and~(\ref{eq:inversionEigs})).  In this notation, $\rho_{i}$ indicates the i$^{\text{th}}$ irreducible representation of $4mm$, the $\mathcal{I}$- (and $\mathcal{I}\times\mathcal{T}$-) broken unitary subgroup of $4/mmm1'$ (and $4/m'mm$); bars indicate double-valued (co)representations; and the corepresentations of $4/mmm1'$ at $\Gamma$ and at $Z$ are labeled with additional $\pm$ superscripts to indicate whether they have positive or negative parity ($\mathcal{I}$) eigenvalues.  Specifically, the numbering for $\rho_{i}$ is chosen to match the order of irreducible representations displayed in the~\textsc{REPRESENTATIONS: DBG} tool on the BCS~\cite{QuantumChemistry,Bandrep1} for the $\Gamma$ point of the unitary subgroup of SG 99 $P4mm1'$, which is isomorphic to $4mm$.

At $\Gamma$ and $Z$, in the absence of SOC, the bonding and antibonding states nearest the Fermi energy are characterized by the single-valued corepresentations $\rho_{1}^{+}$ and $\rho_{5}^{-}$, respectively~\cite{Bandrep1,Bandrep2,BilbaoPoint} (Fig.~\ref{fig:DFT}(a)).  In terms of the corepresentations of point group~\cite{BilbaoPoint} $4/mmm1'$:
\begin{equation}
\rho_{1}^{+}\equiv A_{1g},\ \rho_{5}^{-} \equiv E_{u}.
\end{equation}
In first-principles calculations incorporating the effects of SOC, SOC drives a band inversion in KMgBi at $\Gamma$ between states labeled by double-valued corepresentations with opposite inversion characters (parity eigenvalues).  Specifically, after this SOC-driven band inversion, the states just above and below $E_{F}$ at $\Gamma$ are characterized by $\bar{\rho}_{6}^{-}$ and $\bar{\rho}_{7}^{+}$, respectively (Fig.~\ref{fig:DFT}(b)), whose $C_{4z}$ and $\mathcal{I}$ inversion characters are respectively given by:
\begin{equation}
\chi_{\bar{\rho}_{6}^{-}}(C_{4z})=\frac{-1+i}{\sqrt{2}} + \frac{-1-i}{\sqrt{2}}=-\sqrt{2},\ \chi_{\bar{\rho}_{7}^{+}}(C_{4z})=\frac{1+i}{\sqrt{2}} + \frac{1-i}{\sqrt{2}}=\sqrt{2},\ \chi_{\bar{\rho}_{6}^{-}}(\mathcal{I}) = -2,\ \chi_{\bar{\rho}_{7}^{+}}(\mathcal{I})=2, 
\label{eq:67rep1}
\end{equation}
where $\chi_{\rho}(g)$ is the character of the unitary symmetry $g$ in the irreducible representation $\rho$, and is equivalent to the trace of the matrix representation of $g$ (\emph{i.e.}, the sum of the symmetry eigenvalues of $g$ in $\rho$).  In terms of the corepresentations of point group~\cite{BilbaoPoint} $4/mmm1'$:
\begin{equation}
\bar{\rho}_{6}^{-}\equiv \bar{E}_{2u},\ \bar{\rho}_{7}^{+} \equiv \bar{E}_{1g}.
\label{eq:67equiv1}
\end{equation} 
Because, SOC drives bands with opposite parity eigenvalues to become inverted at $\Gamma$, the Hamiltonian of the $k_{z}=0$ plane of KMgBi is topologically equivalent to a 2D TI, as occurs in many other topological semimetals~\cite{NagaosaDirac}, such as Na$_3$Bi~\cite{NaDirac,SchnyderDirac,KargarianDiracArc1,KargarianDiracArc2,ZJDirac2,SYDiracSurface}, both the room- ($\alpha$) and intermediate-temperature ($\alpha''$) phases of Cd$_3$As$_2$ (Refs.~\onlinecite{WeylReview,AshvinDiracSurface,NagaosaDirac,ZJDirac,ZJSurface} and Appendix~\ref{sec:cadmium}), and WC~\cite{WCarc}.  Specifically, incorporating the effects of SOC in KMgBi, the product of the parity eigenvalues per Kramers pair up to the Fermi energy is positive at $\Gamma$ and negative at $X$, $X'$, and $M$.  The Hamiltonian of the $k_{z}=0$ plane is therefore $\mathbb{Z}_{2}$-nontrival by the Fu-Kane parity index~\cite{FuKaneInversion}.  As shown in Fig.~\ref{fig:DFT}(b), because the bands that cross along $\Gamma Z$ are labeled by $\bar{\rho}_{6}^{-}$ and $\bar{\rho}_{7}^{+}$ at the TRIM points, then the Hamiltonian of the $k_{z}=0$ plane \emph{also} exhibits the same fourfold rotation eigenvalues as a QI in $p4m$ (Appendices~\ref{sec:bandrep} and~\ref{sec:TIboundary}).  This indicates that, at intermediate values of $k_{z}$, bands must be labeled by $\bar{\rho}_{6,7}$, and implies that their crossing (Dirac) points must exhibit HOFA states when projected to the hinges of fourfold-symmetric nanorods (Appendices~\ref{sec:double} and~\ref{sec:SGs}).

\begin{table}[h]
\caption{The parameters used to fit the bands of the $k\cdot p$ theory in Eq.~(\ref{eq:ZJ}) to the first-principles electronic structure of KMgBi in the vicinity of the $\Gamma$ point (Fig.~\ref{fig:DFT}(c)).}
\begin{tabular}{c|c|c|c|c|c|c|c|c}
 $C_0$ (eV)& $C_1$ (eV \AA$^{-2}$) & $C_2$ (eV \AA$^{-2}$) & $A$ (eV \AA$^{-1}$) & $M_0$ (eV) & $M_1$ (eV \AA$^{-2}$) & $M_2$ (eV \AA$^{-2}$) & $B_ 1$ (eV \AA$^{-3}$) & $B_2$ (eV \AA$^{-3}$) \\
\hline
-0.06595 & 31.58273 & 7.89568 & 2.51327 & -0.06595  & -31.58273 & -13.42266 & -124.0251 &  0    \\
\end{tabular}
\label{tb:param}
\end{table}

To demonstrate the presence of hinge-localized HOFAs in KMgBi, we form a $p4m$, $\mathcal{I}$-, and $\mathcal{T}$-symmetric, four-band $k\cdot p$ theory near the $\Gamma$ point.  We choose the basis in which bands characterized by $\bar{\rho}_{7}^{+}$ are labeled $\ket{s,\uparrow}$ and $\ket{s,\downarrow}$, and bands characterized by $\bar{\rho}_{6}^{-}$ are labeled $\ket{p_x+ip_y,\uparrow}$ and $\ket{p_x-ip_y,\downarrow}$.  It is important to note that, because we are here using $p_{x,y}$ orbitals, and not $p_{z}$ orbitals, then the spinful states labeled by $\ket{p_x+ip_y,\uparrow}$ and $\ket{p_x-ip_y,\downarrow}$ exhibit the same fourfold rotation eigenvalues (but not the same parity eigenvalues) as spinful $d_{x^{2}-y^{2}}$ orbitals (Appendices~\ref{sec:bandrep},~\ref{sec:TIboundary}, and~\ref{sec:TItoTrivial}); therefore, as shown in Appendix~\ref{sec:TIboundary}, we expect Eq.~(\ref{eq:ZJ}) to exhibit the same HOFA states as the $p-d$-hybridized Dirac semimetal in Appendix~\ref{sec:pd}.  Using these four states, we formulate a $k\cdot p$ Hamiltonian that is the same to quadratic order as the one introduced in Ref.~\onlinecite{ZJDirac} for the Dirac points in the centrosymmetric phases of Cd$_3$As$_2$ (in Appendix~\ref{sec:cadmium}, we detail calculations showing HOFA states in $\alpha''$-Cd$_3$As$_2$):
\begin{eqnarray*}
  H_\Gamma({\bf k}) & = & \epsilon_0({\bf k})+\left(\begin{array}{cccc}
      M({\bf k}) & Ak_{+} & 0 & B^*({\bf k})\\
      Ak_{-} & -M({\bf k}) & B^*({\bf k}) & 0 \\
      0 &B({\bf k}) & M({\bf k}) & -Ak_{-}\\
      B({\bf k}) & 0 & -Ak_{+} & -M({\bf k})
\end{array}\right),
\label{eq:ZJ}
\end{eqnarray*}
where:
\begin{eqnarray}
\epsilon_0({\bf k}) &=& C_{0}+C_{1}k_{z}^{2}+C_{2}(k_x^{2}+k_y^2),\ k_{\pm}=k_{x}\pm ik_{y},\ B({\bf k})=B_1k_z^ck_+^2+B_2k_z^ck_-^2,\nonumber \\
M({\bf k}) &=& M_{0}-M_{1}k_{z}^{2}-M_{2}(k_x^{2}+k_y^2). 
\end{eqnarray}
We chose $M_0,M_1, M_2<0$ to reproduce the band inversion, and set $\epsilon_0({\bf k})\rightarrow 0$ for simplicity.  Using our first-principles calculations, we predict that the two Dirac points in KMgBi are located at $(0,0,\pm k_z^c)$ with $k_z^c=0.06085~(2\pi/c)=0.046$~$\AA^{-1}$.  Using the energy ordering of the irreducible representations $\bar{\rho}_{6,7}$ (Appendices~\ref{sec:bandrep} and~\ref{sec:double}) and the nested Wilson loop~\cite{multipole,WladTheory} (Fig.~\ref{fig:DFT}(d)), we deduce that $k_{z}$ slices with topological quadrupole moments and hinge-localized HOFA states occur between $0<|k_z|<k_z^c$.  Fitting the model in Eq.~(\ref{eq:ZJ}) to the calculated electronic structure (Fig.~\ref{fig:DFT}(c)), we obtain the $k\cdot p$ parameters listed in Table~\ref{tb:param}.

\begin{figure}[h]
\centering
\includegraphics[width=0.4\textwidth]{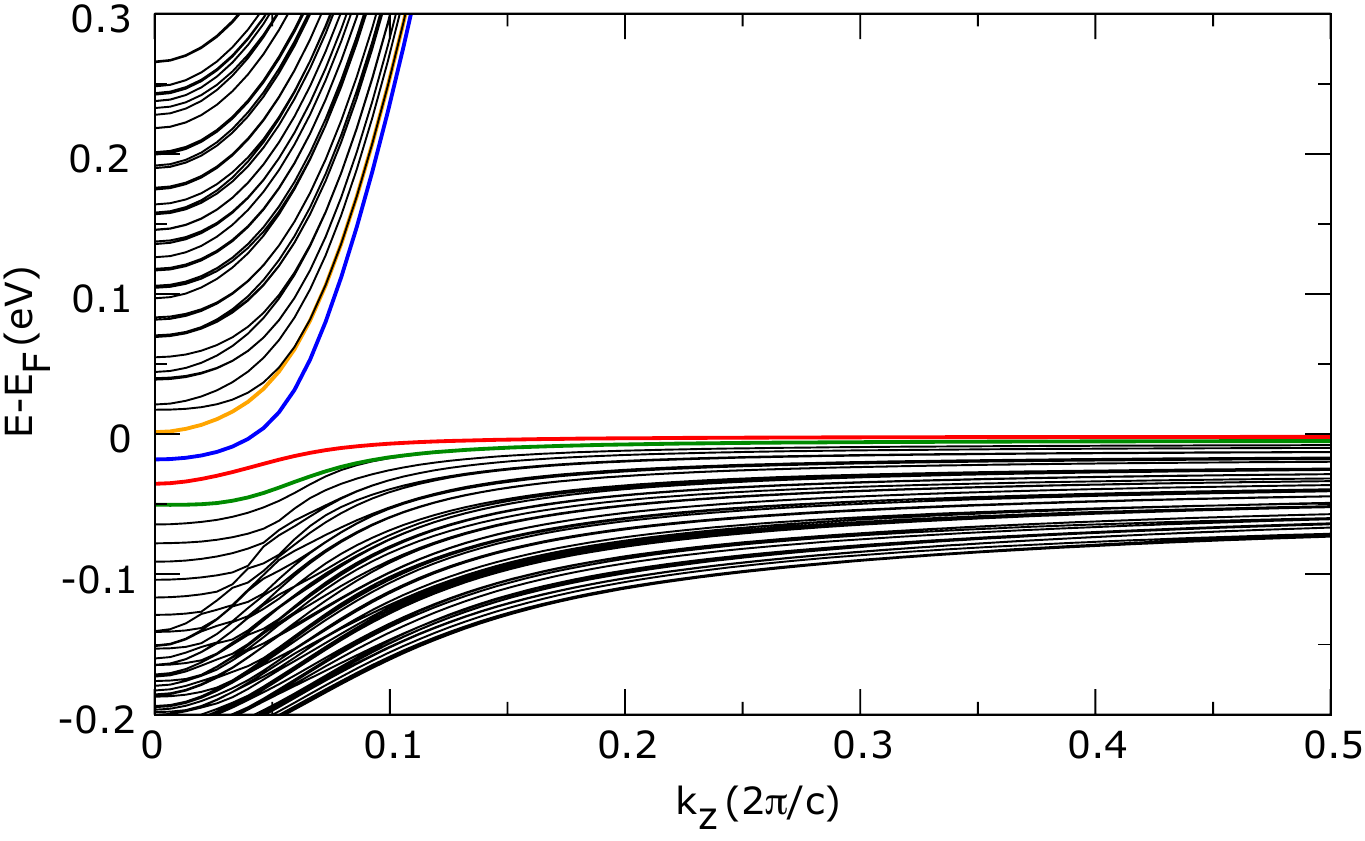}
\caption{The band structure of a square, $z$-directed rod of a tight-binding model obtained from the $k\cdot p$ theory in Eq.~(\ref{eq:ZJ}) and fit to the first-principles electronic structure of KMgBi in the vicinity of the $\Gamma$ point (Fig.~\ref{fig:DFT}(c)) using the parameters in Table~\ref{tb:param}.  Because the rod preserves spinful $\mathcal{I}\times\mathcal{T}$ symmetry (or alternatively, because the rod preserves the spinful $x$ and $y$ reflection symmetries of Eq.~(\ref{eq:ZJ}), whose representations anticommute), the bands are doubly degenerate~\cite{BigBook,WiederLayers}.  Fixing the rod filling to that of the bulk Dirac points, we label the two highest valence (lowest conduction) bands in red (blue).  Though the hinge projection of the bulk Dirac point at $k_{z}^{c}=0.06085~(2\pi/c)$ is split by finite-size effects, we still observe four, half-filled HOFA states at $0<|k_z|<k_z^c$ (which are split into two occupied (red) and two unoccupied (blue) HOFA states by finite-size effects).}
\label{fig:kmgbiRod}
\end{figure}

We then map this fitted $k\cdot p$ theory to a lattice tight-binding model, following the procedure employed in Ref.~\onlinecite{ZJDirac}.  Next, we cut the lattice tight-binding model into a fourfold-symmetric (square) rod that is finite in the $x$ and $y$ directions and infinite in the $z$ direction and calculate its bulk, surface, and hinge states (Fig.~\ref{fig:kmgbiRod}); in particular, we observe that the four hinge states closest to the Fermi energy are half-filled, in agreement with the characterization of the corner modes of a QI in Fig.~\ref{fig:cornerModes}.  However, because the spinful rod groups do not host symmetry-stabilized four-dimensional corepresentations away from rod TRIM points~\cite{BigBook,Bandrep1}, then there is no crystal symmetry that can force the bands corresponding to the bulk Dirac points and HOFA hinge states to appear in strict fourfold degeneracies in the rod bands in Fig.~\ref{fig:kmgbiRod}; instead, their fourfold degeneracy is only restored in the thermodynamic limit that states along the rod do not hybridize.  Therefore, because of finite-size effects, the hinge projections of the bulk Dirac points and the HOFA hinge states split into two sets of states in the rod bands in Fig.~\ref{fig:kmgbiRod}.

Next, to visualize the HOFA state that appears on a single hinge of a macroscopic, $z$-directed sample of KMgBi, we perform a hinge Green's function calculation, the results of which are shown in Fig.~\ref{fig:DFTmain}(d) of the main text.  To isolate the surface and corner (hinge) states,  we first form a slab of the tight-binding model in Eq.~(\ref{eq:ZJ}) that is infinite in the $z$ direction, semi-infinite in the $x$ direction, and large ($\sim 60$ unit cells) in the $y$ direction.  We then employ \emph{hinge} Green's functions to calculate the hinge states localized on just a single edge of the semi-infinite slab (Fig.~\ref{fig:DFTmain}(d) of the main text).  We observe clear HOFA states connecting the hinge projections of the bulk 3D Dirac points to the hinge projections of the 2D topological surface cones of the $\mathbb{Z}_{2}$-nontrivial bulk plane at $k_{z}=0$ (Fig.~\ref{fig:DFTmain}(d) of the main text).  Unlike in our previous rod calculation (Fig.~\ref{fig:kmgbiRod}), the semi-infinite slab used to calculate the hinge Green's function exhibits strongly broken fourfold rotation symmetry, and is therefore not strictly required to exhibit topological HOFA states, which may be removed through surface gap closures, analogous to the corner modes of the $C_{4z}$-broken QI in Ref.~\onlinecite{multipole}.  Nevertheless, in the Green's function of a single hinge (Fig.~\ref{fig:DFTmain}(d) of the main text), we still observe the same, isolated, half-filled HOFA state as is present on each of the four hinges in the fourfold-symmetric rod tight-binding calculation (Fig.~\ref{fig:kmgbiRod}) that we performed to confirm the presence of intrinsic HOFA states in KMgBi.  We postulate that this is because the slab still has $90$-degree corners, like in the square rod calculation (Fig.~\ref{fig:kmgbiRod}), and because it is very (infinitely) large in the $xy$-plane compared to the hoppings in Eq.~(\ref{eq:ZJ}); therefore, at most (if not all) $k_{z}$ points with HOFA states in the hinge Green's function calculation (Fig.~\ref{fig:DFTmain}(d) of the main text), a very large suface potential may be required to change the sign of the surface (edge) gap from its value in the fourfold-symmetric rod calculation (Fig.~\ref{fig:kmgbiRod}).

\subsection{HOFA States in $\alpha''$-Cd$_3$As$_2$}
\label{sec:cadmium}

We also find that the criteria for HOFAs in Appendix~\ref{sec:SGs} (Table~\ref{tb:SGrod4Only}) are satisfied by the archetypal Dirac semimetal Cd$_3$As$_2$ in both its room- ($\alpha$) and intermediate-temperature ($\alpha''$) phases (SGs 142 ($I4_{1}/acd1'$) and 137 ($P4_{2}/nmc1'$), respectively)~\cite{CavaDirac1,CavaDirac2,ZJDirac,ZJSurface}.  Because of its simple primitive tetragonal Bravais lattice, we here focus on $\alpha''$-Cd$_3$As$_2$, though the calculations performed in this section could also be adapted to characterize the HOFA states in $\alpha$-Cd$_3$As$_2$ after carefully mapping its body-centered lattice to a primitive tetragonal rod (Appendix~\ref{sec:bodycenter}).

\begin{figure}[h]
\centering
\includegraphics[width=0.95\textwidth]{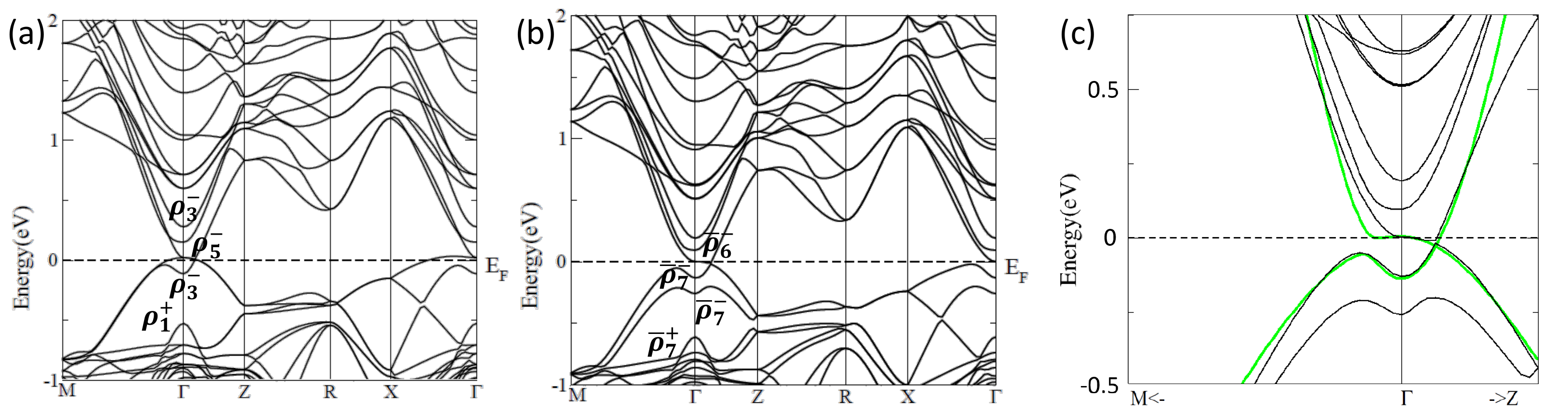}
\caption{The electronic structure of $\alpha''$-Cd$_3$As$_2$ in SG 137 $P4_{2}/nmc1'$ calculated from first principles.  Bands (a) without and (b) with spin-orbit coupling, labeled by the single- and double-valued corepresentations, respectively, of point group $4/mmm1'$ ($D_{4h}$), to which the little co-group at $\Gamma$ is isomorphic.  (c) Bands fitted from the $k\cdot p$ Hamiltonian in Eq.~(\ref{eq:ZJ}) (green) to the bands obtained from first-principles calculations (black).}  
\label{fig:DFTCd}
\end{figure}

The $\alpha''$ phase of Cd$_3$As$_2$ (ICSD~\cite{ICSD} No. 609930) has been extensively studied in theoretical works~\cite{CavaDirac1,CavaDirac2,ZJDirac}, and has been stabilized in experiment in single crystalline form at room temperature and below by 2\% zinc doping~\cite{StableCadmium}.  Furthermore, as Zn is isoelectronic to Cd, this doping should not affect the Fermi level.  In Fig.~\ref{fig:DFTCd}(b,a), we show the electronic structure of $\alpha''$-Cd$_3$As$_2$ calculated from first principles with and without incorporating the effects of SOC, respectively, obtained using the same methodology previously employed to calculate the electronic structure of KMgBi (Appendix~\ref{sec:kmgbi}).  As with KMgBi in Appendix~\ref{sec:kmgbi}, the little co-group of the $\Gamma$ point is isomorphic to point group $4/mmm1'$ ($D_{4h}$) (though unlike in KMgBi, the little co-group of the $Z$ point is here \emph{not} isomorphic to a point group~\cite{BigBook,Steve2D,WiederLayers}, because of the projective action of the fractional lattice translation in the $4_2$ screw symmetry in SG 137 $P4_{2}/nmc1'$).  The representation labels of the bands in Fig.~\ref{fig:DFTCd}(a,b) were obtained from first principles, are given in the convention previously established in Appendix~\ref{sec:kmgbi}, and agree with the results of previous investigations~\cite{CavaDirac1,CavaDirac2,ZJDirac,ZJSurface} of $\alpha''$-Cd$_3$As$_2$.  In terms of the corepresentations of point group~\cite{BilbaoPoint} $4/mmm1'$, the single-valued corepresentations in Fig.~\ref{fig:DFTCd}(a) are related by the equivalences:
\begin{equation}
\rho_{1}^{+}\equiv A_{1g},\ \rho_{3}^{-}\equiv A_{2u},\ \rho_{5}^{-} \equiv E_{u}.
\end{equation}
When the effects of SOC are incorporated, the corepresentations at $\Gamma$ are labeled by $\bar{\rho}_{6}^{-}$, $\bar{\rho}_{7}^{+}$, where the $s_{4_{2}z}$ and $\mathcal{I}$ characters of $\rho_{6}^{-}$ and $\bar{\rho}_{7}^{+}$ were previously given in Eq.~(\ref{eq:67rep1}) ($s_{4_{2}z}$ exhibits the same set of eigenvalues and commutation relations with other spatial symmetries as $C_{4z}$ at the $\Gamma$ point of any SG~\cite{BigBook}), and where the $s_{4_{2}z}$ and $\mathcal{I}$ characters of $\bar{\rho}_{7}^{-}$ are:
\begin{equation}
\chi_{\bar{\rho}_{7}^{-}}(s_{4_{2}z})=\frac{1+i}{\sqrt{2}} + \frac{1-i}{\sqrt{2}}=\sqrt{2},\ \chi_{\bar{\rho}_{7}^{-}}(\mathcal{I})=-2. 
\label{eq:67rep2}
\end{equation}
In terms of the double-valued corepresentations of point group~\cite{BilbaoPoint} $4/mmm1'$:
\begin{equation}
\bar{\rho}_{6}^{-}\equiv \bar{E}_{2u},\ \bar{\rho}_{7}^{+} \equiv \bar{E}_{1g},\ \bar{\rho}_{7}^{-} \equiv \bar{E}_{1u}.
\label{eq:67equiv2}
\end{equation}

Away from the TRIM points $\Gamma$ and $Z$, bands along $\Gamma Z$ cross to form a time-reversed pair of symmetry-stabilized Dirac points at $k_{x}=k_{y}=0$, $k_{z}=\pm k_{z}^{c}$, where $k_{z}^{c}=0.125~(2\pi/c)$ (Fig.~\ref{fig:DFTCd}(b)), where the crossed bands are labeled by $\bar{\rho}_{6,7}$ of $4mm$ with an additional $\mathcal{I}\times\mathcal{T}$ symmetry whose symmetry-representation commutation relations contain phases that reflect the $4_{2}$ screw symmetry in SG 137 $P4_{2}/nmc1'$~(Appendices~\ref{sec:bandrep} and~\ref{sec:pinnedHOFAs}).  Furthermore, $\alpha''$-Cd$_3$As$_2$ crystals in SG 137 are theoretically capable of being shaped into nanowires with fourfold axes, as discussed in the caption of and text surrounding Table~\ref{tb:SGrod4Only} (indeed nanowires~\cite{Cd3As2Nanowire} and $z$- ($c$-axis-) directed samples~\cite{StemmerCadmiumGrowth} of the room-temperature ($\alpha$) phase of Cd$_3$As$_2$ in SG 142 ($I4_{1}/acd1'$) have already been synthesized in experiment).  Therefore, as shown in Appendices~\ref{sec:double} and~\ref{sec:SGs}, a $c$-axis-directed, $4_{2}$-screw-symmetric nanowire of $\alpha''$-Cd$_3$As$_2$ should exhibit HOFA states on its 1D hinges.

To demonstrate the presence of HOFA states in $\alpha''$-Cd$_3$As$_2$, we follow the same procedure previously employed for KMgBi in Appendix~\ref{sec:kmgbi}.  We introduce the $k\cdot p$ Hamiltonian in Eq.~(\ref{eq:ZJ}) to model the electronic structure of $\alpha''$-Cd$_3$As$_2$ calculated from first principles (Fig.~\ref{fig:DFTCd}(b)).  Unlike previously with KMgBi, there is an additional subtlety in using Eq.~(\ref{eq:ZJ}) to model the low-energy electronic structure of $\alpha''$-Cd$_3$As$_2$.  As shown in Fig.~\ref{fig:DFTCd}(b) and discussed in detail in Ref.~\onlinecite{ZJDirac}, the electronic structure of $\alpha''$-Cd$_3$As$_2$ features multiple band inversions and split representations at the $\Gamma$ point: specifically, after incorporating the effects of SOC, the inverted bands closest to the Fermi energy are labeled by corepresentations with different $s_{4_{2}z}$ eigenvalues and the same parity eigenvalues ($\bar{\rho}_{6,7}^{-}$), and there is a second, larger band inversion, also at $\Gamma$, between bands with opposite parity eigenvalues (the valence band of which is labeled with $\bar{\rho}_{7}^{+}$).  To simplify the description of the bands at the Fermi energy, we follow the procedure developed in Ref.~\onlinecite{ZJDirac}, and use a four-band tight-binding model in which the valence states at $\Gamma$ closest to the Fermi energy ($\bar{\rho}_{7}^{-}$) in Fig.~\ref{fig:DFTCd}(b) are replaced with states labeled by $\bar{\rho}_{7}^{+}$, reflecting the summed parity and fourfold rotation eigenvalues of the entire valence manifold at $\Gamma$.  As shown in Ref.~\onlinecite{ZJDirac}, this simplified description still captures the band inversion in $\alpha''$-Cd$_3$As$_2$ between Kramers pairs of Cd $5s$ orbitals and the $m_{j}=\pm 3/2$ subset of spinful As $4p_{x,y}$ orbitals, because spinful $m_{j}=\pm 3/2$ $p$ orbitals (\emph{i.e.} $p_{x} + ip_{y},\uparrow$ and $p_{x} - ip_{y}, \downarrow$ orbitals) were previously shown in Appendix~\ref{sec:kmgbi} to exhibit the same fourfold rotation eigenvalues (but not the same parity eigenvalues) as spinful $d_{x^{2}-y^{2}}$ orbitals (Appendices~\ref{sec:bandrep},~\ref{sec:TIboundary}, and~\ref{sec:TItoTrivial}).

\begin{table}[h]
\caption{The parameters used to fit the bands of the $k\cdot p$ theory in Eq.~(\ref{eq:ZJ}) to the first-principles electronic structure of  $\alpha''$-Cd$_3$As$_2$ in the vicinity of the $\Gamma$ point (Fig.~\ref{fig:DFTCd}(c)).}
\begin{tabular}{c|c|c|c|c|c|c|c|c}
 $C_0$ (eV)& $C_1$ (eV \AA$^{-2}$) & $C_2$ (eV \AA$^{-2}$) & $A$ (eV \AA$^{-1}$) & $M_0$ (eV) & $M_1$ (eV \AA$^{-2}$) & $M_2$ (eV \AA$^{-2}$) & $B_ 1$ (eV \AA$^{-3}$) & $B_2$ (eV \AA$^{-3}$)\\
\hline
-0.066 & 9.8696 & 12.23831 & 0.62832 & -0.07 & -18.16007 & -18.94964 & 124.025106 & 0 \\
\end{tabular}
\label{tb:paramCd}
\end{table}

\begin{figure}[h]
\centering
\includegraphics[width=0.4\textwidth]{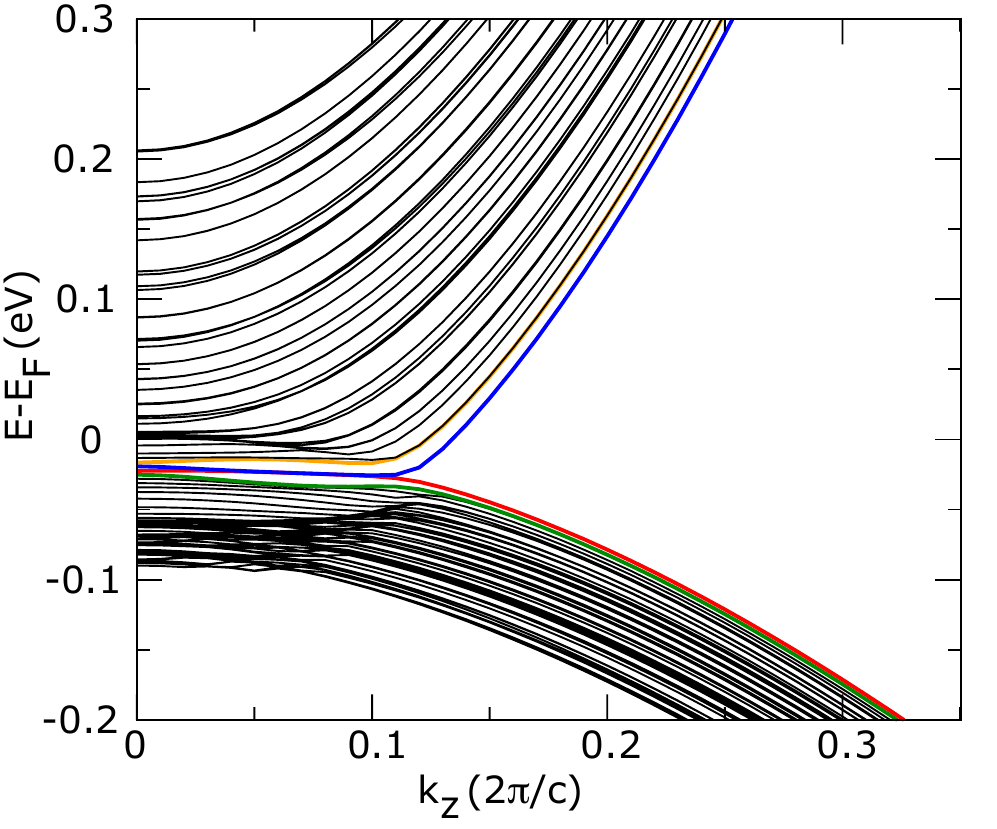}
\caption{The band structure of a square, $z$-directed rod of a tight-binding model obtained from the $k\cdot p$ theory in Eq.~(\ref{eq:ZJ}) and fit to the first-principles electronic structure of $\alpha''$-Cd$_3$As$_2$ in the vicinity of the $\Gamma$ point (Fig.~\ref{fig:DFTCd}(c)) using the parameters in Table~\ref{tb:paramCd}.  Because the rod preserves spinful $\mathcal{I}\times\mathcal{T}$ symmetry (or alternatively, because the rod preserves the spinful $x$ and $y$ reflection symmetries of Eq.~(\ref{eq:ZJ}), whose representations anticommute), the bands are doubly degenerate~\cite{BigBook,WiederLayers}.  Fixing the rod filling to that of the bulk Dirac points at $k_{z}=\pm k_{z}^{c}$, where $k_{z}^{c}=0.125~(2\pi/c)$, we label the two highest valence (lowest conduction) bands in red (blue).  We observe four, half-filled HOFA states connecting the hinge projection of the surface TI cone at $k_{z}=0$ to hinge projection of a bulk Dirac point.}
\label{fig:cdasRod}
\end{figure}

Next, we fit the bands of the simplified $k\cdot p$ theory in Eq.~(\ref{eq:ZJ}) to the first-principles electronic structure of $\alpha''$-Cd$_3$As$_2$ (Fig.~\ref{fig:DFTCd}), obtaining the fitting parameters listed in Table~\ref{tb:paramCd}.  We then map this fitted $k\cdot p$ theory to a lattice tight-binding model, which we place on a fourfold-symmetric rod finite in the $x$ and $y$ directions, and calculate its surface and hinge states (Fig.~\ref{fig:cdasRod}).  In $\alpha''$-Cd$_3$As$_2$, it is well documented that the Hamiltonian of the $k_{z}=0$ plane is equivalent to a 2D TI~\cite{AshvinDiracSurface,ZJSurface,NagaosaDirac}, whose edge states correspondingly manifest as rod surface states at $k_{z}=0$.  In both the complete DFT description of $\alpha''$-Cd$_3$As$_2$ (Fig.~\ref{fig:DFTCd}(b)) and in Eq.~(\ref{eq:ZJ}) with the fitting parameters listed in Table~\ref{tb:paramCd}, the Hamiltonian of the $k_{z}=0$ plane \emph{also} exhibits the same fourfold rotation eigenvalues as a QI in $p4m$ (Appendices~\ref{sec:bandrep} and~\ref{sec:TIboundary}).  As shown in Appendix~\ref{sec:TIboundary}, this indicates that the TI surface states at $k_{z}=0$ will gap into HOFA hinge states away from $k_{z}=0$.  In our rod calculation, we also observe four, half-filled hinge states connecting the hinge projections of the bulk 3D Dirac points to the projections of the 2D TI cones at $k_{z}=0$.  Finally, using the semi-infinite slab construction described in Appendix~\ref{sec:kmgbi}, we use hinge Green's functions to calculate the states on a single hinge of a large, $z$-directed crystal of $\alpha''$-Cd$_3$As$_2$, the results of which are shown in Fig.~\ref{fig:DFTmain}(c) of the main text.  The hinge spectrum exhibits clear HOFA states connecting the hinge projections of the bulk 3D Dirac points to the hinge projections of the 2D topological surface cones of the 2D-TI-equivalent bulk plane at $k_{z}=0$.

\subsection{HOFA States and Fragile Corner Modes in $\beta'$-PtO$_2$}
\label{sec:pto2}

\begin{figure}[h]
\centering
\includegraphics[width=0.95\textwidth]{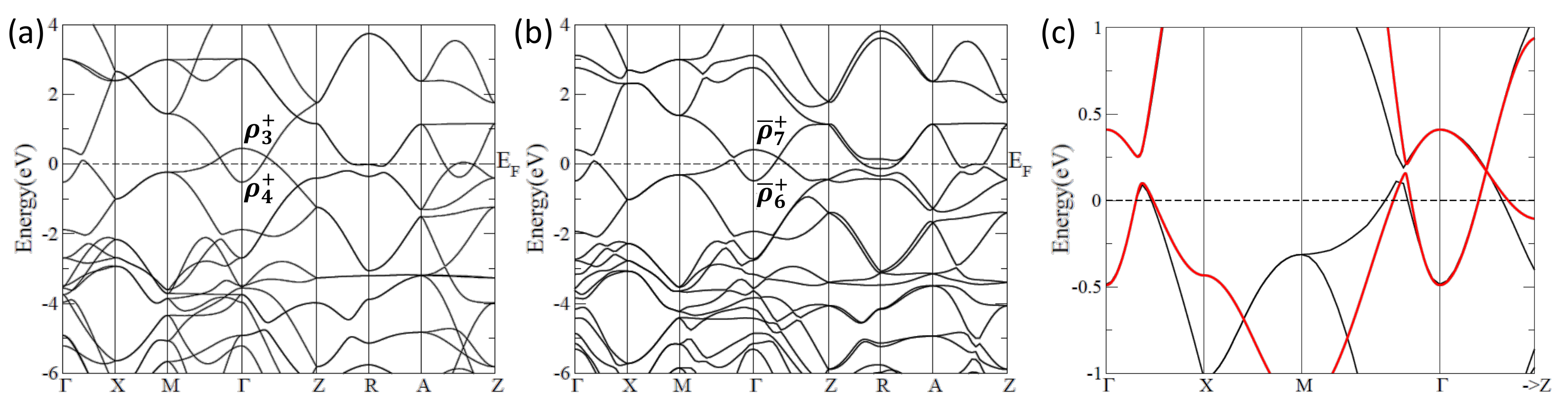}
\caption{The electronic structure of $\beta'$-PtO$_2$ in SG 136 $P4_{2}/mnm1'$.  Bands (a) without and (b) with spin-orbit coupling, labeled by the single- and double-valued corepresentations, respectively, of point group $4/mmm1'$ ($D_{4h}$), to which the little co-group at $\Gamma$ is isomorphic.  (c) Bands fitted from the $k\cdot p$ Hamiltonian in Eq.~(\ref{eq:ZJPt}) (red) to the bands obtained from first-principles calculations (black).  The Dirac points in $\beta'$-PtO$_2$ lie along $\Gamma Z$ in (b,c).}
\label{fig:DFTPt}
\end{figure}

Finally, we also use first-principles and tight-binding calculations to demonstrate the presence of HOFA states and related fragile-phase corner modes in the candidate Dirac semimetal~\cite{PtO21,PtO22,PtO2Supp,PtO2Supp2} PtO$_2$ in its rutile-structure ($\beta'$) phase (SG 136 ($P4_{2}/mnm1'$), ICSD~\cite{ICSD} No. 647316, further details available at~\cite{QuantumChemistry,AndreiMaterials,BCS1,BCS2}~\url{https://topologicalquantumchemistry.org/#/detail/647316}), which satisfies the criteria for HOFA states derived in Appendix~\ref{sec:SGs}.  In $\beta'$-PtO$_2$, the Pt atoms occupy the $2a$ Wyckoff position and the O atoms occupy the $4f$ position.  In Fig.~\ref{fig:DFTPt}(b,a), we respectively show the electronic structure of $\beta'$-PtO$_2$ with and without incorporating the effects of SOC, calculated using the same methodology previously employed for KMgBi (Appendix~\ref{sec:kmgbi}).  Even before incorporating the effects of SOC, spin-degenerate bands in $\beta'$-PtO$_2$ are already strongly inverted at $\Gamma$.  Specifically, the valence and conduction bands in $\beta'$-PtO$_2$ (Fig.~\ref{fig:DFTPt}(a,b)) only cross in the vicinity of inverted bands at the $\Gamma$ point, though there are also additional electron and hole pockets from bands that do not connect across the gap nearest the Fermi energy.  In the absence of SOC, the inverted states at $\Gamma$ are labeled by the single-valued corepresentations $\rho^{+}_{3,4}$ of the little co-group of the $\Gamma$ point of SG 136, which is isomorphic to point group $4/mmm1'$ ($D_{4h}$), and where corepresentations are labeled using the convention previously established in Appendix~\ref{sec:kmgbi}.  In terms of the corepresentations of point group~\cite{BilbaoPoint} $4/mmm1'$, the single-valued corepresentations in Fig.~\ref{fig:DFTPt}(a) are related by the equivalences:
\begin{equation}
\rho_{3}^{+}\equiv A_{2g},\ \rho_{4}^{+} \equiv B_{2g}.
\end{equation} 
When the effects of SOC are incorporated (Fig.~\ref{fig:DFTPt}(b)), the nodal lines near $\Gamma$ split into a time-reversed pair of Dirac points located at ${\bf k}=(0,0,\pm k_z^c)$, where $k_z^c=0.1663~(2\pi/c)=0.33386$~\AA$^{-1}$.  In the electronic structure of $\beta'$-PtO$_2$ incorporating SOC, the bands closest to the Fermi energy at $\Gamma$ are labeled by $\bar{\rho}_{6,7}^{+}$, whose $s_{4_{2}z}$ and $\mathcal{I}$ characters were previously listed in Eqs.~(\ref{eq:67rep1}) and~(\ref{eq:67rep2}), and which are related to the corepresentations of $4/mmm1'$ through the equivalences:
\begin{equation}
\bar{\rho}_{6}^{+}\equiv \bar{E}_{2g},\ \bar{\rho}_{7}^{+} \equiv \bar{E}_{1g},
\label{eq:67equiv3}
\end{equation} 
again noting, as we did previously in Appendix~\ref{sec:cadmium}, that $s_{4_{2}z}$ exhibits the same set of eigenvalues and commutation relations with other spatial symmetries as $C_{4z}$ at the $\Gamma$ point of any SG~\cite{BigBook}.

\begin{table}[h]
\caption{The parameters used to fit the bands of the $k\cdot p$ theory in Eq.~(\ref{eq:ZJPt}) to the first-principles electronic structure of $\beta'$-PtO$_2$ in the vicinity of the $\Gamma$ point (Fig.~\ref{fig:DFTPt}(c)).}
\begin{tabular}{c|c|c|c|c|c|c|c|c}
 $\epsilon_0$ (eV)& $\epsilon_1$ (eV \AA$^{-2}$) & $\epsilon_2$ (eV \AA$^{-2}$) & $M_0$ (eV) & $M_1$ (eV \AA$^{-2}$) & $M_2$ (eV \AA$^{-2}$) & $A$ (eV \AA$^{-1}$) &  $B$ (eV \AA$^{-1}$) & $C$ (eV \AA$^{-1}$)\\
\hline
-0.04 & 2.3 & 4.023& 0.45 & -4.90  & -8.046& 1.6 & 1.3 &  0.4  \\
\end{tabular}
\label{tb:paramPt}
\end{table}

Unlike in the previous HOFA Dirac semimetals KMgBi and $\alpha''$-Cd$_3$As$_2$ (Appendices~\ref{sec:kmgbi} and~\ref{sec:cadmium}, respectively), because the Dirac points in $\beta'$-PtO$_2$ originate from a single band inversion between bands with the same parity eigenvalues (Eqs.~(\ref{eq:67rep1}) and~(\ref{eq:67rep2})), the Hamiltonian of the $k_{z}=0$ plane is not equivalent to a 2D TI.  Instead, because the inverted bands exhibit different fourfold rotation eigenvalues (and the same parity eigenvalues), the Hamiltonian of the $k_{z}=0$ plane of $\beta'$-PtO$_2$ is equivalent to a $C_{M_{z}}=2$ mirror TCI~\cite{ChenBernevigTCI,ChenTCI,NagaosaDirac,PtO22} (Eq.~(\ref{eq:MirrorChernIndicator})).  As in the model of an $s-d$-hybridized 3D Dirac semimetal introduced in this work (Eqs.~(\ref{eq:my2Dquad}) and~(\ref{eq:hinge})), the nontrivial mirror Chern number $C_{M_{z}}=2$ of the Hamiltonian of the $k_{z}=0$ plane necessitates the presence of two, twofold degenerate TCI cones at $k_{z}=0$ on $M_{z}$-preserving surfaces (Fig.~\ref{fig:pd}(b,c)).  We note that the Hamiltonian of the $k_{z}=0$ plane \emph{also} exhibits the same fourfold rotation eigenvalues as a QI in $p4m$ (Appendices~\ref{sec:bandrep} and~\ref{sec:TCIBoundary}); as shown in Appendix~\ref{sec:TCIBoundary}, this indicates that the TCI surface states at $k_{z}=0$ will gap into HOFA hinge states away from $k_{z}=0$.

To calculate the HOFA states in $\beta'$-PtO$_2$, we first form a $4\times 4$ $k\cdot p$ Hamiltonian of the bands closest to the Dirac points at the Fermi energy.  Because the inverted bands in $\beta'$-PtO$_2$ have different parity and fourfold rotation eigenvalues than in KMgBi and $\alpha''$-Cd$_3$As$_2$, then we cannot employ the previous model of Eq.~(\ref{eq:ZJ}).  Instead, we begin by defining a different four-band basis in which the two states labeled with $\bar{\rho}_{7}^{+}$ ($\bar{\rho}_{6}^{+}$) are denoted as $\ket{\pm \frac{1}{2}}$ ($\ket{\pm\frac{3}{2}}$), which we summarize in a Pauli-matrix notation in which $\tau$ indexes orbital components (\emph{i.e.} $J=\frac{1}{2},\frac{3}{2}$) and $\sigma$ indexes spin components (\emph{i.e.}, $\sgn(m_{j}) = \sgn(\pm\frac{1}{2})$ or $\sgn(\pm\frac{3}{2})$).  In this basis, the four-band $k\cdot p$ Hamiltonian of the bands closest to the Fermi energy in $\beta'$-PtO$_2$ takes the form:
\begin{equation}
H_{\Gamma}({\bf k}) = \epsilon({\bf k})\mathds{1}_{\tau\sigma} + M({\bf k})\tau^{z} + A (k_x^2-k_y^2)\tau^{x} + \tau^{y} (Bk_xk_y\sigma^z+Ck_yk_z\sigma^x+Ck_xk_z\sigma^y), 
\label{eq:ZJPt}
\end{equation}
where $\mathds{1}_{\tau\sigma}$ is the $4\times 4$ identity and:
\begin{equation}
\epsilon({\bf k}) = \epsilon_{0}+\epsilon_{1}k_{z}^{2}+\epsilon_{2}(k_x^{2}+k_y^2),\ M({\bf k}) = M_{0}+M_{1}k_{z}^{2}+M_{2}(k_x^{2}+k_y^2). 
\end{equation}
In particular, Eq.~(\ref{eq:ZJPt}) is identical to the $k\cdot p$ Hamiltonian of the $\Gamma$ point of the four-band model of an $s-d$ hybridized Dirac semimetal in SG 123 $P4/mmm1'$ introduced in this work (Eqs.~(\ref{eq:my2Dquad}) and~(\ref{eq:hinge})), and therefore satisfies the symmetry representation used throughout this work (Table~\ref{tb:symsMain} of the main text).  To model the electronic structure of $\beta'$-PtO$_2$, we map the $k\cdot p$ theory in Eq.~(\ref{eq:ZJPt}) to a lattice tight-binding model, which we then fit to the electronic structure of $\beta'$-PtO$_2$.  From this, we obtain the fitting parameters listed in Table~\ref{tb:paramPt}.

\begin{figure}[h]
\centering
\includegraphics[width=0.8\textwidth]{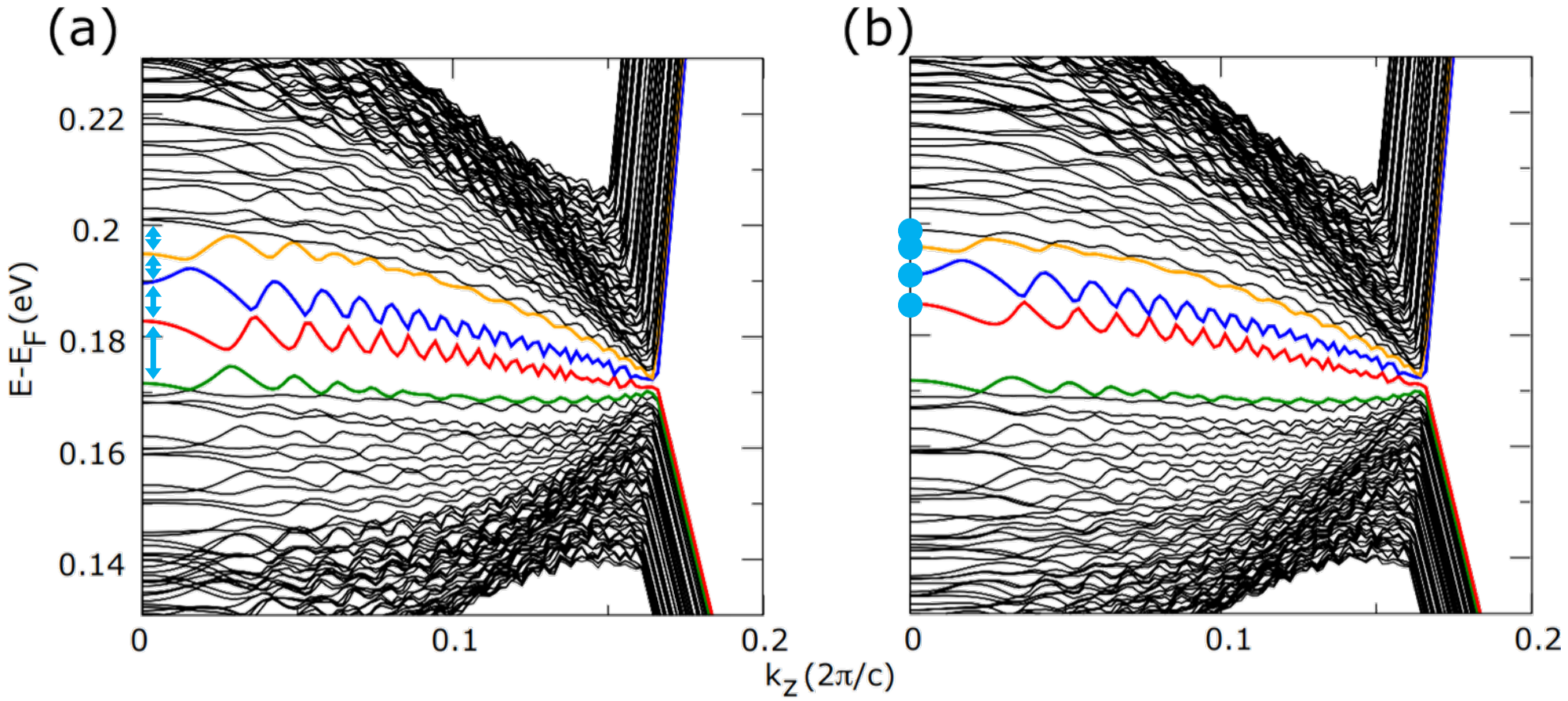}
\caption{The band structures of square, $z$-directed rods of tight-binding models obtained from the $k\cdot p$ theory in Eq.~(\ref{eq:ZJPt}) and fit to the first-principles electronic structure of $\beta'$-PtO$_2$ in the vicinity of the $\Gamma$ point (Fig.~\ref{fig:DFTPt}(c)) using the parameters in Table~\ref{tb:paramPt}.  The bands in (a,b) were respectively calculated in the absence and presence of an external electric field modeled by setting $D=0.04$ (eV \AA$^{-1}$) in Eq.~(\ref{eq:Efield}).  Because the rod in (a) preserves spinful $\mathcal{I}\times\mathcal{T}$ symmetry (or alternatively, because the rod in (a) preserves the spinful $x$ and $y$ reflection symmetries of Eq.~(\ref{eq:ZJPt}), whose representations anticommute), the bands in (a) are doubly degenerate~\cite{BigBook,WiederLayers}; in (b), even though $\mathcal{I}\times\mathcal{T}$ is broken by the external electric field, the anticommuting $x$ and $y$ reflections are still preserved, and the rod bands remain doubly degenerate.  In (a,b), fixing the rod filling to that of the bulk Dirac points, we label the two highest valence (lowest conduction) bands in red (blue), and label the next two highest valence (lowest conduction) bands in green (orange).  In (a), the TCI surface states at $k_{z}=0$ are split by finite-size effects (blue arrows in (a) at $k_{z}=0$), but we still observe eight hinge-localized surface states spanning from $k_{z}=0$ to the hinge projection of the bulk Dirac point at $k_z^c=0.1663~(2\pi/c)$; these eight bands correspond to the bands labeled in red, blue, and orange, as well as the black pair of conduction bands immediately above the orange pair of bands.  Of these eight states, only two (the red bands) are filled.  When the effects of an external field are incorporated in (b), the TCI surface states split in the hinge spectrum calculated through hinge Green's functions (Fig.~\ref{fig:DFTfragileMain}(c) of the main text), but the rod spectrum remains nearly identical to (a) ((a) and (b) are only distinguished by the growing gap at $k_{z}=0$ between the red and green pairs of bands, and by the shrinking gap at $k_{z}=0$ between the orange pair of bands and the lowest black pair of conduction bands).  In (b), we interpret the eight HOFA states identified in (a) as connecting to four, quarter-filled Kramers pairs of states at $k_{z}=0$ (blue circles in (b) at $k_{z}=0$) whose eightfold degneracy is split in the rod spectrum by finite-size effects, and which correspond to the Kramers pairs of hinge states at $k_{z}=0$ in the hinge Green's function shown in Fig.~\ref{fig:DFTfragileMain}(c) of the main text.}
\label{fig:pto2Rod}
\end{figure}

Next, to identify and characterize the hinge states in $\beta'$-PtO$_2$ we follow the procedure detailed in Appendix~\ref{sec:kmgbi}.  First, we calculate the bands of a fourfold-symmetric, $z$-directed rod of the lattice tight-binding model obtained from Eq.~(\ref{eq:ZJPt}).  In agreement with analysis performed throughout this work, when we fix the system filling to that of the Dirac points, we observe in Fig.~\ref{fig:pto2Rod}(a) a single, half-filled set of four HOFA states connecting the hinge projections of the bulk Dirac points to the projections of the 2D TCI cones at $k_{z}=0$ (which have been gapped by finite-size effects, as indicated by the blue arrows in Fig.~\ref{fig:pto2Rod}(a)), as well as a second set of empty HOFA states above them in energy.  As previously discussed in Appendix~\ref{sec:kmgbi}, because the spinful rod groups do not host symmetry-stabilized four-dimensional corepresentations away from rod TRIM points~\cite{BigBook,Bandrep1}, then there is no crystal symmetry that can force the HOFA hinge states to appear in strict fourfold degeneracies in the rod bands in Fig.~\ref{fig:pto2Rod}; instead, their fourfold degeneracy is only restored in the thermodynamic limit that states along the rod do not hybridize.  Therefore, because of finite-size effects, the eight HOFA hinge states in Fig.~\ref{fig:pto2Rod}(a) split into four sets of two states, which are labeled, in increasing energy, with red, blue, orange, and black.

To isolate the states on a single hinge, we then calculate the hinge Green's function of a semi-infinite {slab} (Appendix~\ref{sec:kmgbi}), which we plot in Fig.~\ref{fig:DFTfragileMain}(b) of the main text.  At each hinge $k$ point between $|k_{z}|=0,k_{z}^{c}$ we observe two narrowly split hinge states with an overall one-quarter filling.  This agrees with the analysis performed in Appendix~\ref{sec:TCIBoundary}.  Specifically, the hinge spectrum at each $k$ point with HOFA states in Fig.~\ref{fig:DFTfragileMain}(b) of the main text represents the particle-hole conjugate of the particle-hole-broken QI corner spectrum in Fig.~\ref{fig:TCIcorners}(c).

Crucially, unlike the surface 2D TI cones in KMgBi and $\alpha''$-Cd$_3$As$_2$ (Appendices~\ref{sec:kmgbi} and~\ref{sec:cadmium}, respectively), the surface 2D TCI cones in $\beta'$-PtO$_2$ are only protected by $M_{z}$ symmetry, and therefore can be gapped without breaking $\mathcal{T}$ symmetry.  As shown in the main text and discussed in detail in Appendices~\ref{sec:fragile} and~\ref{sec:TCIBoundary}, breaking $M_{z}$ while keeping fourfold rotation, in-plane reflection, and $\mathcal{T}$ symmetries gaps the surface cones of a $C_{M_{z}}=2$ 2D TCI formed from band inversion at $\Gamma$ (such as the $k_{z}=0$ plane of $\beta'$-PtO$_2$) into the Kramers pairs of corner modes of a fragile topological phase.  Because the corner states are a manifestation of an anomalous absence of ($2+8n$ or $6+8n$) states from the valence manifold in the $k_{z}=0$ plane, and are a property of the inverted (fragile) bands near the Fermi energy (Appendix~\ref{sec:TCIBoundary}), then they \emph{remain} present even when trivial bands below the Fermi energy are added to trivialize the fragile valence manifold.  To gap the surface TCI cones in our model of $\beta'$-PtO$_2$, we introduce a term that breaks $M_{z}$ and $\mathcal{I}$ symmetries while preserving fourfold rotation, in-plane reflection, and $\mathcal{T}$ symmetries:
\begin{equation}
H^{el}_{\Gamma}({\bf k}) =H_{\Gamma}({\bf k})+D\tau^z(\sigma^xk_y-\sigma^yk_x),
\label{eq:Efield}
\end{equation}
where $H_{\Gamma}({\bf k})$ is given in Eq.~(\ref{eq:ZJPt}).  The $D$ term in Eq.~(\ref{eq:Efield}) can be induced in experiment by directing an external electric field that is spatially constant (or slowly varying on the scale of the lattice spacing) along the $z$- ($c$-) axis of a $4_{2}$-screw-symmetric nanorod of $\beta'$-PtO$_2$.  To demonstrate the presence of fragile-phase corner charges in $\beta'$-PtO$_2$ in an external field, we set $D=0.04$ (eV \AA$^{-1}$) in Eq.~(\ref{eq:Efield}) and again calculate the bands of a fourfold-symmetric rod and the hinge Green's function of a semi-infinite slab (Appendix~\ref{sec:kmgbi}).  In the rod calculation (Fig.~\ref{fig:pto2Rod}(b)), we observe that the electric field term in Eq.~(\ref{eq:Efield}) shifts in energy the four Kramers pairs of hinge states (blue circles in Fig.~\ref{fig:pto2Rod}(b) at $k_{z}=0$) that resulted from gapping the TCI surface states at $k_{z}=0$ (the surface states were already largely split by finite-size effects in (a)); only two of the eight hinge states at $k_{z}=0$ (the red states) are filled when the system filling is fixed to that of the Dirac points.  Even though our rod calculation preserves fourfold rotation and $\mathcal{T}$ symmetry, because the spinful rod groups do not host symmetry-stabilized eight-dimensional corepresentations~\cite{BigBook,Bandrep1}, then there is no crystal symmetry that can force the bands corresponding to the eight Kramers pairs of fragile-phase corner modes at $k_{z}=0$ in Fig.~\ref{fig:pto2Rod}(b) to appear in a strict eightfold degeneracy; instead, their eightfold degeneracy is only restored in the thermodynamic limit that states along the rod do not hybridize.

Finally calculating the hinge Green's function of a $z$-directed macroscopic sample of $\beta'$-PtO$_2$ in the presence of a $z$-directed electric field (Fig.~\ref{fig:DFTfragileMain}(c) of the main text), we observe well-isolated, quarter-filled fragile-phase corner modes at $k_{z}=0$ connected to narrowly split pairs of half-filled and fully unoccupied HOFA states, in agreement with the analysis in Appendix~\ref{sec:TCIBoundary}.  Specifically, the quarter-filled corner states at $k_{z}=0$ in Fig.~\ref{fig:DFTfragileMain}(c) of the main text represent the particle-hole conjugates of the three-quarters-filled fragile-phase corner modes observable at $k_{z}=0$ in the rod bands of the model of an $s-d$-hybridized, noncentrosymmetric, fragile topological Dirac semimetal introduced in this work (Fig.~\ref{fig:HingeSMmain}(h) of the main text and Appendix~\ref{sec:TCIBoundary}).


\end{appendix}
\bibliography{HingeBibs}
\end{document}